\pdfoutput=1
\documentclass[11pt,twoside,a4paper,cmspaper,final,collab]{cms-tdr}
\def\svnVersion{359f535}\def\svnDate{2022/06/13}\def\cmsCernNoTag{CERN-EP-2021-203}\def\cmsCernDate{\today}\def\cmsMessage{Published in Nuclear Instruments and Methods A as \href{http://dx.doi.org/10.1016/j.nima.2022.166795}{\doi{10.1016/j.nima.2022.166795}.}}
\begin{document}\cmsNoteHeader{TRK-20-001}

\newlength\cmsTabSkip\setlength{\cmsTabSkip}{1ex}
\newcommand{\HIPPY} {{\textsc{HipPy}}\xspace}
\newcommand{\pp}{\ensuremath{\Pp\Pp}}
\newcommand{\pthat}{\ensuremath{\hat{p}_\mathrm{T}}\xspace}
\newcommand{\Ztomm}{\PZ\to\Pgm\Pgm}
\newcommand{\mmumu}{\ensuremath{m_{\Pgm\Pgm}}}
\newcommand{\pdpd}[2]{\ensuremath{\frac{\partial #1}{\partial #2}}}
\newcommand{\pdpdinline}[2]{\ensuremath{\partial #1/\partial #2}\xspace}
\newcommand{\dxy}{\ensuremath{d_{xy}}}
\newcommand{\dz}{\ensuremath{d_z}}
\newcommand{\phimup}{\ensuremath{\phi_{\mu_+}}}
\newcommand{\sigmaijm}{\ensuremath{\sigma_{ij}^\mathrm{m}}}
\newcommand{\sigmaalign}{\ensuremath{\sigma_\text{align}}}

\cmsNoteHeader{TRK-20-001} 
\title{Strategies and performance of the CMS silicon tracker alignment during LHC Run~2}

\author*{The Tracker Alignment group}

\date{\today}

\abstract{
    The strategies for and the performance of the CMS silicon tracking system alignment during the 2015--2018 data-taking period of the LHC are described. 
    The alignment procedures during and after data taking are explained. Alignment scenarios are also derived for use in the simulation of the detector response.
    Systematic effects, related to intrinsic symmetries of the alignment task or to external constraints, are discussed and illustrated for different scenarios.
}

\hypersetup{%
pdfauthor={CMS Collaboration},%
pdftitle={Run 2 Tracker Alignment Performance},%
pdfsubject={CMS},%
pdfkeywords={CMS, detector, alignment, tracker, Legacy reprocessing, MillePede-II, HipPy}}

\maketitle 
\tableofcontents

\section{Introduction}
The innermost subdetector of the CMS experiment constitutes the largest silicon tracker in the world, both in terms of the total surface area and the number of sensors.
To benefit from the excellent resolution of the silicon sensors for measuring the trajectories of charged particles, the position and orientation of each sensor must be precisely measured. 
During the installation procedure, a mechanical alignment yields a precision in the position of the tracker of $\mathcal{O}(0.1\unit{mm})$, which is much larger than the design hit resolution of $\mathcal{O}(0.01\unit{mm})$. 
Therefore, a further correction to the position, orientation, and surface deformations of the sensors needs to be derived. This correction is commonly referred to as the spatial alignment of the tracker or simply the tracker alignment. We will refer to the parameters of this correction as the tracker alignment constants.
To maintain the targeted precision, the alignment constants must be updated regularly to include effects such as the ramping of the magnetic field or temperature variations.
The approach employed by CMS consists in determining the alignment constants by performing track fits with the corresponding track parameters unconstrained. 

Previous publications~\cite{Chatrchyan:2009sr,Chatrchyan:2014wfa}, covering the 2010--2012 LHC data-taking period (Run~1), showed how the roughly 200\,000 parameters necessary to describe the alignment of the tracker modules were determined using track-based methods.
Based on the same techniques, this article describes the strategies and recent developments utilized for the tracker alignment during the LHC data-taking period between 2015 and 2018, which we refer to as Run~2. We also quantify the performance of the alignment that is achieved at various stages based on observables sensitive to tracking. 

After a short description of the CMS detector in Section~\ref{sec:CMS}, the concept of track-based alignment is illustrated in Section~\ref{sec:concepts}, followed by a description of the data sets used to perform and validate the alignment in Section~\ref{sec:datasets}. 
We present studies that identify and minimize systematic distortions, which can be left uncorrected by the track-based alignment, in Section~\ref{sec:WMs}.
Two algorithms, \MILLEPEDE-II and \HIPPY, are used for the tracker alignment. The recent software developments of these algorithms are summarized in Section~\ref{sec:software}, and the derivation of the alignment constants during data taking is described in Section~\ref{sec:datataking}, taking as representative examples the start-up of Run~2 and the first alignments of the pixel detector in 2017 after the Phase-1 upgrade~\cite{CMS:2012sda,Adam_2021}.
Section~\ref{sec:Legacy} focuses on the strategies developed to provide the best possible alignment calibration for reprocessing the data before using them in physics analyses. Because of the higher intensity of the LHC compared to Run~1, a dedicated strategy was successfully developed to include the fast changes in the local reconstruction conditions. 
In Section~\ref{sec:MC}, the derivation of an alignment scenario for simulation is discussed, along with a comparison of the tracking performance between data and simulation. 
Section~\ref{sec:Summary} summarizes the strategies, observations, and results.

\section{The CMS detector} \label{sec:CMS}
The central feature of the CMS apparatus is a superconducting solenoid of 6\unit{m} internal diameter, providing a magnetic field of 3.8\unit{T}. Within the solenoid volume are a silicon pixel and strip tracker, a lead tungstate crystal electromagnetic calorimeter, and a brass and scintillator hadron calorimeter, each composed of a barrel and two endcap sections. Forward calorimeters extend the pseudorapidity coverage provided by the barrel and endcap detectors. Muons are measured in gas-ionization detectors embedded in the steel flux-return yoke outside the solenoid.

Events of interest are selected using a two-tiered trigger system. The first level (L1), composed of custom hardware processors, uses information from the calorimeters and muon detectors to select events at a rate of around 100\unit{kHz} within a fixed latency of about 4\mus~\cite{Sirunyan:2020zal}. The second level, known as the high-level trigger (HLT), consists of a farm of processors running a version of the full event reconstruction software optimized for fast processing, and reduces the event rate to around 1\unit{kHz} before data storage~\cite{Khachatryan:2016bia}. 

During the 2016 (2017 and 2018) LHC running periods, the silicon tracker consisted of 1440 (1856) silicon pixel and 15\,148 silicon strip detector modules. 
After the 2016 data-taking period, the pixel detector was upgraded to its Phase-1 configuration. The upgraded pixel detector features one more layer in the barrel, and one more disk in each of the forward pixel endcaps, than the pixel detector that was in use up to the end of 2016 (Phase-0 pixel detector).
This extended the acceptance of the tracker from a pseudorapidity range $\abs{\eta} < 2.5$ to $\abs{\eta} < 3.0$, and improved the impact parameter resolution. 
Before the Phase-1 upgrade, the track resolutions were typically 1.5\% in transverse momentum~(\pt) and 25--90 (45--150)\mum in the transverse (longitudinal) impact parameter for nonisolated particles of $1 < \pt < 10\GeV$ and $\abs{\eta} < 1.4$~\cite{Chatrchyan:2014fea}. 
For nonisolated particles of $1 < \pt < 10\GeV$ and $\abs{\eta} < 3.0$, the track resolutions are typically 1.5\% in \pt and 20--75\mum in the transverse impact parameter for data recorded after the Phase-1 upgrade~\cite{CMS-DP-2020-049}. 

The mechanical structure of the silicon tracker consists of several high-level structures: two half barrels in the barrel pixel tracker (BPIX), four half cylinders in the two forward pixel tracker regions (FPIX), two half barrels in the strip tracker inner barrel (TIB) and in the strip tracker outer barrel (TOB), two endcaps in the tracker inner disks (TID) and in the tracker endcaps (TEC).
For the Phase-0 (Phase-1) detector, the half barrels in the BPIX consist of three (four) layers and the half cylinders in the FPIX consist of two (three) disks, separated in the plane parallel to the beam axis and perpendicular to the LHC plane.
In the barrel, groups of eight pixel modules are mounted on rods arranged in cylindrical layers. The rods are mounted such that the modules of two adjacent ladders are rotated by 180 degrees around~$z$ with respect to each other, thus having the silicon surface pointing inwards or outwards. In the FPIX, modules are supported by blades arranged in a turbine-like geometry, each hosting two modules mounted back-to-back, pointing in opposite directions.

A more detailed description of the CMS detector, together with a definition of the coordinate system used and the relevant kinematic variables, can be found in Ref.~\cite{Chatrchyan:2008zzk}.

\section{General concepts of alignment} \label{sec:concepts}
In the track-based alignment approach, the alignment parameters $\mathbf{p}$ are derived by minimizing the following~$\chi^2$ function:
\begin{equation} \label{eq:eqn}
    \chi^2(\mathbf{p},\mathbf{q}) = \sum_j^\text{tracks} \sum_i^\text{hits} \left( \frac{m_{ij} - f_{ij}(\mathbf{p},\mathbf{q}_j) }{\sigmaijm} \right)^2,
\end{equation}
where
\begin{itemize}
    \item   $\mathbf{p}$ represents the alignment parameters (also called alignables),
    \item   $\mathbf{q}$ represents the track parameters (\eg parameters related to the track curvature and the deflection by multiple scattering~\cite{Chatrchyan:2014fea}),
    \item   $m$ represents the measurements (\eg hits) and $f$ for the predictions, and
    \item   $\sigma^\mathrm{m}$ represents the uncertainty in the measurements (\eg local hit resolution, alignment uncertainty).
\end{itemize}
The number of alignables varies depending on the desired granularity of the alignment.
For an alignment of the large mechanical structures in the pixel tracker, six parameters for the position and orientation of each of the structures would typically be used, leading to 36 parameters in total (see Section~\ref{sec:PCL}).
In contrast, the alignment of every single module of the whole tracker, including eight or nine parameters per module describing corrections to the position, orientation, and surface deformations, as well as additional parameters to account for changes over time, would require several hundreds of thousands of alignment parameters (Section~\ref{sec:UL}).

With both track and alignment parameters, the $\chi^2$ can potentially include millions of parameters. To still be able to minimize it in that case, given an approximate set of alignment parameters $\mathbf{p_0}$ and the corresponding track parameters $\mathbf{q_0}$, we can first linearize the prediction term~$f$ in the $\chi^2$:
\begin{linenomath}
\ifthenelse{\boolean{cms@external}}
{
\begin{multline} \label{eq:linearization}
    \chi^2(\mathbf{p_0} + \Delta \mathbf{p},\mathbf{q_0} + \Delta \mathbf{q}) =\\ 
    \sum_j^{\text{tracks}} \sum_i^{\text{hits}} \left(\frac{1}{\sigmaijm}\right)^2 \left(m_{ij} - f_{ij}(\mathbf{p_0},\mathbf{q_0}_j) \right.\\ 
    \left.- \Delta \mathbf{p} \pdpd{f_{ij}}{\mathbf{p}}(\mathbf{p_0},\mathbf{q_0}_j) - \Delta \mathbf{q}_j \pdpd{f_{ij}}{\mathbf{q}_j}(\mathbf{p_0},\mathbf{q_0}_j) \right)^2.
\end{multline}
}
{
\begin{equation} \label{eq:linearization} 
    \chi^2(\mathbf{p_0} + \Delta \mathbf{p},\mathbf{q_0} + \Delta \mathbf{q}) = \sum_j^\text{tracks} \sum_i^\text{hits} \left( \frac{m_{ij} - f_{ij}(\mathbf{p_0},\mathbf{q_0}_j) - \Delta \mathbf{p} \pdpd{f_{ij}}{\mathbf{p}}(\mathbf{p_0},\mathbf{q_0}_j) - \Delta \mathbf{q}_j \pdpd{f_{ij}}{\mathbf{q}_j}(\mathbf{p_0},\mathbf{q_0}_j) }{\sigmaijm} \right)^2.
\end{equation} 
}
\end{linenomath}
After some manipulation, this $\chi^2$~minimization is reformulated into a system of tens or hundreds of thousands of linear equations, and treated like a matrix inversion problem:
\begin{equation} \label{eq:matrixinvfull}
    \mathbf{C} \times \begin{pmatrix} \Delta\mathbf{p} \\ \Delta\mathbf{q} \end{pmatrix} = \mathbf{b},
\end{equation}
where $\mathbf{C}$ is a correlation matrix whose components are functions of $\pdpdinline{f_{ij}}{\mathbf{p}}$, $\pdpdinline{f_{ij}}{\mathbf{q}_j}$, and $\sigmaijm$, and $\mathbf{b}$ is a source term whose components are functions of $\pdpdinline{f_{ij}}{\mathbf{p}}$, $\pdpdinline{f_{ij}}{\mathbf{q}_j}$, $\sigmaijm$, and of the $m_i$.
The size of $\mathbf{C}$ corresponds to the total number of track and alignment parameters.
The matrix $\mathbf{C}$ must be inverted; however, since this matrix is sparse and we are only interested in the alignment parameters, it is not necessary to perform a full inversion.
The steps of the matrix inversion related to the determination of the track parameters themselves are not necessary for determining the alignment parameters. The only requirement is that their correlations with the alignment parameters are taken into account.
Using block matrix algebra, the problem posed in Eq.~(\ref{eq:matrixinvfull}) is simplified by first focusing on the blocks related to the track parameters, and then modifying the large block related to the alignment parameters, as well as the source term.
Equation~(\ref{eq:matrixinvfull}) is then reduced to a system of linear equations including the alignment parameters, and keeping all the correlations from the tracks~\cite{Blobel:2002ax}:
\begin{equation} \label{eq:matrixInversion}
    \mathbf{C'} \times \Delta\mathbf{p} = \mathbf{b'},
\end{equation}
where $\mathbf{C'}$ ($\mathbf{b'}$) is obtained from $\mathbf{C}$ ($\mathbf{b}$) with a significantly smaller size. 
If the size of the matrix to invert is reduced to $\mathcal{O}(10\,000)$~parameters, as is typical for the alignment of the pixel tracker, it can be inverted exactly. If the size of the matrix is larger, as is typical for the alignment of the whole pixel and strip tracker, alternative numerical approaches are used to perform an approximate matrix inversion.
The two implementations of the track-based alignment used at CMS are discussed in Section~\ref{sec:software}.

Track-based alignment may suffer from different types of systematic biases inherited from the tracking algorithm, \ie in~$f$ in Eq.~(\ref{eq:eqn}), such as changes of conditions not included in the model.

During operation of the detector, changes in running conditions, such as changes of the magnetic field or changes in temperature, are sometimes unavoidable. These changes happen a few times a year and may affect the alignment procedure.
In general, it is essential to define the interval of validity~(IOV) of a set of alignment constants.
For instance, after ramping down and then ramping up the magnet (magnet cycle), movements of the high-level structures of $\mathcal{O}(1\mm)$ have been observed. 
In data-taking mode, the tracker is cooled down to temperatures close to $-15\,{}^\circ\text{C}$ ($-20\,{}^\circ\text{C}$) for the 2015--2017 (2018) data-taking period. For maintenance purposes, typically during a year-end technical stop (YETS), cooling may be interrupted. This can potentially cause movements of the modules of~$\mathcal{O}(10\mum)$.

\begin{figure*}[!ht]
    \centering

    \includegraphics[width=\textwidth]{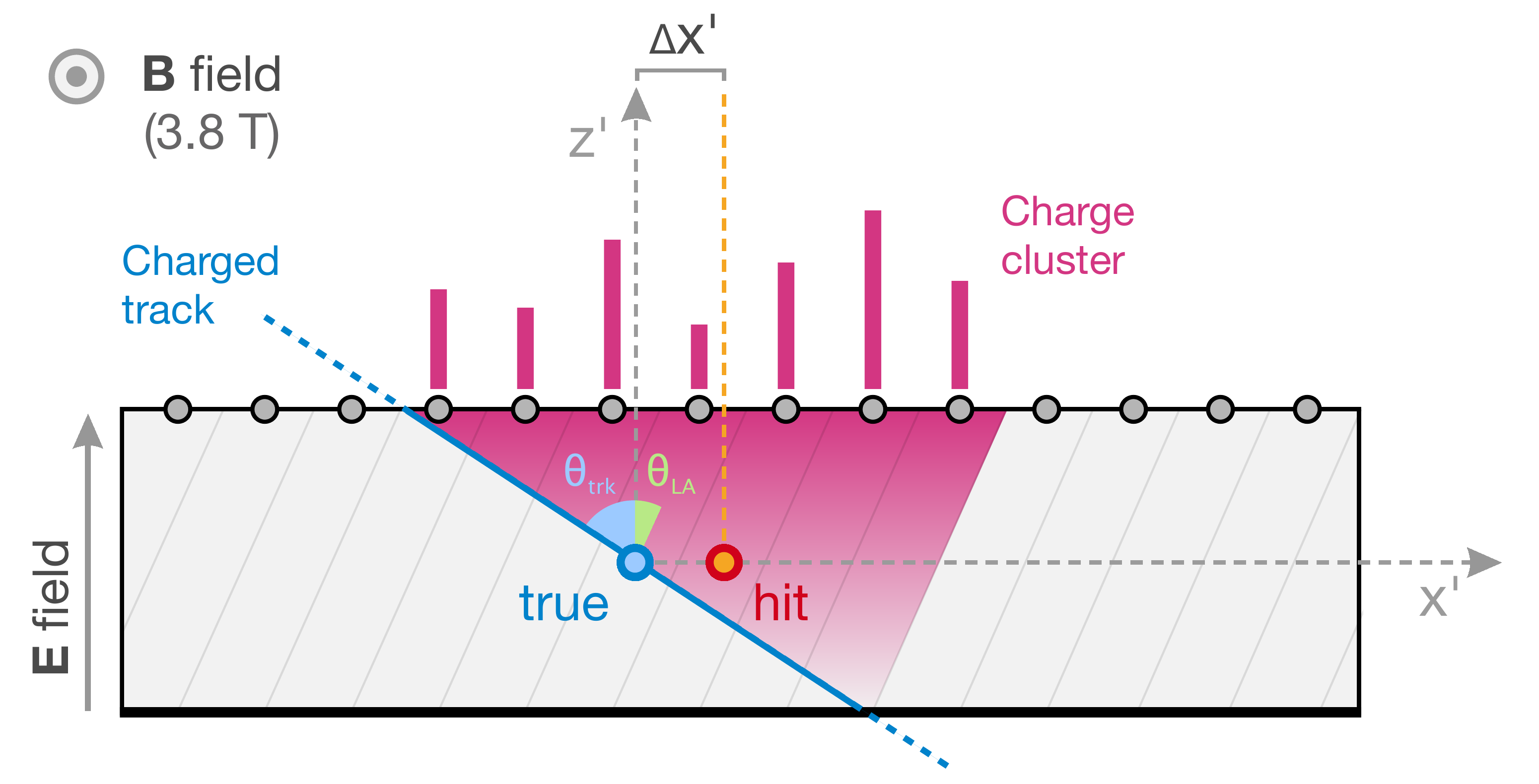}

    \caption{
        Sketch showing the transverse view of a silicon module working in a magnetic field~$\mathrm{B}$, with the backplane of the module located at the bottom.
        Here, $x'$ and $z'$ are the local coordinates of the module.
        The grey lines in the shaded rectangle indicate the direction of the Lorentz drift, forming an angle~$\theta_\text{LA}$ with the $z'$~axis.
        The blue line represents a charged particle traversing the module with incident angle~$\theta_\text{trk}$, and the magenta shaded area represents the volume in which charge carriers released by the ionization drift towards the electrodes at the top of the module.
        The blue-cyan (orange-red) dot represents the reconstructed hit if the Lorentz drift is (is not) included in the reconstruction.
        An example of a reconstructed charge cluster is shown by the vertical magenta bars above the module.
    }

    \label{fig:lorentz_angle}
\end{figure*}

\begin{figure}[!ht]
    \centering

    \includegraphics[width=0.5\textwidth]{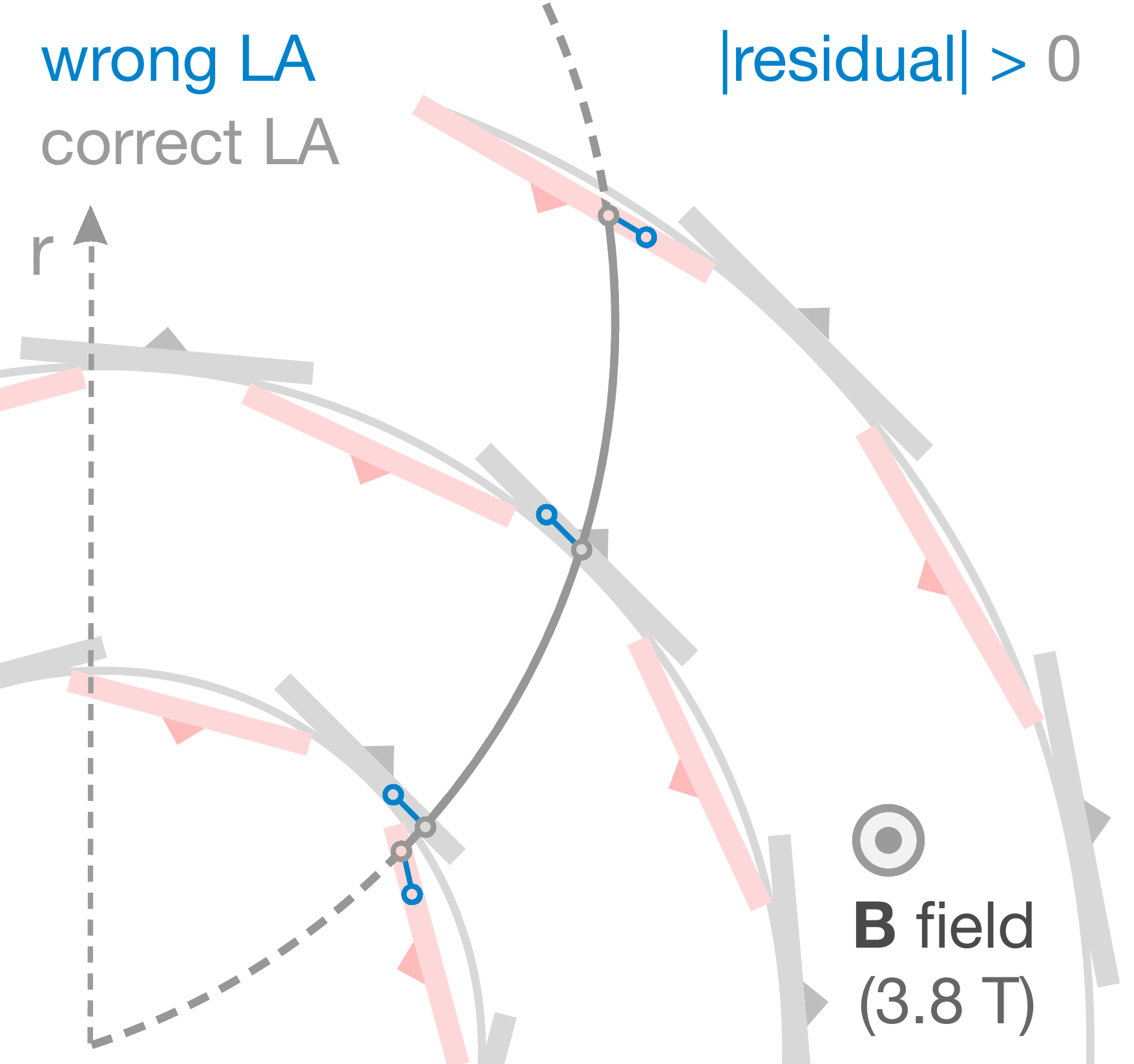}

    \caption{
        Sketch showing the transverse view of the Phase-0 barrel pixel subdetector, made of successive layers of silicon modules. The alternating orientation of the modules within each layer is indicated by the triangles.
        The blue (grey) circles represent the reconstructed hit positions using incorrect (correct) Lorentz angles in the presence of a magnetic field~$\mathrm{B}$. The grey curve corresponds to a track built from the hits that were reconstructed with the correct Lorentz angles.
        Hits reconstructed with incorrect Lorentz angles are displaced in a direction defined by the orientation of the module, increasing the residual distance between the hits and the track.
    }

    \label{fig:lorentz_drift_hits}
\end{figure}

Furthermore, the modules operate in a high-radiation environment, which affects their performance over time.
One quantity that is sensitive to the irradiation dose and plays a role in the alignment calibration is the Lorentz drift, as illustrated in Fig.~\ref{fig:lorentz_angle}.
It corresponds to the lateral drift of the charge carriers in the silicon induced by the external magnetic field, and is orthogonal to the electric field direction.
Aside from the magnetic field, the magnitude of the drift depends on the electric field, the mobility of the charge carriers, and the thickness of the active zone.
Since these quantities are not constant, the measured hit position changes over time by $\Delta x' \propto \tan \theta_\text{LA}$, where $\theta_{\text{LA}}$ is the Lorentz angle.
The sign of the shift depends on the orientation of the electric field, so that the shift in the hit position in modules pointing inward is opposite with respect to this shift in outward-pointing modules. 
As a consequence of the higher irradiation dose close to the interaction point, the mobility of the charge carriers changes faster in the pixel detector than in the strip detector.
These changes are corrected using a dedicated calibration method and residual effects are corrected in the alignment procedure.
The impact of the irradiation on the track reconstruction, and therefore on the alignment procedure, is illustrated in Fig.~\ref{fig:lorentz_drift_hits} and will be discussed in Section~\ref{sec:UL}.

To account for changes over time, either the alignment parameters should be updated regularly or additional parameters should be included.
In the latter case, a hierarchy of the parameters is introduced such that the absolute position and orientation of the chosen mechanical structures (for example high-level mechanical structures in the strip tracker, or ladders and blades in the pixel tracker) are allowed to change with time, whereas the relative position and orientation of the modules are assumed to be constant over time.

Another class of systematic biases arises from the internal symmetries of the alignment problem, such as the cylindrical symmetry of the detector, or the fact that most tracks originate from a single region of space.
This results in nonphysical geometrical transformations, also known as weak modes~(WMs).
Systematic distortions will be further discussed in Section~\ref{sec:WM}.

\section{Data sets} \label{sec:datasets}
Different types of data sets are used in the alignment procedure and in its validation.
In this section, we first describe data sets from collision events, then data sets from cosmic ray muons.
We generated corresponding simulated data samples; these are not expected to exactly describe the observed data. However, it is important that the simulated samples cover the same phase space as the observed data, with similar event topologies and numbers of tracks, for the derivation of alignment scenarios in the simulation.

\subsection{Proton-proton collisions}
To achieve the desired statistical precision of track-based alignment, a large track sample of at least several million tracks accumulated in the proton-proton ($\pp$) physics run is indispensable. These events have tracks propagating outwards from the interaction point, which therefore correlate detector elements radially.

\subsubsection{Inclusive L1 trigger}
\begin{figure*}
	\centering			

    \includegraphics[width=0.4\textwidth]{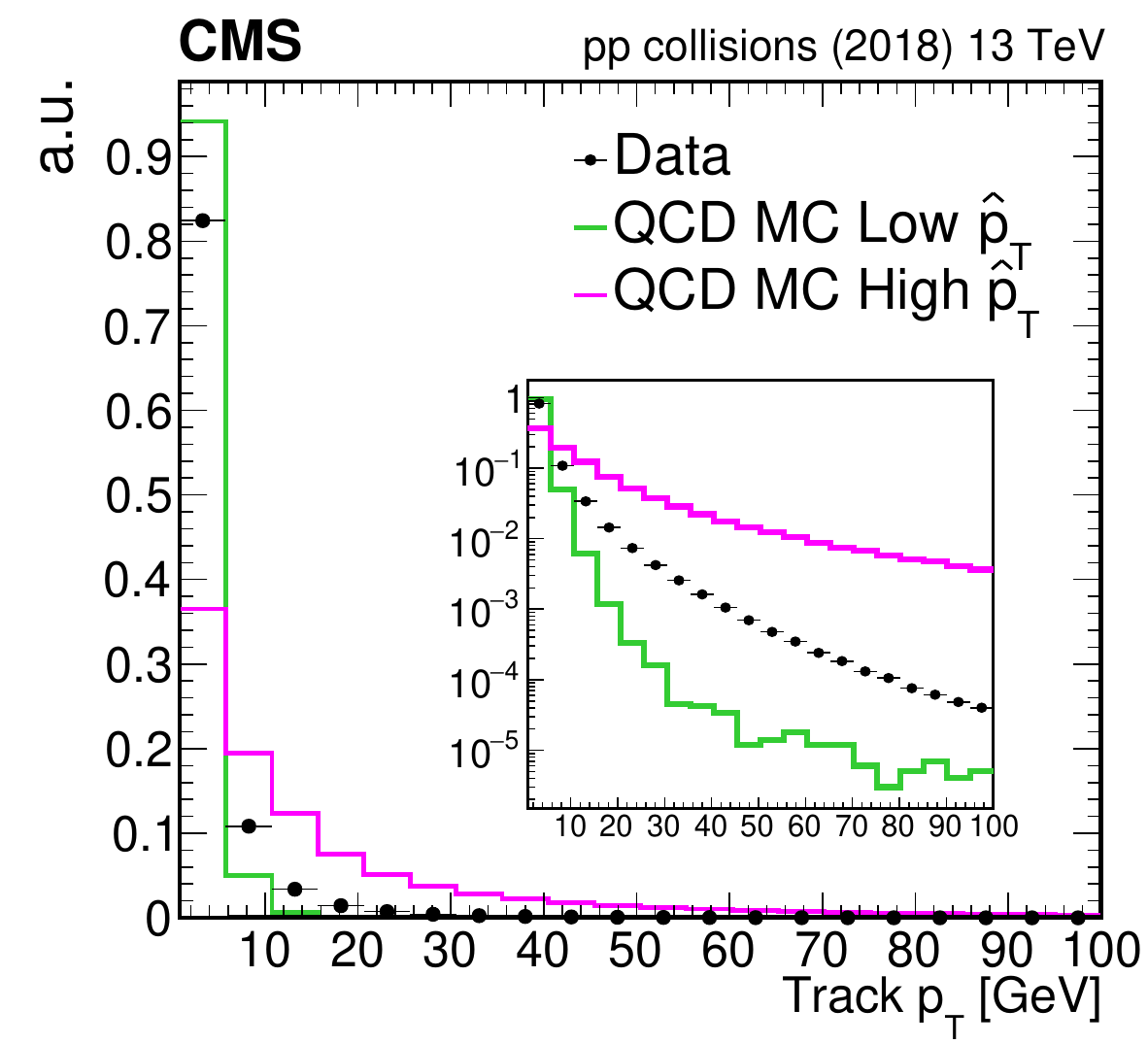}~\includegraphics[width=0.4\textwidth]{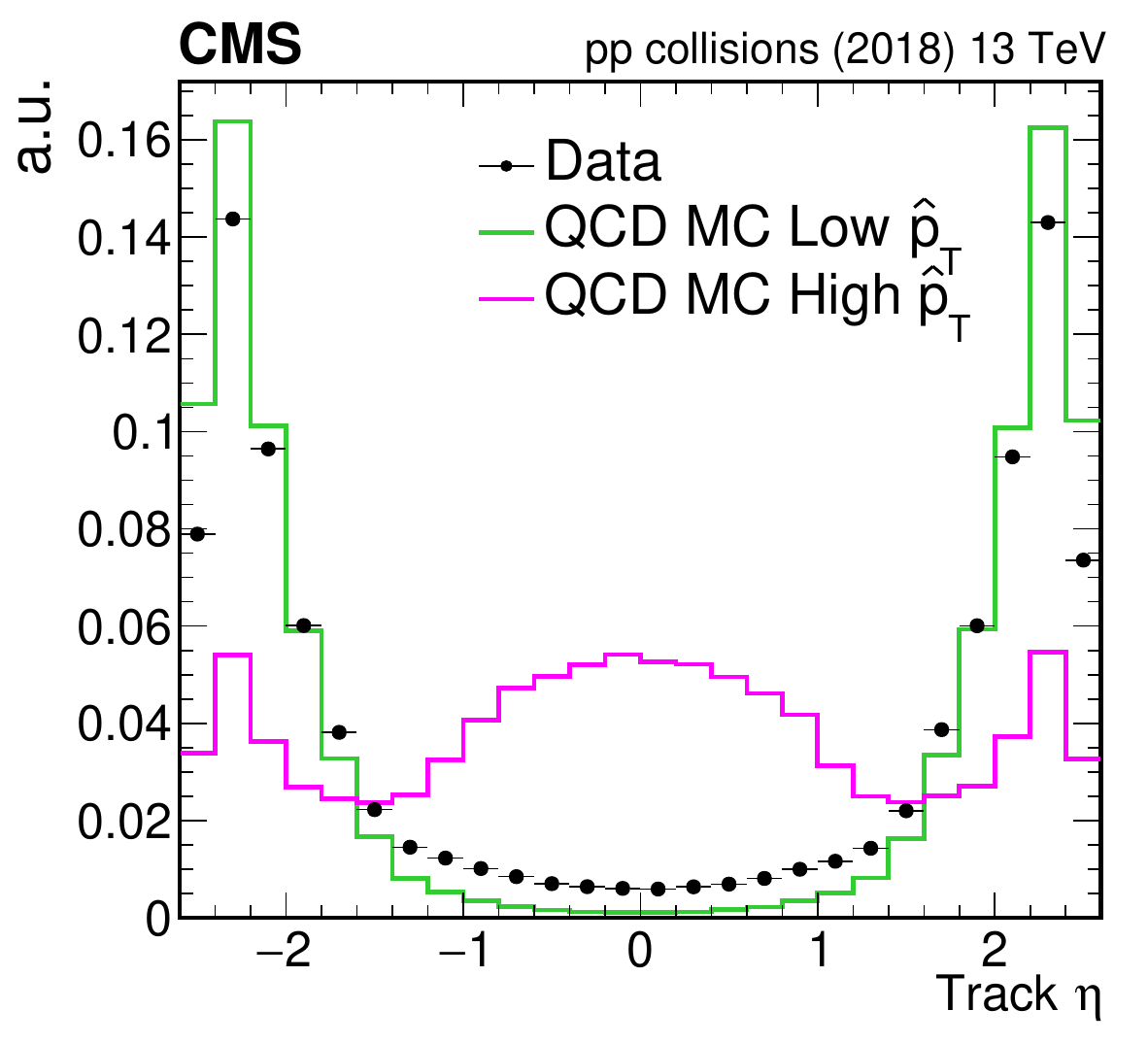}

    \includegraphics[width=0.4\textwidth]{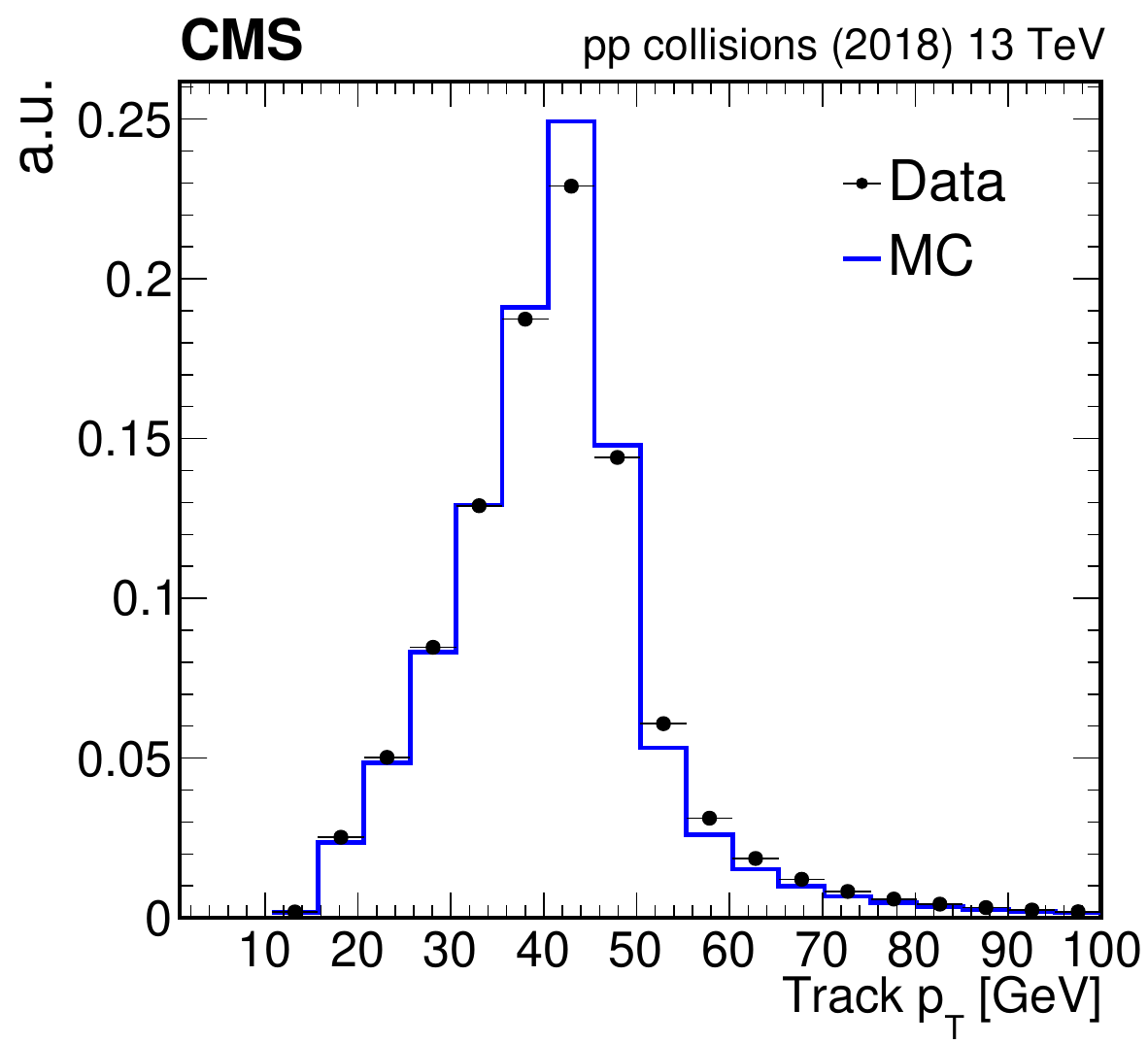}~\includegraphics[width=0.4\textwidth]{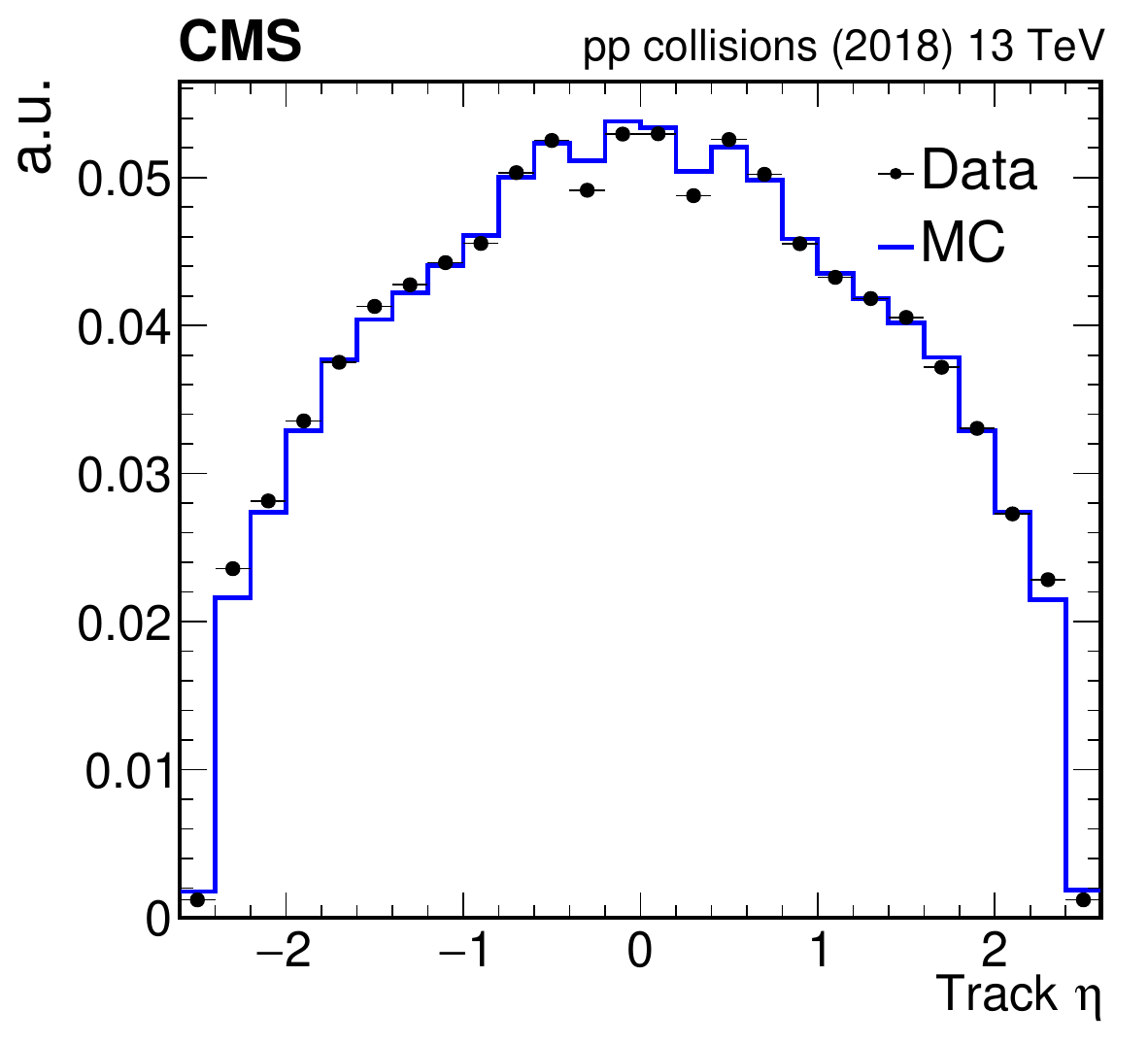}

    \includegraphics[width=0.4\textwidth]{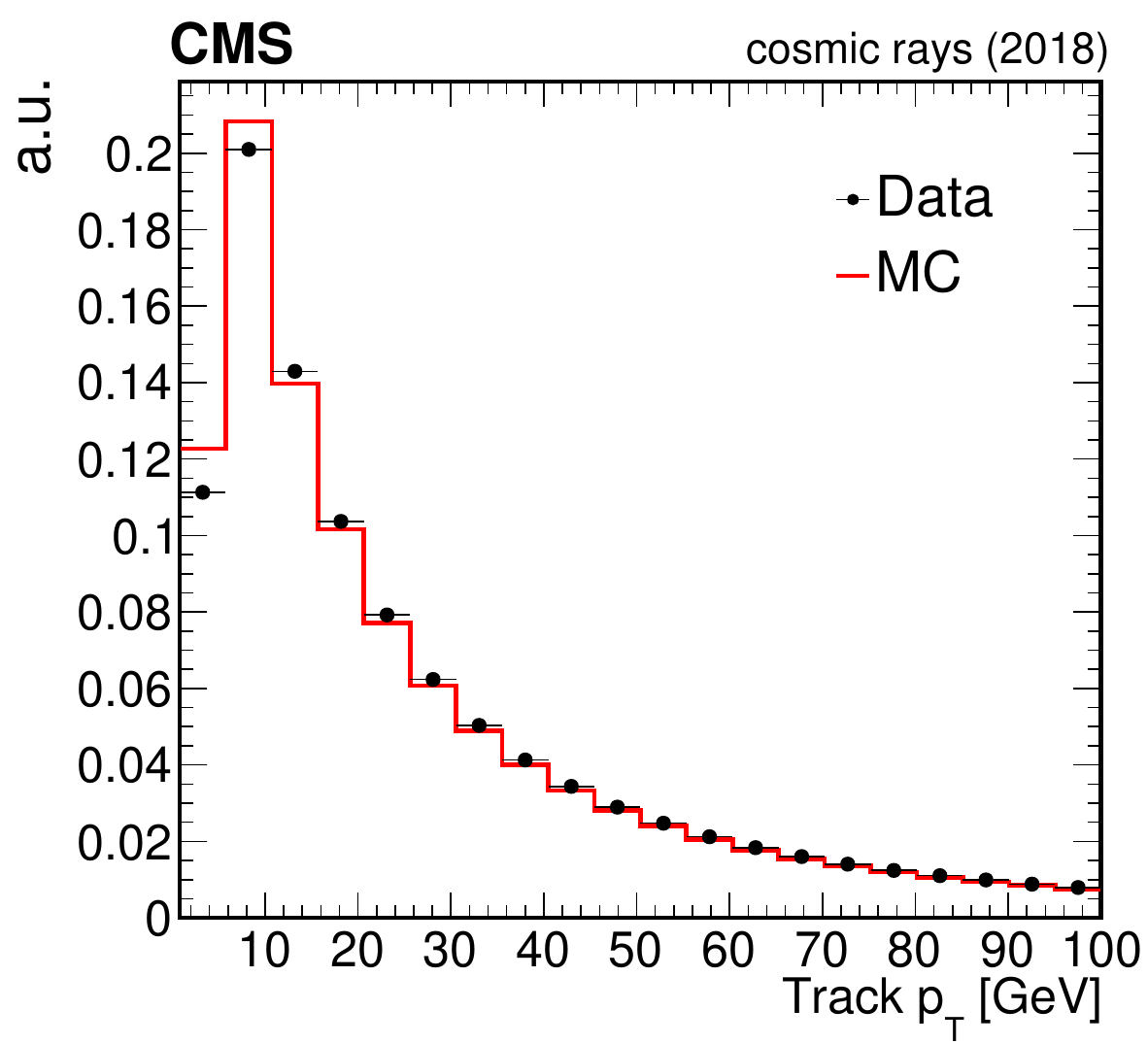}~\includegraphics[width=0.4\textwidth]{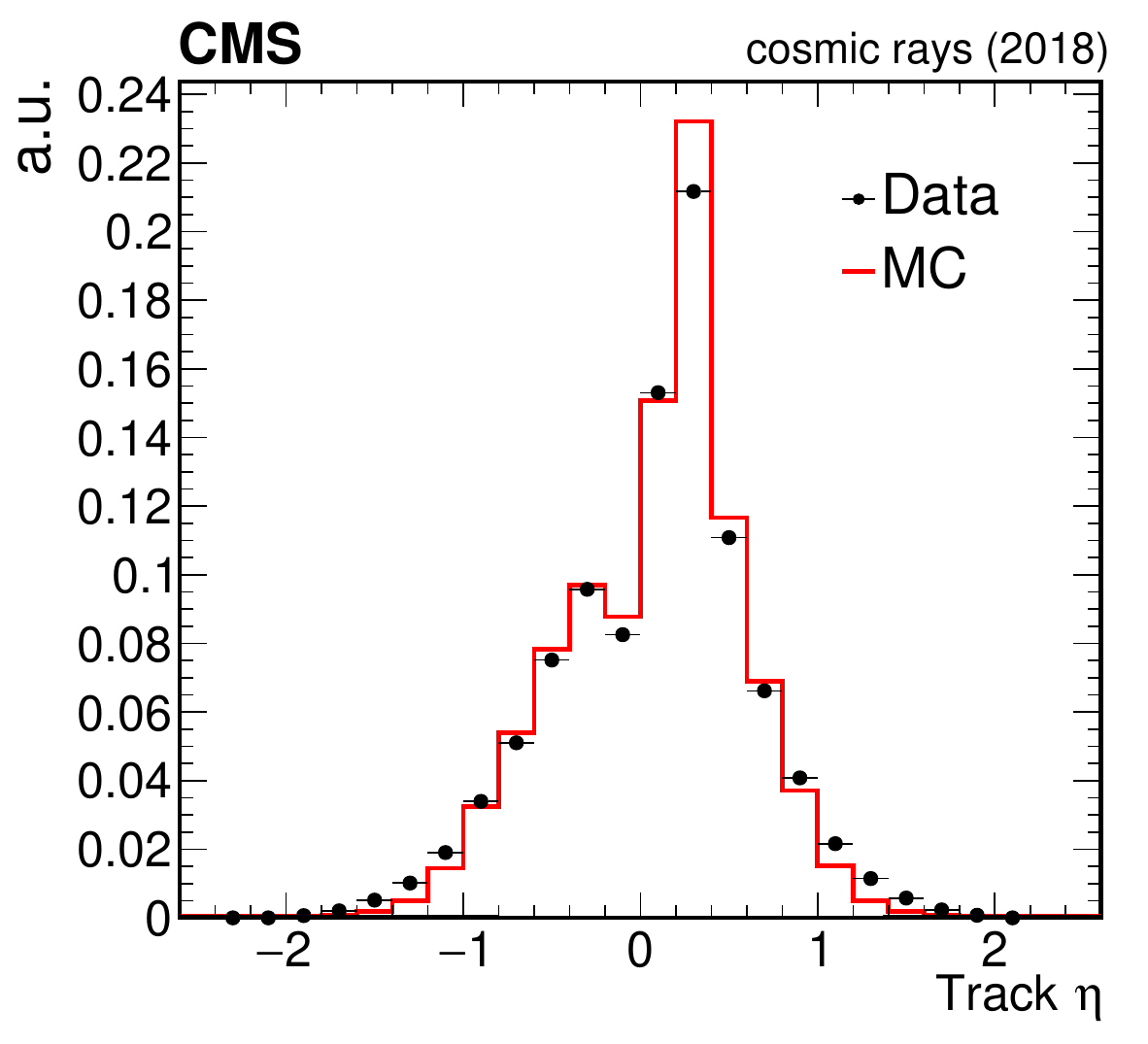}

    \caption{Normalized track \pt (left) and $\eta$ (right) distributions for the inclusive L1 trigger (top), $\Ztomm$ (middle), and interfill cosmic ray muon (bottom) data sets in arbitrary units.
        Data collected with the CMS detector in 2018 and used for the final alignment in that year (solid black circles) are compared with the simulation (solid coloured lines).
        Distributions in data are obtained from a sample of $11\times 10^6$, $55\times 10^6$, and $3.4\times 10^6$ tracks for the inclusive L1 trigger, $\Ztomm$, and cosmic ray muon data sets, respectively.
        For the inclusive L1 trigger data set, data are compared with two sets of simulated QCD events with different ranges of transverse momentum transfers $\pthat$.
        The green line corresponds to $\pthat$ between 15 and 30\GeV, whereas the magenta line corresponds to $\pthat$ between 1000 and 1400\GeV.
        The inset in the \pt distribution of the inclusive L1 trigger data set (top left) shows the same distribution with a logarithmic scale for the $y$ axis.
        No correction for the limited modelling of the trigger efficiency in the simulation has been applied for the $\Ztomm$ data set. 
        The statistical uncertainty is smaller than the symbol size and therefore imperceptible.
    }
	\label{fig:DatasetDistributions}
\end{figure*}

Events recorded with loose triggering conditions are referred to as belonging to the inclusive L1 trigger data set.
It consists of a sample of randomly chosen events passing an L1 trigger~\cite{Sirunyan:2020zal}.
Because of their large production rate, these events are particularly important during low-luminosity runs, as well as in the early stages of data taking for providing a sufficient amount of tracks for the alignment procedure.  
The track selection requires tracks to be reconstructed from a set of at least ten hits in the tracker.
The tracks must have a momentum $p>8\GeV$ and $\pt > 1 \GeV$. 
The vast majority of the final-state particles have low~\pt, and their tracks are concentrated in the high-$\eta$ region, as shown in Fig.~\ref{fig:DatasetDistributions} (top row).  
This figure shows the \pt and $\eta$ distributions for tracks from the inclusive L1 trigger data set collected by the CMS detector in 2018, after applying the track selection described above.  
The data are compared with Monte Carlo (MC) simulation for both low-$\pthat$ and high-$\pthat$ interactions, where $\pthat$ is the scale in the $2\to2$ matrix element calculation of the hard process.
The events are simulated using \PYTHIA 8.240~\cite{Sjostrand:2006za,Sjostrand:2014zea} with the CP5 tune~\cite{cp5tune}.
The low- and high-$\pthat$ samples, which correspond to events with $\pthat$ in the range of 15--30\GeV and 1000--1400\GeV, respectively, are used as two opposite reference points and the data naturally fall in between.
The small fraction of events coming from the high-$\pthat$ interactions produces the harder \pt spectrum and the more central $\eta$ distribution observed in data with respect to the simulation.

\subsubsection{Isolated muons}
Another suitable data set for the alignment procedure consists of isolated high-\pt muons from leptonic decays of \PW bosons, since they are recorded with very high efficiency and their track parameters can be measured very precisely in the detector. This data set consists of events passing the selection of at least one among several single-muon triggers. These triggers require the presence of an isolated muon and differ in the \pt threshold applied. 
Tracks of muon candidates reconstructed both in the silicon tracker and in the muon spectrometer, termed global muons, are selected if they have at least ten hits in the tracker, including at least one in the pixel detector. 
Events must have exactly one isolated muon candidate with $\pt>5 \GeV$.
An isolation condition is imposed on the muon by requiring it be separated from the axis of any jet candidate by $\Delta R > 0.1$, where $\Delta R = \sqrt{\smash[b]{(\Delta\eta)^2+(\Delta\phi)^2}}$ and $\Delta\eta$ and $\Delta\phi$ are differences in the pseudorapidity and the azimuthal angle, respectively.
Isolated muons cover a different phase space with respect to collision tracks from the inclusive L1 trigger data set, because they are characterized by a harder \pt spectrum and hence a more central $\eta$ distribution.

\subsubsection{Dimuon resonances}
This data set is formed by events passing the selection of a collection of double-muon triggers with different muon~\pt and isolation requirements. Tracks from the decay products of well-known dimuon resonances are particularly valuable for alignment purposes, because additional information from vertex and invariant mass constraints can be added to Eq.~(\ref{eq:eqn}) to constrain certain kinds of systematic distortions, especially those that bias the track momentum. 
By considering different resonances we can connect different groups of modules, because the difference in $\Delta\phi$ between the two tracks depends on the boost of the mother particle. 
For this reason, muon tracks from both \PgU meson and \PZ~boson decays are included.
To target $\PgU\to\Pgm\Pgm$ events, the applied selection requires track $\pt>3 \GeV$ and a dimuon invariant mass in the range $9.2<\mmumu<9.7\GeV$. To select muon pairs from \PZ~boson decays these requirements are
$\pt>15 \GeV$ and $85.8<\mmumu<95.8\GeV$.
The \pt and $\eta$ distributions for tracks recorded in 2018 that satisfy the $\Ztomm$ event selection are shown in Fig.~\ref{fig:DatasetDistributions}. 
The MC events for comparison with data are generated using \MGvATNLO 2.6.0~\cite{madgraph}, which is interfaced with \PYTHIA 8.240 to simulate parton showering and hadronization.
The lower event yields in data in the central $\eta$ region are due to a known trigger inefficiency, which is not included in the simulation.

\subsection{Cosmic ray muons}
\label{Cosmic-rays}
Cosmic ray muons, referred to as cosmics or cosmic events, recorded by the CMS detector are used for detector commissioning and calibration.
Before turning on the magnetic field, events are recorded during the \emph{Cosmic RUns at ZEro Tesla}~(CRUZET). 
Cosmic ray muon tracks are also recorded in the 3.8\unit{T} magnetic field provided by the CMS solenoid, during the \emph{Cosmic Runs At Four Tesla}~(CRAFT). 
Tracks from cosmic ray muons are crucial for the derivation of the alignment constants for two main reasons. First, they can be recorded before the start of LHC collisions, and are therefore employed to derive the first alignment corrections after a shutdown period, as described in Section~\ref{sec:startup}.
Second, they have a very different topology compared with collision tracks.
Unlike collision tracks, tracks from cosmic ray muons cross the whole detector and connect modules located in the top and bottom halves of the tracker. 
This breaks the cylindrical symmetry typical of collision tracks and helps to constrain several classes of systematic distortions. 
Figure~\ref{fig:DatasetDistributions} (bottom row) shows the \pt and $\eta$ distributions for cosmic ray muons in which the asymmetry in $\eta$  is attributed to the location of the CMS cavern shaft. 
The MC events in the figure, used for comparison with data, are simulated using the cosmic muon generator CMSCGEN~\cite{CosmicMuonGenerator}.

Throughout Run~2, cosmic ray muon events were recorded before the start of LHC collisions in dedicated commissioning runs and in the time intervals between two LHC fills (interfill runs). 
Since 2018, it has been possible to record cosmic events during collision data taking. Figures in this paper that contain data from cosmic events are indicated with the label ``cosmic rays'' instead of ``$\pp$ collisions''.

\subsubsection{Commissioning and interfill runs}
\begin{figure}[ht!]
    \centering
    \includegraphics[width=0.5\textwidth]{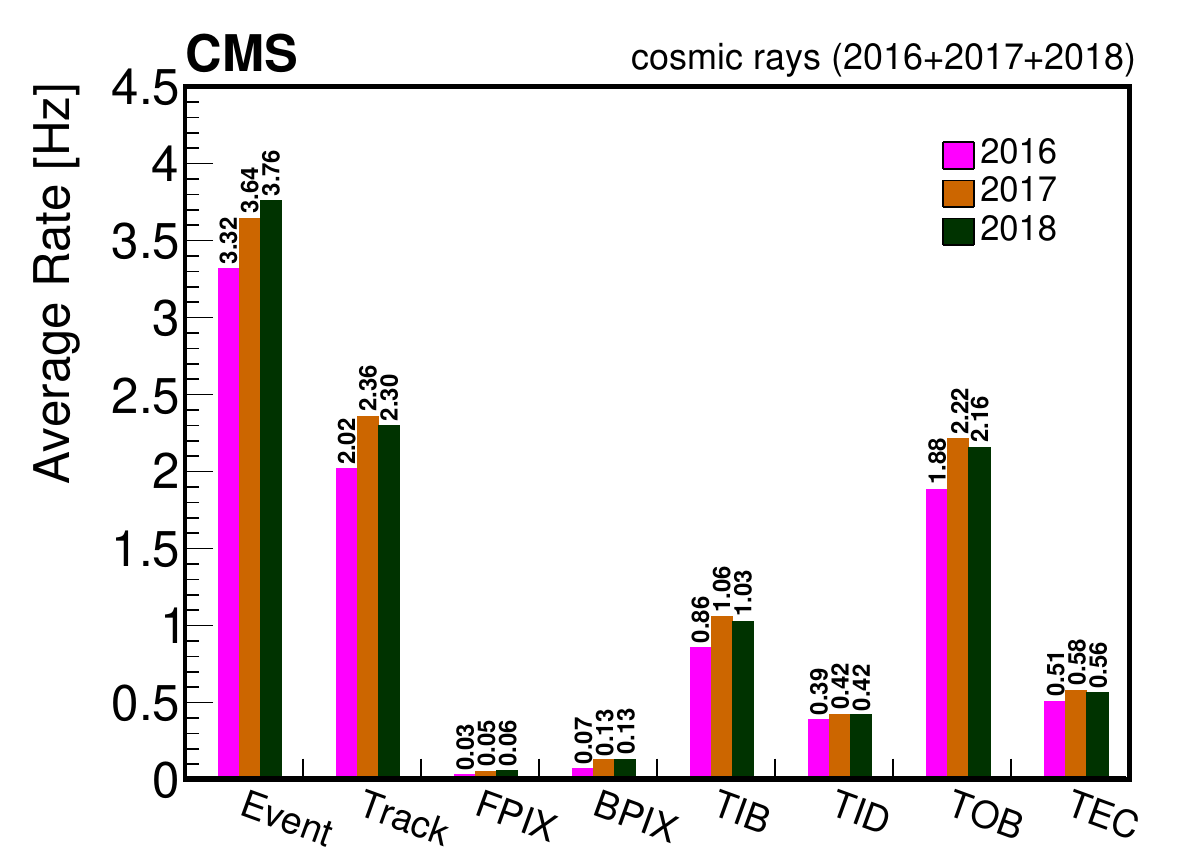}
    \caption{Average event rates of cosmic ray muon data recorded with the CMS tracker during the years 2016, 2017, and 2018, obtained as explained in the text.
        The statistical uncertainty in the measured rates is negligible and is not shown in the figure.}
    \label{fig:Cosmic_Rate_plot}
\end{figure}

A total of $1.72\times 10^7$ and $1.33\times 10^7$ cosmic ray muon events were collected with the CMS detector during commissioning and interfill runs, respectively, from 2016 to 2018. These events are selected using an unprescaled single-muon trigger with no $\pt$ threshold applied. Figure~\ref{fig:Cosmic_Rate_plot} shows the average rate of cosmic ray muon events recorded by the CMS detector in this time period. Cosmic events are recorded using dedicated muon triggers. The tracks are reconstructed using three different algorithms, which are described in Ref.~\cite{1742-6596-119-3-032030}: combinatorial track finder, cosmic track finder, and road search.
Reconstructed tracks obtained by the combinatorial track finder algorithm are used for the alignment procedure, and are required to have at least seven hits, of which at least two must be in either the pixel detector or in stereo module pairs. Stereo module pairs consist of two strip modules mounted back-to-back, with their strips aligned at a relative angle to provide a measurement of both the $r$-$\phi$ and the $r$-$z$ coordinates. The average track rates after this selection are also shown in Fig.~\ref{fig:Cosmic_Rate_plot}, both inclusively and separately for tracks with at least one valid hit in a given tracker partition. 
A systematic study of the average track rates is essential for estimating the duration of cosmic data collection during commissioning and of the interfill runs needed to accumulate a sufficient number of tracks for the alignment procedure.
Additional quality requirements are applied to the tracks used in the alignment fit; events with more than one track are rejected, as are tracks with $p<4\GeV$. 

\subsubsection{Cosmics during \texorpdfstring{$\pp$}{pp} collisions}
To increase the number of cosmic ray muon tracks used in the alignment procedure, an effort was made to collect cosmics during collisions~(CDC).
Dedicated trigger sequences, which rely on the longer trajectory of muons from cosmic rays inside the detector with respect to muons being produced in the $\pp$~interaction region, were developed for this. 
The typical time of flight of a cosmic ray muon passing through the whole detector is ${\approx}30\unit{ns}$, larger than the interval of 25\unit{ns} between two consecutive bunch crossings. 
A cosmic ray muon candidate can be identified by requiring two consecutive signals of the global muon trigger in a back-to-back topology.
In the L1 trigger, only muon candidates with $\pt>3\GeV$ in the central region, $\abs{\eta}<1.2$, of the detector are retained to reduce the background from low-\pt muon tracks from $\pp$~interactions. 
A larger fraction of low-\pt background tracks is rejected after the muon reconstruction is performed by the HLT, and the kinematic requirements are tightened to keep the trigger rate below a threshold of~$\mathcal{O}(5-10\unit{Hz})$. 
Owing to the dynamical dependence of the trigger rate on the number of additional $\pp$~interactions from the same or nearby bunch crossing (pileup), two CDC triggers were introduced in the HLT. 
A \pt~threshold of 10\,(5)\GeV is required in the main (low-pileup) trigger. 

These dedicated CDC triggers were deployed in July 2018 and were active until the end of the $\pp$~collision run of that year.
Approximately 700\,000 tracks collected with the CDC triggers passed the alignment selection criteria for cosmic ray muon tracks.

\section{Systematic misalignments} \label{sec:WMs}\label{sec:WM} \label{sec:systDistort}
Systematic shifts of the assumed positions of the silicon modules of the tracker, when compared with the actual positions of the active elements, can occur. Such shifts will be called systematic distortions, or misalignments, of the tracker geometry. 
These systematic misalignments may cause biases in the track reconstruction and this can have a negative impact on physics measurements. 
Therefore, a dedicated programme of studies of such systematic distortions was developed.
The WMs mentioned in Section~\ref{sec:concepts} form a particular class of systematic distortions. These are transformations that change a set of valid tracks into another set of valid tracks and satisfy $\Delta \chi^2 \approx 0$. They may arise when the alignment parameters of all modules are free in the alignment fit without additional constraints.
Although such a transformation does not affect the individual track parameter fit performance in the alignment procedure, it may affect certain topologies of the tracks or correlations between tracks that are later used in physics measurements.

The most obvious example of a WM is a global movement of the whole detector, but more subtle effects are possible.
The fact that all collision tracks come from the centre of the detector and that the detector is symmetric around the beam axis may cause certain WM biases that leave the $\chi^2$ of the individual collision tracks invariant. 
Such a systematic distortion is not necessarily a WM, but the effect may be especially large in the direction with the weakest constraints.

In this section, we first present the methods used to detect the presence of systematic distortions in the alignment constants, then we review nine canonical systematic distortions.
At this stage, only pure systematic distortions are discussed, without considering any alignment procedures.

\subsection{Validation of systematic distortions} 
\label{sec:systematic-validation}
Several validations are used to check the effect of misalignments and determine whether a particular set of alignment constants performs well.  
A validation is essentially a measurement of a variable of interest that is also a metric for the alignment performance.  
The quantities we choose to study typically have a known value under perfectly aligned conditions.  
For example, a distribution of residuals is expected to peak at 0 with a given width. 
The difference in parameters between two halves of a cosmic ray track is also expected to be 0 on average.  
The mass of a reconstructed \PZ~boson should be around 91.2\GeV.  
By detecting deviations from these expected values, especially deviations as functions of the track location or direction, we can search for biases.

\subsubsection{Geometry comparison} 
Once the alignment fit has been performed, the new geometry is compared with a reference geometry, such as the design geometry or a previously aligned geometry.
Systematic differences in such a comparison may reveal distortions in the tracker geometry.
Although it is not possible to assess the validity of systematic shifts in the module positions from geometry comparisons alone, they may serve as a guide and visualization of possible effects in the tracker.
Certain distortions may be known to be unphysical from the detector design constraints, and would form an early warning of biases in the alignment procedure prior to more detailed tests with the reconstructed track data. 
The geometry comparison validation was derived from the tools developed for the optical survey constraint within the \HIPPY algorithm discussed in Section~\ref{sec:software}.
These tools match two geometries by translating and rotating certain structures before the differences between the geometries are calculated.
These differences are treated as survey residuals in the alignment algorithm~\cite{Chatrchyan:2009sr}.
The global shift and rotation of large structures are removed and the module displacements $\Delta z$, $\Delta r$, and $\Delta\phi$ are measured with respect to the reference geometry as a function of $z$, $r$, and $\phi$.
The other coordinates ($x$, $y$, and $z$) and three angular rotations can be visualized in this manner as well. 

Geometry comparison validation is performed without any reconstructed track data and can be applied with reference to any prior geometry, such as design, survey, or previously aligned track-based geometry. 

\subsubsection{Cosmic ray muon track validation} 
The track parameter resolutions can be validated by independently reconstructing the upper and lower portions of cosmic ray muon tracks that cross the tracker and comparing the track parameters at the point of closest approach to the nominal beamline. We will refer to this procedure as the cosmic ray muon track split validation.
This method is powerful because we know that the two halves of a given cosmic ray track should have the same parameters at the point closest to the nominal beamline, while each half of a track mimics a regular collision track originating from that point.
Systematic differences between the track halves can indicate a misalignment.

\begin{figure*}[htb!]
    \centering
    \includegraphics[width=0.9\textwidth]{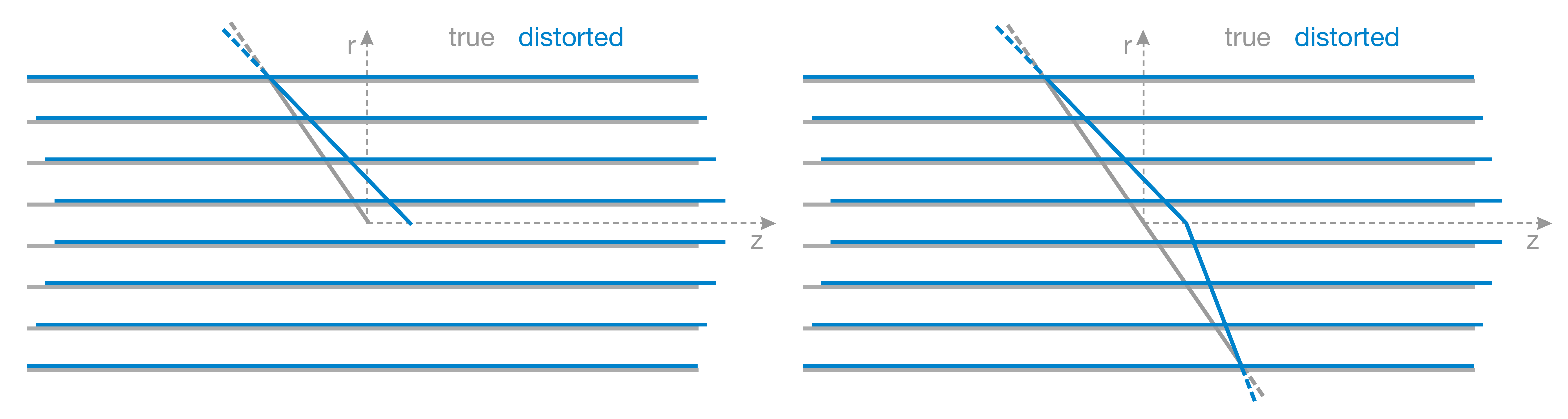}
    \caption{
        Diagram demonstrating distortions of the tracker geometry that may not affect the consistency of a reconstructed collision track with the measured hits (left), but introduce a kink when reconstructing the two halves of a cosmic ray muon track (right), leading to an inconsistency.
        This illustrates the telescope effect from Table~\ref{table:weakModes}.
    }
    \label{fig:CosmicDiagram}
\end{figure*}

Cosmic ray muon tracks and collision tracks have different topologies.
Therefore, systematic distortions that may appear as WMs with collision tracks may be well constrained or visible with cosmics.
In particular, the fact that these tracks do not originate at the centre of the detector means they connect the top and bottom halves of the detector directly through a single track. 
Such a connection is not possible with tracks originating from the beam collision point.
This effect is shown in Fig.~\ref{fig:CosmicDiagram}. 
Since cosmic ray muons leave predominantly vertical tracks, they primarily constrain horizontal modules along the $z$-axis and have limited sensitivity for aligning vertical modules in the endcap regions or modules connecting different parts of the detector in the horizontal direction.

Cosmic ray muon track validation can be performed without data from beam collisions and serves as an early validation of the detector geometry before LHC operation starts.
It remains a powerful tool during collision data taking because of the unique topology of the cosmic ray muon tracks. 

\subsubsection{Overlap of hits within the same layer of modules}
\begin{figure}[htb!]
     \centering
     \includegraphics[width=0.5\textwidth]{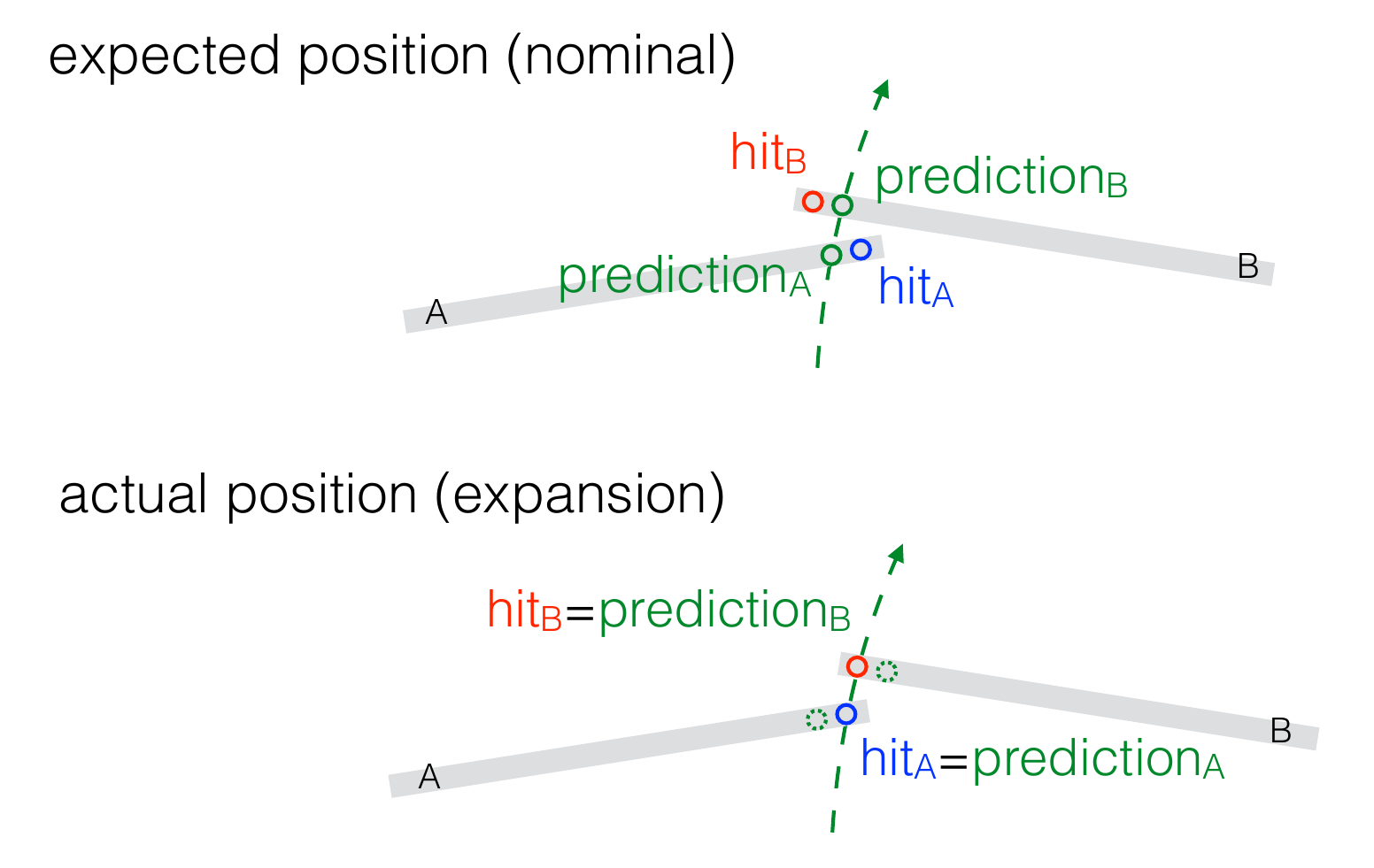}
       \caption{
               Diagram demonstrating the overlap regions of two representative modules $A$ and $B$.
               In the upper diagram, the predicted track impact points (green circles) and the actual hits with charge depositions (red and blue circles)
               do not coincide because of a wrong prediction of the module positions. In the lower diagram, the actual module positions are shown
               for the geometry with radial expansion, and the predicted impact points and hits coincide. 
               Uncertainties due to track propagation are ignored in this illustration, but are greatly
               reduced in the difference of residuals as discussed in the text.
               The green dashed circles in the lower diagram indicate predicted impact points from the nominal geometry in the upper diagram.
       }
       \label{fig:OverlapDiagram}
\end{figure}

The overlap validation monitors the alignment by using hits from tracks passing through regions where modules overlap within a layer of the tracker.
It can be performed either with the cosmic data or with data from beam collisions. 
Tracks are required to have two hits in separate modules within the same layer.
In this method we take advantage of the small distance between the two hits, and therefore the small uncertainty in the track parameter propagation between the two modules.
The double difference in estimated and measured hit positions is very sensitive to systematic deformations.
Unexpected deviations between the reconstructed hits and the predicted positions can indicate a misalignment.
This is characterized by a nonzero mean of the difference of residuals.
An illustration of the overlap measurements is shown in Fig.~\ref{fig:OverlapDiagram}. 
The quantity of interest is the difference of residuals calculated as 
\begin{equation}
(\text{hit}_A-\text{prediction}_A)-(\text{hit}_B-\text{prediction}_B),
\end{equation}
where $\text{hit}_{A,B}$ refers to the position of a hit in module $A$ or $B$, and $\text{prediction}_{A,B}$ refers to the position of a predicted impact point of the track in module $A$ or $B$ derived from its fit using measurements in other modules.
The advantage of the overlap method is that most uncertainties in the track propagation are cancelled in the difference $(\text{prediction}_A-\text{prediction}_B)$. 
The difference of residuals is expected to be zero on average for a perfectly aligned detector. 
A positive shift in the mean is expected for expansion and a negative shift for contraction.

\begin{figure}[htb!]
    \centering
    \includegraphics[width=0.9\linewidth]{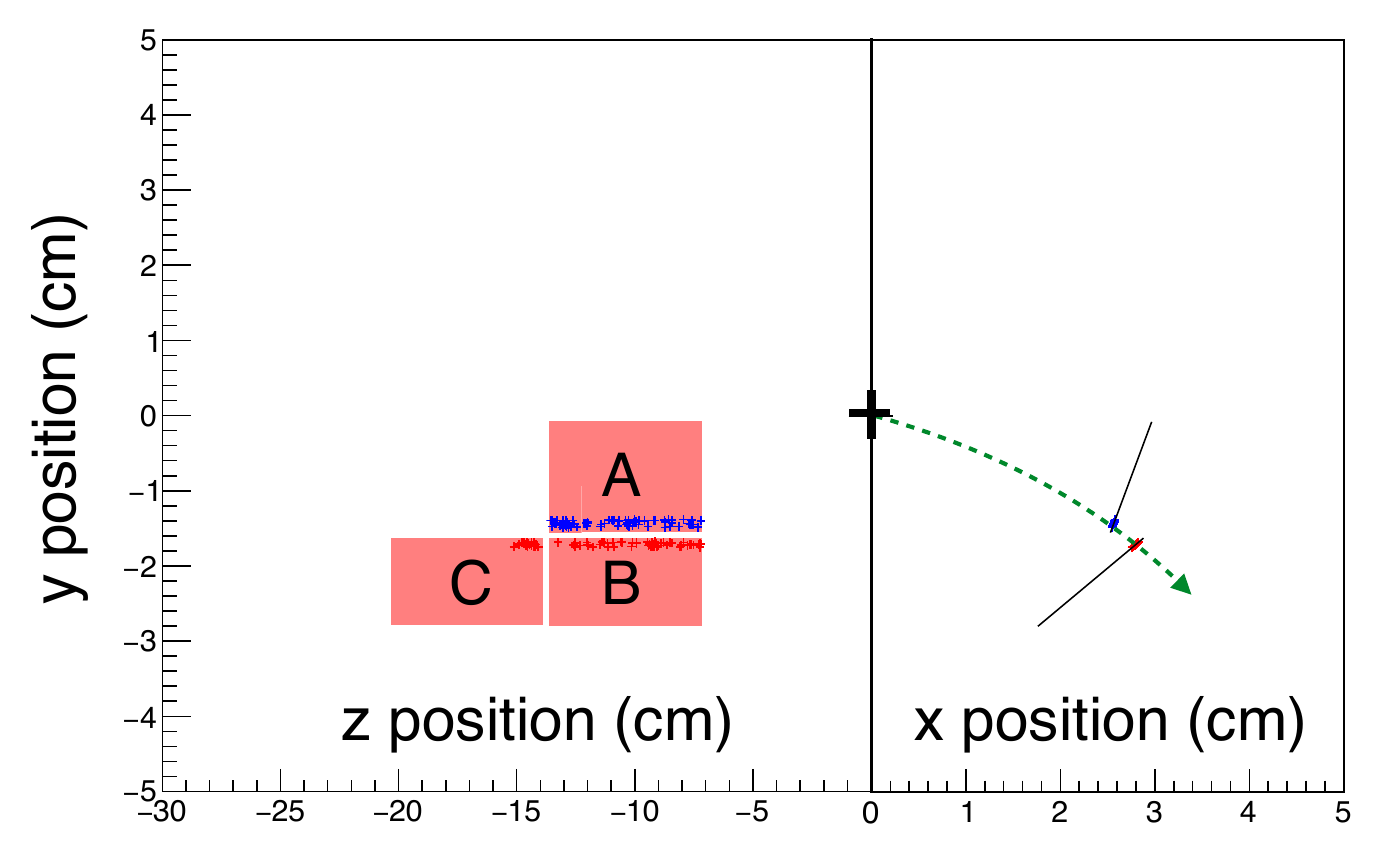}
    \caption{
        Diagram demonstrating the overlap regions of three representative modules $A$, $B$, and $C$ in the first layer 
        of the barrel pixel detector. The $y$-$z$ view (left) and $y$-$x$ view (right) are shown for the same modules.         
        The overlap hits are indicated with the blue (inner) and red (outer) crosses and appear in tracks with hits 
        in two consecutive modules in the same layer of the detector.
        The black cross represents the interaction point.
        The overlap between modules $A$ and $B$ constrains the distance between modules in the $\phi$ direction,
        whereas the overlap between modules $A$ and $C$ constrains the distance between modules in the $z$ direction.
     }
    \label{fig:OverlapDiagram_2}
\end{figure}

In Fig.~\ref{fig:OverlapDiagram}, the overlap between modules $A$ and $B$ constrains the 
circumference of the detector, and therefore its radial scale, when measured for all pairs of modules.
Figure~\ref{fig:OverlapDiagram_2} shows an example of the module overlaps in the $\phi$ and $z$~directions 
for three representative modules in the first layer of the BPIX.
The overlap between modules $A$ and $B$ constrains the circumference of the detector, as in Fig.~\ref{fig:OverlapDiagram}.
The overlap between modules $A$ and $C$ constrains the distance between modules in the $z$~direction, and
therefore the longitudinal scale. 

\subsubsection{Dimuon validation}
\begin{figure}[htb!]
    \centering
    \includegraphics[width=0.9\linewidth]{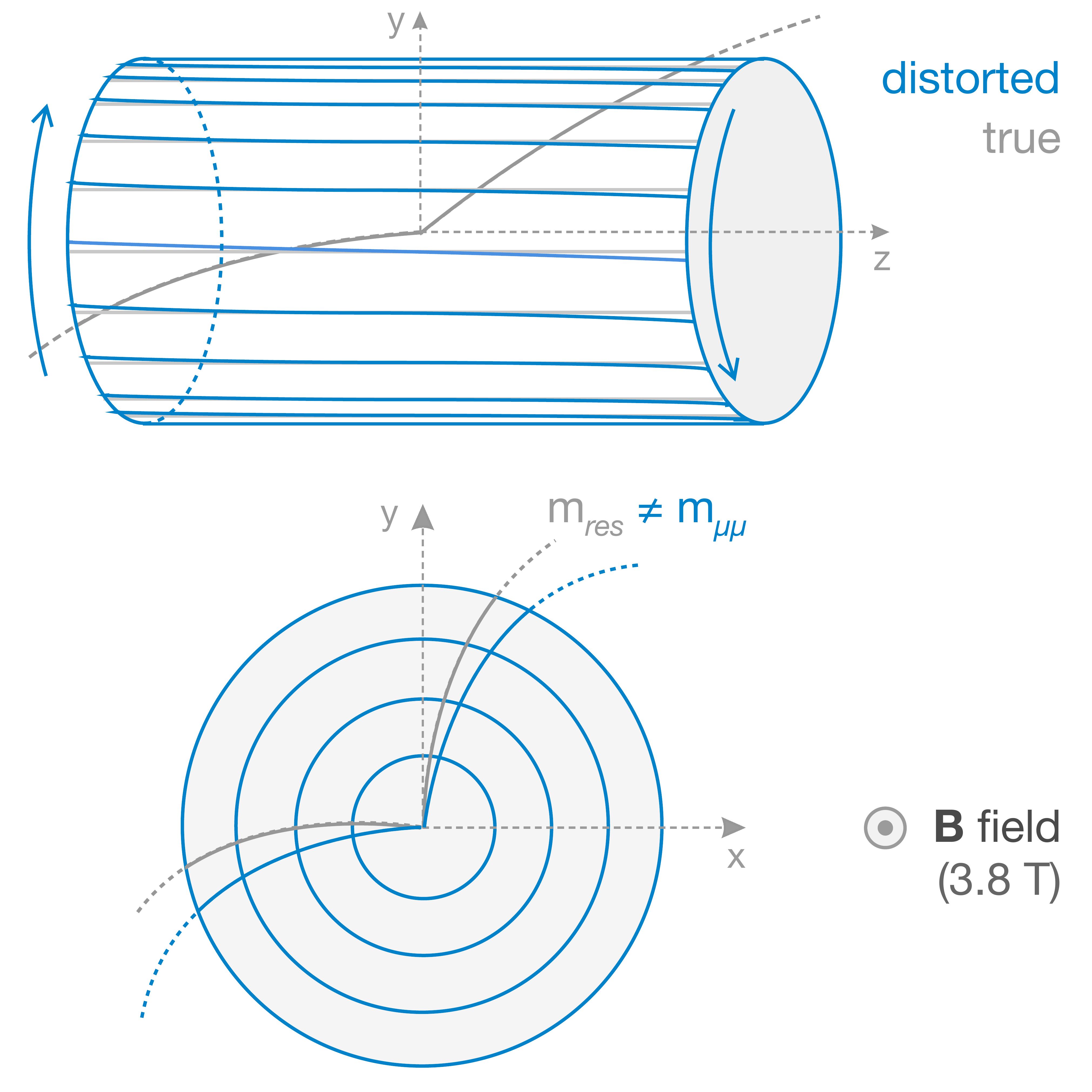}
    \caption{
        Diagrams demonstrating distortions of the tracker geometry in the $r$-$z$ view (upper) and in the $x$-$y$ view (lower) with the reconstructed muon pair from a $\Ztomm$ decay. The invariant mass of the pair of muons deviates from the expected value and becomes a function of the track parameters.
        This illustrates the twist effect from Table~\ref{table:weakModes}.
    }
    \label{fig:ZmumuDiagram}
\end{figure}

In an ideally aligned tracker, the reconstructed $X\to\Pgm\Pgm$ invariant mass should be minimally dependent on where in the detector the muons travel.
Therefore, the quality of the set of alignment constants can be assessed by looking for biases in the reconstructed mass of a known resonance $X$.
Any resonance can be used, but in practice we primarily consider \PZ~boson decays into muons.
This is because \PZ~bosons are often produced with a relatively small boost, which results in the two muons passing through opposite ends of the tracker.
Figure~\ref{fig:ZmumuDiagram} shows an example of a systematic distortion to which $\Ztomm$ decays are very sensitive (twist distortion, described in Section~\ref{sec:modandval}).

Each selected event, with its reconstructed mass, is placed into a bin depending on the $\eta$ and $\phi$ of the muons.
The mass distribution of each bin is then fit with a Gaussian function, and the mean of this function is recorded as the reconstructed mass in that bin.
The bins are then used to construct profiles of the invariant mass as a function of $\eta$ or $\phi$.
Misalignment in the tracker may be detected if the mean reconstructed mass strays from the expected value of 91.2\GeV, either uniformly or as a function of $\eta$ and $\phi$.

Dimuon validation is performed with the LHC collision data after a sufficiently
large sample of $\Ztomm$ events has been accumulated. Therefore, this validation is powerful in stable operating conditions.

\subsection{Modelling and validation of global systematic distortions} 
\label{sec:modandval}
To study systematic distortions, and WMs in particular, we introduce nine first-order deformations natural for the cylindrical geometry of the CMS tracker and parameterize them with simple models described by a single parameter $\epsilon$ for each distortion.
The systematic displacements from the reference geometry in $\Delta z$, $\Delta r$, and $\Delta\phi$ 
are functions of $z$, $r$, and $\phi$, with an overall scaling given by $\epsilon$.
The functional forms used to generate each systematic misalignment are listed in Table~\ref{table:weakModes}.

\begin{table*}[ht]
	\centering
	\topcaption{The nine basic systematic distortions in the cylindrical system, with the names of each systematic misalignment, the function by which the misalignment is generated, and a validation type sensitive to the misalignment.  
		The parameter $z_0={271.846}\cm$ is half of the length of the CMS tracker, and $\phi_0$ is an arbitrary constant phase. 
		}
	\begin{tabular}{cccc}
		&$\Delta z$  &$\Delta r$  &$\Delta\phi$  \\ \hline
                 & \textit{$z$~expansion} & \textit{bowing} & \textit{twist} \\        
                vs.~$z$ &$\Delta z =\epsilon z$ & $\Delta r=\epsilon r(z_0^2-z^2)$ & $\Delta\phi=\epsilon z$\\
                 & overlap & overlap & $\Ztomm$ \\ [\cmsTabSkip]
                & \textit{telescope} & \textit{radial} & \textit{layer rotation}\\
                vs.~$r$ & $\Delta z=\epsilon r$ & $\Delta r=\epsilon r$ & $\Delta \phi = \epsilon r$ \\
                & cosmics & overlap & cosmics \\ [\cmsTabSkip]
                & \textit{skew} & \textit{elliptical} & \textit{sagitta}\\
                vs.~$\phi$ & $\Delta z = \epsilon\cos(\phi+\phi_0)$ & $\Delta r = \epsilon r \cos(2\phi+2\phi_0)$ & $\Delta\phi = \epsilon\cos(\phi+\phi_0)$ \\
                & cosmics & cosmics & cosmics \\
	\end{tabular}
	\label{table:weakModes}
\end{table*}

The sign of $\epsilon$ is critical in the description of its value for misalignments.
To save computing time, MC simulations are always performed using the ideal geometry, and the track reconstruction is performed with a possibly misaligned geometry.
That is, the detector position remains fixed to the ideal geometry, and the geometry used in the reconstruction changes.
When discussing data, the opposite convention is more natural: the geometry used in the reconstruction is initially fixed and the detector itself moves.
Taking the radial misalignment as an example, a value of $\epsilon>0$ means that the geometry used for reconstruction is expanded in the $r$~direction with respect to the geometry used during data taking.
If this happens in data, we call it a radial contraction, because the detector has moved with respect to the expected position.

\begin{figure*}
	\centering

    \includegraphics[width=.32\textwidth]{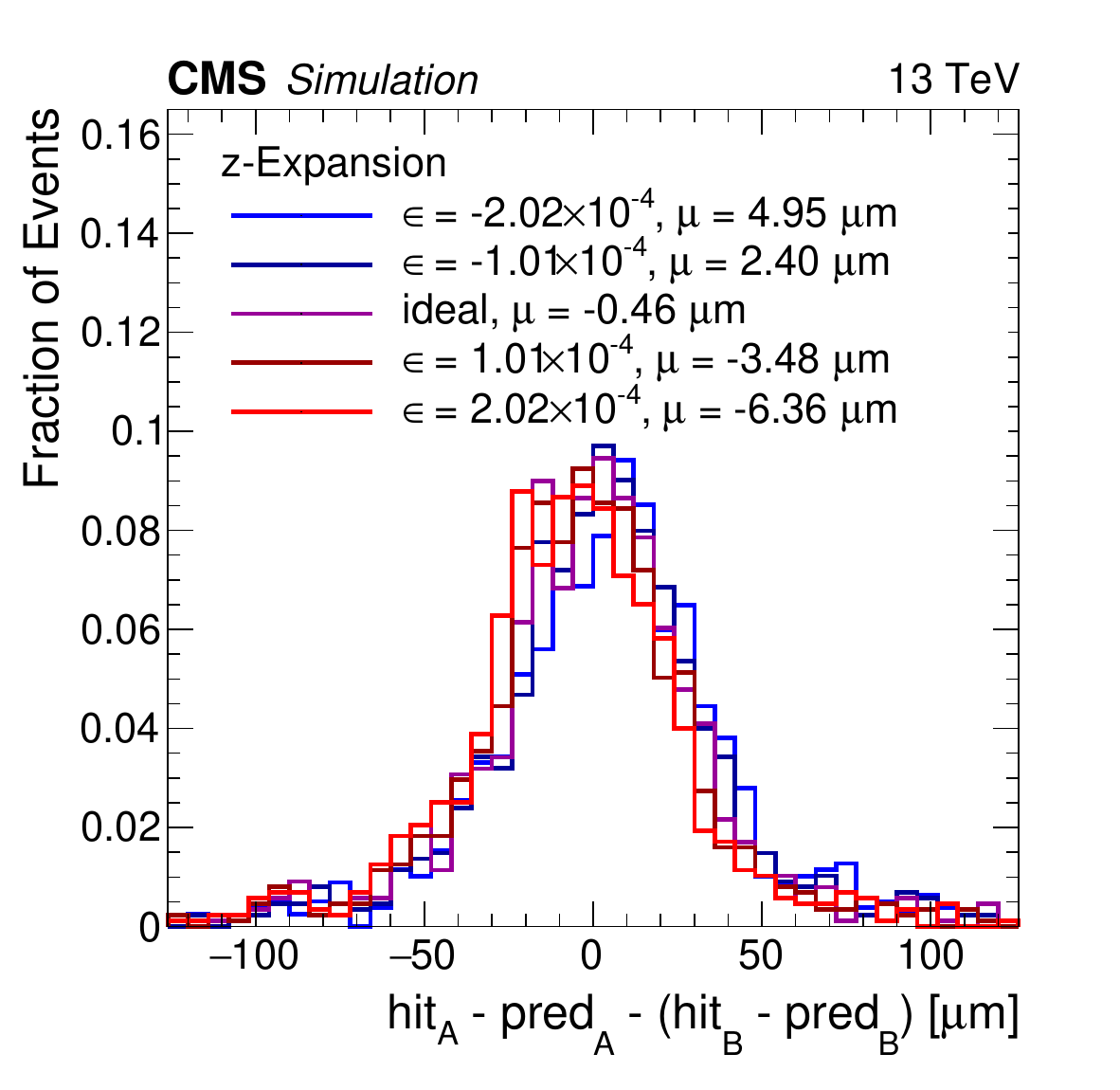}
	\includegraphics[width=.32\textwidth]{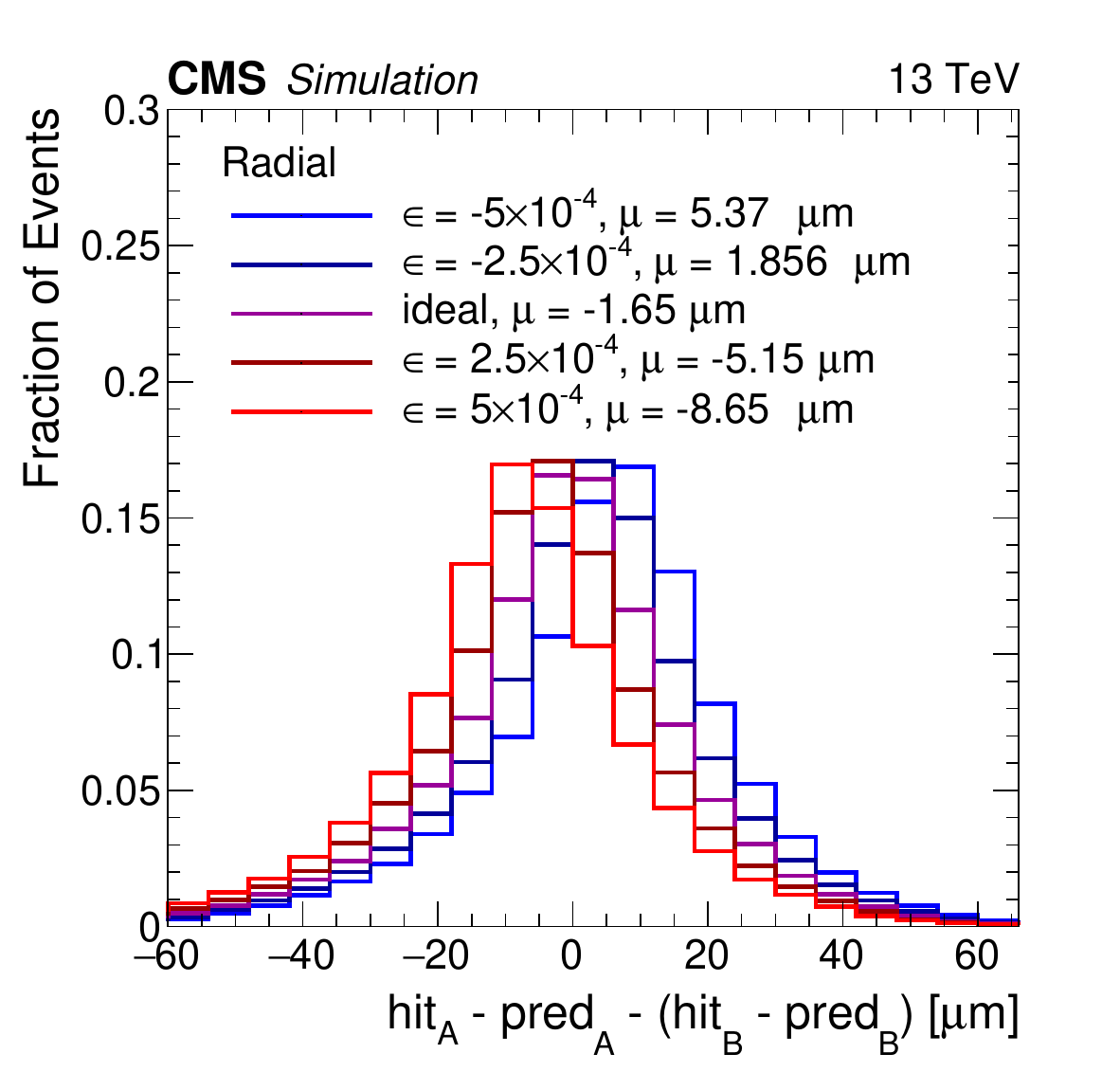}
    \includegraphics[width=.32\textwidth]{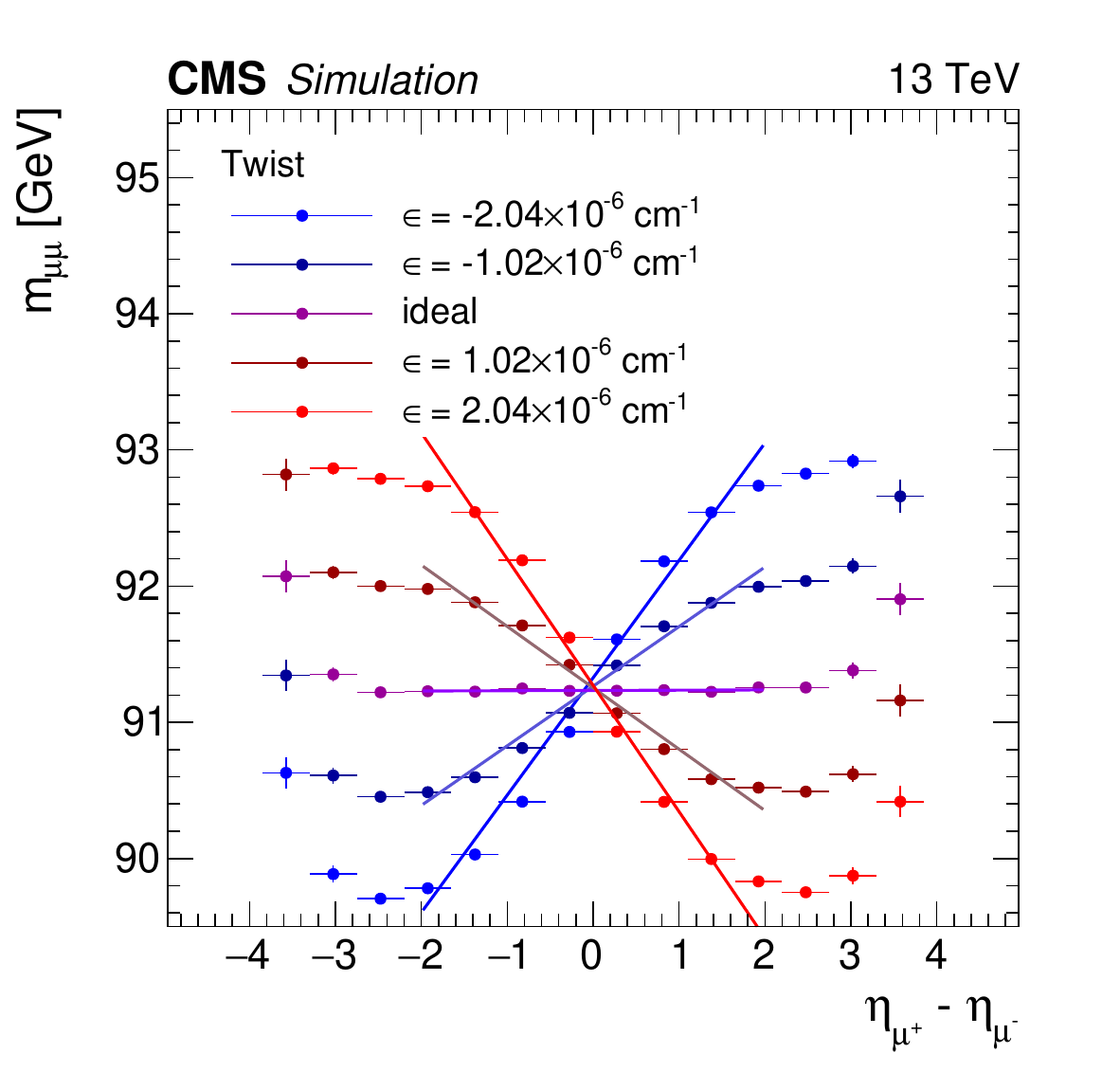}

    \includegraphics[width=.32\textwidth]{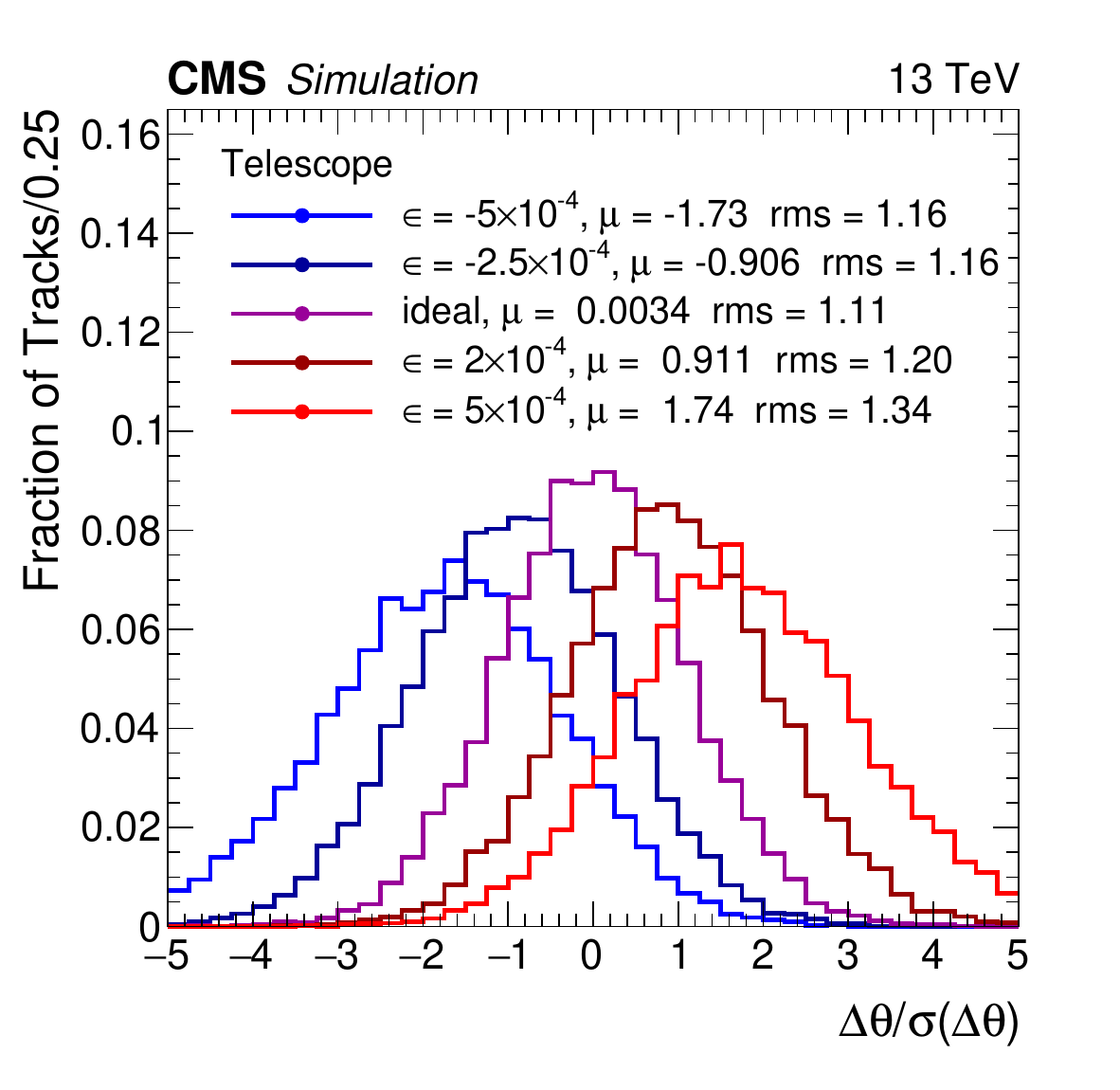}
    \includegraphics[width=.32\textwidth]{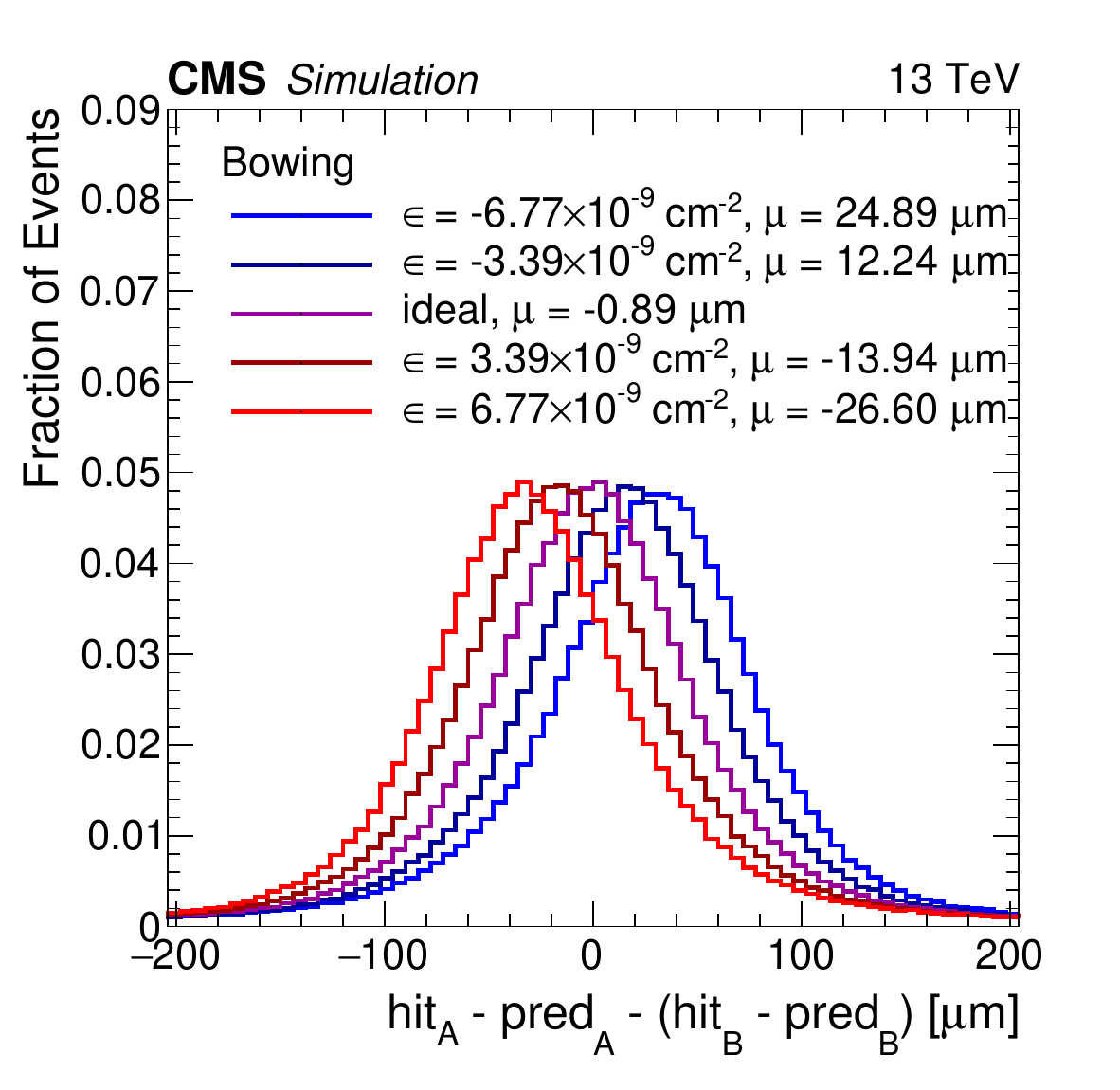}
	\includegraphics[width=.32\textwidth]{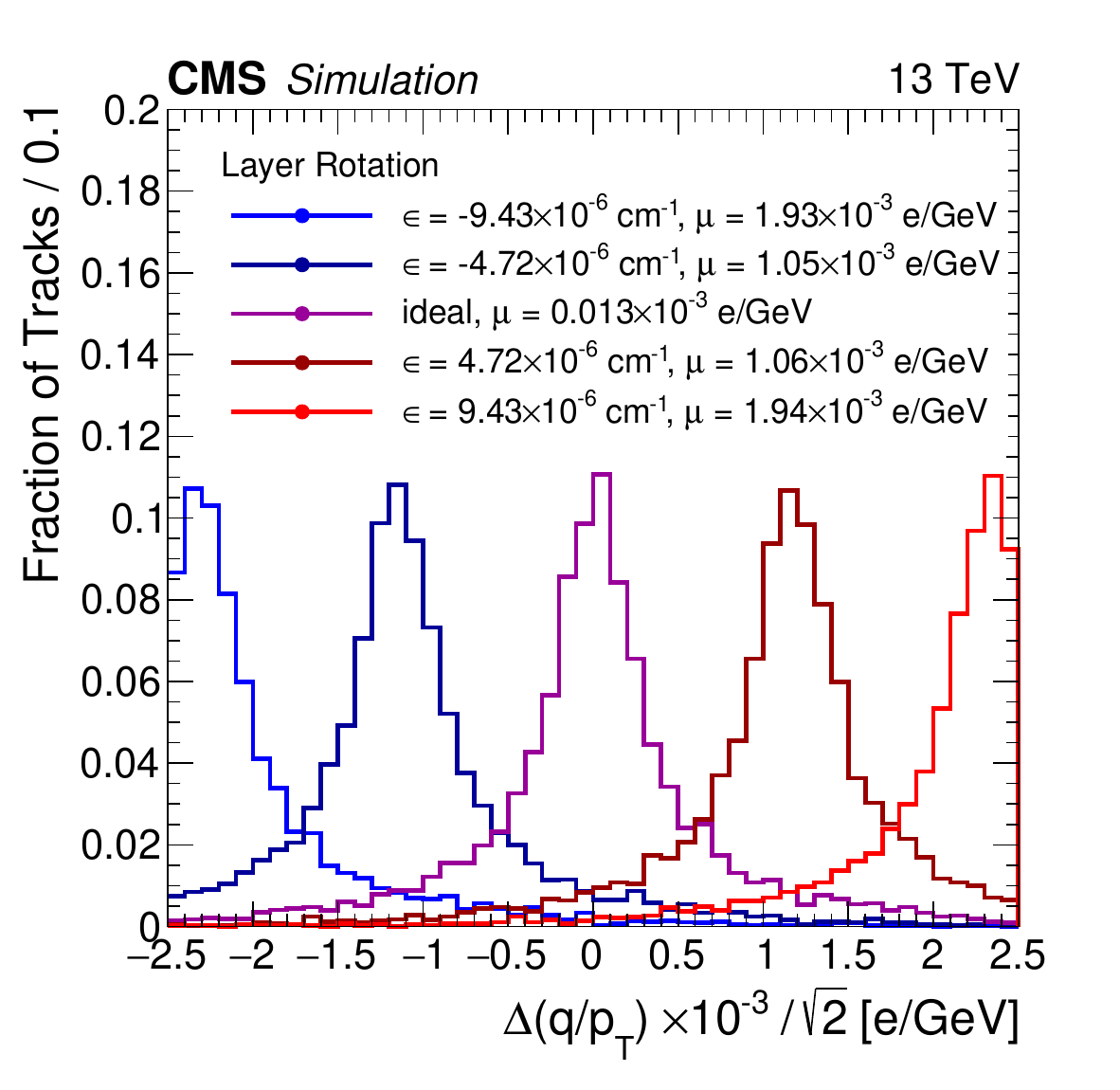}

    \includegraphics[width=.32\textwidth]{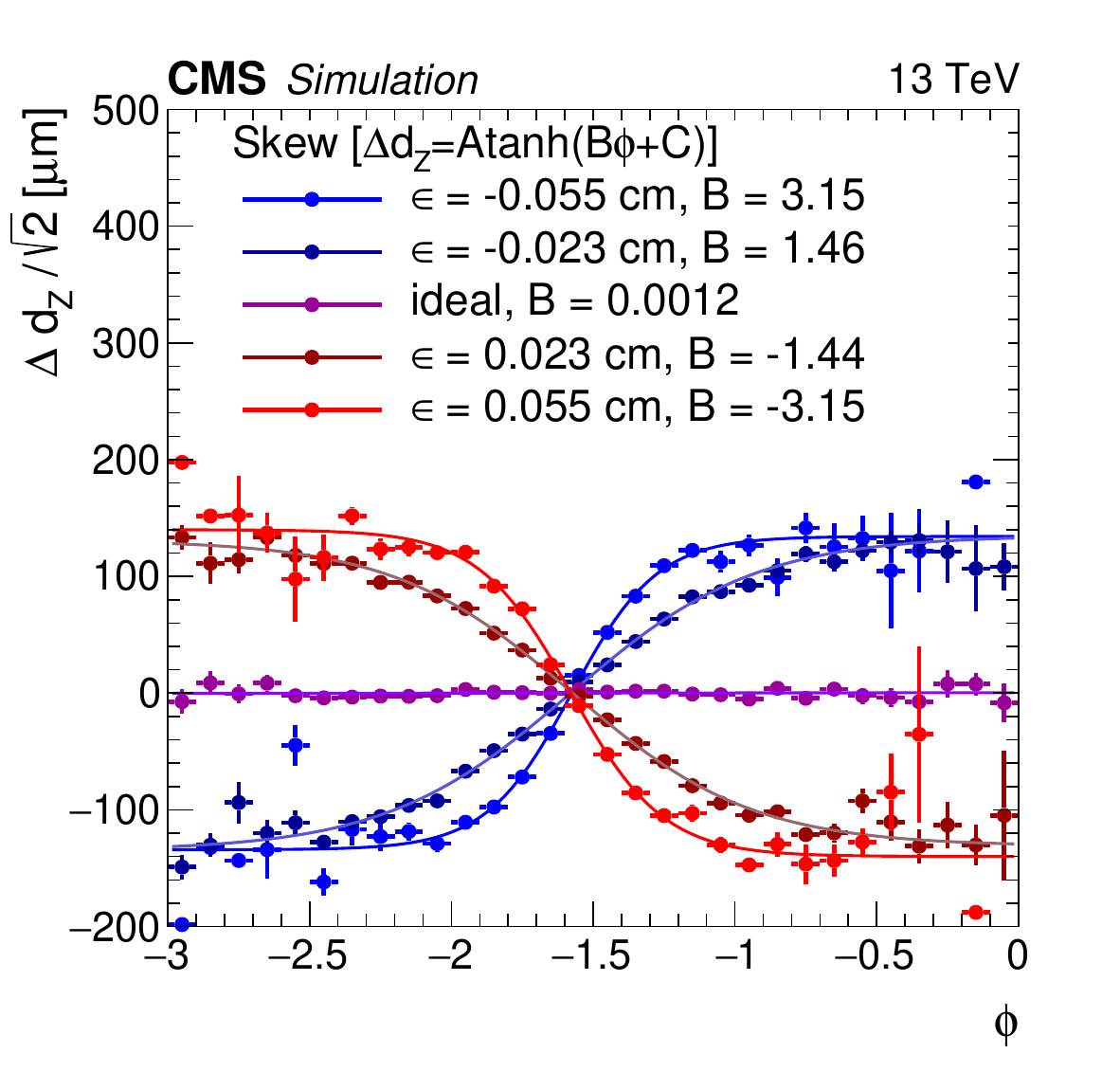}
	\includegraphics[width=.32\textwidth]{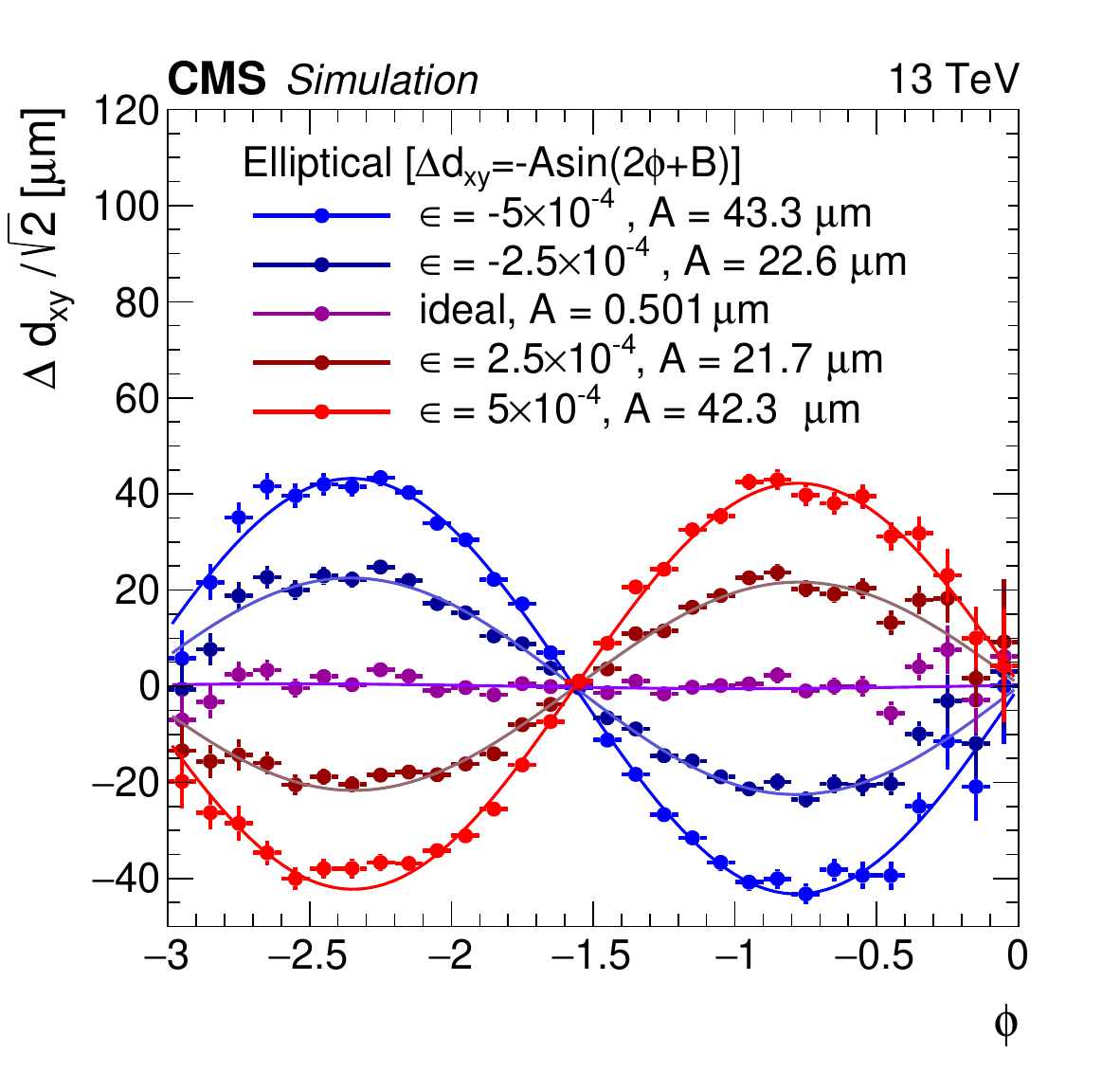}
	\includegraphics[width=.32\textwidth]{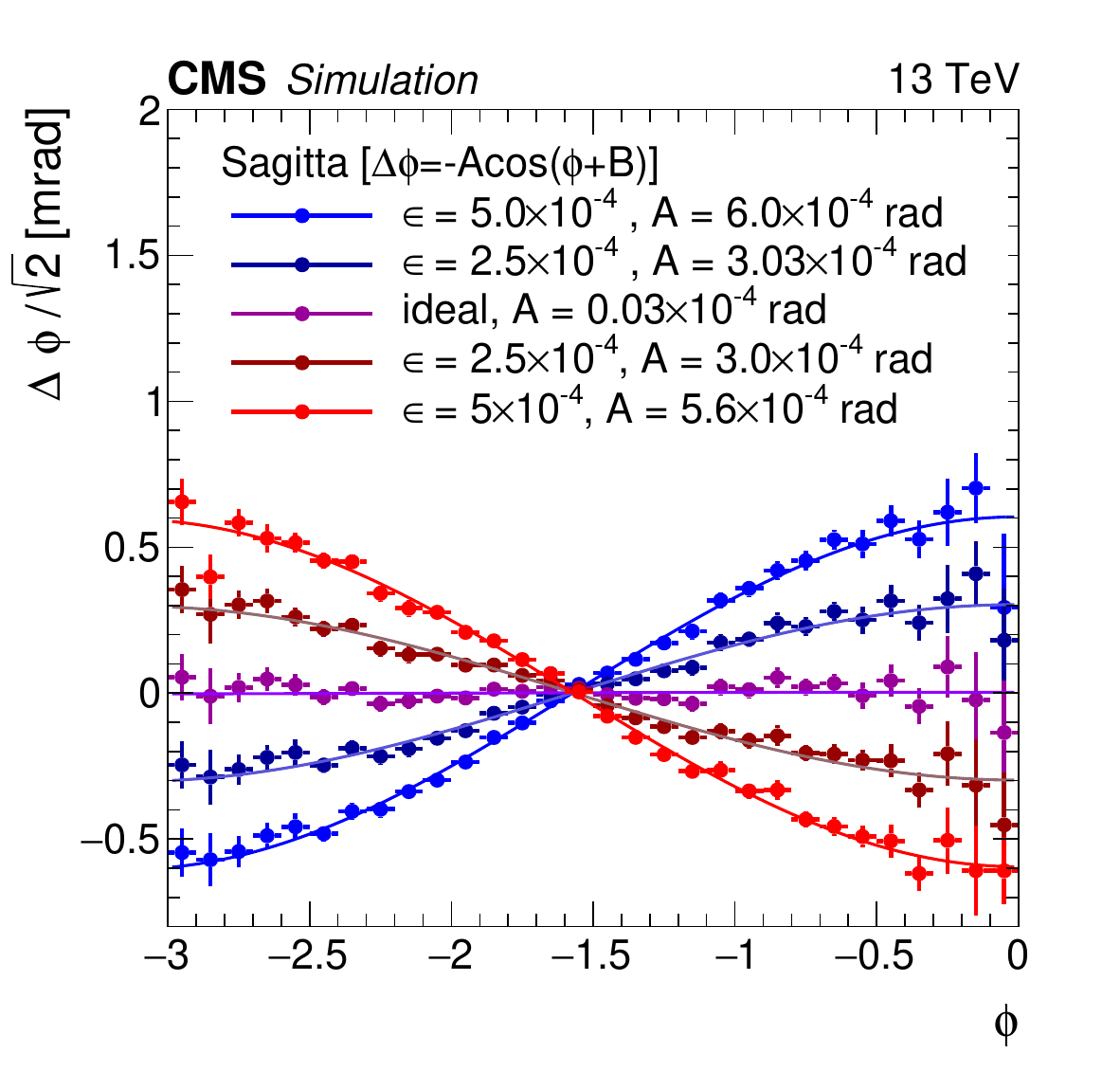}

	\caption{Validation of the nine basic systematic distortions summarized in Table~\ref{table:weakModes}
	using reconstructed MC simulations with five variations of the 
	misalignment parameter $\epsilon$ in each case.
	The ideal geometry in MC simulation corresponds to $\epsilon=0$. The horizontal lines show the uncertainty on the average of a measurement in a given bin. 
 	The most sensitive validation out of cosmic ray muon track, overlap, or dimuon validation is employed in each case, 
	as discussed in more detail in the text and as indicated in Table~\ref{table:weakModes}.
	In the bottom row, the formulae indicate the functional form of the fit used to extract the parameter quoted
	in the legend, which can be used to quantify the distortion. 
	The convention for the sign of $\epsilon$ is discussed in the text and corresponds to a distortion in the geometry 
	used for the reconstruction of MC events.
	This is opposite to the sign of the distortion if it were to be introduced in simulation of the detector components
	traversed by the charged particles. 
	}
	\label{fig:weakModes} \label{fig:twist}
\end{figure*}

The nine basic systematic distortions summarized in Table~\ref{table:weakModes} are not necessarily WMs when considering all possible topologies of tracks. We found that cosmic ray muon track, overlap, and dimuon validation 
are sufficient to detect these global, coherent movements of modules. This is illustrated in Fig.~\ref{fig:weakModes}, 
where for each of the nine misalignments a representative validation using one of the three techniques is shown.
In all cases, the five distributions 
are constructed using an MC simulation where the true positions of the modules
are known. The distributions correspond to $\epsilon=0$ (no misalignment) and two nonzero values of $\epsilon$, both positive and negative. Where appropriate, the mean ($\mu$) and root-mean-square (RMS) of the
distributions are also given in the figure legends. 
The results of this section are crucial when deriving the alignment parameters in $\pp$ collision data, as discussed in Section~\ref{sec:UL}.

The uniform misalignment of the tracker in the $z$~direction is known as $z$~expansion (or contraction).
In the BPIX, $z$~expansion can be detected using overlapping sensors in the same layer. 
This validation is not possible with the silicon strip modules because there is no precise measurement of the $z$~coordinate. 
We find that a change in $\epsilon$ causes a shift in the mean of the distributions for overlaps in the $z$~direction. 
The design of the silicon pixel detector does not provide a $z$~overlap for modules at the same azimuthal angle, but it is possible for modules that are near in~$\phi$. 
Figure~\ref{fig:weakModes} (top row, left) shows the distribution of differences of residuals in the overlapping modules in the $z$~direction with modules overlapping in the $z$~direction in the BPIX for cosmic muon events in MC simulation.
The $z$~expansion misalignment is tested with $\epsilon = -2.02\times10^{-4}$, $-1.01\times10^{-4}$, 0, $1.01\times10^{-4}$, {and} $2.02\times10^{-4}$.
Constraining the $z$~expansion in the strip detector is a more challenging task. 
For example, the global $z$~position of the silicon modules in the endcap detectors is weakly constrained.
Although the distribution of material has been studied extensively and is well described in Ref.~\cite{CMS:2018wqs}, certain biases in the track reconstruction may appear if inactive material is not fully included in the detector model. This may lead to distortions in the detector geometry appearing in the form of a $z$~expansion.

Radial expansion (or contraction) is the uniform misalignment of the tracker in the $r$~direction as a function of~$r$ ($r\to r + \epsilon r$).
Because of the uniform and symmetric nature of this misalignment, it is not easily detected with cosmic ray muon track splitting or
$\Ztomm$ decays. However, it is easily detected using the overlap validation, since in the case of a radial
expansion, modules that overlap in the radial direction will move apart uniformly. Therefore, the difference between the true
and the predicted hit locations in two overlapping modules is a good indicator of a radial expansion or contraction.
The linear relationship between the mean of the overlap validation figures and the magnitude of the radial misalignment
are used to categorize the presence of radial expansion or contraction in $\pp$~collision data.
Figure~\ref{fig:weakModes} (top row, middle) shows the distribution of overlaps in the $\phi$~direction for modules overlapping in the $\phi$~direction in the
BPIX for collision events in MC simulation.
The MC events are simulated with $\epsilon = 5\times10^{-4}$, $2.5\times10^{-4}$, 0, $-2.5\times10^{-4}$, and $-5\times10^{-4}$.

Twist is the misalignment of the tracker in the $\phi$~direction as a function of~$z$.
As such, twist shows up clearly in the $\Ztomm$ validation, and also in the overlap validation.
The parameter used is the slope of the invariant mass~$\mmumu$ vs. $\Delta\eta_{\Pgm\Pgm}$ distribution. It ranges from $\Delta\eta_{\Pgm\Pgm}=-2$ to $+2$, as the distribution becomes nonlinear for larger values of $\Delta\eta_{\Pgm\Pgm}$.
Figure~\ref{fig:weakModes} (top row, right) shows the profile of invariant mass $\mmumu$ vs. $\Delta\eta_{\Pgm\Pgm}$ for $\Ztomm$ events in MC simulation.
The MC events are simulated with $\epsilon = 2.04\times10^{-6}$,  $1.02\times10^{-6}$, 0,  $-1.02\times10^{-6}$, and $-2.04\times10^{-6}\cm^{-1}$.

The telescope effect is the uniform misalignment of the tracker in the $z$~direction as a function of $r$ $(z\to z + \epsilon r)$.
This creates concentric rings that are offset in the $z$~direction, and this misalignment can be visualized by imagining an actual telescope.
Because of its $z$ dependence, the telescope effect is identified primarily using the reconstruction of cosmic ray muon tracks.
Figure~\ref{fig:weakModes} (middle row, left) shows the distribution of $\Delta\theta/\sigma(\Delta\theta)$ for cosmics in MC simulation.
The MC events are simulated with $\epsilon = 5\times10^{-4}$, $2.5\times10^{-4}$, 0, $-2.5\times10^{-4}$, and $-5\times10^{-4}$.

Bowing is the misalignment of the tracker in the $r$~direction as a function of~$z$.
It is similar to the radial expansion, and differs only by the fact that the bowing effect is a function of $z$.
Figure~\ref{fig:weakModes} (middle row, middle) shows the distribution of overlaps in the $\phi$~direction with modules overlapping 
in the $\phi$~direction in the TOB for cosmic ray muon tracks in MC simulation.
The MC events are simulated with the ideal detector geometry and reconstructed using five geometries, corresponding to the bowing
misalignment with $\epsilon = 6.77\times10^{-9}$, $3.39\times10^{-9}$, 
0, $-3.39\times10^{-9}$, and $-6.77\times10^{-9}\cm^{-2}$.

Layer rotation is the misalignment of the tracker in the $\phi$~direction as a function of~$r$.
The outer layers twist with a different magnitude to that of the inner layers.
This distortion is easily picked up with cosmic ray muon track splitting, since we can see a change in track curvature between the two track halves.  
As such, we take the mean of a value proportional to the curvature for each value of~$\epsilon$. 
Figure~\ref{fig:weakModes} (middle row, right) shows the distribution of $\Delta(q/\pt)$ for cosmic events in MC simulation.
The MC events are simulated with 
$\epsilon=9.43\times10^{-6}$, $4.72\times10^{-6}$, 0, $-4.72\times10^{-6}$, and $-9.43\times10^{-6}\cm^{-1}$.

Skew is the misalignment of the tracker in the $z$~direction as a function of~$\phi$.
Because of the $\phi$~dependency, it can be detected with cosmic ray muon track splitting.
The distribution of $\Delta \dz$ vs. $\phi$ can be fit with a hyperbolic tangent function,
$A\times\tanh(B(\phi + C))$, from which we can extract~$\epsilon$.
Figure~\ref{fig:weakModes} (bottom row, left) shows the profile of $\Delta \dz/\sqrt{2}$ vs. $\phi$ for cosmic events in MC simulation.
The MC events are simulated with 
$\epsilon = 5.5\times10^{-2}$, $2.25\times10^{-2}$, 0, $-2.25\times10^{-2}$, and $-5.5\times10^{-2}\cm$.

Elliptical distortion is the uniform misalignment of the tracker in the $r$~direction as a function of~$\phi$
$(r\to r + r\epsilon\cos(2\phi + \delta))$. Because of its $\phi$ dependency, elliptical distortion is easily detected
with cosmic ray muon track splitting. This misalignment is especially clear in the modulation of the difference in the impact
parameter $\Delta \dxy$ as a function of the azimuthal angle of the track.
We fit a sinusoidal function to this modulation, $\Delta \dxy = -A\times\sin(2\phi+B)$, and find a linear relationship
between $A$ and~$\epsilon$.  Figure~\ref{fig:weakModes} (bottom row, middle) shows the profile of $\Delta \dxy/\sqrt{2}$ vs. $\phi$ 
for cosmic ray muon events in MC simulation. The MC events are simulated with 
$\epsilon = 5\times10^{-4}$, $2.5\times10^{-4}$, 0, $-2.5\times10^{-4}$, and $-5\times10^{-4}$.

Sagitta distortion is the uniform misalignment of the tracker in the $\phi$~direction as a function of $\phi$.
As with the elliptical misalignment, the $\phi$ dependence in the sagitta distortion means it can be detected with cosmic ray muon track splitting validation.
The effect of the misalignment can be seen in distributions of $\Delta\phi$ vs. $\phi$. The distributions of $\Delta\phi$ vs. $\phi$ are fit with a cosine function, $\Delta\phi=-A\cos(\phi+B)$, from which we can extract $\epsilon$.
Figure~\ref{fig:weakModes} (bottom row, right) shows the distribution of $\Delta\phi$ vs. $\phi$ for cosmic events in MC simulation.
The MC events are simulated with $\epsilon = 5\times10^{-4}$, $2.5\times10^{-4}$, 0, $-2.5\times10^{-4}$, and $-5\times10^{-4}$.

As the above studies show, various systematic distortions in the tracker geometry can be detected using combinations of 
different types of tracks and hits. Therefore, it is essential to combine all this information in the alignment procedure, which 
will be discussed in the next section. Balanced information in the input to the alignment procedure would ensure that 
such distortions are not present in the tracker geometry prepared for the reconstruction of tracks. 

\section{Alignment algorithms} \label{sec:software}
The CMS Collaboration uses two independent implementations of the track-based alignment, \MILLEPEDE-II and \HIPPY.
In the mathematical formulation presented in Section~\ref{sec:concepts}, \MILLEPEDE-II also performs a global matrix inversion, whereas \HIPPY neglects the blocks relating the alignment parameters to the track parameters and iterates to improve this approximation.
Furthermore, the two algorithms follow different strategies. They are developed, maintained, and used independently; thus facilitating independent cross-checks.
The earlier implementations of both algorithms, including techniques such as vertex and mass constraints, were described in Refs.~\cite{Adam:2009aa,Chatrchyan:2009sr,Chatrchyan:2014wfa}.
The algorithm that was used will be addressed in the following sections when investigating practical cases.
In this section, we outline the improvements in the two algorithms motivated by the needs of the alignment procedure of the CMS tracker during Run~2.

\subsection{\MILLEPEDE-II}
\label{millepede}
The \MILLEPEDE-II~algorithm~\cite{terascale-wiki,Blobel:2002ax,BLOBEL20065} has been discussed in the context of CMS in Ref.~\cite{Chatrchyan:2014wfa}.

It is still being developed further to meet the growing user needs; in addition to CMS, the Belle-II experiment~\cite{belle2align} is a main user driving the developments.
In this section, we review the main algorithms implemented in the software and describe recent improvements.
The \MILLEPEDE-II algorithm allows determination of the position, the orientation, and the curvature of the tracker modules.
The algorithm consists of two steps: 
\begin{description}
\item[\textsc{Mille}] 
This program has to be integrated into the track fitting software of the specific experiment. For each track the independent residuals with errors and the derivatives of the track (local) and module (global) parameters from Eq.~(\ref{eq:linearization}) have to be calculated and stored in custom binary files. The track fitting method has to fit all hits simultaneously, providing the complete covariance matrix of all track parameters. Although a solution based on the standard Kalman filter~\cite{Fruhwirth:1987fm} is also possible~\cite{Hulsbergen2009,Amoraal:2012qn}, only the general broken lines method~\cite{GBL,terascale-wiki-gbl} has been implemented for the track fit in \MILLEPEDE-II.
This is a refit of the trajectory defined by the track parameters from the Kalman filter at one given hit, \eg the first. The output to binary \textsc{mille} files contains the subset of the trajectory attributes that are needed by \MILLEPEDE-II. Only the global derivatives have to be added. 
\item[\textsc{Pede}]
This is an experiment-independent Fortran program that builds and solves the linear equation system from Eq.~(\ref{eq:matrixInversion}).
It reads a text file with steering information and the tracks with the hit information from the \textsc{mille} binary files to perform the local (track) fits to
construct the global matrix $\mathbf{C'}$. This symmetric matrix is stored in full (lower triangular part) or sparse (only nonzero parts) mode. 
Several solution methods are implemented. An overview is given in Table~\ref{tab:MPalgos}.
\end{description}

\begin{table*}[!ht]
    \centering
    \topcaption{
        List of the main solution methods implemented in \MILLEPEDE-II. The computation time is given as a function of the number of parameters $n$ and the number of internal iterations $n_\text{it}$ if applicable. The type of solution delivered by the algorithm is also shown.
    }
    \begin{tabular}{l l l c}
        Method                                       & Computing time                & Solution type & Error calculation \\
        \hline                                                                                       
        Inversion (Gauss--Jordan)                        & ${\sim}n^3 $                   & Exact         & Yes               \\
        Cholesky decomposition                          & ${\sim}n^3 $                   & Exact         & Skipped (for speed)    \\
        \textsc{MinRes}~\cite{choi2011minres,Choi2013ALGORITHMD} & ${\sim}n^2 \times n_\text{it}$ & Approximate   & No                \\
    \end{tabular}

    \label{tab:MPalgos}
\end{table*}

Compared with the version used previously by CMS~\cite{Chatrchyan:2014wfa}, the most important technical improvements used for the alignment fits described in this paper are:
\begin{enumerate}
    \item The migration from Fortran~77 to Fortran~90 allowing for dynamic memory management.
    \item The implementation of the solution of problems with constraints by elimination, in addition to Lagrange multipliers. Especially for large problems where an approximate solution is obtained~\cite{choi2011minres}, elimination shows superior numerical performance. 
    \item The analysis of the input (global parameters and constraints) for optional factorization of a large problem into smaller ones using block matrix algebra.
    \item Alignable objects are, in general, described by several global parameters. These global parameters appear together in the binary files, and are now split into groups by relying on the adjacent global (user-defined) labels with which the parameters appear. This means the global matrix is organized as a collection of block matrices instead of a collection of single values. The size corresponds to the two contributing parameter groups. By arranging the matrix in this way, operations on the global matrix are sped up using the caching and vectorization options available on modern processors. This helps especially in the case of sparse storage, since typically 10--30\% of the elements of the global matrix are nonzero.
\end{enumerate}
In Table~\ref{table:PedeWallTime}, we illustrate the amount of running time used by \textsc{Pede} to solve the linear equation system from Eq.~(\ref{eq:matrixInversion}); a record typically corresponds to a track or to a pair of tracks coming from a resonance decay.
These results were obtained with a test machine at DESY; in practice, for the alignment fits presented in this paper, similar machines at CERN were used.

\begin{table*}[!ht]
    \centering
    \topcaption{Examples of \textsc{Pede} wall time (time taken from start of the program to end) for some larger alignment campaigns using \textsc{MinRes} on a dedicated test machine 
        (Intel Xeon E5-2667 @ 3.2\unit{GHz}, 256\unit{GB} memory @ 51\unit{GB/s}).}
    \begin{tabular}{c c c c c}
        Number of & Number of  & Number of  & Matrix size [GB] & Wall time [s] \\
        global parameters & constraints & records & (sparse) & (10 threads) \\
        \hline
        217500 & 138 & $4.46\times 10^7$ & 44 & $8.4\times 10^3$ \\
        213900 & 1782 & $2.90\times 10^7$ & 85 & $6.8\times 10^3$ \\
        576000 & 942 & $5.20\times 10^7$ & 218 & $4.4\times 10^4$ \\
    \end{tabular}
    \label{table:PedeWallTime}
\end{table*}

No numerical problems have been observed. 
Either rank deficits are detected, or the matrix is inverted correctly.

\subsection{\HIPPY} \label{HipPy}
The \HIPPY algorithm is based on the hits-and-impact-points algorithm~\cite{Karimaki:2003bd,HIP} with additional
features introduced using the constraints developed for the BaBar track-based alignment~\cite{Brown:2008ccb}. 
It has been used extensively during commissioning of the CMS tracker~\cite{Adam:2009aa} and during the CMS start-up period 
in Run~1~\cite{Chatrchyan:2009sr,Chatrchyan:2014wfa}. Further improvements were introduced during Run~2,
as described below. The improved algorithm is now named hits-and-impact-points-past-year-1 (\HIPPY). 

The main distinguishing feature of the \HIPPY algorithm, compared with \MILLEPEDE-II, is its local nature. The position 
and orientation of each sensor are determined independently of the other sensors. This approach has advantages and disadvantages
compared to \MILLEPEDE-II.
One disadvantage is that multiple iterations of running the algorithm are required to solve correlations
between the sensor parameters. The number of iterations can be several dozen up to a hundred.
This means multiple runs of the CPU-expensive track fits are needed, which limits the practical application of
this algorithm. Advantages include the native integration with CMS software, immediately providing features such as the CMS Kalman filter code for track propagation without additional development.
As a result, any constraint, such as 
mass or vertex constraints, implemented in the CMS software can be incorporated in the algorithm. Each iteration of the algorithm is a very 
simple application of a small matrix inversion. This simplicity and dependence on the CMS software makes the \HIPPY 
algorithm complementary to \MILLEPEDE-II. 

Since our previous publications, the most important technical improvements in the \HIPPY alignment fit algorithm are:
\begin{enumerate}
    \item The inclusion of alignment parameters beyond the three position and three orientation coordinates of each sensor,
    namely the curvature of the sensors.     
    \item The possibility to apply a weight to certain types of input to balance statistical and systematic uncertainties. 
    \item The option to perform sequential, hierarchical alignment over multiple time periods, when the time stability of the structures differs among the hierarchical levels. 
    \item The inclusion of possible mass and/or vertex constraints in certain types of events with known physics process. 
\end{enumerate}

\begin{figure*}[htbp]
	\centering
    \includegraphics[width=0.75\textwidth]{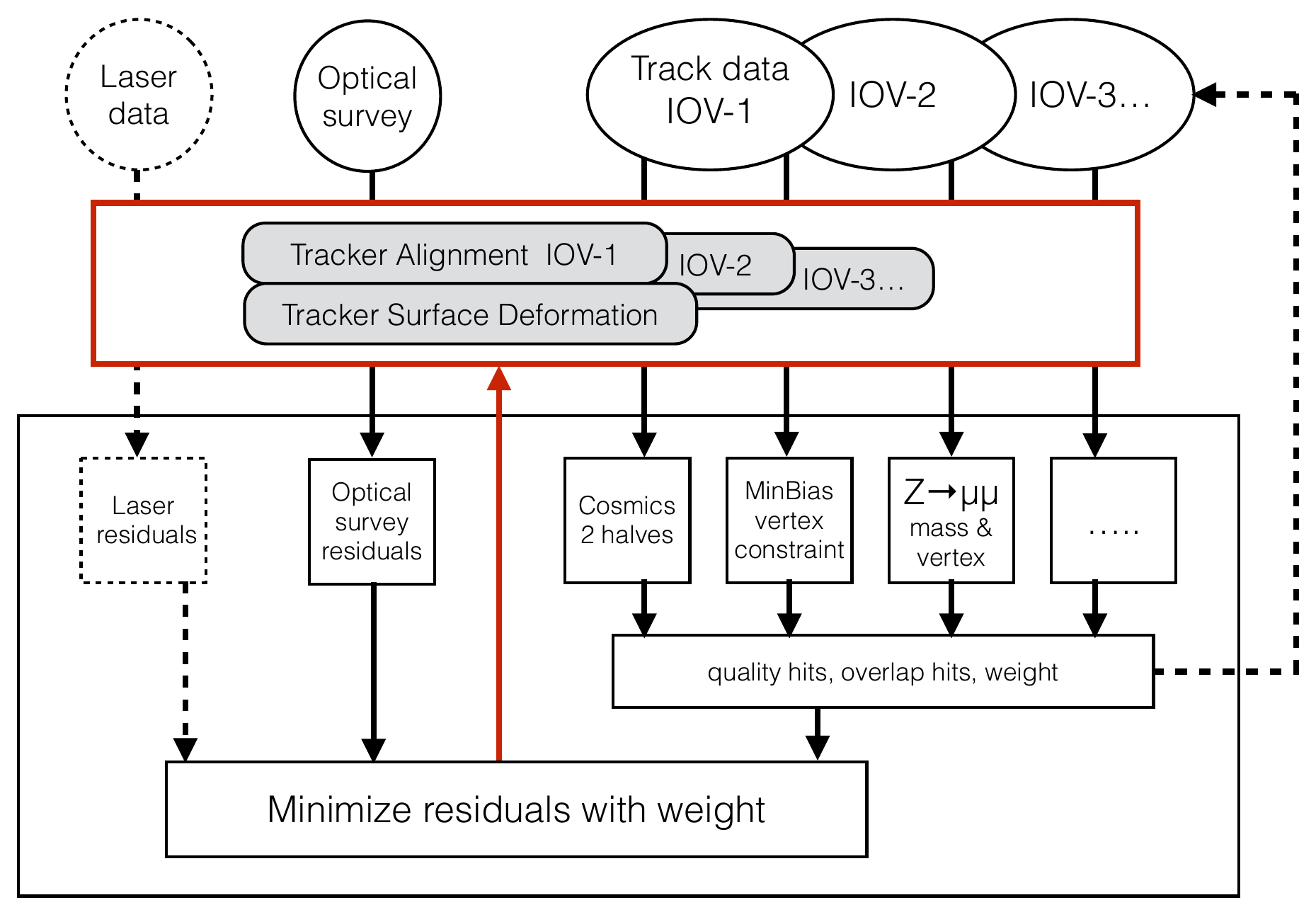}
		\caption{Diagram of the \HIPPY algorithm design with the
		sequence for event, track, and hit selection, including application of the weight factors and constraints. 
		Not all features were used in the Run~2 alignment procedure, as described in the text.
		The algorithm operates in iterative mode, indicated with the arrows. 
		}
	\label{fig:HipPy_chart}
\end{figure*}

The diagram in Fig.~\ref{fig:HipPy_chart} shows the design of the \HIPPY algorithm with
the sequence for the event, track, and hit selection, including the application of the weight factors and constraints.
The arrows indicate the flow of information and the dashed arrows indicate features that are not used in this work.
The track data are categorized in several paths (vertical arrows pointing down) and the corresponding 
constraints are applied in each event during the track fit with a given set of alignment conditions
(indicated by the red box). During the minimization procedure, different weights are assigned to different 
types of input, and a new set of alignment conditions is passed back to the track fitting procedure 
(vertical arrow pointing up). The process is repeated until convergence is reached. 
The inclusion of the laser calibration data~\cite{Adam:2009aa,Chatrchyan:2009sr} was developed during Run~1 but
the laser system was not supported anymore in Run~2. 
The optical survey data constraint was used during the Run~1 start-up~\cite{Chatrchyan:2009sr} and was used as
a constraint to a prior geometry during the Run~2 start-up.  The main new features of the \HIPPY algorithm discussed above are
indicated in the diagram. These are the multi-IOV reconstructed track data, the sensor surface deformation database object
in addition to the sensor position, the division of input track data into categories with the corresponding constraints,
and the application of weights to certain types of input in the minimization process.
Further details can be found in the description of Fig.~10 in Ref.~\cite{Brown:2008ccb}.

\section{Alignment during data taking} \label{sec:datataking}
The changes observed in the tracker require it to be realigned several times during the year.
In this section, we focus on the two typical cases:
\begin{enumerate}
    \item   Restarting the detector with limited statistical power after a technical shutdown; we discuss the beginning of Run~2 (Section~\ref{sec:2015}) and the Phase-1 upgrade~\cite{Schroder:2019fbv} (Section~\ref{sec:2017}).
    \item   Running an automated, unsupervised alignment during data taking, with limited degrees of freedom, illustrated for the period 2016--2018 (Section~\ref{sec:PCL}).
\end{enumerate}

\subsection{Start-up alignments} \label{sec:startup}
The general strategy during the start-up periods has been to run a series of alignment fits, where each time the starting point for the alignment fit is the set of alignment constants obtained in the previous fit.
In case of large misalignments in the starting geometry, the linearization approximation made in the alignment algorithms is not valid anymore, and several iterations are required to achieve convergence.
To further simplify the alignment problem, we usually only gradually increase the number of degrees of freedom of the fit. This means the high-level mechanical structures are aligned first and the individual modules are aligned in a later iteration.
In addition to the convergence aspects, this strategy also allows for early alignments with a small number of recorded tracks that do not sufficiently cover all modules, and would therefore be insufficient for a full module-level alignment.
Thus, a continuous improvement of the alignment precision can be achieved during the start-up period as data are collected.
In view of the potentially large misalignments to be expected during the start-up period, a particular effort has been made to run the \HIPPY and \MILLEPEDE-II alignment algorithms independently to cross-check each other.
In some cases, the two algorithms have also been run successively to refine the alignment.
For a specific alignment task, the performances of the various algorithmic results were compared and were typically very similar, increasing confidence in the results.
The alignment with the best performance was selected, and is presented in the following.

The first alignment constants after the extended shutdowns during Run~2 were derived ahead of the $\pp$ collision run, using exclusively cosmic ray muon tracks recorded by CMS during commissioning runs.
This corrected for the large misalignments that are often observed after the extraction and subsequent reinstallation of the pixel detector.
To retain tracking efficiency in view of potentially large misalignments, large alignment parameter uncertainties (APUs) were assumed for the track reconstruction and fit. This is affordable only at very low event rates, which is the case during cosmic data taking.
A further refinement of the alignments derived with only cosmic ray muon tracks was performed after recording a sufficiently large sample of collision tracks to be able to derive the alignment corrections with a higher granularity.    
In the following, the alignment strategies and their performance with early 2015 and 2017 data are discussed.

\subsubsection{Start-up of Run~2} 
\label{sec:2015}
During the shutdown period between Runs~1 and~2 of the LHC, extensive maintenance and repair work was performed on the tracking detector.
The pixel tracker was removed to replace the beam pipe.
The pixel detector half barrel on the $+x$ side was repaired, including replacements of several modules.
During reinstallation, the barrel pixel detector was centred around the beam pipe by displacing it upward by 3.4\mm and horizontally by 1.3\mm relative to its previous position.
The strip detector was kept in place, but the cooling system was partially replaced.

\begin{figure*}[!htb]
  \centering
  \begin{tabular}{cc}
      \includegraphics[width=0.48\textwidth]{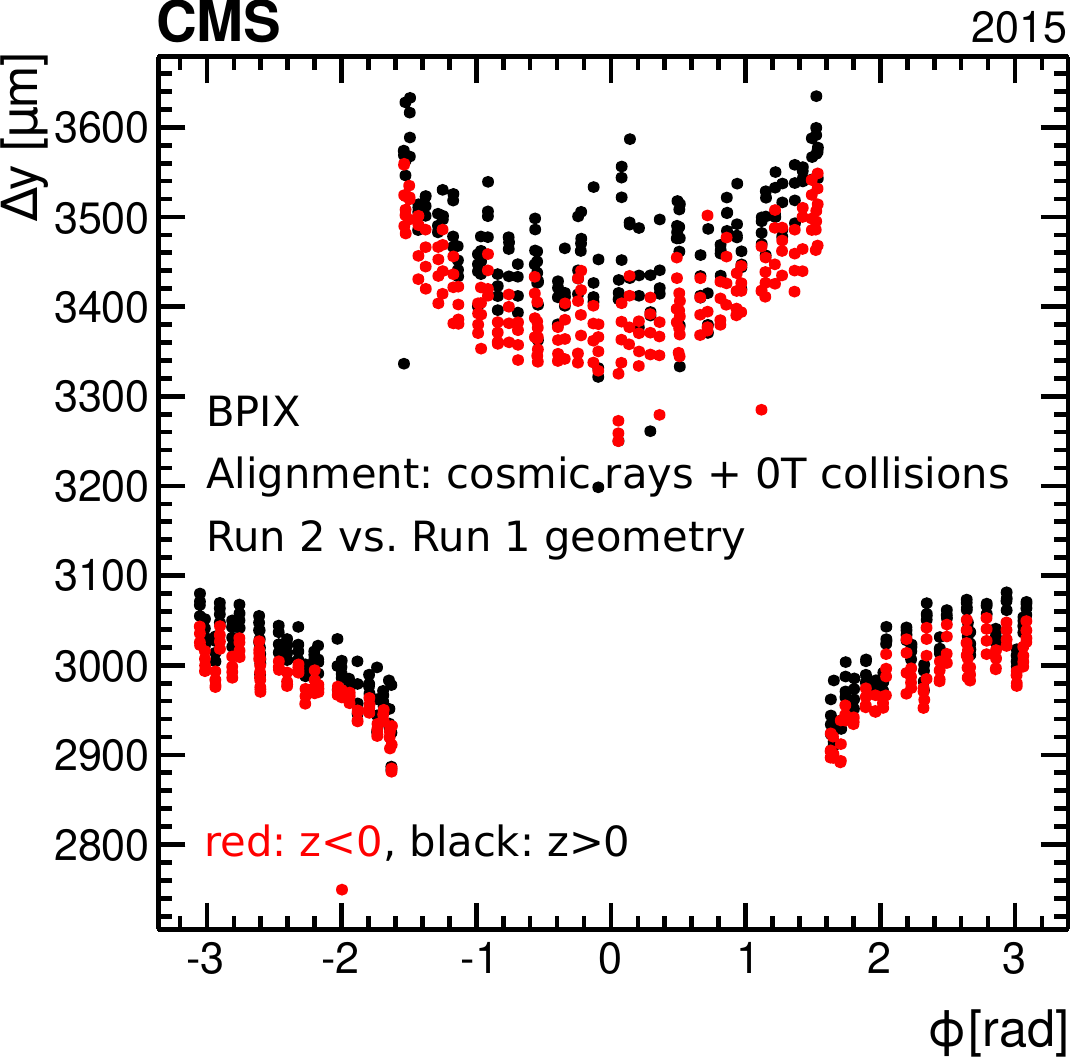}
      \includegraphics[width=0.475\textwidth]{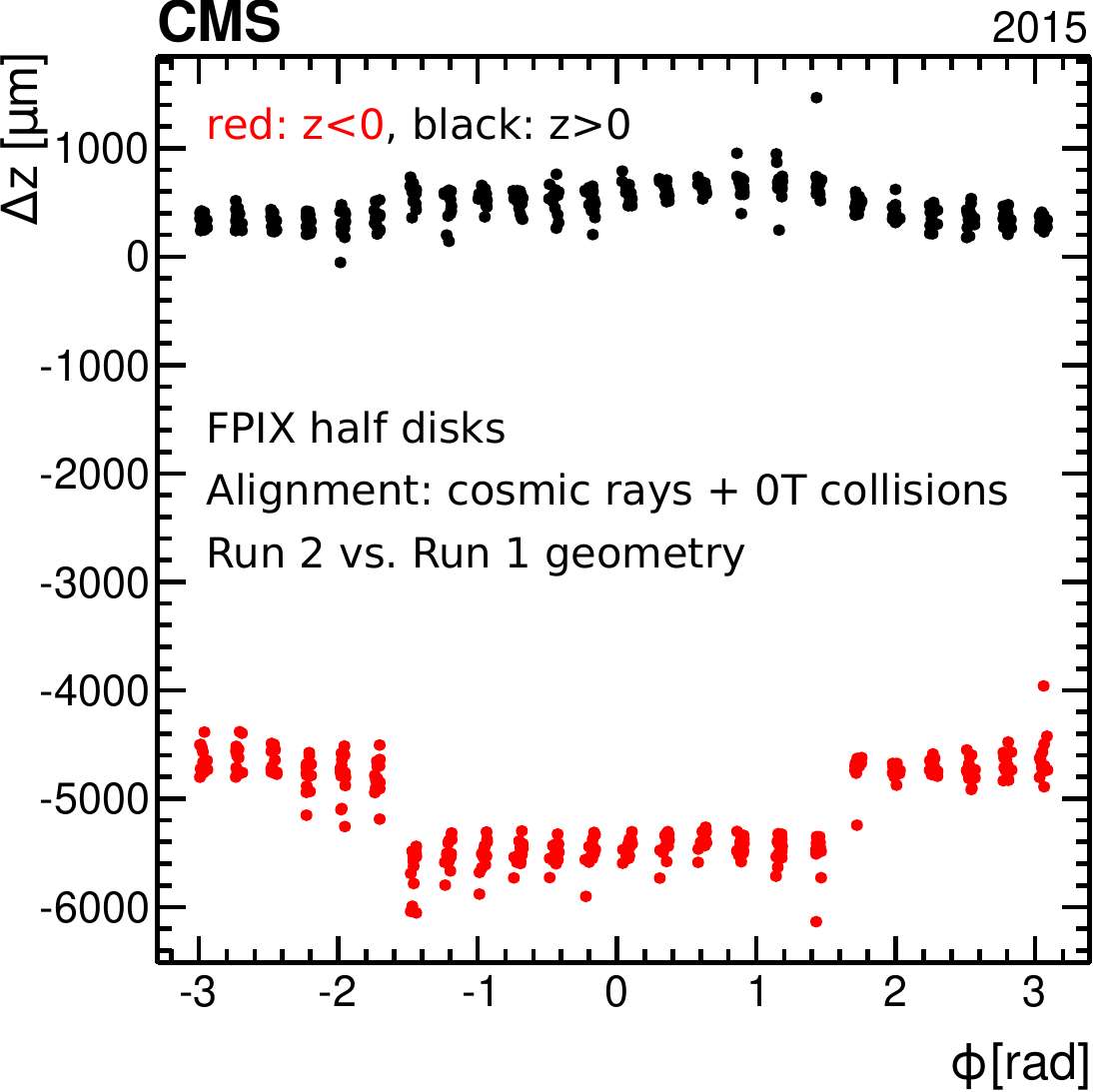}
  \end{tabular}
  \caption{
    Comparison of the positions of the modules in the first IOV of Run~2 and the last IOV of Run~1 in the BPIX (left) and FPIX (right) detectors, determined using cosmic data collected with 0\unit{T} and 3.8\unit{T} magnetic field in the solenoid and collision data at 0\unit{T}~\cite{CMS-DP-2015-029}.
    The differences in position $\Delta y$ (Run~2 $-$ Run~1) and $\Delta z$ (Run~2 $-$ Run~1) of the sensor modules are shown as a function of $\phi$ in global coordinates.
    Modules on the $-z$ side are shown in red, modules on the $+z$ side are shown in black.
  }
  \label{fig:Run2startup:gc}
\end{figure*}

Initially, the latest available geometry of the Run~1 detector was used to reconstruct the data. 
An observed asymmetry in the track rate in the forward pixel endcaps quickly invalidated the assumed alignment constants and provided hints of a large initial misalignment of the FPIX.
This was confirmed by the first alignment constants derived with the \MILLEPEDE-II algorithm using 0\unit{T} cosmic data to determine the relative positions and rotations of the high-level mechanical structures of the pixel and strip detectors.   
The several million cosmic ray muon tracks collected at 3.8\unit{T} were used to further improve the alignment configuration up to the level of single modules, by performing several successive alignment fits with an increasing number of degrees of freedom using the \MILLEPEDE-II and \HIPPY algorithms in sequence.
Figure~\ref{fig:Run2startup:gc} shows the shifts, relative to the Run~1 geometry, of the pixel module positions measured after the module-level alignment fit.
The BPIX as a whole was subject to a 3--3.5\mm shift, attributed to the aforementioned recentring procedure of the BPIX around the beam pipe during the reinstallation.
In addition, a relative shift between the two half barrel structures $-\pi/2<\phi<\pi/2$ and $\phi<-\pi/2$, $\phi>\pi/2$ was visible, which is attributed to the extensive repair and replacement work in the $-\pi/2<\phi<\pi/2$ half barrel.
The FPIX half disks on the $-z$ side were displaced by $-4.5\mm$ ($\phi<-\pi/2$, $\phi>\pi/2$) and $-5.5\mm$ ($-\pi/2<\phi<\pi/2$) compared with the Run 1 position.
Much smaller relative movements of up to 200\mum were observed between the half disks on the $+z$ side.

\begin{figure*}[!htb]
  \centering
  \begin{tabular}{cc}
    \includegraphics[width=0.48\textwidth]{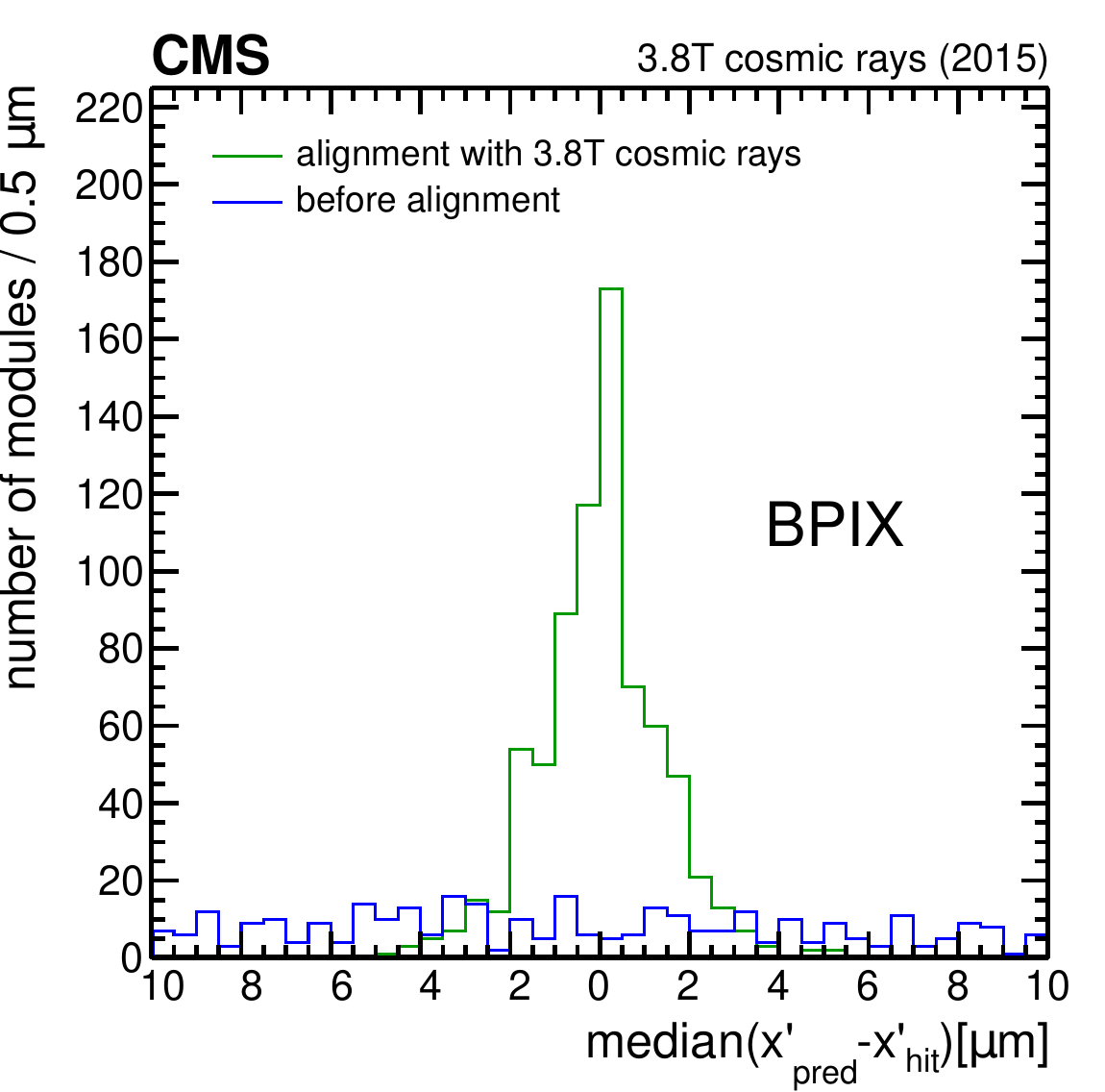} &
    \includegraphics[width=0.48\textwidth]{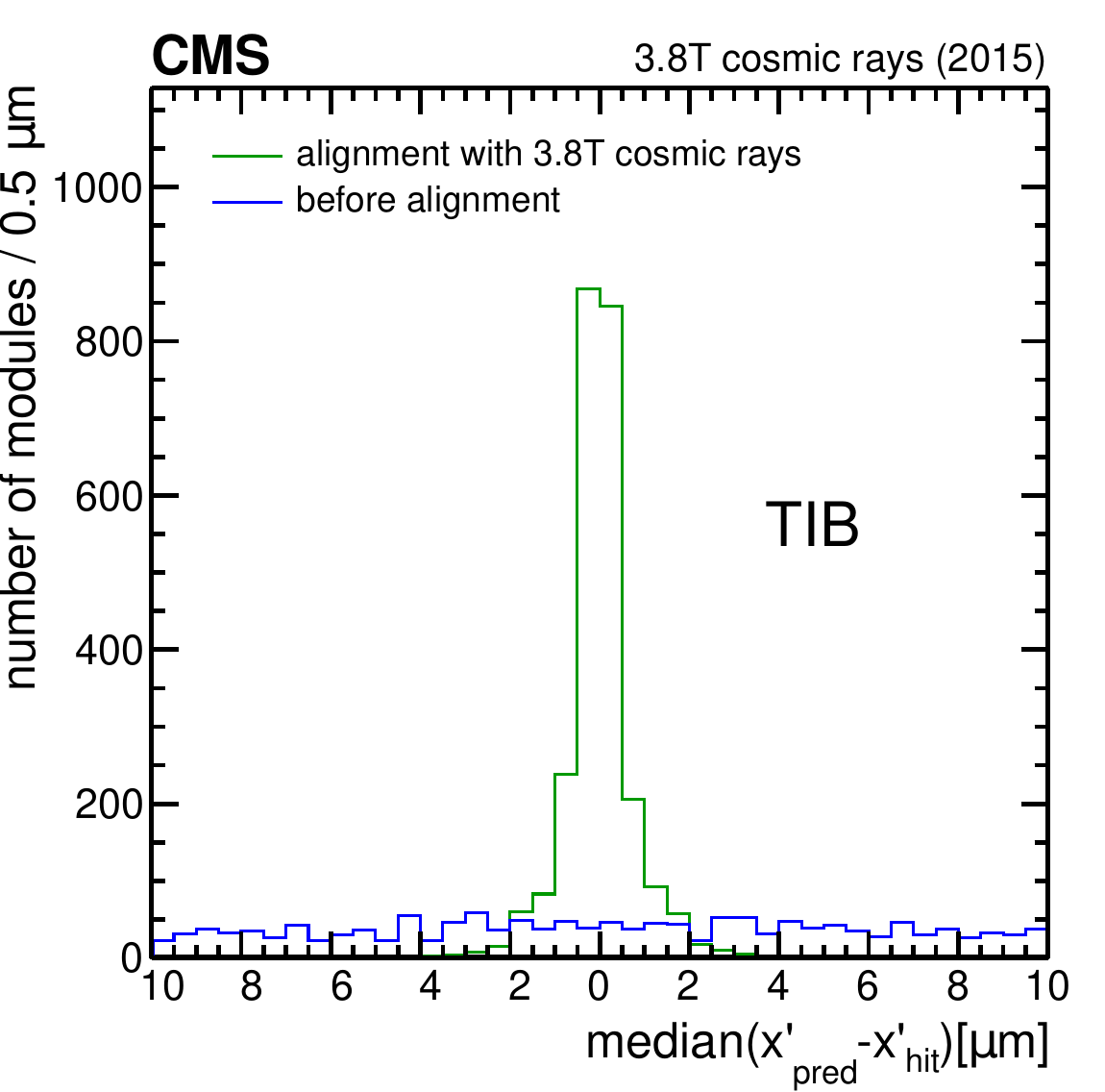} \\
    \includegraphics[width=0.48\textwidth]{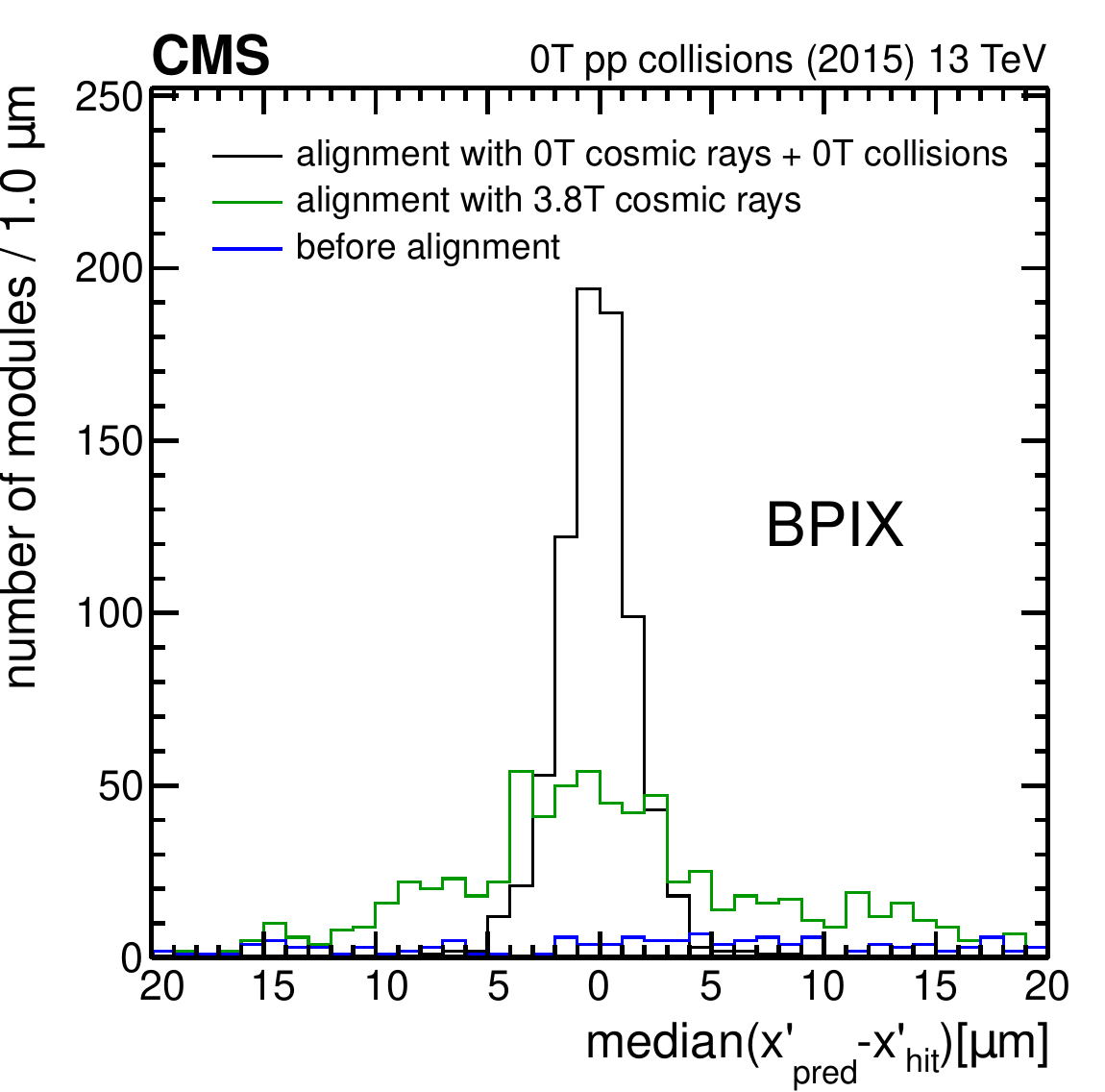} &
    \includegraphics[width=0.48\textwidth]{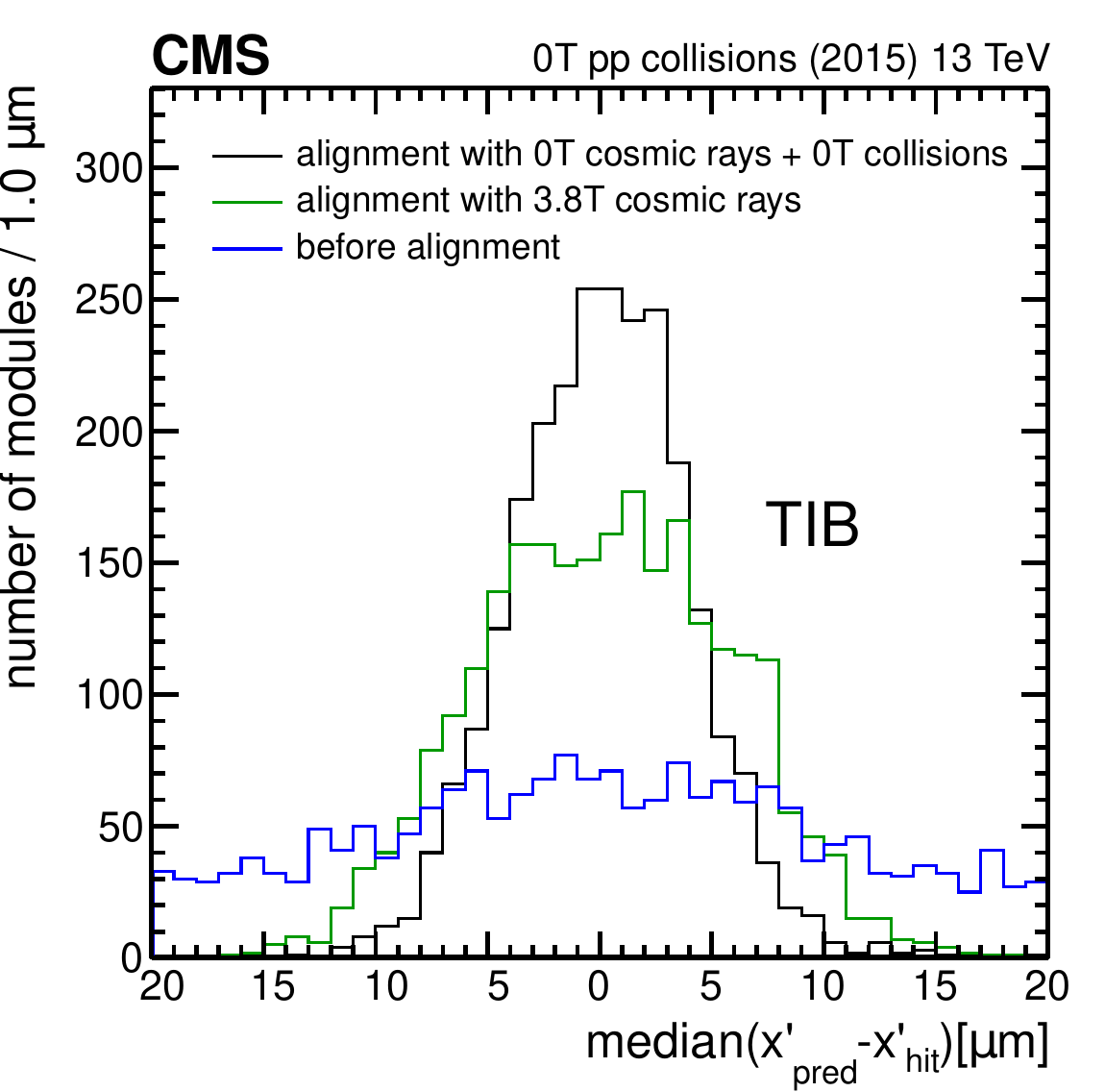} 
  \end{tabular}
  \caption{
    DMRs for the local~$x$ coordinate in the BPIX (left) and TIB (right), evaluated using $2\times 10^6$ cosmic ray muon tracks collected at 3.8\unit{T} (top) and $1.8\times10^6$ tracks from collision data at 0\unit{T} (bottom).
    The alignment constants used to fit the tracks were determined successively from cosmic data at 3.8\unit{T} (green line) and, after the magnet was switched off, from 0\unit{T} cosmics and collision data (black line) as described in the text.
    Because of the detector movements caused by the change in the magnetic field, the alignment constants derived with 3.8\unit{T} data (green line) are not optimal for the track fits in the 0\unit{T} data (bottom row).
    The blue line shows the DMR computed assuming the Run~1 geometry, which is no longer valid for Run~2 data. 
    }
  \label{fig:Run2startup:dmr}
\end{figure*}

An example of the early alignment performance is depicted in Fig.~\ref{fig:Run2startup:dmr}, showing the distributions of median track-hit residuals (DMRs) per module for BPIX and TIB modules. 
To avoid biasing the measurement, the track is refitted removing the hit under consideration. 
The DMRs represent a measurement of the local alignment precision based on data. 
In the case of perfect alignment, the distribution is expected to be centred at zero. 
The width of the DMR is influenced by the residual random misalignment of the modules, as well as by an intrinsic statistical component due to the number of tracks used to construct the distribution itself. 
For this reason, the DMRs for two different sets of alignment constants can be compared only if they are obtained with the same number of tracks. 

The local alignment precision of the pixel and the strip modules achieved in Run~2 with the fit using the cosmic data is compared to the one obtained for the Run~1 geometry (Fig.~\ref{fig:Run2startup:dmr}, top row). The latter is not expected to describe the detector well because of movements during the shutdown period, but was the best available geometry description before realignment.
Although the alignment corrections for the strip detector were smaller than for the pixel detector, the misalignment of the pixel tracker had a noticeable effect on the performance of the strip detector by reducing the accuracy of the tracks.

\begin{figure*}[!htb]
  \centering
  \begin{tabular}{cc}
    \includegraphics[width=0.48\textwidth]{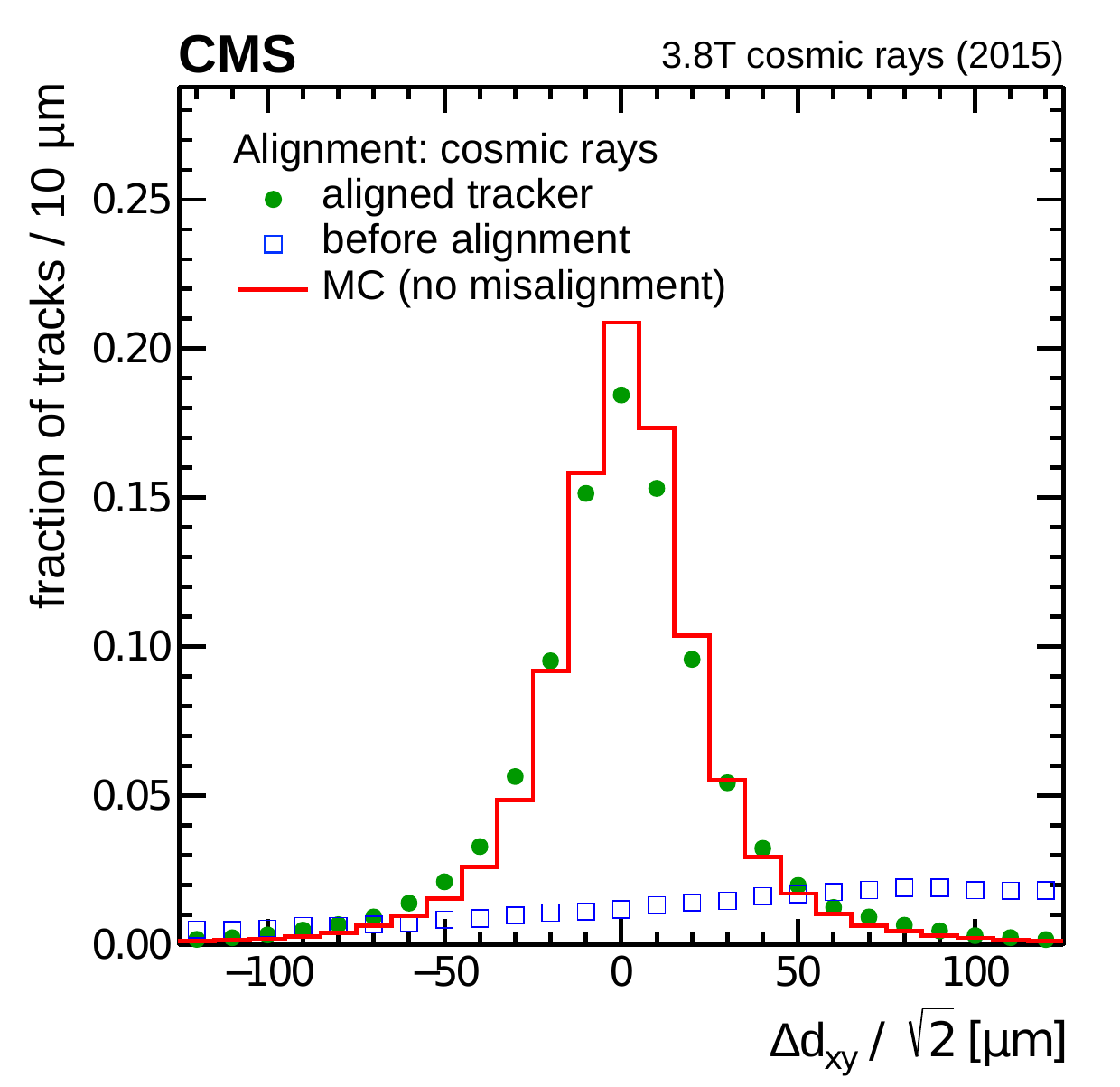} &
    \includegraphics[width=0.48\textwidth]{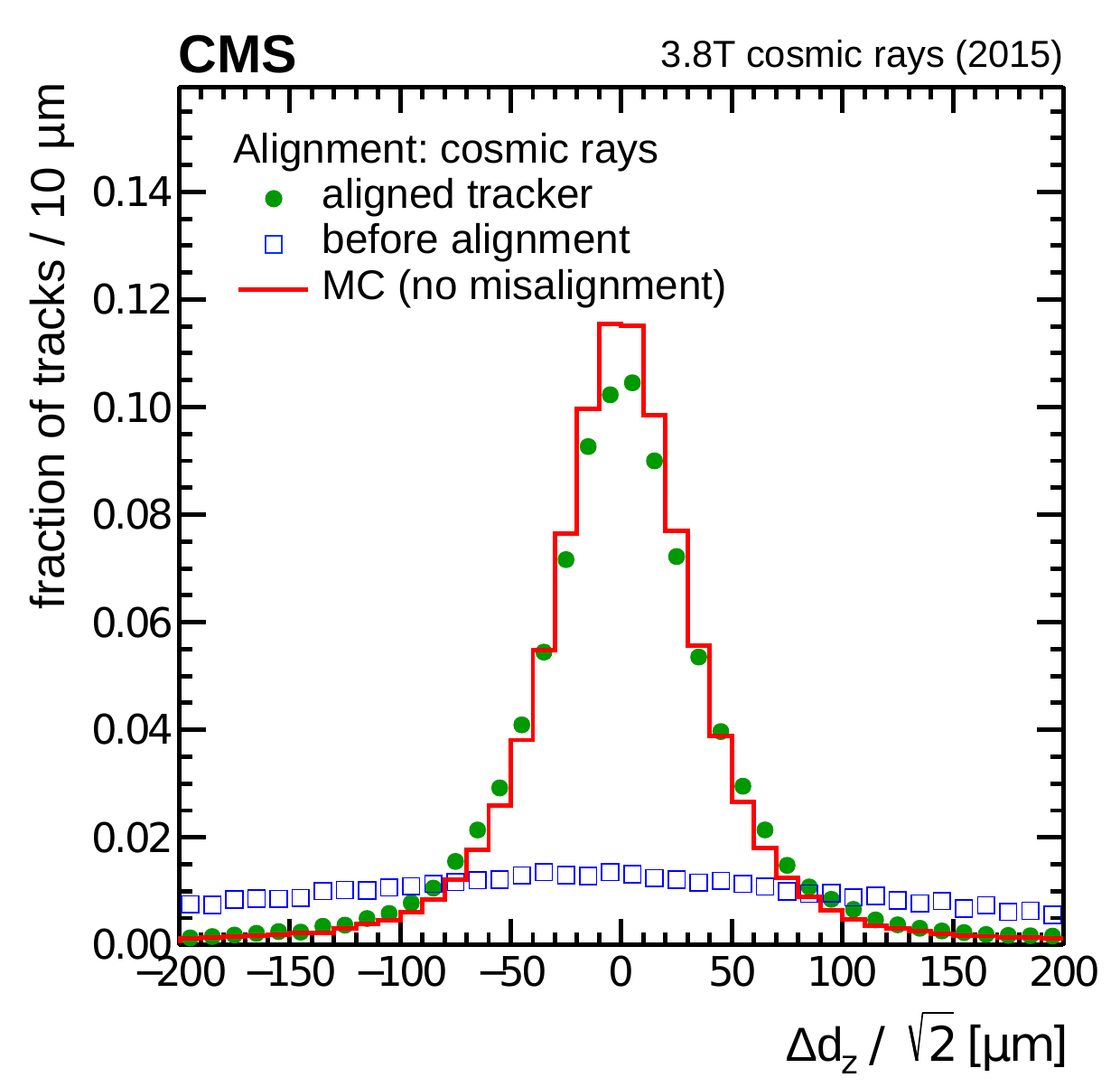}
  \end{tabular}
  \caption{
    Distribution of the difference between two halves of a cosmic ray muon track, scaled by a factor $\sqrt{2}$ to account for the two independent measurements.
    The track is split at the point of closest approach to the interaction region, in the $x$-$y$ (left) and $z$ (right) distance between the track and the origin.
    The tracks are fit using the alignment constants determined with cosmics at 0\unit{T} and 3.8\unit{T} (green circles) and using the Run~1 geometry (blue squares), which is no longer valid for Run~2 data.
    Vertical error bars represent the statistical uncertainty due to the limited number of tracks; they are smaller than the marker size.
    For comparison, the case of perfect alignment and calibration obtained from simulated events is shown (red line)~\cite{CMS-DP-2015-029}.
  }
  \label{fig:Run2startup:tracksplit}
\end{figure*}

The track performance is also measured using the cosmic ray muon track split validation, as described in Section~\ref{sec:WM}.
The differences between the transverse and longitudinal track impact parameters of the two track halves are shown in Fig.~\ref{fig:Run2startup:tracksplit}.
The observed precision using the aligned geometry comes close to that expected from the simulation for the case of perfect alignment. 

Because of problems with the cooling system, the superconducting CMS magnet was switched off shortly before the first collisions.  
As a consequence, the tracker geometry changed between the CRAFT data and the first collisions, as seen by a widening of the DMR evaluated with the 0\unit{T} collision data when the alignment constants obtained from the 3.8\unit{T} cosmic ray data are used to fit the tracks (Fig.~\ref{fig:Run2startup:dmr}, bottom row, green line).
Because of the way the mechanical structures are mounted, a change of the magnetic field generally moves all of the high-level structures of the pixel detector, particularly in the global $z$~direction. The relative positions of the individual modules remain mostly unchanged.
As expected, these effects are mainly apparent in the pixel detector, which is installed on wheels and has some freedom to move relative to the rest of the detector, whereas the strip tracker position is relatively stable against magnetic field changes.

Starting from the alignment constants obtained using the 3.8\unit{T} cosmic data that produced the green line in Fig.~\ref{fig:Run2startup:dmr}, a new set of alignment constants was derived with the \HIPPY algorithm using the first 0\unit{T} collision data and cosmic data taken between collision runs. 
The pixel detector was aligned at the level of single modules using a large sample of collision tracks, whereas for the strip detector only the positions of the high-level structures were updated. 
A module-level alignment fit of the full detector with only 0\unit{T} collision data would have been vulnerable to WMs, \eg to radial expansions of the entire detector, given the radial symmetry of the tracks.
The fixed position of the strip modules relative to their high-level structures provided a reference system.
Furthermore, the addition of cosmic ray muon tracks, with their different topology, helped to avoid WMs.

The new alignment procedure improves the tracker performance, which is apparent from the narrower DMR seen in Fig.~\ref{fig:Run2startup:dmr} (bottom row, black line).
The alignment precision is limited by the poorer track resolution at 0\unit{T}, which is caused by the poor description of multiple-scattering effects at 0\unit{T}, because the track momentum cannot be measured in the absence of a magnetic field.
A track momentum of 5\GeV is assumed in the reconstruction at 0\unit{T}, motivated by the typical track momentum in the cosmic ray data.

\begin{figure*}[!htb]
  \centering
  \begin{tabular}{cc}
      \includegraphics[width=0.48\textwidth]{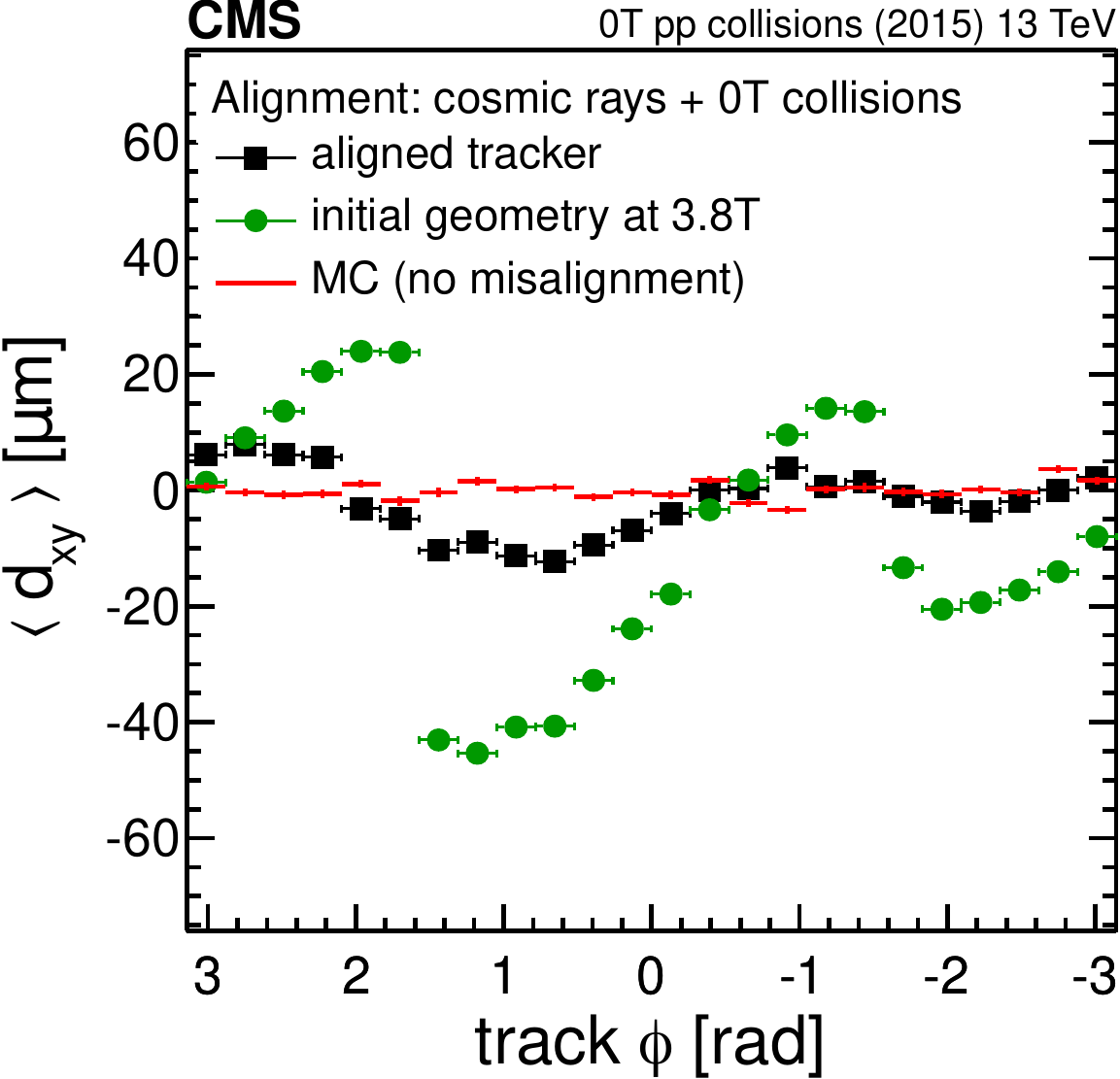} &
      \includegraphics[width=0.48\textwidth]{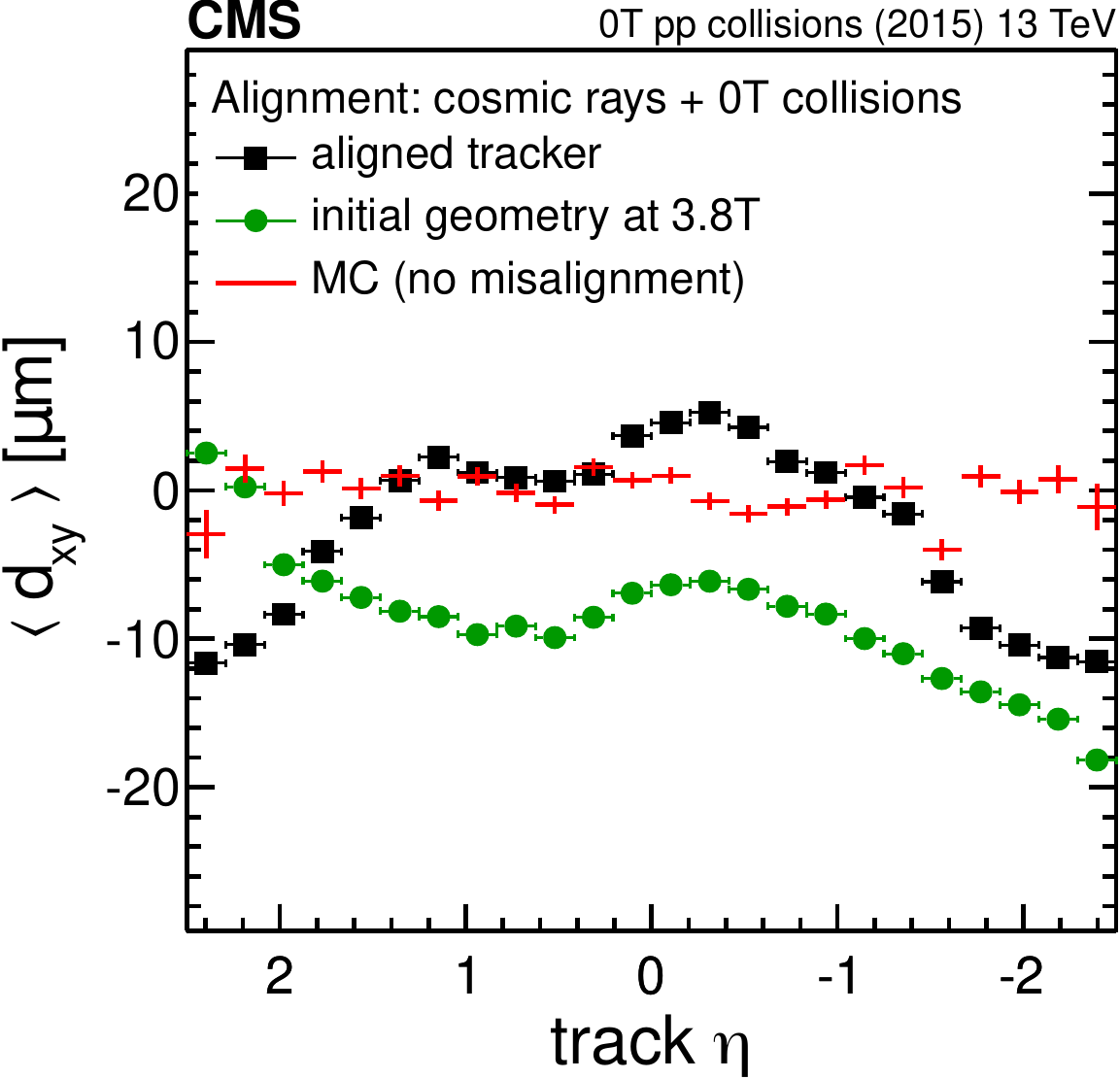}
  \end{tabular}
  \caption{
    Mean distance in the transverse plane of the track at its closest approach to a refit unbiased PV as a function of the track $\phi$ (left) and $\eta$ (right), measured in approximately $5.5\times 10^6$  collision events collected at 0\unit{T} magnetic field.
    Two different alignments are used to fit the tracks:
    the alignment constants obtained during the commissioning phase with cosmic ray muon tracks at 3.8\unit{T} prior to collision data taking (green circles)
    and the alignment constants determined subsequently with 0\unit{T} collision data (black squares).
    For comparison, the case of perfect alignment and calibration obtained from simulated data is shown (red line).
    Vertical error bars represent the statistical uncertainty due to the limited number of tracks; for the data, they are smaller than the size of the markers~\cite{CMS-DP-2015-029}.
  }
  \label{fig:Run2startup:pv}
\end{figure*}

The alignment quality is also measured by its effect on the physics object performance.
In particular, the reconstruction of the primary vertex~(PV), \eg the vertex belonging to the track with the highest \pt, is driven by the pixel detector since it is the detector closest to the interaction point and has the best intrinsic hit position resolution.
The unbiased track-vertex residuals (track impact parameters) provide a measurement of the vertex reconstruction performance based on data.
For each track, the PV position is reconstructed, excluding the track under scrutiny.
A deterministic annealing clustering algorithm is used to make the method robust against pileup~\cite{Rose:1998dzq,Chatrchyan:2014fea}.
The mean values of the distributions of the unbiased track-vertex residuals are shown in Fig.~\ref{fig:Run2startup:pv}.
In the case of perfect alignment and calibration, mean values of zero are expected.
Random misalignments of the modules affect only the resolution of the unbiased track-vertex residuals, increasing the width of the distributions without biasing their mean.
Systematic misalignments of the modules, however, bias the distributions in a way that depends on the nature and size of the misalignment.
The structures of the green curve, which is obtained when fitting the tracks with the alignment constants derived during the commissioning phase with cosmic data at 3.8\unit{T}, indicate relative movements of the half barrels of the pixel detector during the decrease of the magnetic field.
A clear improvement of the vertex performance is observed for the subsequent alignment with 0\unit{T} collision data (black curve).
Residual biases are attributed to suboptimal coverage of modules due to the limited number of tracks and their incidence angle, in addition to the aforementioned WMs and generally poorer track resolution at 0\unit{T}.

The tracker geometry changed again when the magnetic field was turned back on.
Immediately after the first 3.8\unit{T} collision data were taken, a fast alignment fit was performed with a limited number of tracks, updating only the positions of the pixel tracker high-level structures.
After collecting a larger data set, a full module-level alignment fit of the entire tracking detector was performed. This also included the available updates of the hit position calibration and further improved the tracking performance.

\subsubsection{Phase-1 upgrade} \label{sec:2017}
During the extended YETS starting at the end of 2016, the innermost component of the silicon tracker was replaced with a new upgraded pixel detector~\cite{Adam_2021}.
The first data with the new pixel detector were recorded in spring 2017, prior to the restart of the LHC $\pp$ run. 
After the first period of detector commissioning to derive the initial calibrations, CRUZET and CRAFT data were collected for alignment purposes. 
In general, the alignment fits presented in this subsection were performed with \MILLEPEDE-II. As a linearization point, the best alignment of the strip detector in 2016 was used, and the pixel detector was assumed to correspond to an ideal tracker.

\begin{figure*}[htbp]
	\centering
        \includegraphics[width=0.45\textwidth]{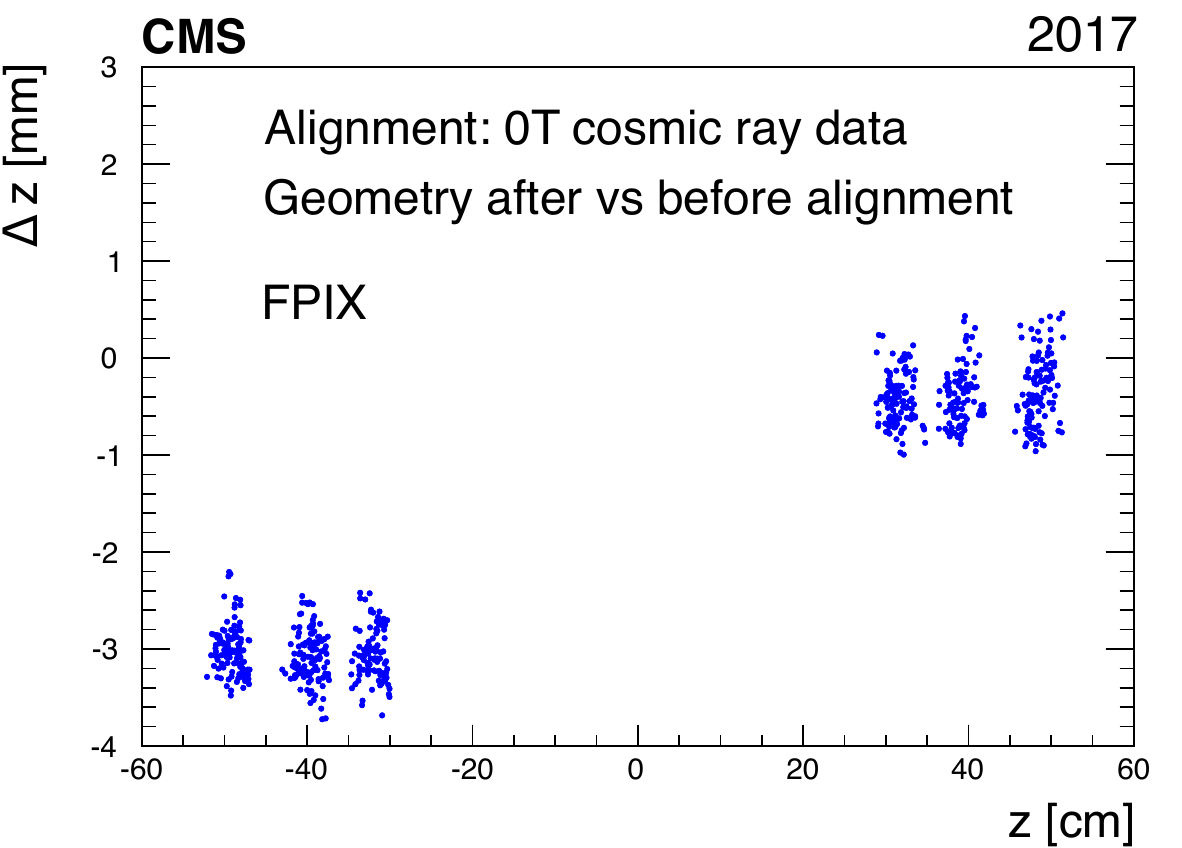}
        \includegraphics[width=0.45\textwidth]{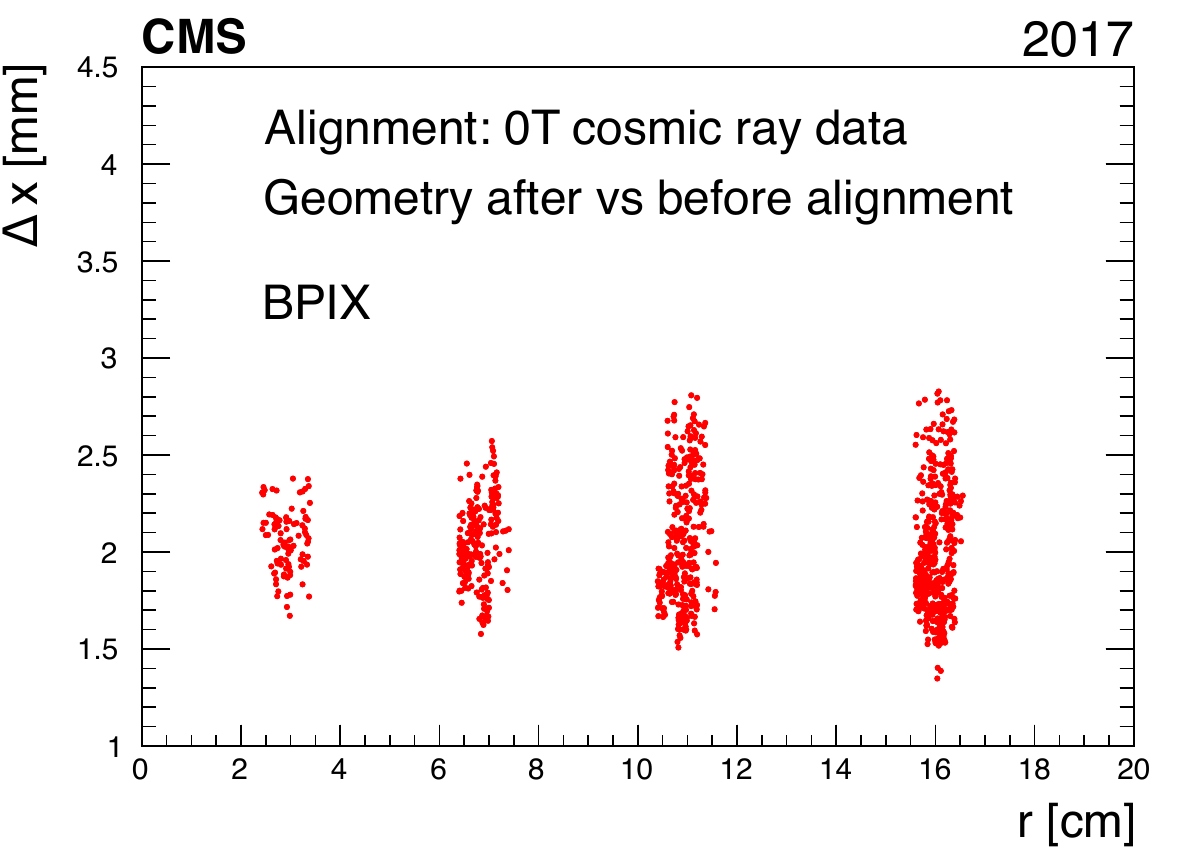}
	\caption{Corrections after the first high-level structures alignment of the pixel detector in 2017, for the three FPIX disks (left) and the four BPIX layers (right).
        Shown are the differences of the module positions after the module-level alignment of the pixel detector, with respect to the ones considered before performing any alignment, as a function of the design positions.
        The alignments were performed with 0\unit{T} cosmic ray muon tracks.
    }
	\label{fig:GC_pixelHL}
\end{figure*}

First, only the forward pixel subsystem was included in data taking. 
Before performing any alignment fit, the asymmetric track rate that was observed between the two FPIX endcaps already provided a hint of a large initial misalignment for this subdetector.
The first alignment fit of the FPIX high-level structures, namely the four half cylinders, was performed using \mbox{$1.5\times10^6$} reconstructed cosmic ray muon tracks collected at $0\unit{T}$.
The high-level structures of the strip detector were included in the alignment procedure as well, to avoid introducing any bias in the alignment constants for the pixel detector due to misalignment in the strip tracker.
For the first track-based alignment after the installation, APUs of 500\mum and 20\mum for the pixel and strip modules, respectively, were used in the track reconstruction and fit. 
The largest measured correction to the assumed geometry was a shift of around 2.8\mm of the $-z$ endcap of the forward pixel detector in the longitudinal direction, further away from the interaction point, as shown in Fig.~\ref{fig:GC_pixelHL} (left).
Correcting this large misalignment eliminated the asymmetric track rate between the two forward pixel sides.
The strip tracker was not substantially misaligned.
The magnitude of the alignment corrections was typically 10\mum for most of the substructures, with the exception of the tracker endcaps, which moved by around 100\mum in the longitudinal direction. 

Later, the barrel pixel detector was also included in data taking, and alignment corrections for the positions and orientations of the two half barrels were derived. This alignment used around 50\,000 cosmic ray muon tracks, recorded at 0\unit{T}.
The largest geometry correction was a shift of around 2\mm in the $x$~direction of the whole barrel, as shown in Fig.~\ref{fig:GC_pixelHL} (right). Since the detector was also rotated with respect to the geometry assumed before alignment, an increasing spread of $\Delta x$ with the radius~$r$ is visible.

\begin{figure*}[htbp]
	\centering
	\includegraphics[width=0.49\textwidth]{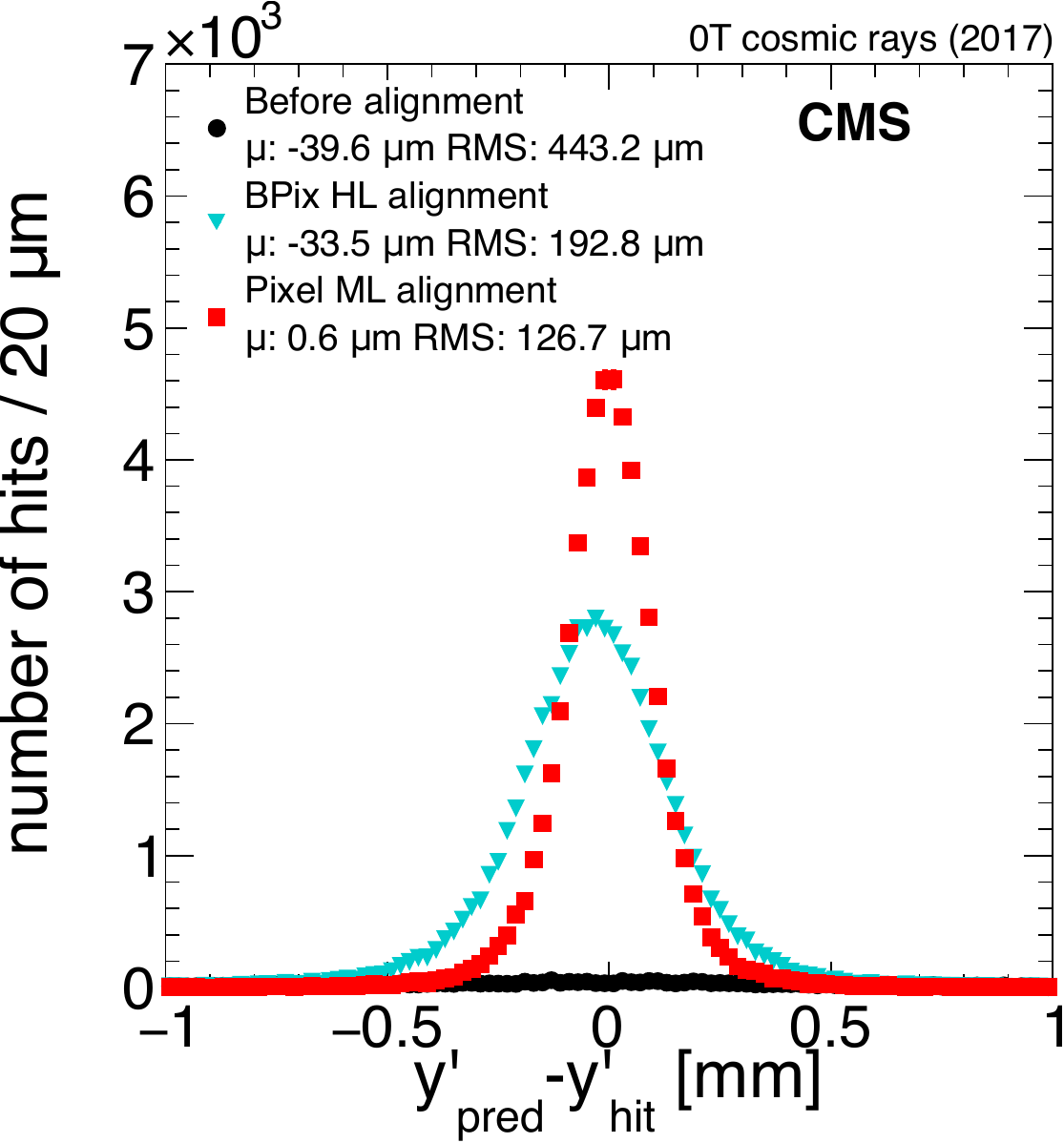}
	\includegraphics[width=0.49\textwidth]{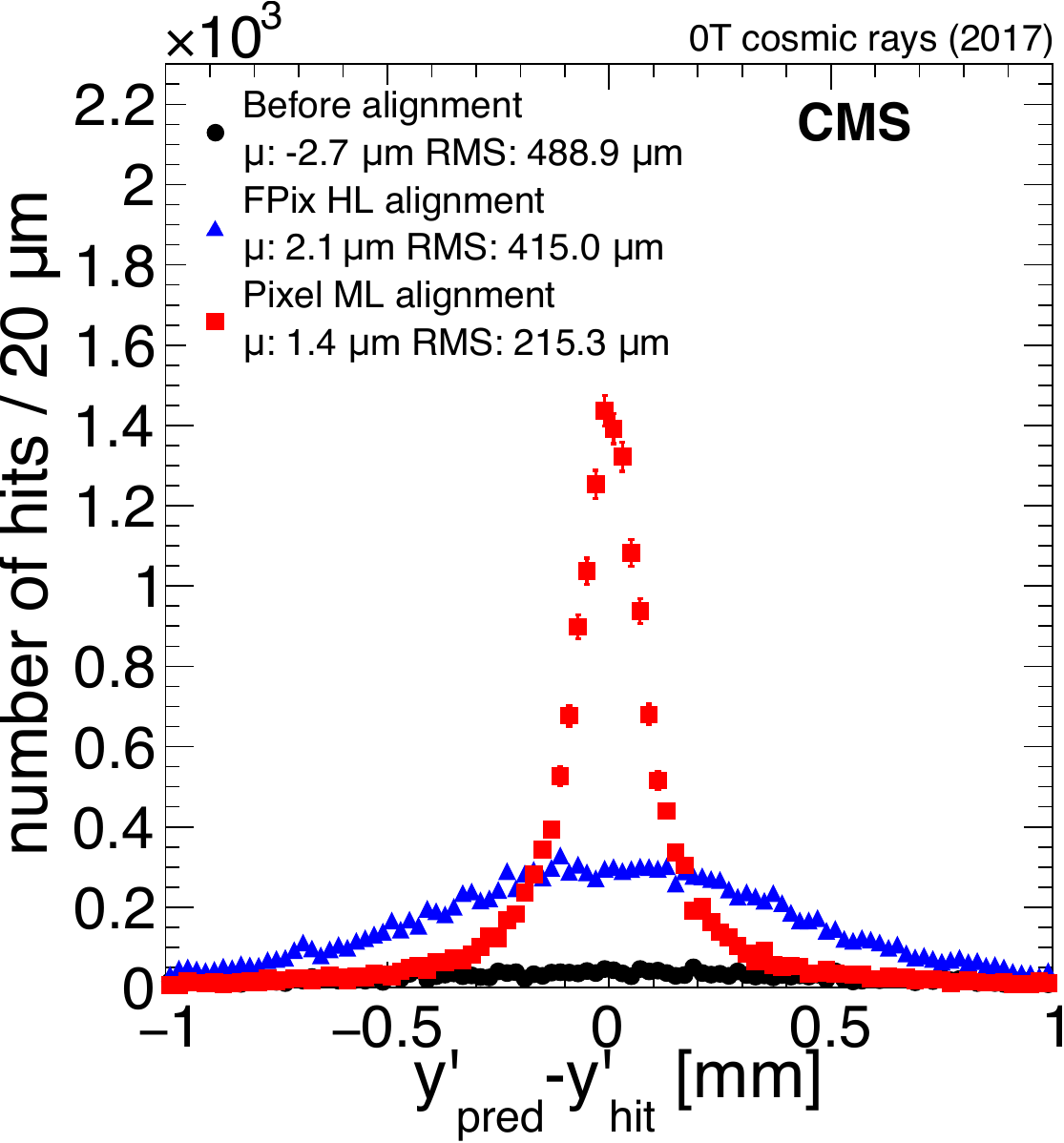}
	\caption{Unbiased track-hit residuals in the BPIX (left) and FPIX (right), in the local~$y$ ($y'$) coordinate.
        The distributions are shown for the different alignment iterations that were performed: the black circles indicate the geometry assumed before performing any alignment fit, the blue and cyan triangles show the high-level (HL) structures alignment of the barrel and forward pixel detectors, and the red squares represent the module-level (ML) alignment of the pixel detector.  The mean ($\mu$) and RMS of the distributions are given in the legend. Vertical error bars represent the statistical uncertainty; they are smaller than the marker size in most of the cases.
    }
	\label{fig:residuals_cruzet}
\end{figure*}

The module-level alignment of the pixel detector was performed using the full data set collected during the CRUZET run, with around $3.2\times 10^6$ tracks used for the alignment.
For the pixel modules, the APUs were reduced to 100\mum. 
The distributions of the unbiased track-hit residuals in the local $x$ and $y$~coordinates for the BPIX and FPIX are shown in Fig.~\ref{fig:residuals_cruzet}, for the two alignment iterations and for the initial geometry assumed before performing any alignment.
After each alignment iteration, the bias and the width of the residual distributions decrease, demonstrating the reduction of systematic misalignment effects and an improved local precision. 

\begin{figure*}[htbp]
	\centering
    \includegraphics[width=0.45\textwidth]{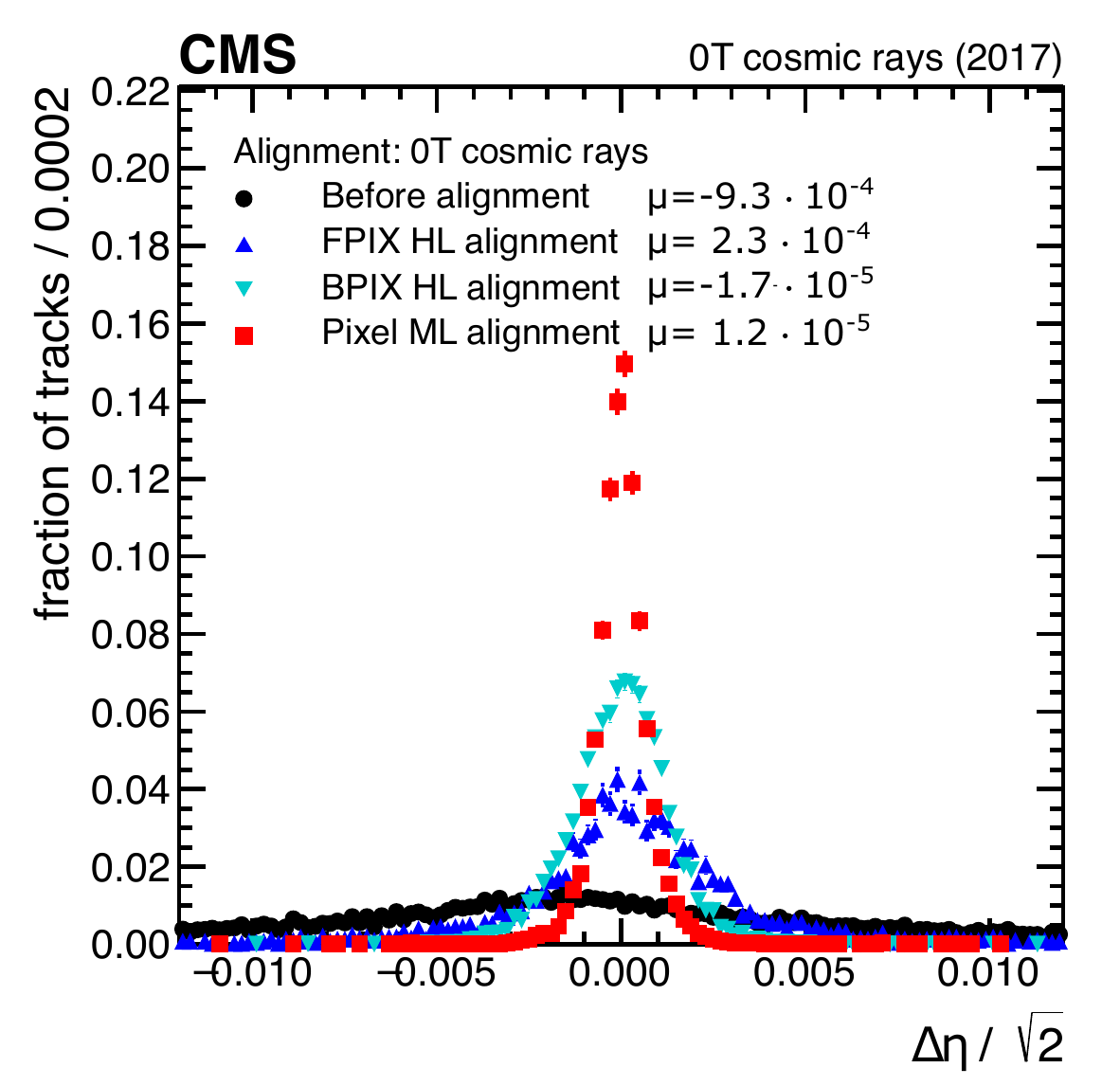}
	\includegraphics[width=0.45\textwidth]{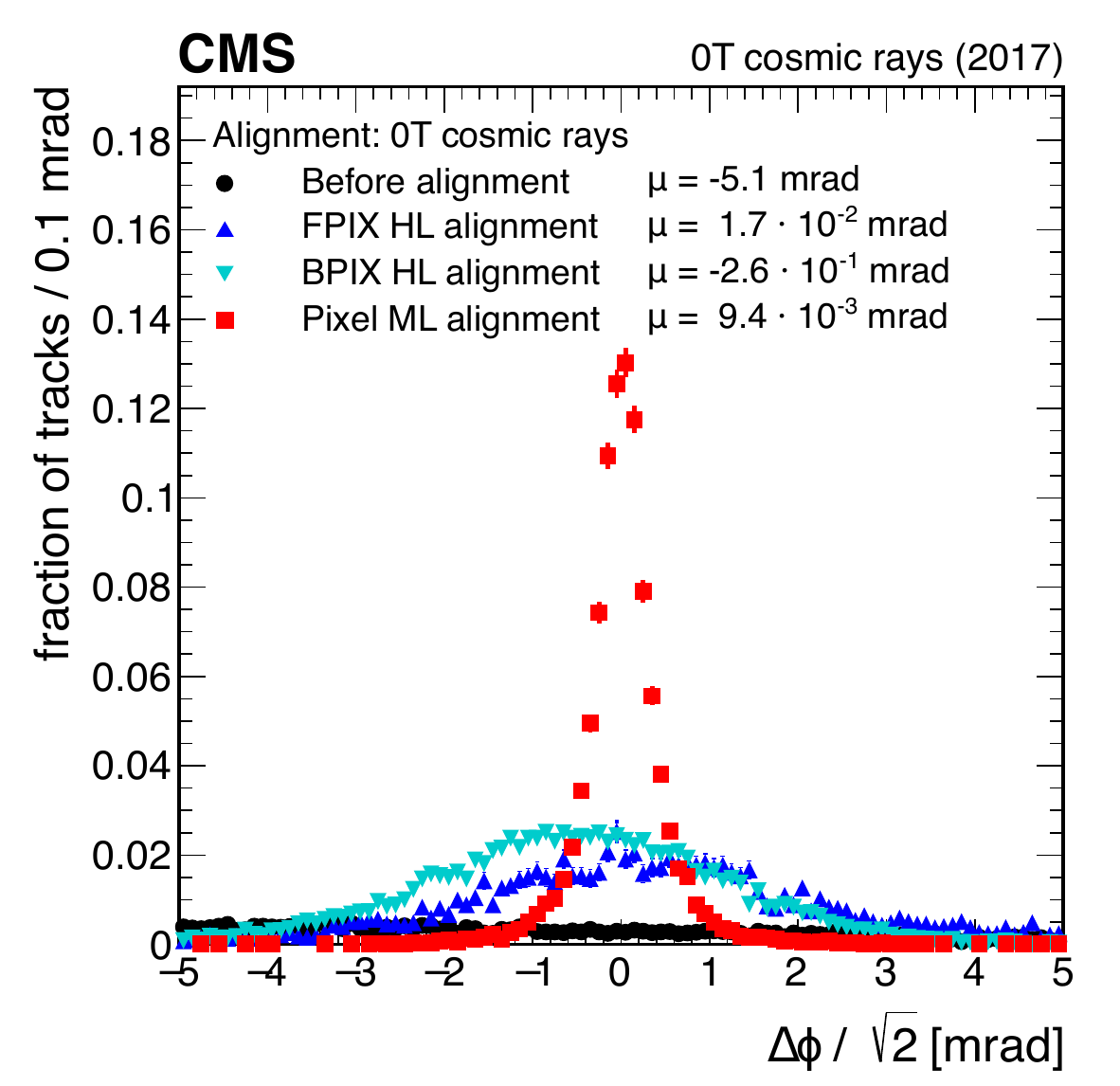}
	\caption{
	Distributions of the difference in reconstructed parameters between the two halves of a cosmic ray muon track, split at the point of closest approach to the interaction region: track $\eta$ (left) and $\phi$ (right) from a dataset recorded at 0\unit{T}. The difference is scaled by a factor $\sqrt{2}$ to account for the two independent measurements.
        The black circles show the geometry assumed before performing any alignment, the triangles show the FPIX and BPIX high-level (HL) structures alignment, and the red squares represent the module-level (ML) alignment of the pixel detector. The mean of the distributions ($\mu$) is given in the legend. Vertical error bars represent the statistical uncertainty; they are smaller than the marker size in most of the cases.
    }
	\label{fig:split_cruzet}
\end{figure*}

The improvements achieved by the different alignment iterations are visible in the cosmic ray muon track split validation as well. Figure~\ref{fig:split_cruzet} shows the differences in the $\eta$ and $\phi$ parameters of the tracks at the different stages of the pixel detector realignment. The improvement achieved after each alignment iteration can be observed, since both the bias and the width of the distributions are considerably reduced.

\begin{figure*}[!ht]
	\centering
    \includegraphics[width=0.45\textwidth]{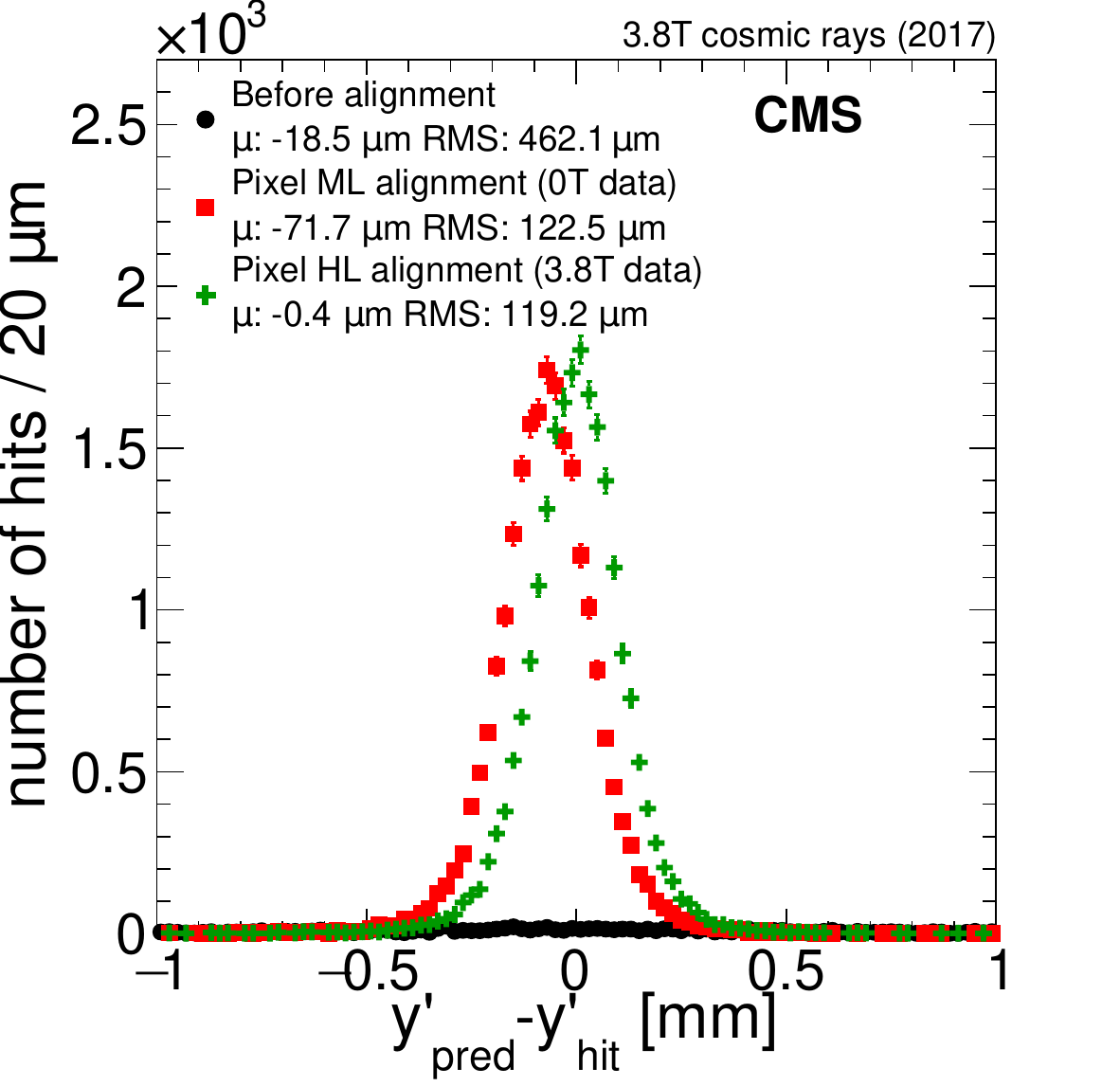}
	\includegraphics[width=0.45\textwidth]{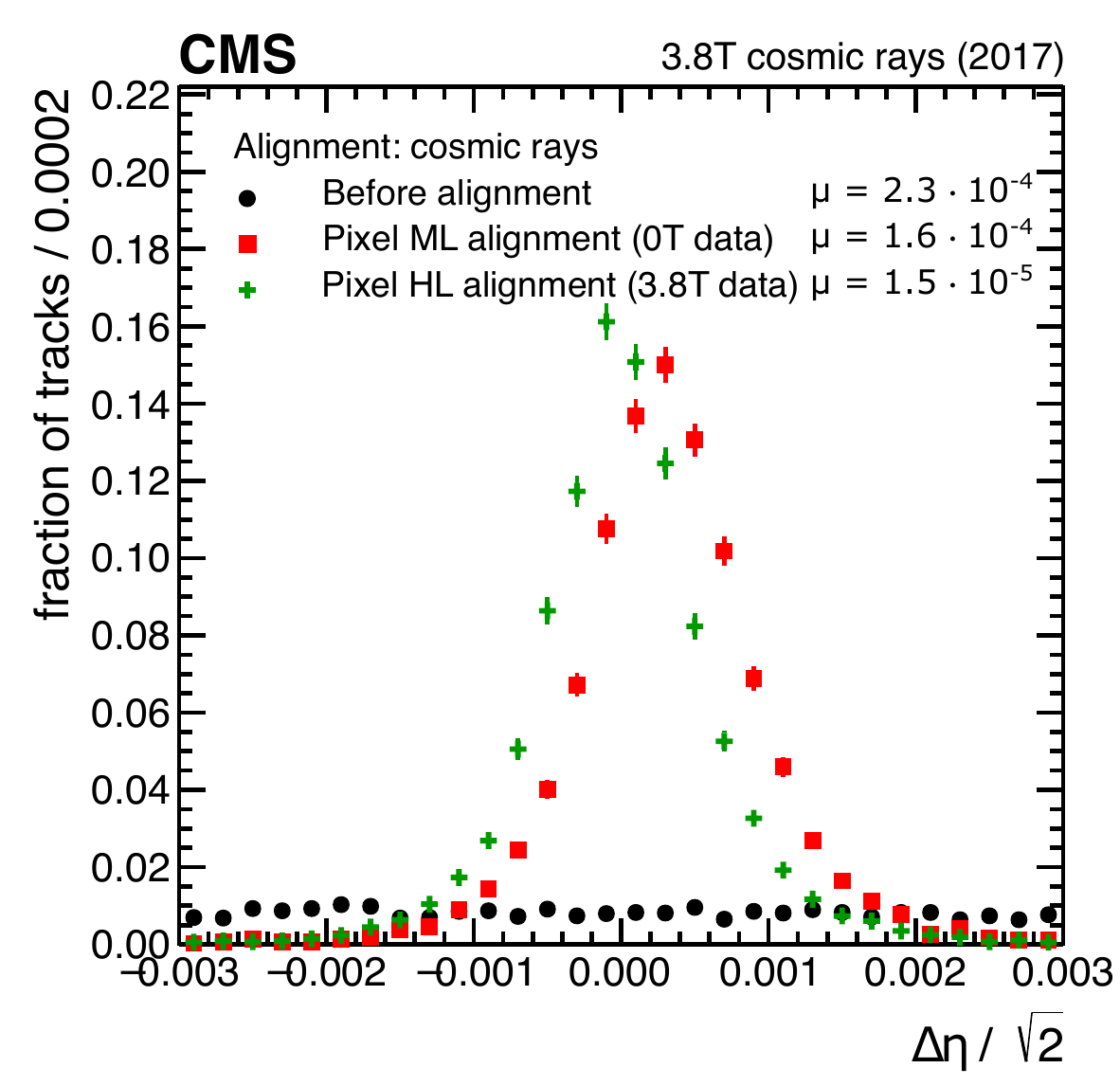}
	\caption{Performance of the 2017 CRAFT alignment fit (green crosses) compared with the geometry obtained after the alignment fit with 0\unit{T} cosmic ray muon tracks (red squares), and with the assumed geometry before performing any alignment (black circles).
        As an example, the track-hit residual distributions in the local~$y$ coordinate for the BPIX (left) and the difference in track $\eta$ from the cosmic ray muon track split validation (right) are shown. The mean ($\mu$) of the distributions is given in the legend. For the track-hit residual distributions, also the RMS is indicated. Vertical error bars represent the statistical uncertainty; they are smaller than the marker size in most of the cases.
    }
	\label{fig:cruzet_craft}
\end{figure*}

An additional alignment was performed after the CMS magnetic field was turned on, using the tracks recorded in around 300\,000 CRAFT events. The alignment constants needed to be rederived to account for movements of the mechanical structures induced by the magnet ramp-up, causing biases in the distributions for the 0\unit{T} alignment fit. 
This is visible in the track-hit residual distributions, as well as in the cosmic ray muon track split validation shown in Fig.~\ref{fig:cruzet_craft} (red squares).
Despite the limited size of the data set, corrections for the strip and pixel detector high-level structures were derived. These corrections removed the observed biases as shown in Fig.~\ref{fig:cruzet_craft} (green crosses). 

\subsection{Automated alignment} \label{sec:PCL}
During data taking the different components of the pixel detector may shift because of changes in the magnetic field or the temperature.
To account for these shifts, an automated alignment procedure was implemented for Run~2 so that fast updates of the alignment parameters can be provided within 48 hours.
This alignment procedure runs as part of the prompt calibration loop~(PCL)~\cite{Cerminara:2121263}, which processes several routines to control and automatically update different detector-related parameters.

\begin{figure*}[!ht]
	\centering
    \includegraphics[width=0.3\textwidth]{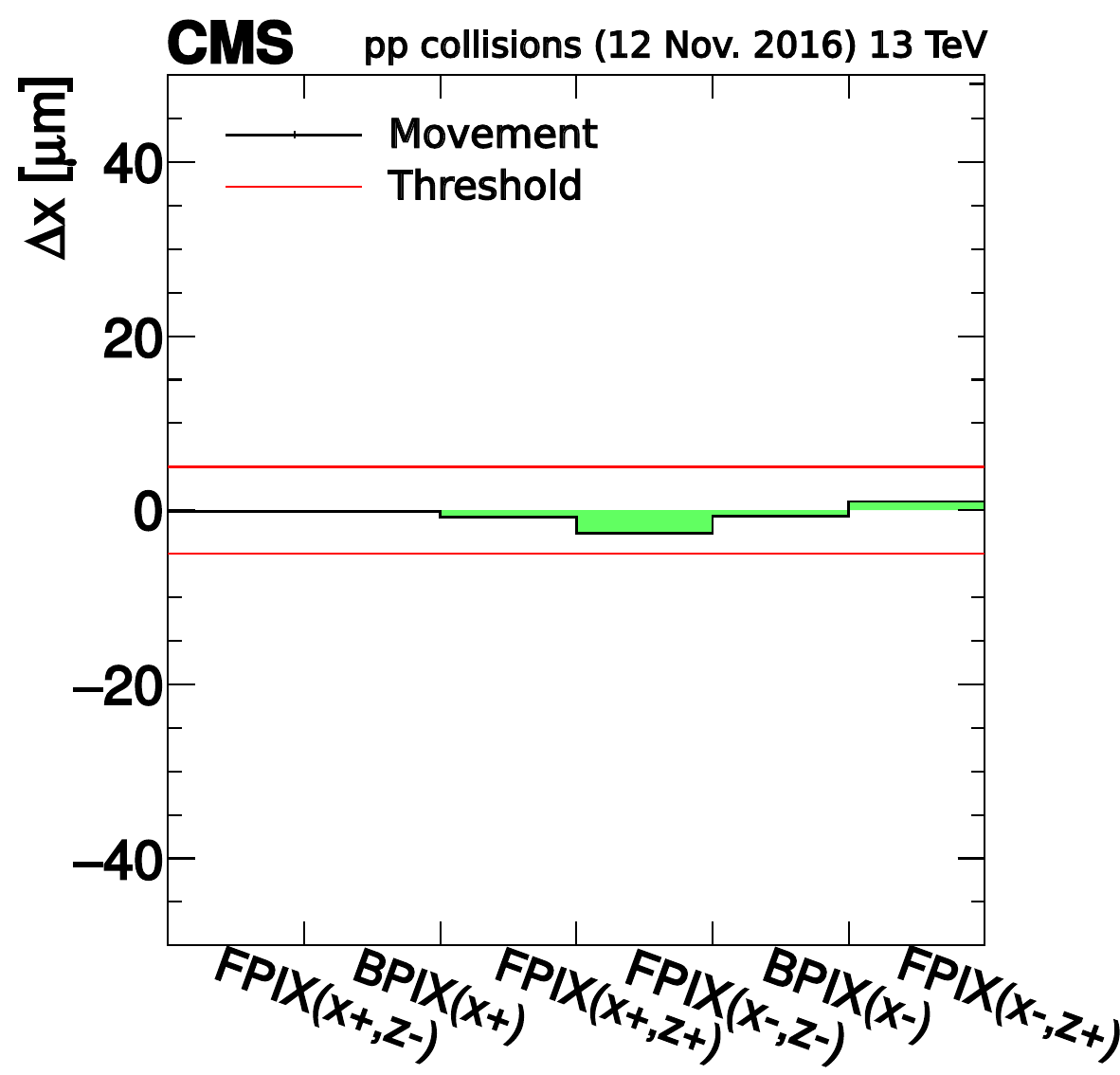}
	\includegraphics[width=0.3\textwidth]{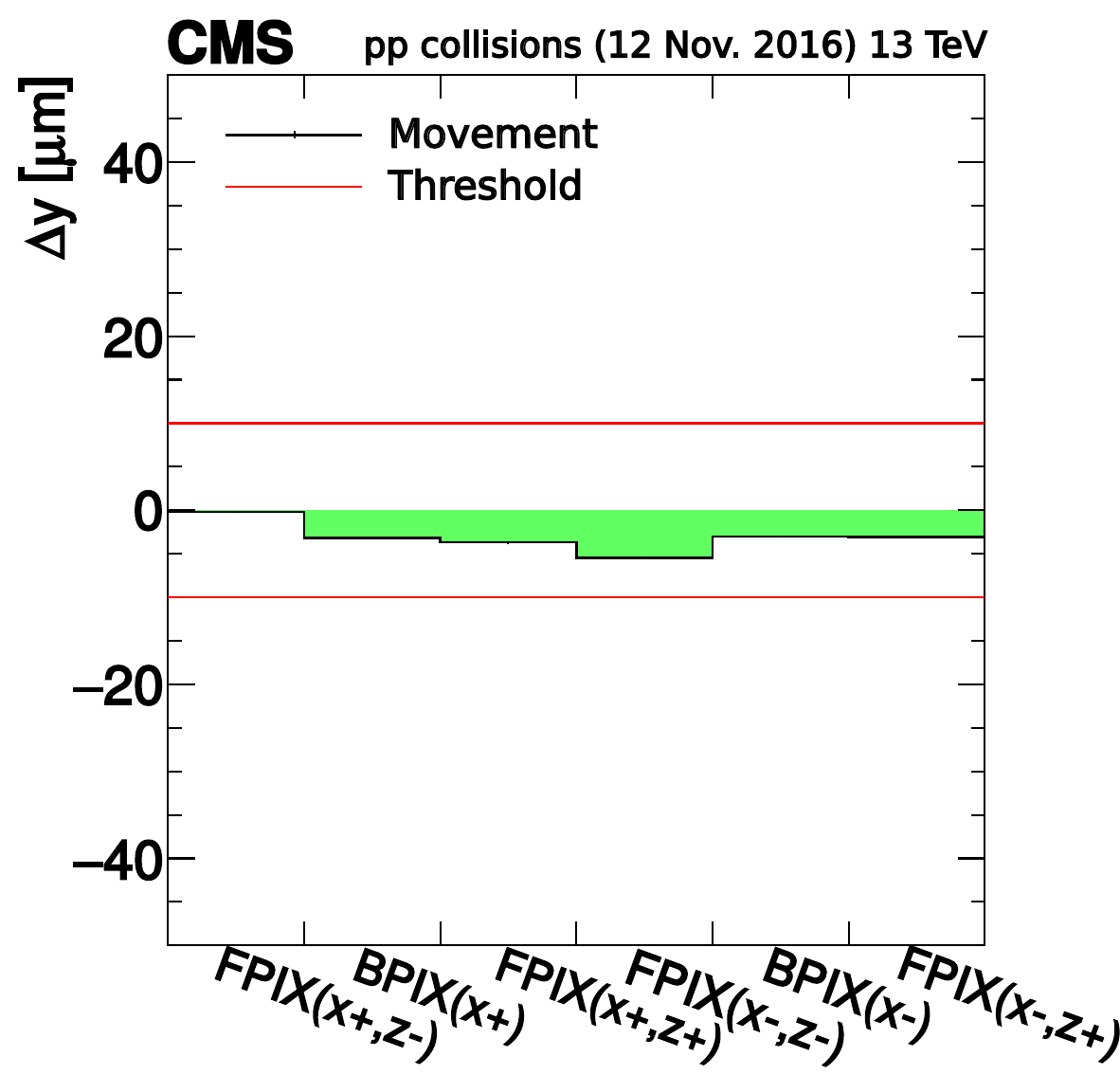}
	\includegraphics[width=0.3\textwidth]{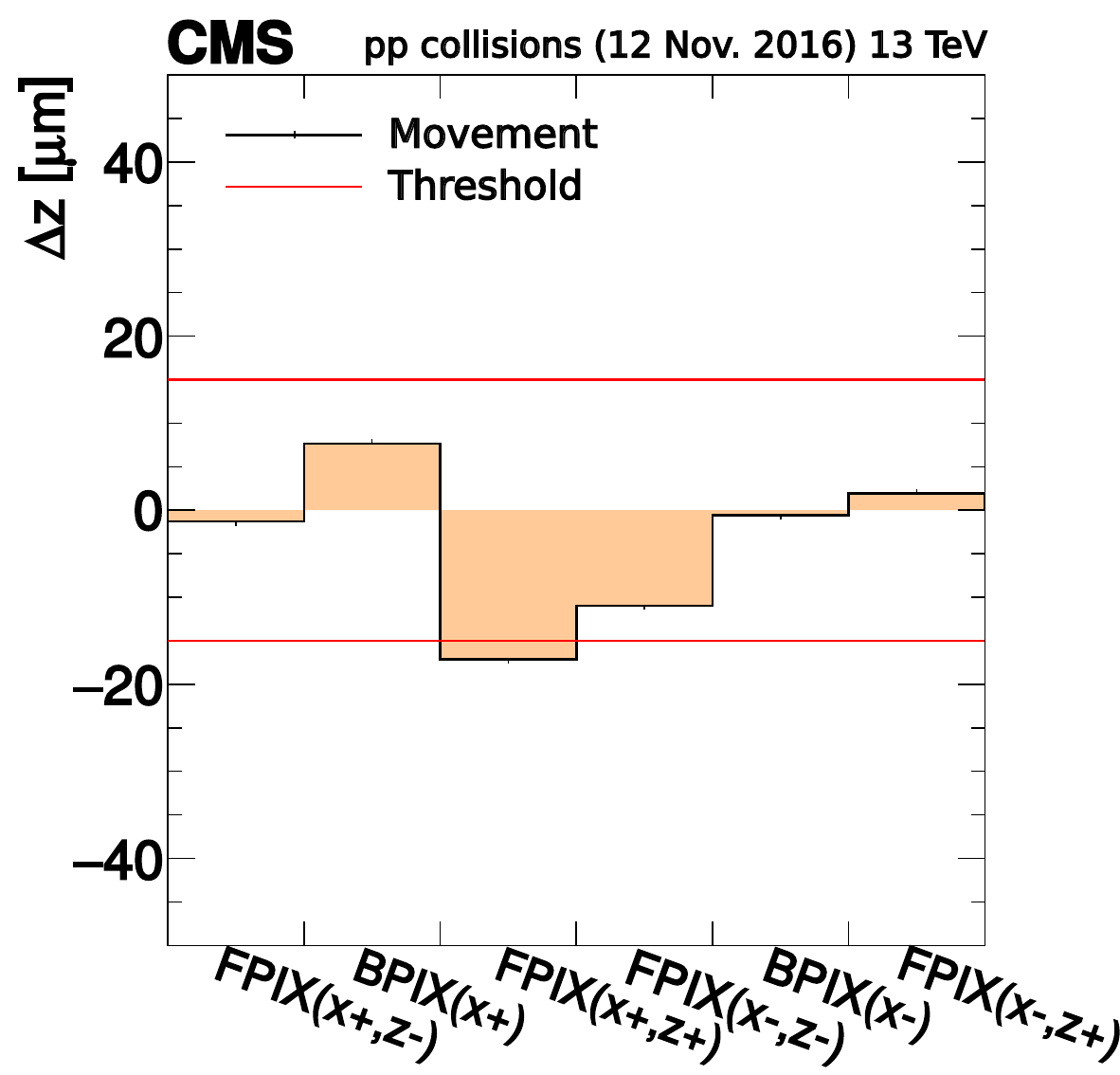}\\
	\includegraphics[width=0.3\textwidth]{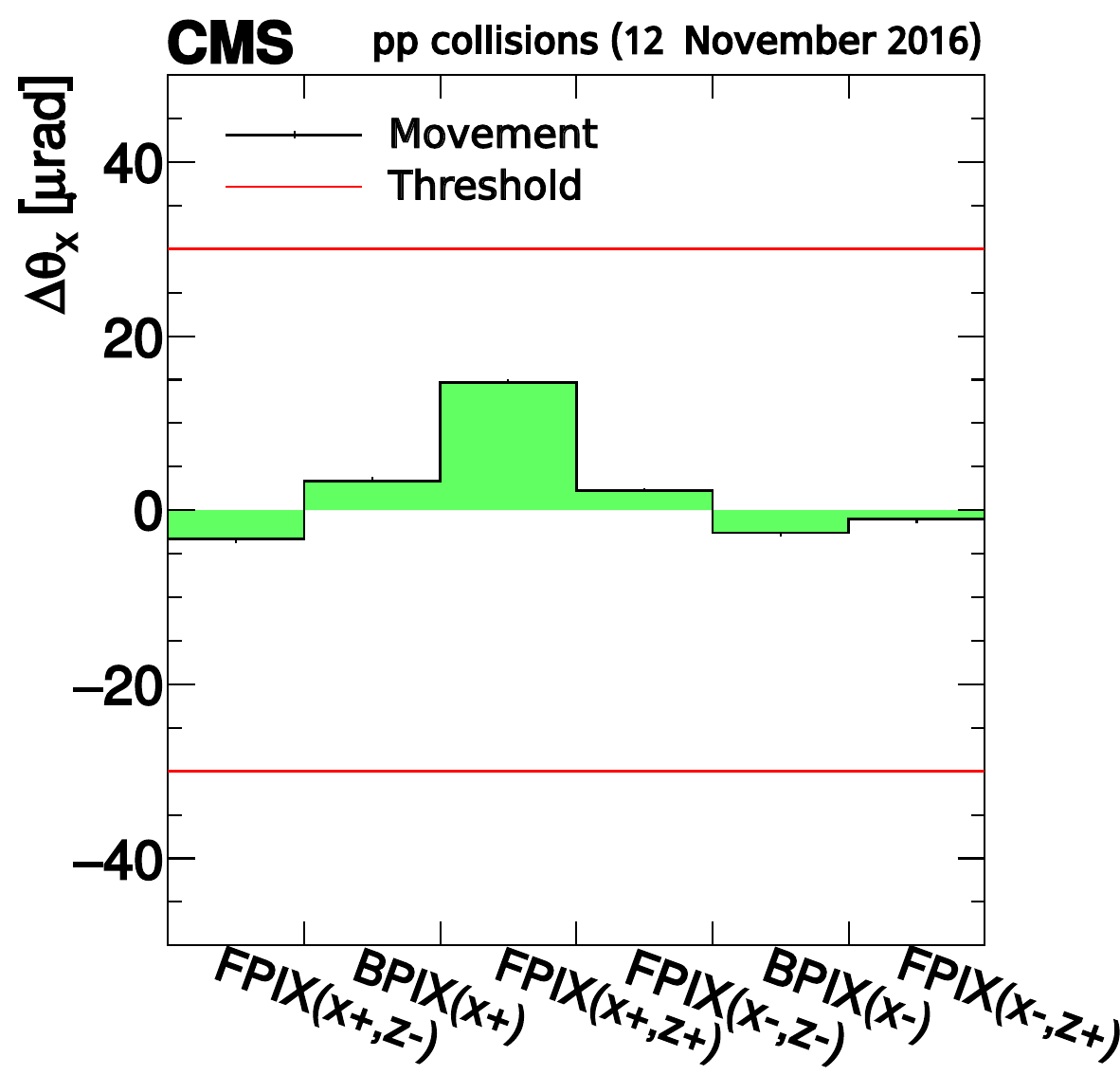}
	\includegraphics[width=0.3\textwidth]{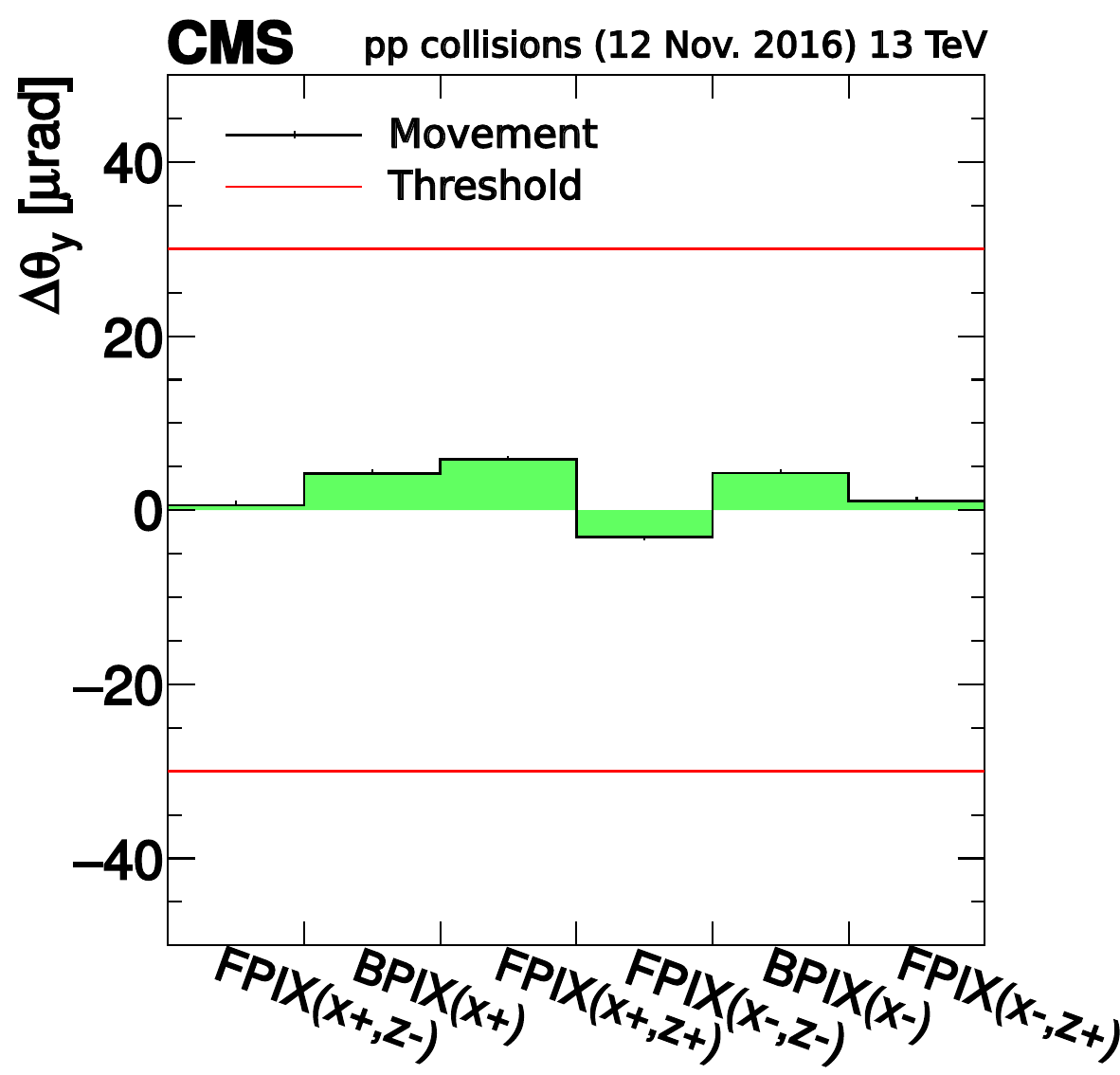}
	\includegraphics[width=0.3\textwidth]{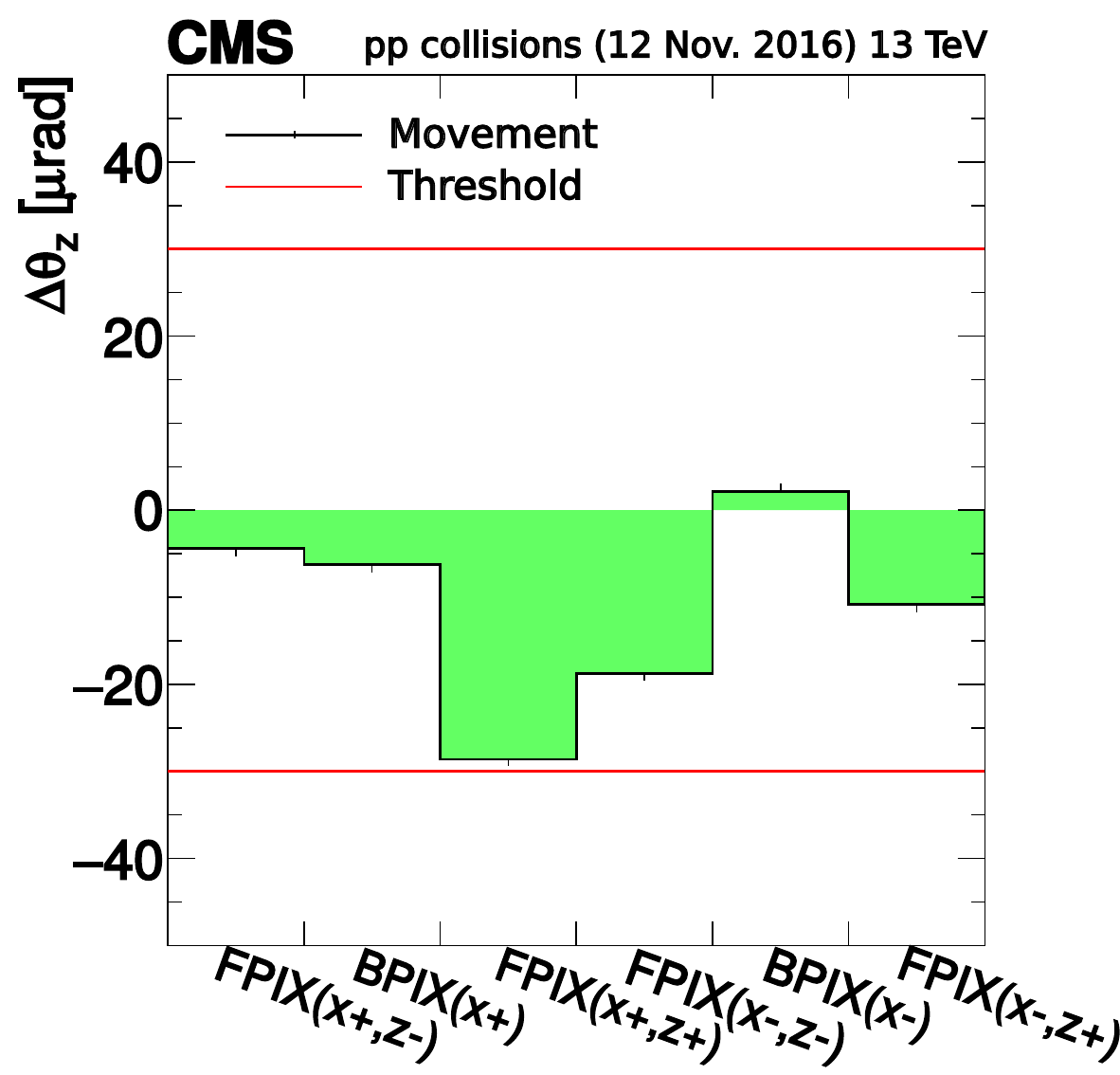}
		\caption{Observed movements of the six high-level structures in the pixel detector from the alignment procedure in the PCL for one arbitrary run in 2016. The two horizontal red lines in each of the figures show the threshold for triggering the deployment of a new set of alignment constants. The error bars represent the statistical uncertainties of the measurements. The orange bars in the figure showing the movement in the $z$~direction indicate that a sufficient movement to deploy a new set of alignment constants was observed.}
	\label{fig:PCL_Run2016}
\end{figure*}

The alignment routine itself is based on a total of 36 degrees of freedom to account for the movement of the high-level structures in the pixel detector, namely two half barrels and two half cylinders in each of the two endcaps (six substructures in total).
For each of these structures, corrections for the positions~($x,y,z$) as well as the rotations~($\theta_x,\theta_y,\theta_z$) are derived using the \MILLEPEDE-II alignment algorithm (six corrections for each of the six substructures). 
The alignment is performed using tracks from the inclusive L1 trigger data set. 
If the position of a structure changes by 5, 10, or 15\mum in the $x$, $y$, or $z$~direction, respectively, or rotates by 30\unit{$\mu$rad}, the alignment parameters are updated for the prompt reconstruction of the next run.
Figure~\ref{fig:PCL_Run2016} shows the results of the alignment routine for an arbitrary run in the 2016 data set.
For this run, the deployment of a new set of alignment constants was triggered by the movement of one of the endcap half cylinders in the $z$~direction.

\begin{figure*}[!ht]
    \includegraphics[width=\textwidth]{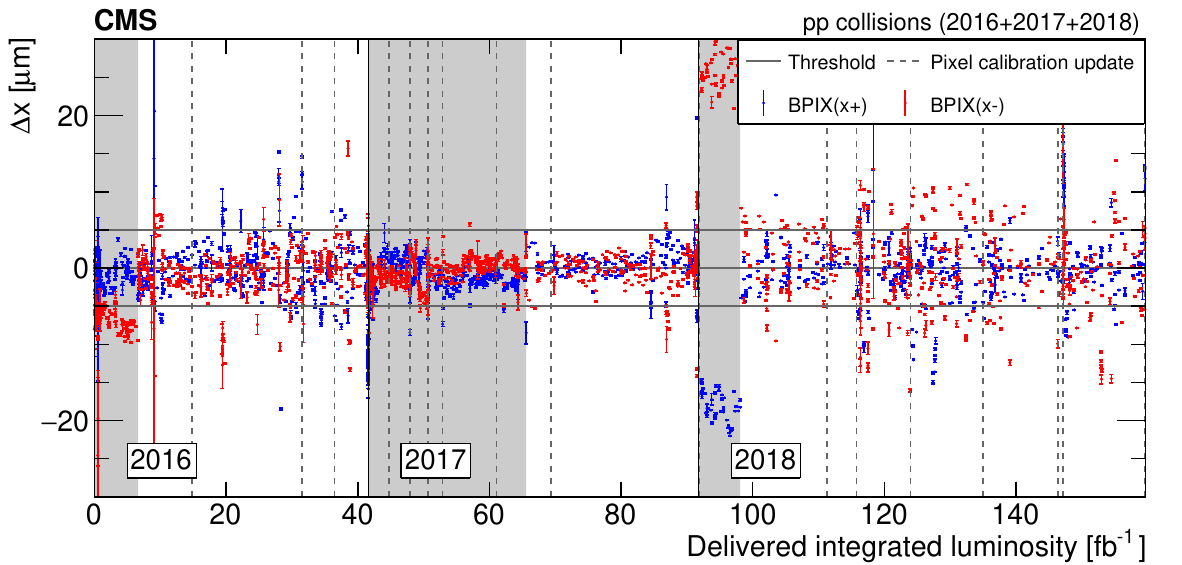}
    \caption{Observed movements in the $x$~direction of the two BPIX half cylinders, as functions of the delivered integrated luminosity, from the alignment procedure in the PCL. Each point corresponds to a single run and shows the movement proposed by the alignment fit with respect to the last deployed alignment. The alignment is only updated for future data taking if at least one of the six movements surpassed its corresponding threshold, otherwise it is not changed. The vertical bars on each point represent the statistical uncertainty of the measurement. The vertical black solid lines indicate the first processed runs for the 2016, 2017, and 2018 data-taking periods, respectively. The vertical dashed lines illustrate updates of the pixel detector calibration. The two horizontal lines show the threshold for triggering the deployment of a new set of alignment constants. The grey bands at the beginning of each year indicate runs where the automated alignment updates were not active.}
    \label{fig:PCL_trend}
\end{figure*}

Figure~\ref{fig:PCL_trend} shows the movement of the two BPIX half cylinders in the $x$~direction as a function of the delivered integrated luminosity during Run~2.
In general, a stable performance of the automated alignment fit was observed over the full range of Run~2.
Compared with 2017, a few larger movements can be observed in 2018 related to the presence of residual misalignments not covered by the degrees of freedom used in the alignment during data taking.
The beginning of data taking in each of the three years shows a period where the automated deployment of the alignment constants was not active.
In 2016 and 2017 these periods show only small movements.
The movements in the corresponding period of 2018 are at a stable high level, indicating that the alignment was not corrected by an automated update.

\begin{figure*}[!ht]
    \centering

    \includegraphics[width=0.76\textwidth]{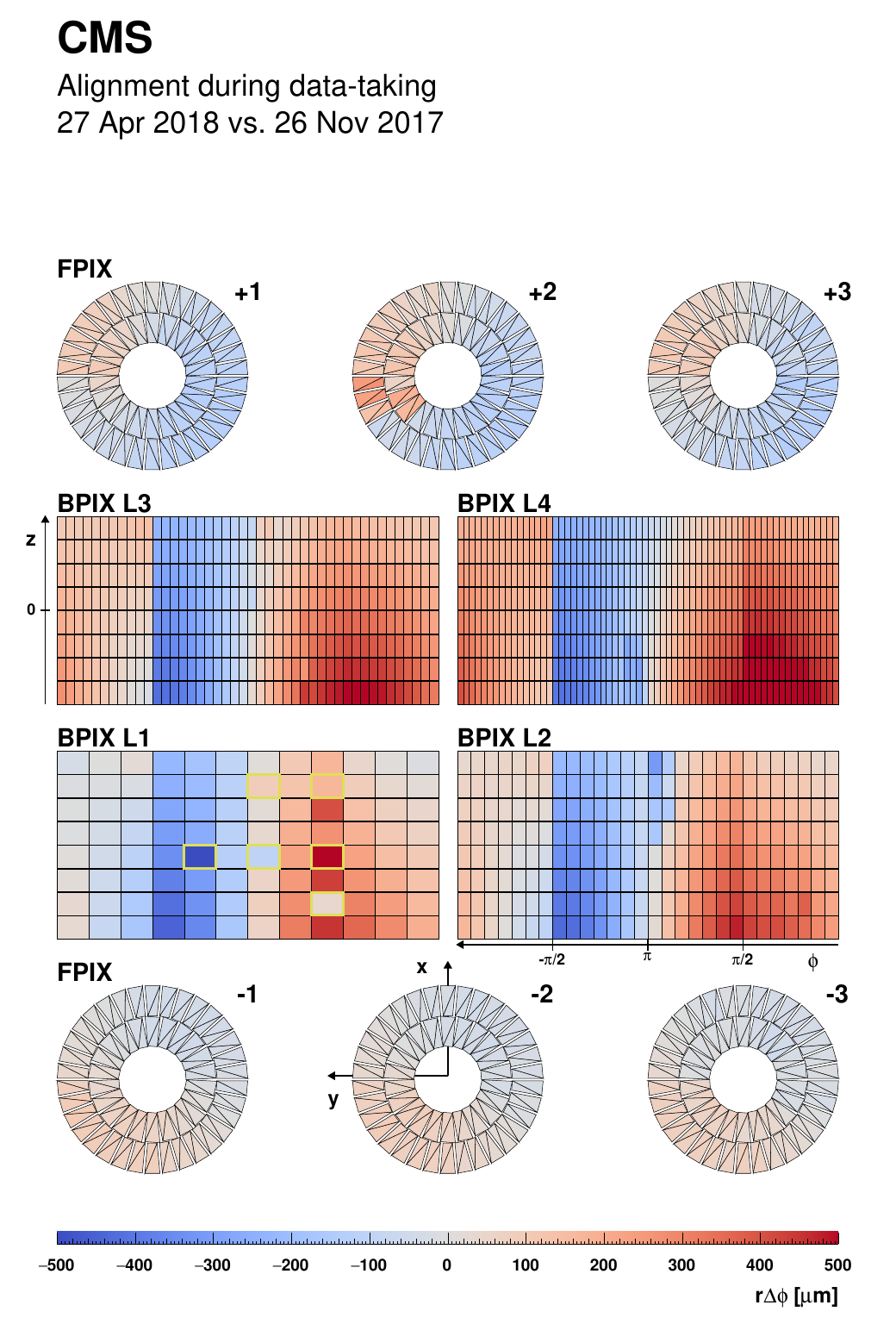}
    \caption{The value of the product $r\Delta\phi$ for each module in the pixel detector, comparing the alignment parameters of the alignment during data taking on 26 November 2017 and on 27 April 2018. These runs correspond to the last run of 2017 and the first run of 2018 after commissioning.
        During this transition, several modules of layer 1 of the BPIX were replaced (indicated in yellow frames), which explains the large movements with respect to their neighbouring modules.
        For each of the detector components, $r$ and $\phi$ correspond to the global coordinate, and $\Delta\phi$ is the shift in $\phi$ across the two alignments resulting in the physical shift $r\Delta\phi$ in the detector.
    }

    \label{fig:TkAlMap_2017to2018}
\end{figure*}

During the transition between 2017 and 2018, a few detector modules of layer 1 of the BPIX were replaced. The entire pixel detector was opened for this procedure. 
This resulted in large movements in the detector, as the PCL alignment correctly indicated, and a manual alignment procedure was performed. Figure~\ref{fig:TkAlMap_2017to2018} shows the movements of all the individual pixel modules during this transition period. 
Large movements in the entire pixel detector, in addition to the relative movements of the replaced parts, are clearly visible in this figure.
In addition to the alignment automatically derived in the PCL, manually derived alignments are also used during data taking.
These alignments are based on a higher granularity. This is necessary, for example, after updates of the pixel detector calibration, as indicated in Fig.~\ref{fig:PCL_trend}.
Unlike the PCL alignment procedure, which is based on high-level structures, the higher granularity of the additional alignment fits further enables the correction of the radiation effects introduced in Section~\ref{sec:concepts}.

\begin{figure*}[!ht]
	\centering
    \includegraphics[width=0.45\textwidth]{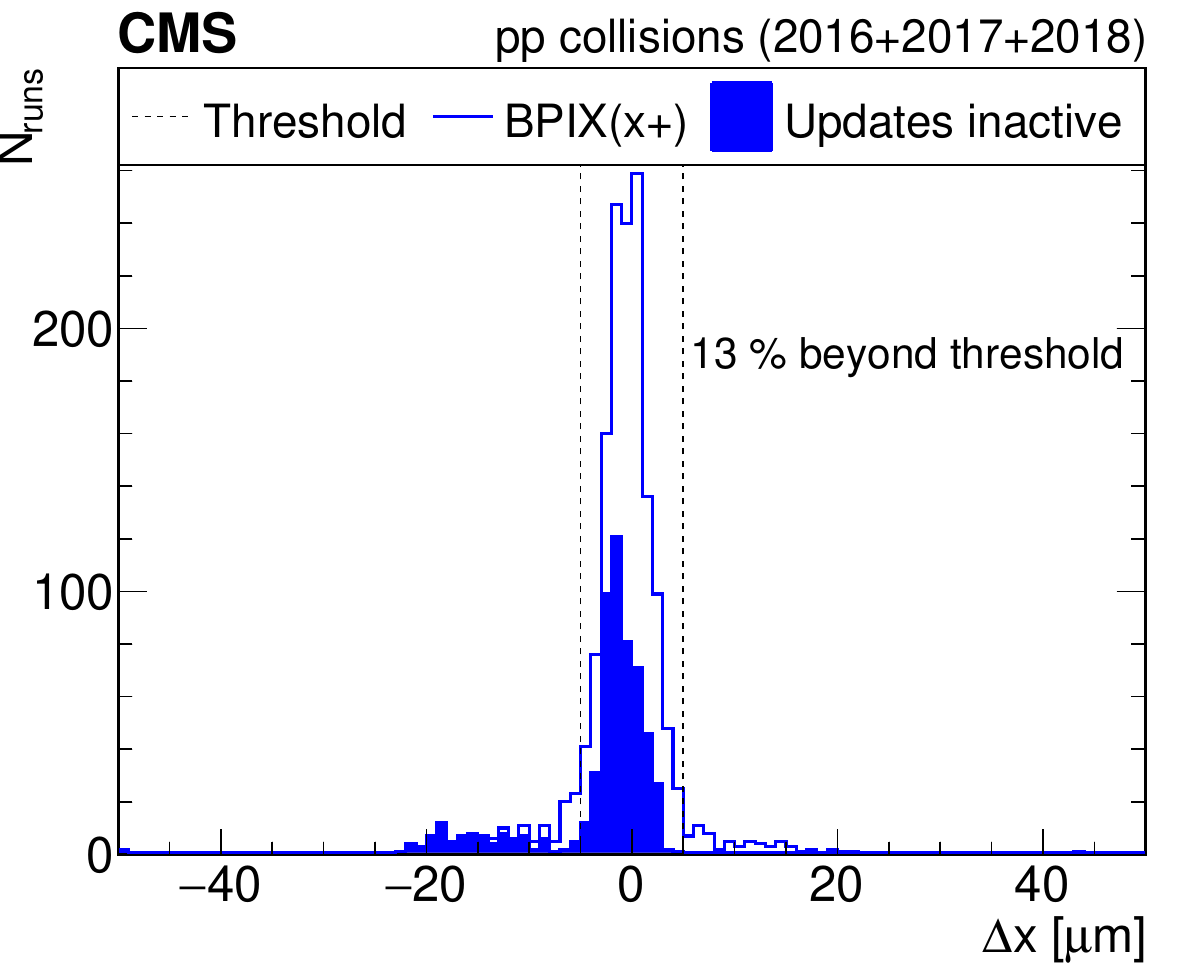}
	\includegraphics[width=0.45\textwidth]{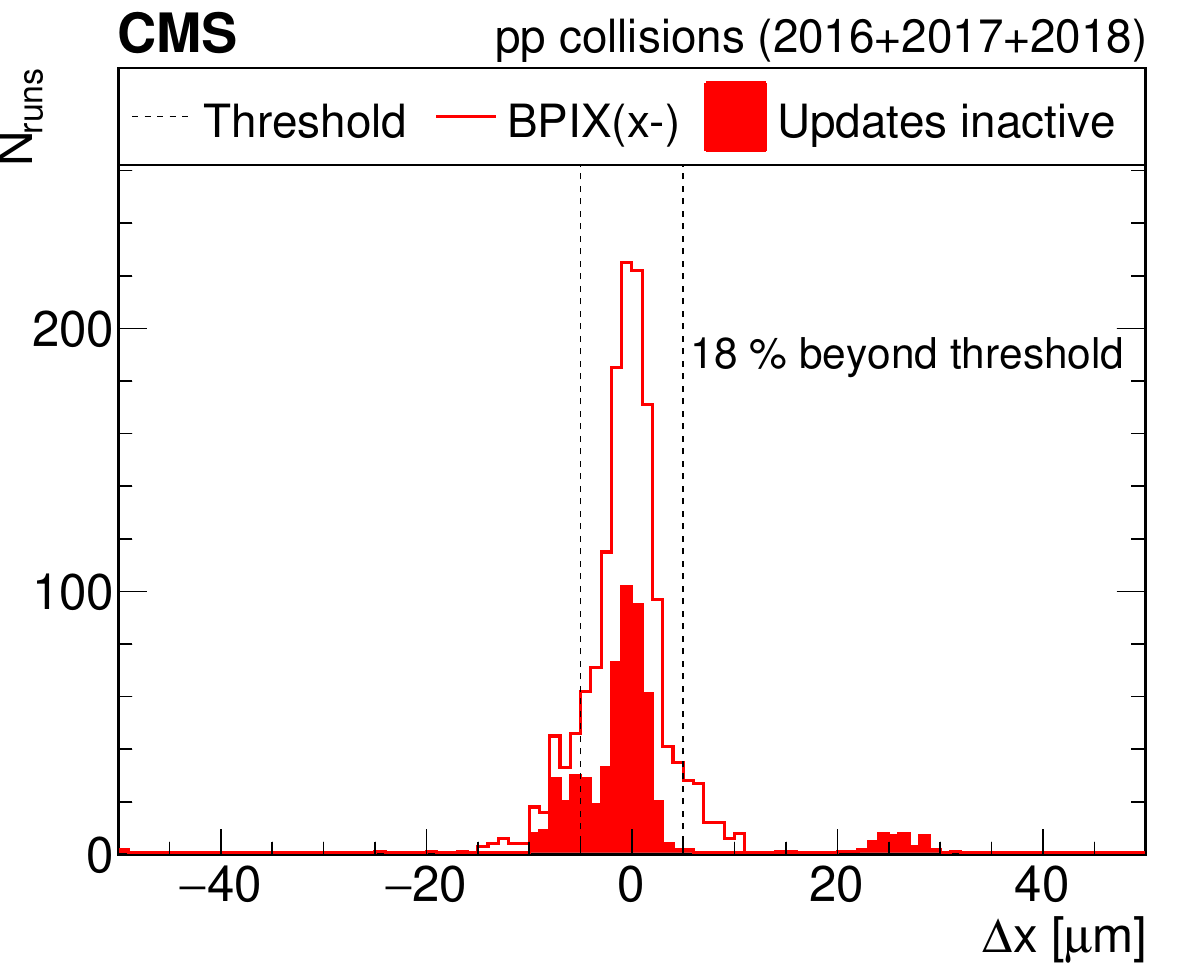}
    \caption{Observed movements in the $x$~direction of the two BPIX half cylinders from the PCL alignment. The two vertical lines show the threshold for a new alignment to be triggered. The filled entries correspond to runs at the start of each year, where the automated updates of the alignment were not active. The percentage of the runs for which a new alignment was triggered by the movement in the $x$~direction is displayed below the legend in both figures. In the calculation of this percentage, the filled entries are not included.}
	\label{fig:PCL_hist}
\end{figure*}

The distribution of the movement of the two BPIX half cylinders in the $x$~direction, which corresponds to the projection of Fig.~\ref{fig:PCL_trend}, is shown in Fig.~\ref{fig:PCL_hist}.
For both of the half cylinders, the majority of the runs show movements within the given thresholds and, therefore, do not trigger an update of the alignment.
In both figures, the larger movements at the beginning of 2018 discussed above are visible as smaller clusters.
Without including runs where the deployment of the new alignment constants was inactive, a total of 13--18\% of the runs yielded a sufficient movement in the $x$~direction of one of the half cylinders to trigger an alignment update.

\begin{figure*}[!ht]
    \includegraphics[width=\textwidth]{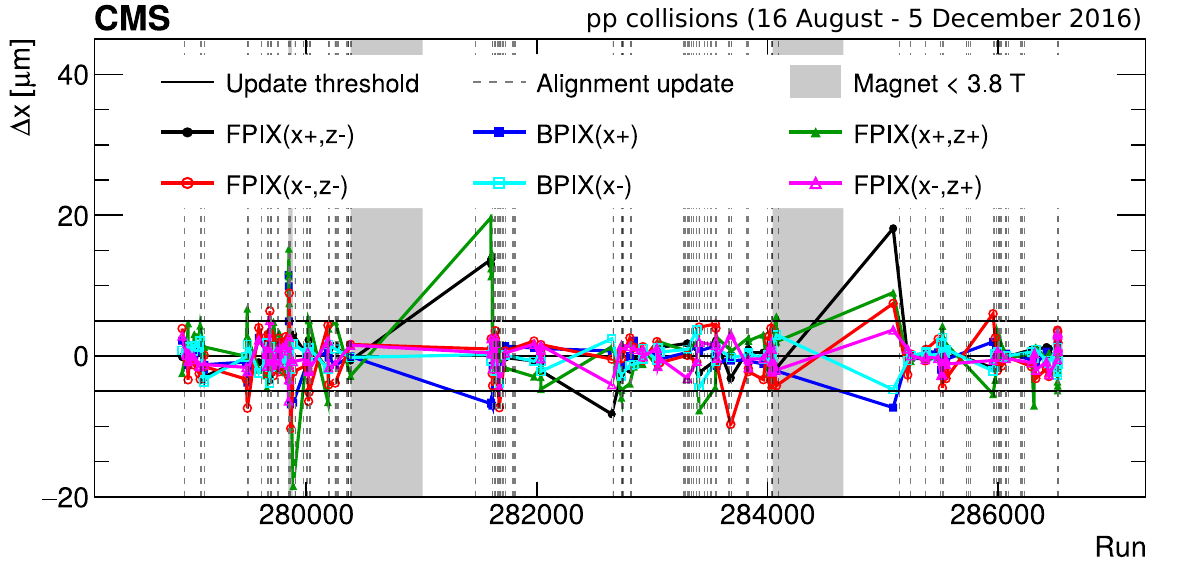}
    \caption{Observed movements in the $x$~direction of the six high-level structures, as functions of the run number from the alignment procedure in the PCL for the data-taking period between 16~August and 5~December 2016. Each point corresponds to a run that triggered an update of the alignment parameters caused by a sufficient change in one of the three positions or rotations. The movements are shown without uncertainties. The vertical dashed lines illustrate the deployments of new sets of alignment constants. The two horizontal lines show the threshold for a new alignment to be triggered. The grey shaded regions indicate runs during which the magnet was not at 3.8\unit{T} (magnet cycle). After each of the two magnet cycles a large movement is observed for the very first run after the cycle.}
    \label{fig:PCL_trend_2016}
\end{figure*}

The necessity, as well as the success, of the automated alignment during data taking is also visible in Fig.~\ref{fig:PCL_trend_2016}.
Here, for a short period in 2016, where the magnetic field was changed twice, the movement in the $x$~direction of all six high-level structures is shown for each run that triggered an update of the alignment parameters.
In both cases, large movements were observed in the first run after the change in the magnetic field.
These movements triggered an update of the alignment, which was able to correct for the changes introduced by the magnetic field.

\section{Alignment for physics analysis} \label{sec:EOY} \label{sec:UL} \label{sec:Legacy}
This section describes the determination of the alignment constants in view of performing physics analyses with the data taken from 2016 to 2018.

The derivation of alignment constants for physics analysis requires large samples to determine the positions, orientations, and surface deformations of all sensors in the pixel and strip detectors.
Therefore, it is typically performed in the middle or at the end of the year. This is referred to as the ``end-of-year (EOY) reconstruction'' and it is shown in red in the figures in this section. In addition, after the completion of Run 2, a new set of alignment constants was derived for the three years. This is referred to as ``legacy reprocessing'' and is shown in green in the figures in this section.
Throughout this section, we will describe how we have obtained these two new sets of alignment constants and compare them with the set of alignment constants used during data taking. This is labelled ``alignment during data taking'' and is shown in blue in the figures in this section.

One important feature discussed in this section is the impact of radiation on the hit reconstruction and consequently on the alignment procedure.
This, together with the aim of controlling potential WMs, drives the strategy.

First we will describe the strategy that was employed to account for the time dependence. Then we will compare the performance of the EOY reconstruction and of the legacy reprocessing with the performance of the alignment during data taking.
Finally, we will discuss special data-taking periods, such as runs with low pileup, runs at a centre-of-mass energy of 5.02\TeV, and heavy ion (HI) runs.

\subsection{General strategy}
\label{sect:ULstrat} \label{sec:ULstrat}
Because of the large changes that occur during a YETS, as illustrated in Fig.~\ref{fig:TkAlMap_2017to2018}, each data-taking year is aligned separately.
To maximize the statistical power of the cosmic ray muon track and dimuon resonance data sets, and to prevent systematic distortions from arising, the data collected during an entire year are combined to perform the alignment fit.
Temporal changes within a year are taken into account by introducing a hierarchy in the alignment fit. The positions of certain sets of modules are aligned over short periods of data taking corresponding to an integrated luminosity of around $1\fbinv$ (a few days); such sets of modules can correspond to mechanical structures, since a whole mechanical structure can move coherently. The positions, orientations, and surface deformations of all the sensors are aligned relative to these high-level alignables.

During Run~1, the high-level alignables usually corresponded to the high-level structures. During Run~2 this was still true for the strip detector, but smaller mechanical structures were chosen in the pixel detector to absorb the effects of radiation damage accumulating over time.
This strategy was applied for the first time in 2016, where ladders and blades, rather than the high-level structures, were treated as the high-level alignables. In 2018, with the increased level of radiation, a better performance was obtained by considering ladders in the BPIX and the modules in the FPIX as the high-level parameters.
This approach is necessary to cope with residual effects that are not covered by the dedicated calibration. These effects are of the order of a few microns at most, but are not constant over time. Without this additional freedom, tensions would appear in the alignment fit and lead to unphysical results.

\label{sect:IOV} \label{sec:IOV}

The IOVs for each set of alignment constants are determined from several sources:
\begin{itemize}
    \item   magnet cycles and known changes of temperature;
    \item   changes in the hit reconstruction (including changes of the local calibration, changes of voltage, and ageing due to the irradiation);
    \item   changes to the distributions of the impact parameters in the transverse plane~$\langle \dxy\rangle$ and on the longitudinal axis~$\langle \dz\rangle$, as a function of the track angular variables~$\eta$ and $\phi$ (as introduced in Section~\ref{sec:2015}), observed by eye on a per-run basis (typically corresponding to 1--100\pbinv).
\end{itemize}
These last types of changes correspond, for example, to steps in the distributions matching with the geometrical coverage of the high-level structures (\eg half-barrels in the BPIX). Such changes can also correspond to a scattering of the points as a function of $\phi$, matching with the position of the ladders.

In addition to the IOV boundaries, the level of precision of the alignment must be configured.
Depending on the required precision and on the available data set, one can include more or fewer alignables, \ie release more or fewer degrees of freedom.
Moreover, the alignment procedure and its physics validation are both computationally demanding, and only a limited number of configurations can be run and compared at a time.
As discussed in Section~\ref{sec:software}, the single matrix inversion necessary for each tested configuration can take up to a full day of computing time.
Similarly, the physics validation may take several days.
Different configurations are attempted, and the configuration that provides the best physics performance is then selected.
As a result of the time involved, the number of attempts to understand the alignment is sometimes limited. This was especially the case for the EOY reconstruction. However, for the legacy reprocessing, the investigations spanned up to a few months. The calibration of the pixel detector local reconstruction was also refined, leading to a better physics performance, as will be illustrated throughout this section: we first investigate the tracking and vertexing performance (Sections~\ref{sec:tracking}-\ref{sec:vertexing}), then the presence of systematic distortions (Sections~\ref{sec:ZmmVal}--\ref{sec:overlapVal}).
The derivation of a final set of alignment constants for one data-taking year takes several weeks, and involves two to five people for running the programs and comparing the different configurations.

For the 2016 EOY reconstruction, the IOV boundaries were the same in the whole tracker. Ladders and blades were used as high-level alignables in the pixel detector. A global fit was first performed with \MILLEPEDE-II, and was further refined with \HIPPY using the same IOV boundaries to test the stability of the solution.
For the 2017 EOY reconstruction, the alignment constants were derived independently using either \HIPPY or \MILLEPEDE-II, but without time dependence within each period.
For the 2018 mid-year reconstruction (labelled `EOY reconstruction' in the figures), only \MILLEPEDE-II was used. Ladders and blades were used as high-level alignables in the pixel detector. In addition, the alignment fit was performed in two steps: first the entire tracker was aligned using $\sim$ 10 IOVs, corresponding to important changes. Then, the module parameters of the strip detector were fixed and only the module parameters of the pixel detector were fit using $\sim$ 80 IOVs.
Note that there is no EOY reconstruction for the data corresponding to the last 33\fbinv of the delivered integrated luminosity, because the derivation of the alignment for the legacy reprocessing started at the end of the year 2018, before the end of the data taking.

For the legacy reprocessing, \MILLEPEDE-II was used exclusively.
The definitions of the IOVs and of the alignables were unchanged for 2016 (Phase~0) with respect to the EOY alignment constants. For 2017 and 2018 (Phase~1), different high-level alignables were chosen for the pixel detector: ladders were used in BPIX and modules in FPIX. In addition, different IOV boundaries were chosen in the pixel and strip detectors. Additional IOVs were introduced in the pixel detector to absorb the bias from the hit reconstruction in the alignment constants.
Adapting the strategy based on the results of the studies described in Section~\ref{sec:WMs}, systematic deformations were avoided or reduced, except for the 2018 data-taking year. In the 2018 data, even though the dimuon vertex and invariant mass constraints had been included in the alignment procedure, the resulting alignment fit contained a residual but significant twist deformation. Therefore, in addition to performing the alignment fit with \MILLEPEDE-II, an inverted twist transformation was manually applied to the geometry to compensate for this systematic distortion. This is shown in Fig.~\ref{fig:twist}.

The different years required different strategies to obtain adequate performance for various reasons.
First, the different phases of the pixel detector imply different levels of precision. This leads to the observation of different tensions in the alignment. With more modules and being closer to the interaction point, the Phase-1 pixel detector was more demanding in terms of alignment precision and suffered more from the effects of radiation than the Phase-0 detector.
Similarly, the alignment for 2018 was more demanding than for 2017, likely because of the increase in instantaneous luminosity and therefore radiation levels.

\subsection{Performance}
In this subsection, the performance for different sets of alignment constants in different data-taking periods is illustrated.
The validation procedures already introduced in the text were repeated to compare the quality of the different sets of alignment parameters. 
In addition, the stability of the performance is investigated as a function of the integrated luminosity delivered by the LHC per run or per IOV. This is helpful to understand the ageing of the modules with the accumulation of radiation effects.

\subsubsection{Tracking performance} \label{sec:tracking}
\begin{figure*}[!ht]  
    \centering
    \includegraphics[width=0.5\textwidth]{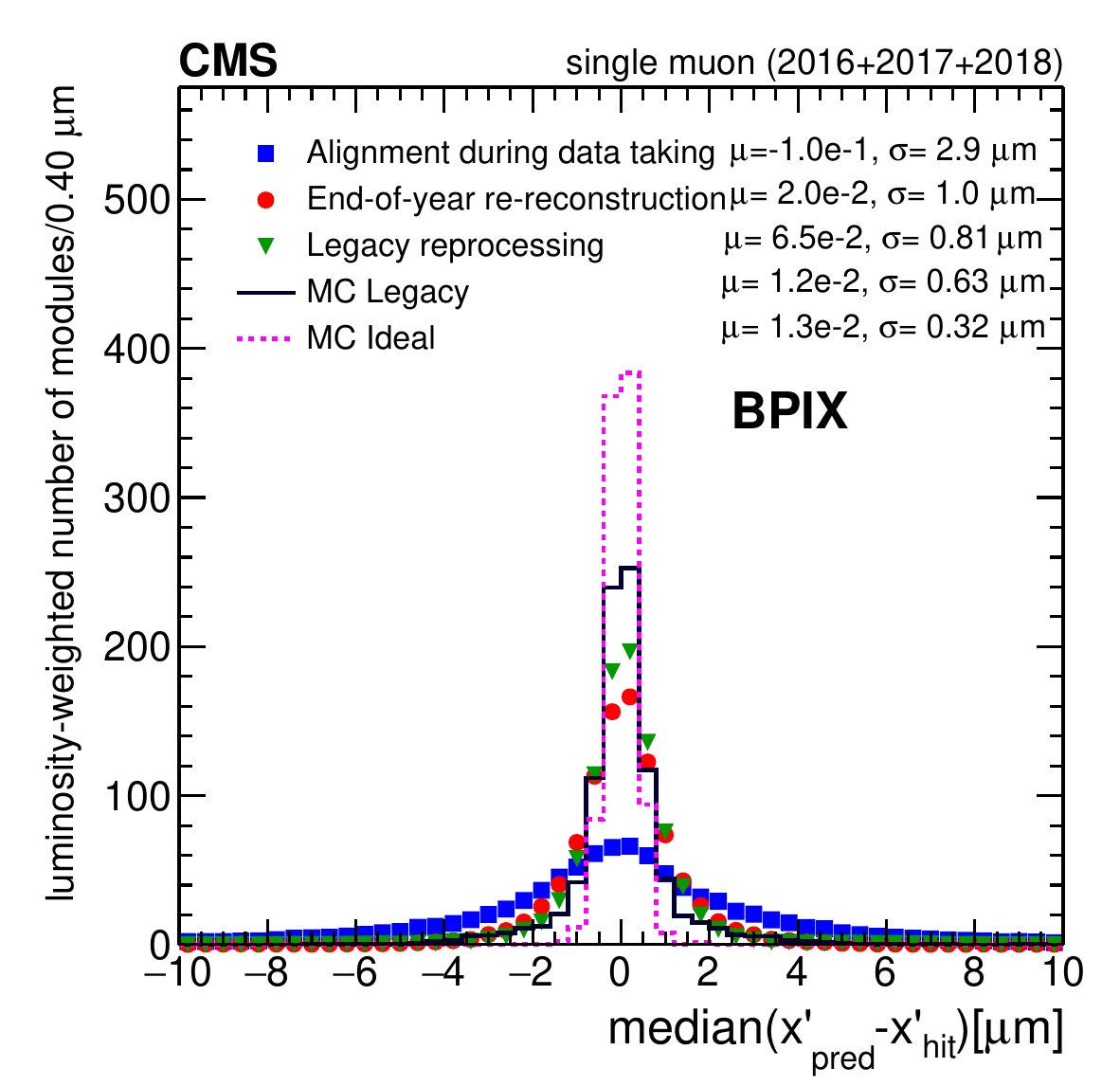}~
    \includegraphics[width=0.5\textwidth]{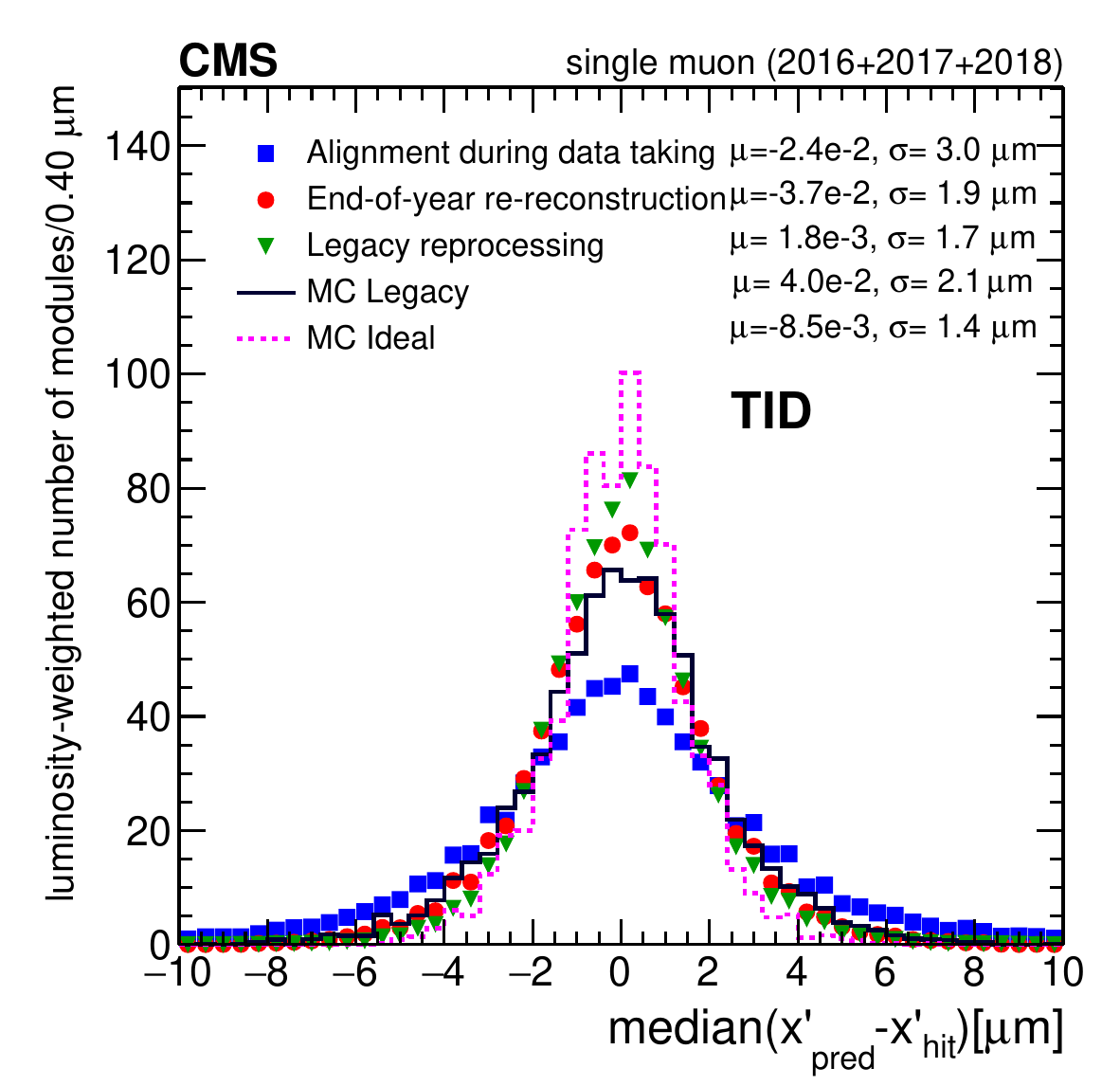}

    \caption{
        Distributions of per-module median track-hit residuals in the local~$x$ ($x'$) coordinate, for two subdetectors (BPIX and TID), produced with the single-muon data set. The distributions are averaged over all IOVs, where each IOV is weighted with the corresponding delivered integrated luminosity. The DMRs are shown for three different geometries in data. They are compared with the realistic MC scenario (black line) and the design MC scenario (magenta) evaluated in simulated isolated-muon events.
        The quoted means~$\mu$ and standard deviations~$\sigma$ are the parameters of a Gaussian fit to the distributions.
    }
    \label{fig:DMRperformance}
\end{figure*}

Figure~\ref{fig:DMRperformance} shows DMRs (introduced in Section~\ref{sec:2015}) along the local~$x$ coordinate of the modules belonging either to the BPIX or to the TID. These distributions are produced with the single-muon data set instead of the L1~inclusive sample that is usually used for DMRs. This is done to compare the performance in the data and simulated samples with similar event topologies, which is achieved by requiring exactly 100 hits per module in the simulation and in each IOV. Pixel detector (strip detector) modules with fewer than 100 hits per IOV over a period of data taking corresponding to an integrated luminosity of more than 2 (7)\fbinv were excluded from this study. Approximately 91\% of all functioning modules passed this condition. The distributions are then averaged over all IOVs weighted by the corresponding delivered integrated luminosity. A comparison is made to the realistic MC scenario derived for each year as well as the design MC scenario. These scenarios will both be introduced in Section~\ref{sec:MC}.
For an ideally aligned tracker, the DMRs are expected to peak at zero. This is visible for the MC simulation with the ideal alignment conditions, shown as the dashed magenta line in Fig.~\ref{fig:DMRperformance}. A nonzero average may indicate a systematic shift of the structure under scrutiny.

The position of the pixel detector is known to be very sensitive to changes of the conditions, \eg\ the temperature and the magnetic field.
Furthermore, the width of the DMRs is determined by several properties: the topology of the tracks; the number of hits per module, which is held constant; and the quality of the alignment. To focus on the performance related to the alignment, the DMRs are calculated using the exact same sample of tracks in each case.
The mean and the width of the DMRs are extracted using a Gaussian fit.
Figure~\ref{fig:DMRperformance} illustrates the improvement of the performance with the large number of parameters in the global fit. There were more IOVs in the legacy reprocessing than in the EOY reconstruction, and smaller time-dependent structures were used.
The improvement is largest for the legacy reprocessing.

\begin{figure*}[!ht]
    \centering

    \includegraphics[width=\textwidth]{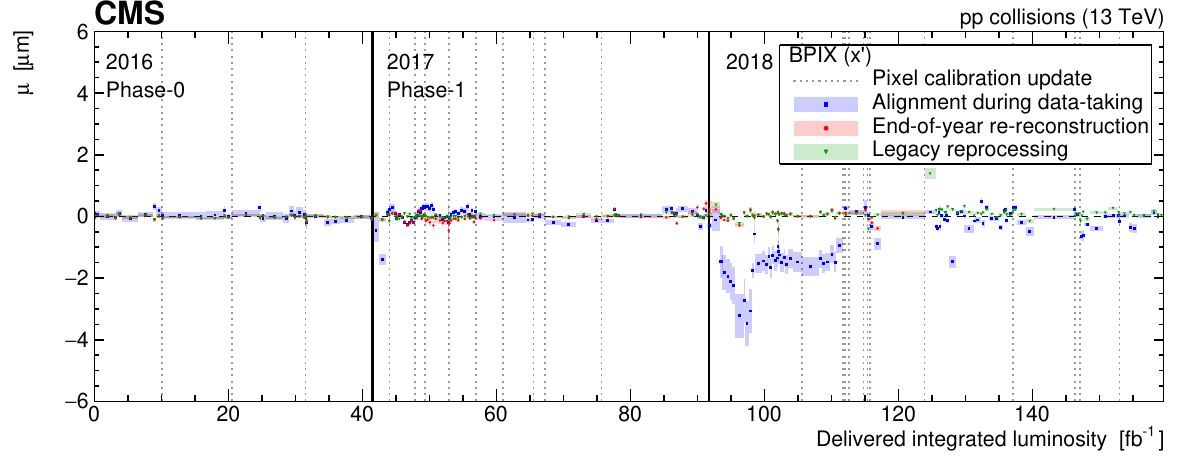}

    \includegraphics[width=\textwidth]{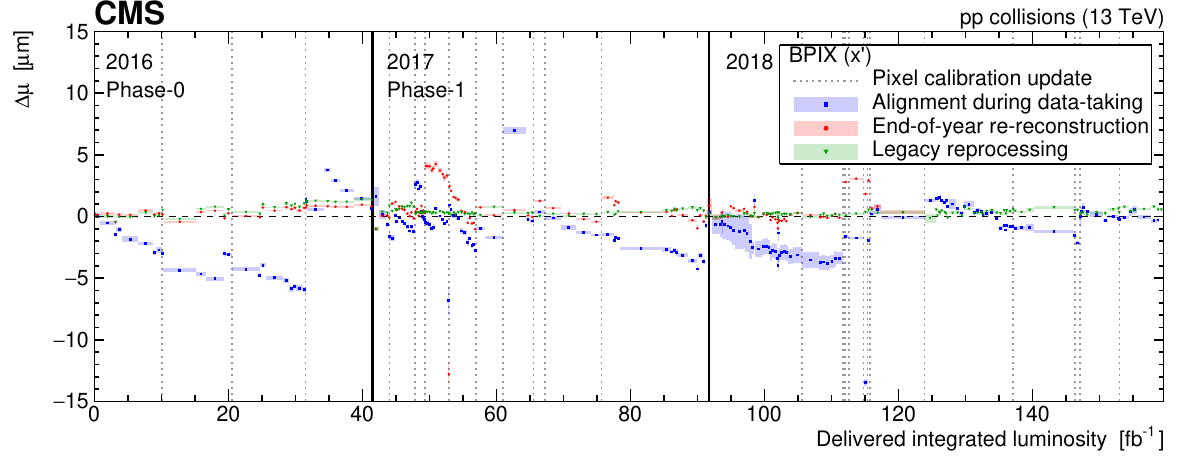}

    \caption{The DMR trends for the years 2016--2018, as functions of the delivered integrated luminosity, evaluated with a sample of data recorded by the inclusive L1 trigger.
        The upper figure shows the mean value of the distribution of median residuals for the local~$x$ ($x'$) coordinate in the BPIX detector.
        The lower figure shows the difference between the mean values $\Delta\mu$ obtained separately for the modules with the electric field pointing radially inwards or outwards. This quantity is also shown in the $x'$~coordinate.
        The shaded band indicates one standard deviation from the Gaussian fit of the corresponding DMR.
    }

    \label{fig:DMRtrend}
\end{figure*}

After the dedicated alignment for the Run~2 legacy reprocessing, the mean value~$\mu$ of the distribution of median residuals is shifted closer to zero and the mean difference shows improved stability.
The top panel of Fig.~\ref{fig:DMRtrend} shows $\mu$ extracted for the three different geometries, evaluated with a sample of data recorded by the inclusive L1 trigger, for each IOV as a function of the delivered integrated luminosity.

Because the direction of the Lorentz drift depends on the orientation of the modules, we also produce DMRs in each subdetector for the inward- and outward-pointing modules separately, and calculate the differences between the means of the DMRs, $\Delta \mu$.
A varying value is a hint of residual biases due to the accumulated effects from radiation in the silicon sensors, as already introduced in Section~\ref{sec:concepts}.
The bottom panel of Fig.~\ref{fig:DMRtrend} shows $\Delta \mu$ extracted for the three different geometries for each IOV as a function of the delivered integrated luminosity.
The finer granularity of the time dependence in the legacy reprocessing reduces the bias, since the rapidly changing shift from the local reconstruction can be absorbed in the position of the ladders and of the modules.
The improvement observed in Fig.~\ref{fig:DMRperformance} with respect to the EOY reconstruction is also directly related to this feature.

\begin{figure*}[!ht]
    \centering

    \includegraphics[width=0.49\textwidth]{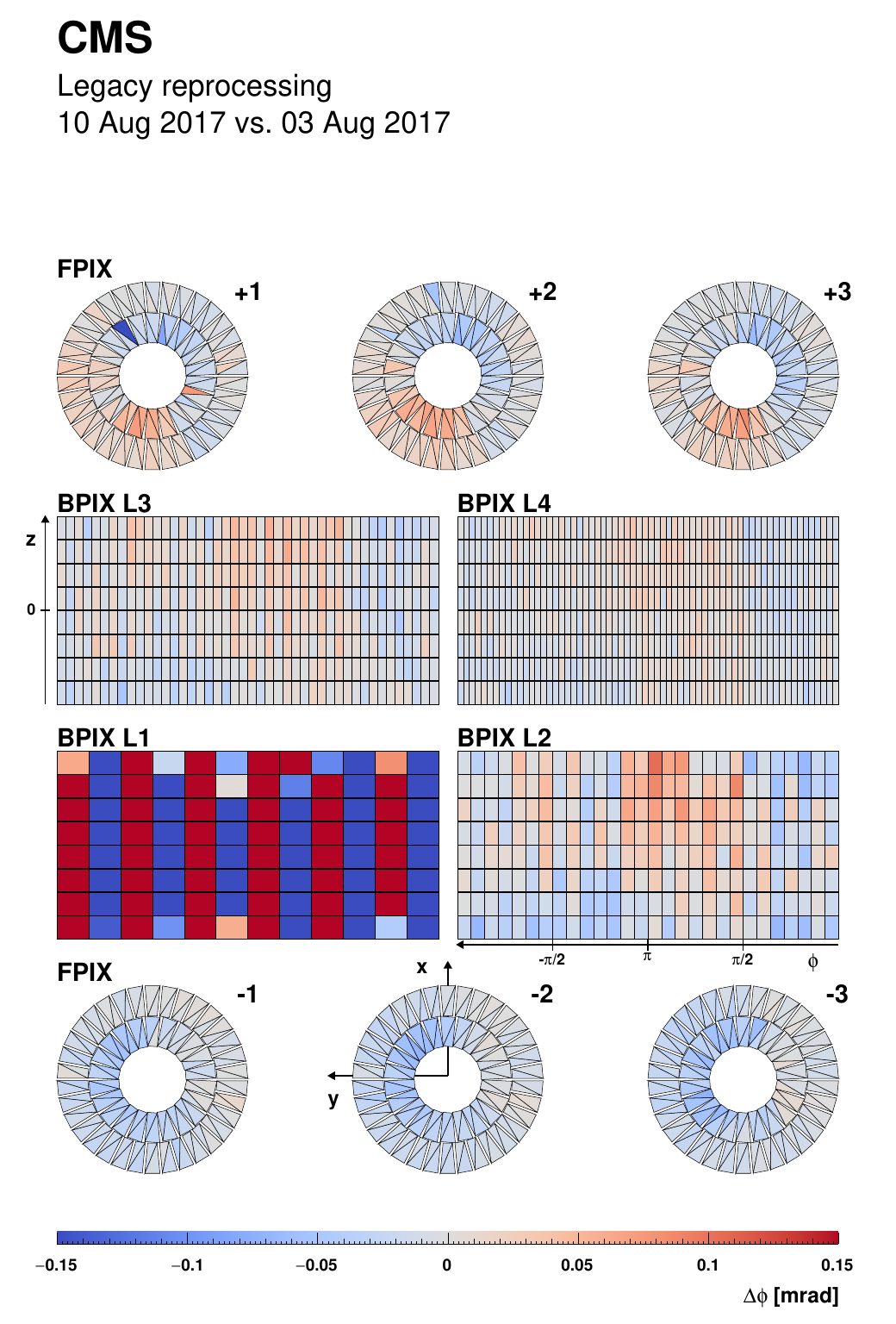}
    \includegraphics[width=0.49\textwidth]{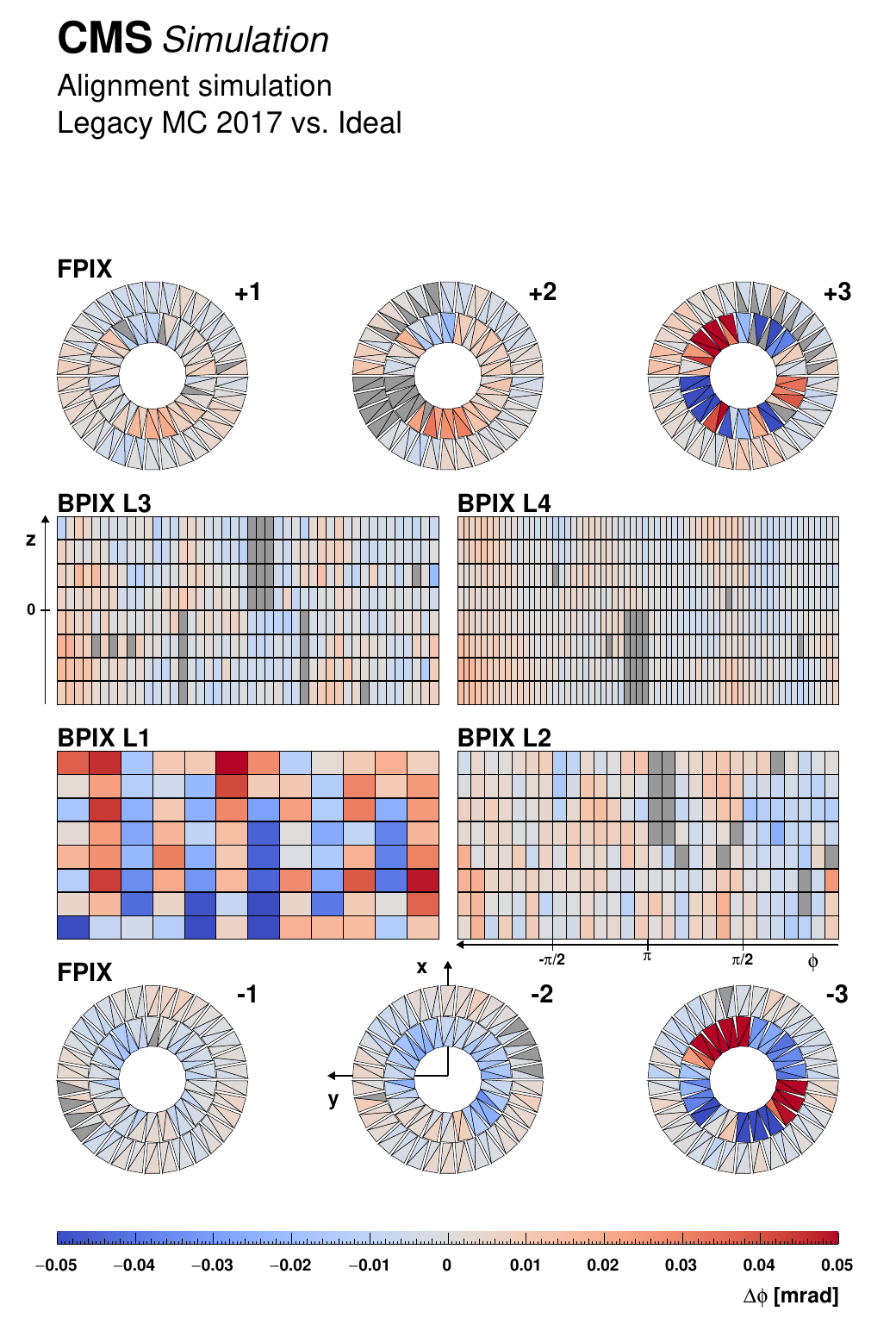}

    \caption{The value of $\Delta\phi$ for each module in the pixel detector, comparing the alignment parameters of the legacy reprocessing on 3 and 10 August 2017 (left) and comparing the alignment based on MC simulation for 2017 and the ideal detector (right).
        For each of the detector components, $\phi$ corresponds to the global coordinate.
        The colours denote the value of the $\Delta\phi$ movement, as shown by the bar at the bottom.
        These values are capped between $-0.15$ and 0.15\unit{mrad} for the alignments in data (left) and between $-0.05$ and 0.05\unit{mrad} for simulation (right).
        In the figure on the right, modules that were inactive in the simulation are indicated in dark grey.
        The alternating pattern visible in layer~1 of the BPIX is caused by radiation damage that is absorbed in the modules between local calibration updates of the pixel modules.
        Due to the opposite orientations of the neighbouring ladders, an alternating pattern is created.
        Radiation damage is more severe in the first layer which is closer to the interaction point, making the pattern more visible.
}
    \label{fig:TkAlMap_2017RD}
    \label{fig:TkAlMap_MC2017}
\end{figure*}

This effect of the radiation damage is shown in Fig.~\ref{fig:TkAlMap_2017RD}, where two different IOVs of the legacy reprocessing are compared with one another by showing $\Delta\phi$ for each module of the pixel detector.
The IOVs were chosen such that the local calibration of the pixel modules was constant in the time interval between them. They correspond to the first and last set of alignment parameters of the 5th pixel IOV in 2017, where the $\Delta \mu$ from Fig.~\ref{fig:DMRtrend} indicates large radiation effects.
The legacy reprocessing includes more updates during this period, which means the effects of accumulating radiation damage can be absorbed in the alignment. This results in an alternating pattern in the ladders, since the modules of neighbouring ladders have opposite orientations.
This pattern is especially visible in layer~1 of the BPIX. It is less visible in the next layers, where the effect is reduced due to the distance and  is folded in with other effects. Such other effects include movements of the high-level structures, \eg the half barrels of the BPIX.

The understanding of the interplay of the local reconstruction and of the alignment procedure is a great improvement in the physics performance in the legacy reprocessing.

\subsubsection{Vertexing performance} \label{sec:PV} \label{sec:SplitVertex} \label{sec:vertexing}
\begin{figure*}[!ht]
    \centering
    \includegraphics[width=\textwidth]{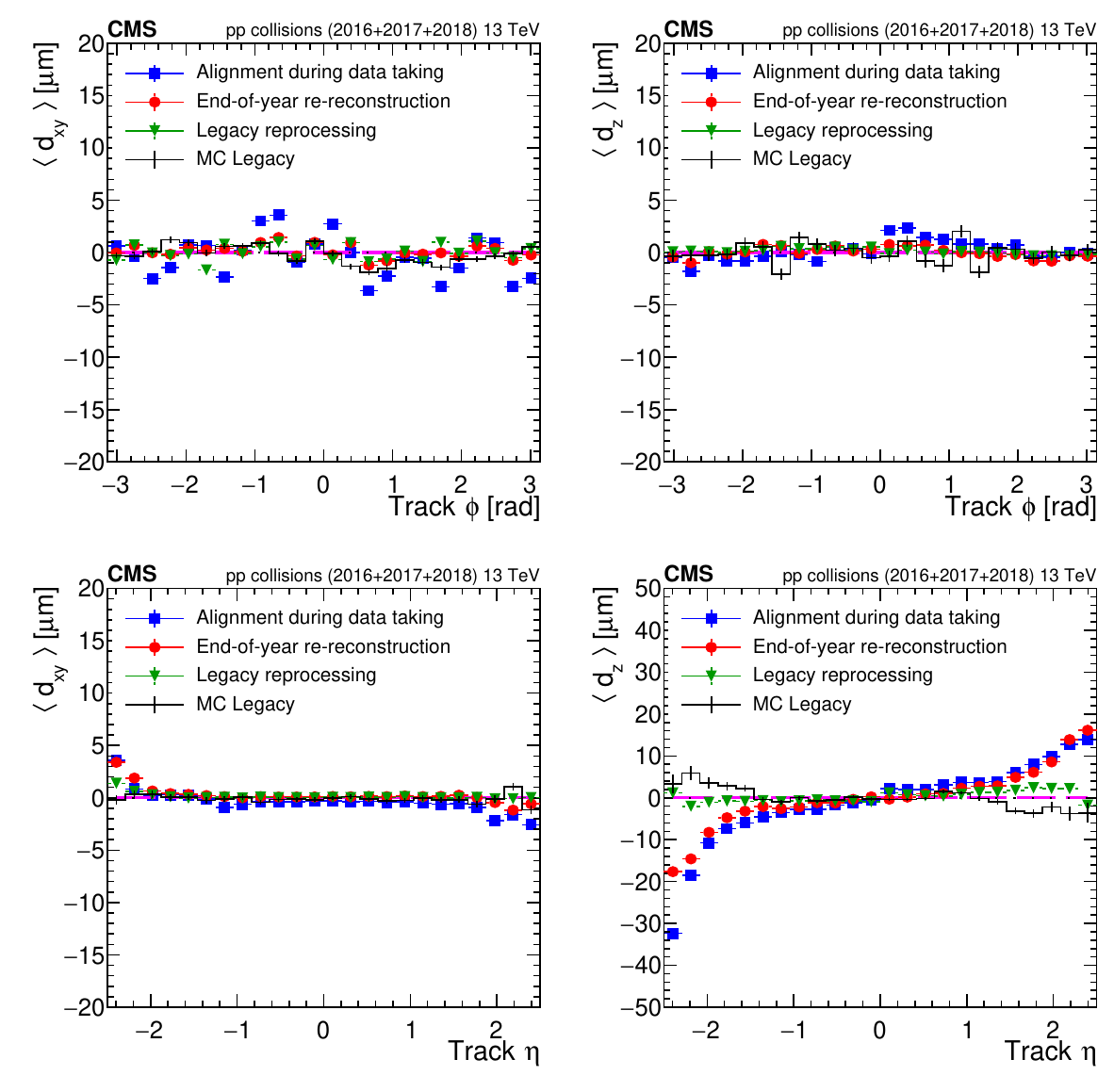}
    \caption{
        Mean track-vertex impact parameter in the transverse plane $\dxy$ (left) and in the longitudinal plane $\dz$ (right), as a function of track $\phi$ (top) and $\eta$ (bottom).
        The impact parameters are obtained by recalculating the vertex position after removal of the track under scrutiny and considering the impact parameter of this removed track.
        Only tracks with $\pt > 3 \GeV$ are considered.
        These distributions are averaged over all runs of 2016, 2017, and 2018 after scaling them with the corresponding delivered integrated luminosity for each run.
        Three alignment geometries in data are compared with the realistic MC alignment scenario evaluated in a sample of simulated inclusive L1 trigger events (black points) scaled to the corresponding luminosity delivered in 2016, 2017, and 2018. The error bars represent the statistical uncertainties due to the limited number of tracks. In case of data points statistical uncertainties are smaller than size of the displayed markers.
    }
    \label{fig:PVperformance}
\end{figure*}

The distributions showing the impact parameters of tracks were already introduced in Section~\ref{sec:2015}.
Figure~\ref{fig:PVperformance} shows the average unbiased track-vertex distance along the beam axis~$\langle \dz \rangle$ and in the transverse plane~$\langle \dxy \rangle$, as a function of the $\eta$ and $\phi$ of the tracks. Only tracks that satisfy $\pt > 3 \GeV$ are considered.
The distributions are averaged over all runs and weighted with the corresponding delivered integrated luminosity per run, normalized to the total integrated luminosity delivered in 2016--2018.
The results are shown for the three geometries and are compared with the realistic MC scenario derived for each year, which will be described in Section~\ref{sec:MC}.
The distributions are expected to be flat and compatible with zero for an ideally aligned tracker.
The impact parameter as a function of $\phi$ shows some improvement with the EOY reconstruction, but the performance as a function of $\eta$ is only improved with the legacy reprocessing.
The remaining residual deviations from the ideal case observed in the legacy reprocessing at high absolute pseudorapidity can be explained by a limited tracking performance related to the pixel detector local reconstruction in the data. The deviation visible in the simulation will be discussed in Section~\ref{sec:MC}.

\begin{figure*}[!ht]
    \centering

    \includegraphics[width=\textwidth]{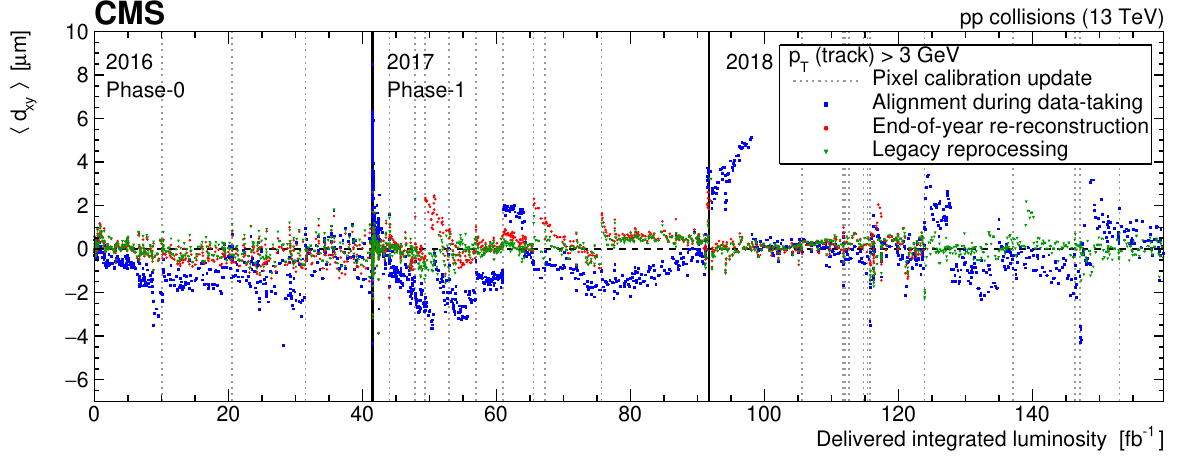}

    \includegraphics[width=\textwidth]{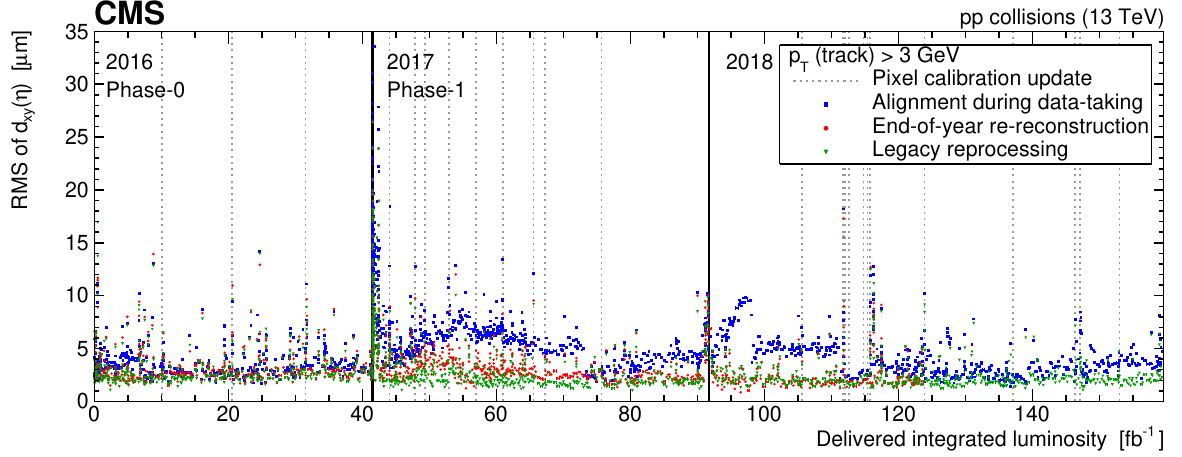}

    \caption{Impact parameter trends in the transverse plane~$\dxy$ as a function of the delivered integrated luminosity.
        Only tracks with $\pt > 3 \GeV$ are considered.
        The upper figure shows the average $\dxy$; the lower figure shows its RMS in bins of the track $\eta$.
    }
    \label{fig:PVtrends}
\end{figure*}

Similarly to the DMR trends, we also calculate impact parameter trends.
Figure~\ref{fig:PVtrends} shows the average and the spread of the distributions on a run-by-run basis, as a function of the delivered integrated luminosity for the three geometries.
The suboptimal tracking performance during the first few inverse picobarns of the 2017 $\pp$~collision run corresponds to the commissioning of the upgraded pixel detector. This is visible in the degraded impact parameter bias around the thick vertical line, which indicates the Phase-1 upgrade.
Apart from this short period, aligning the tracker improves the mean of this distribution.
Short IOVs with a suboptimal configuration of the pixel local reconstruction, \eg~different high-voltage settings or local reconstruction parameters that are inconsistent with the alignment, can give rise to isolated peaks in the trends, especially for the alignment derived during data taking.
The slopes in the RMS trend that are visible between two pixel calibration updates for the alignment during data taking are due to residual radiation effects not included in the dedicated calibration, \eg at the beginning of 2018. This causes the Lorentz drift to change rapidly. This residual effect can be corrected only by aligning with a finer granularity than the automated alignment implements.
In general, the improved performance obtained with the legacy reprocessing is also related to the more granular configuration of the alignment fit in comparison with the alignment during data taking and for the EOY reconstruction. This is especially the case in 2017.

\begin{figure*}
    \centering
    \includegraphics[width=\textwidth]{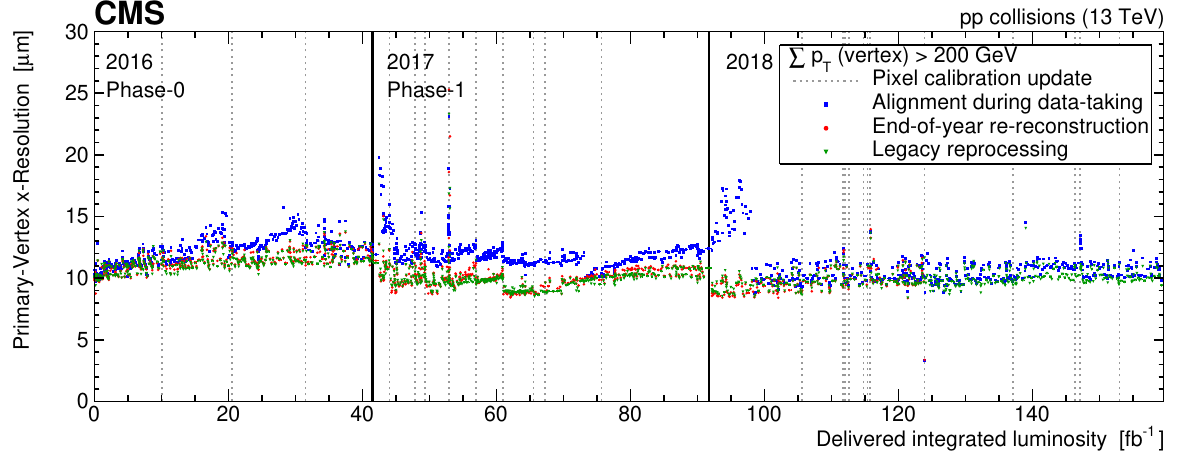}

    \includegraphics[width=\textwidth]{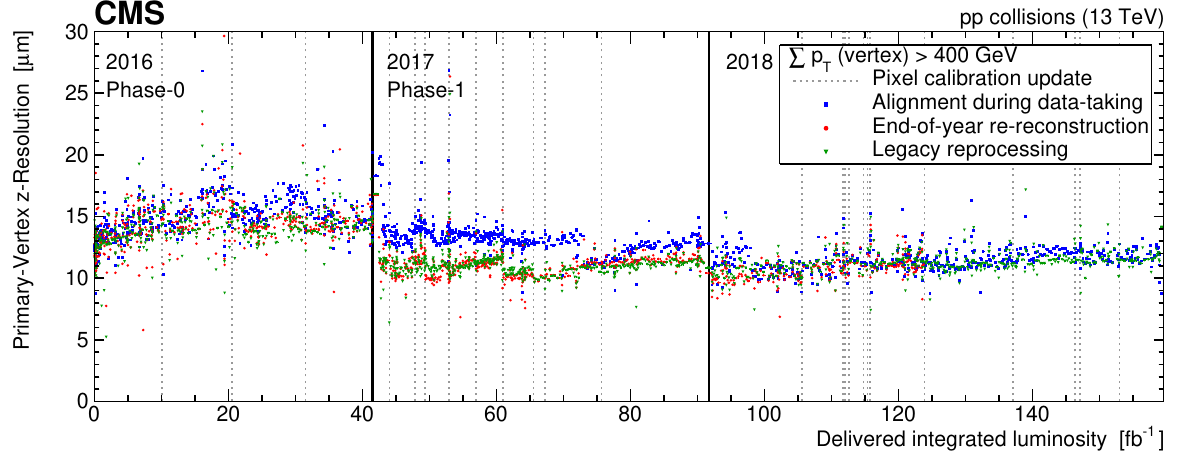}

    \caption{
        Primary vertex resolution in the $x$ (top) and $z$ (bottom) directions, calculated using refitted vertices with $\sum \pt > 200 \GeV$ ($x$) and $\sum \pt > 400 \GeV$ ($z$) in $\pp$~collisions.
        The vertical black lines indicate the first processed runs for the 2016, 2017, and 2018 data-taking periods. The vertical dotted lines indicate changes in the pixel tracker calibration.
    }
    \label{fig:SplitVertexOverview}
\end{figure*}

To obtain information about the PV resolution, the set of tracks assigned to a vertex is split into two independent data sets.
For this, instead of the usual inclusive L1 trigger sample, we use a sample populated with tracks at high~\pt.
The tracks are sorted according to their \pt and assigned to a pair starting from the track with the highest~\pt. From each pair, one track is randomly placed into one of the new samples to ensure similar kinematic attributes. 
For both new data sets, the adaptive vertex fitter is run and the residual of the PV resolution is calculated as the difference between the results.
This is done for several values of $\sum \pt$ for all tracks in the respective data set, and the distribution of residuals is fitted with a Gaussian function.
This scalar \pt sum is used to categorize the vertices, because the resolution improves when a larger overall \pt is assigned to a vertex.
The PV resolution is defined as the mean of the fitted Gaussian distribution.
In Fig.~\ref{fig:SplitVertexOverview}, the resolution for all runs recorded in 2016, 2017, and 2018 is shown using the three usual geometries, for the $x$ and $z$ vertex coordinates. As an illustration, two different selections in terms of the minimum track \pt sum are shown. The maximum resolution for each vertex coordinate is obtained for $\sum \pt > 400 \GeV$, which is approximately 15\% better than in the case of the looser selection $\sum \pt > 200 \GeV$.    
The conclusions are similar to those obtained with the impact parameter trends:
an improvement corresponding to the upgrade of the pixel detector is observed, especially in the $z$~direction, and isolated peaks correspond to short IOVs with a suboptimal configuration of the pixel local reconstruction.
Although the EOY reconstruction shows some improvement with respect to the alignment during data taking, the legacy reprocessing seems to improve only the performance of certain outliers.

\subsubsection{Dimuon invariant mass reconstruction} \label{sec:ZmmVal}
\begin{figure}[!ht]
    \centering

    \includegraphics[width=0.5\textwidth]{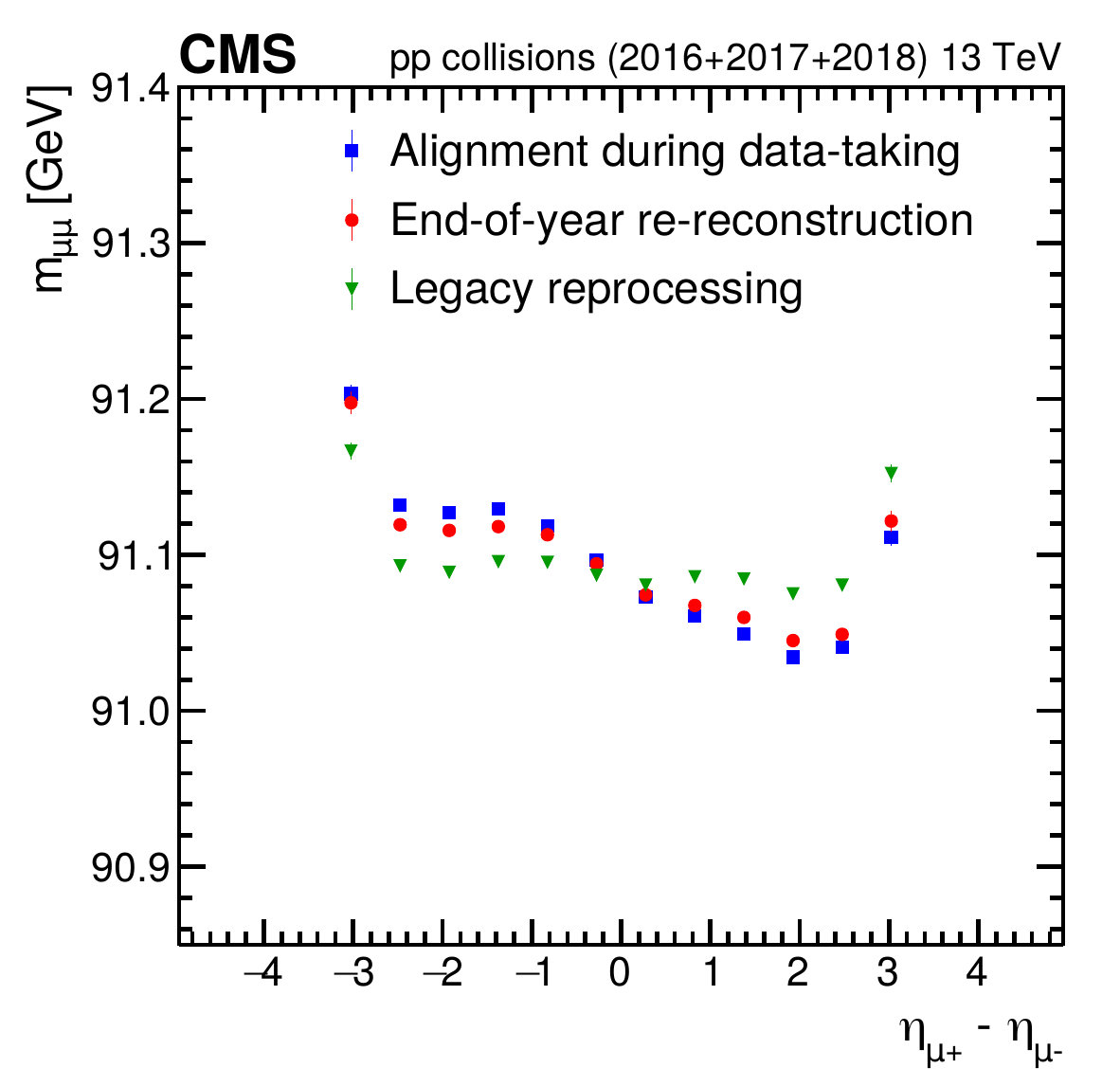}

    \caption{
         Reconstructed \PZ~boson mass as a function of the difference in $\eta$ between the positively and negatively charged muons, calculated from the full sample of dimuon events in the years 2016, 2017, and 2018. The error bars show the standard deviation of the invariant \PZ~boson mass as retrieved from a fit of dimuon mass distribution to a Breit-Wigner convolved with a Crystal Ball function.
    }

    \label{fig:DeltaEtaTwist}
    
\end{figure}

The dimuon mass validation using $\Ztomm$ events was already discussed in Section~\ref{sec:WM}. 
Beyond the improvement of the local precision, another improvement brought by the legacy reprocessing is related to the systematic distortions, as demonstrated in Fig.~\ref{fig:DeltaEtaTwist}.
The sigmoid shape, typical of a twist distortion, is reduced in the legacy reprocessing when compared with the alignment during data taking and to the EOY reconstruction.
Although the EOY reconstruction of the 2016 data did not suffer from such a distortion, that of the 2017 and 2018 data did suffer from it. For 2017, the distortion was improved by performing the alignment fit with \MILLEPEDE-II in the legacy reprocessing. However, the distortion observed in 2018 was not removed by the same procedure and a twist transformation was applied in the opposite direction in addition to running \MILLEPEDE-II for the legacy reprocessing.
The reason for the need for extra processing in 2018 is not understood to date.

\begin{figure*}
    \centering 
    \includegraphics[width=0.5\textwidth]{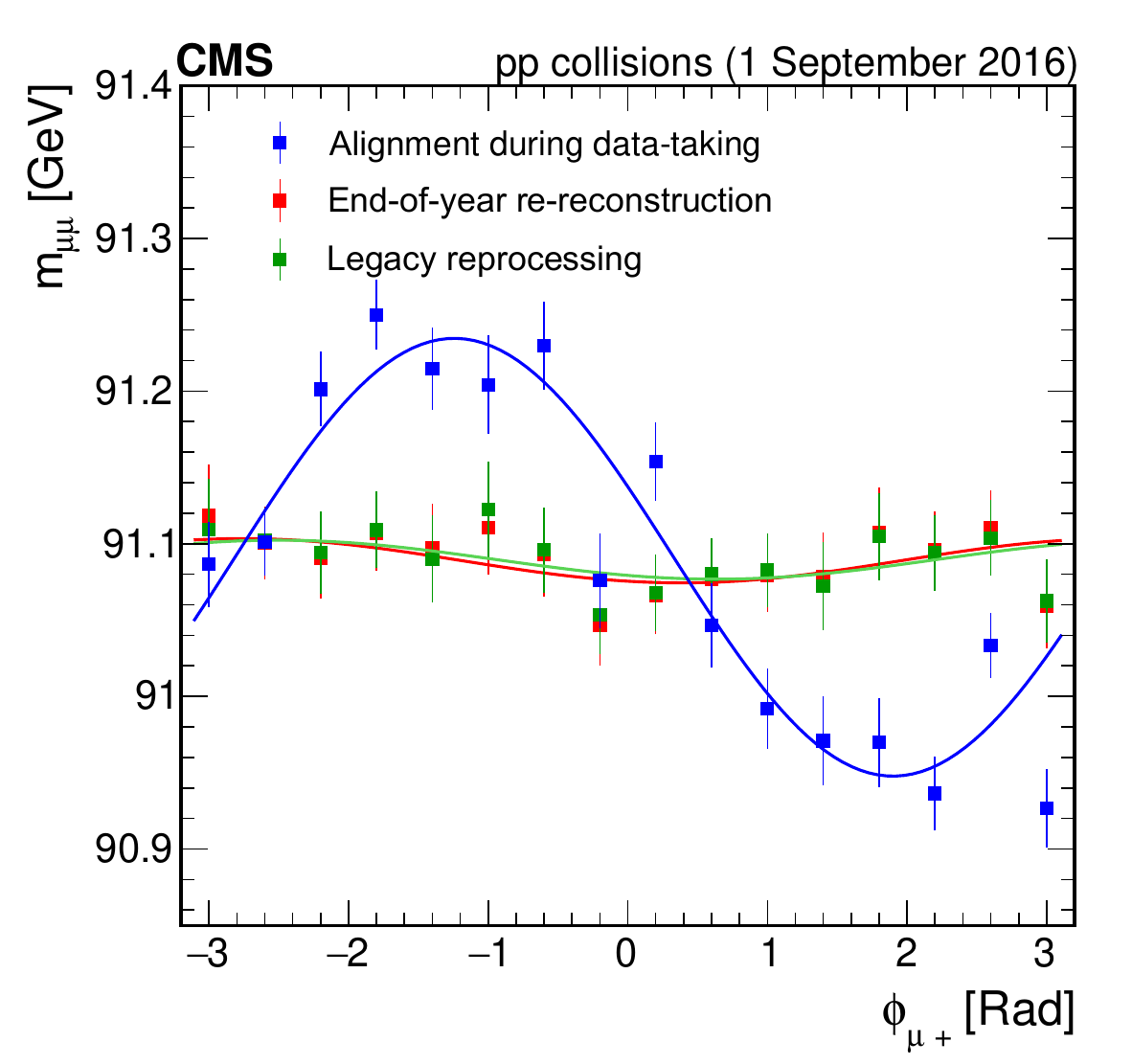}

    \includegraphics[width=\textwidth]{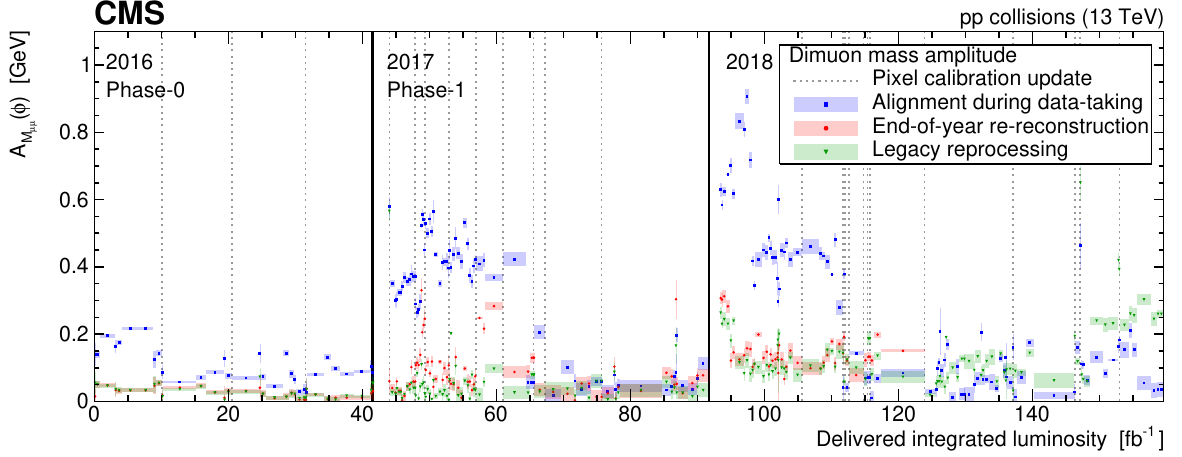}

    \caption{
        The upper figure shows the invariant mass of the dimuon system, as a function of the azimuthal angle of the positively charged track for a single~IOV.
        The lower figure shows the amplitude $A$, obtained by fitting the invariant mass of the dimuon system versus $\phimup$ with a function of the form $A \cos{\left(\phi + \phi_{0}\right)} + b$ as a function of the delivered integrated luminosity.
        The vertical bars on the points in the upper figure show the uncertainty in the average $\mmumu$ of a given $\phimup$ bin. 
	The shaded bands in the lower figure show the uncertainty in the fitted parameters calculated by a $\chi^2$ regression. 
    }

    \label{fig:DiMuon_IOV} \label{fig:Amp_DiMuon_Run2}
    
\end{figure*}

Further systematic distortions have been reduced with the legacy reprocessing.
For instance, the bias in the reconstructed mass as a function of $\phi$ due to a systematic distortion is periodic to first order, and the distributions are easily fit with a cosine function.
As such, the amplitude of the fitted cosine function is a good measure for quantifying the magnitude of the bias in the reconstructed dimuon mass.
The top figure of Fig.~\ref{fig:DiMuon_IOV} demonstrates the periodicity of the reconstructed \PZ~boson mass, whereas the bottom figure of Fig.~\ref{fig:Amp_DiMuon_Run2} shows the amplitude of the reconstructed \PZ~boson mass as a function of the delivered integrated luminosity.
The amplitude shows the average spread of the reconstructed \PZ~boson mass with respect to $\phimup$, which is expected to be zero in a well-aligned detector. A nonzero amplitude indicates that the reconstructed mass has some dependence on the spatial coordinates of the detector.
We observe an improvement in the legacy reprocessing compared with the earlier alignment procedures.
However, the legacy reprocessing still shows a suboptimal performance in the 2018 data-taking year in comparison with the two other years. In addition to this, from the point of view of Fig.~\ref{fig:Amp_DiMuon_Run2}, the legacy reprocessing shows worse performance than the alignment during data taking, especially in the last 30\fbinv. This is suspected to be related to the discrepancies between different IOVs with different configurations while operating the pixel detector, \eg a change of voltage or annealing.

\subsubsection{Cosmic ray muon track reconstruction} \label{sec:cosmics}
\begin{figure*}
    \centering

    \includegraphics[width=0.5\textwidth]{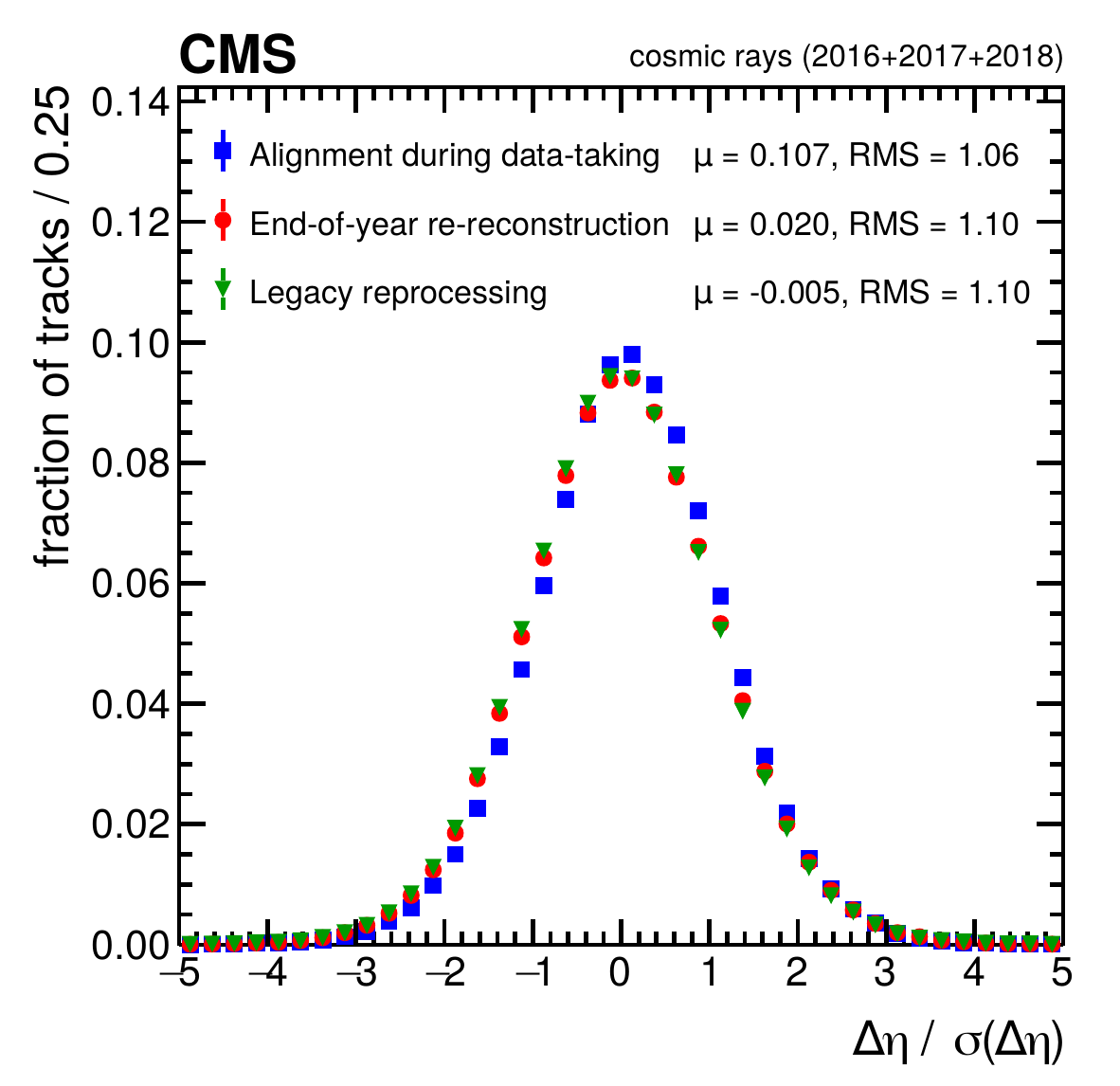}~
    \includegraphics[width=0.5\textwidth]{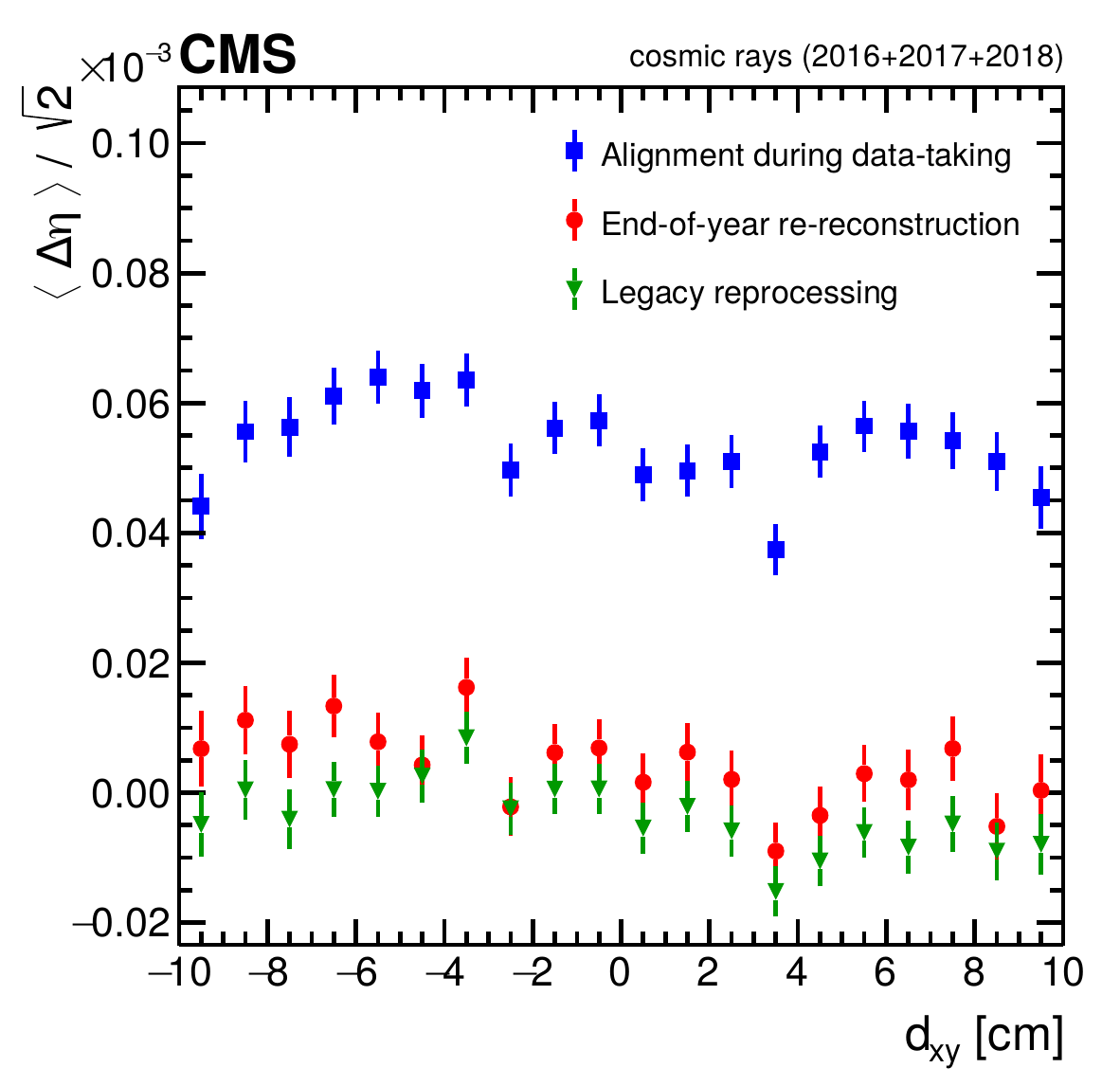}

    \caption{Performance results for cosmic ray muon tracks recorded during commissioning and interfill runs at 3.8\unit{T} during 2016, 2017, and 2018. The top and bottom halves of the cosmic ray track are reconstructed independently and the track parameters are compared at the point of closest approach to the interaction region.
        The mean and RMS of the distribution of $\Delta\eta$ relative to its uncertainty are shown in the figure on the left.
        The mean $\eta$~difference between the two tracks is presented as a function of $\dxy$ on the right, scaled down by $\sqrt{2}$ to account for the two independent measurements. The error bars show the statistical uncertainty related to the limited number of tracks.
    }

    \label{fig:MTS}
\end{figure*}

As was mentioned in Section~\ref{sec:WM}, cosmic ray muon tracks are one of the key ingredients used to control systematic distortions in the alignment procedure.
The quality of the alignment constants, and in particular the presence of certain systematic distortions, can also be better assessed by studying the performance of the reconstruction of cosmic ray muon tracks.
Figure~\ref{fig:MTS} shows $\Delta \eta$ and the difference in the impact parameter in the transverse plane between the two half tracks refitted from the hits of a cosmic ray muon traversing the detector.

From this perspective, the strategy followed in the legacy alignment procedure has led to better performance. 
In particular, the improvement of the distribution of $\Delta \eta$ relative to its uncertainty is related to the improved statistical precision of the alignment fit. The reduction in the difference between the impact parameters is related to the reduction of systematic distortions; in particular, for the left figure, it can be interpreted as a reduction of the telescope WM.

\subsubsection{Overlap validation} \label{sec:overlapVal}
\begin{figure*}
   \centering
   \includegraphics[width=\textwidth]{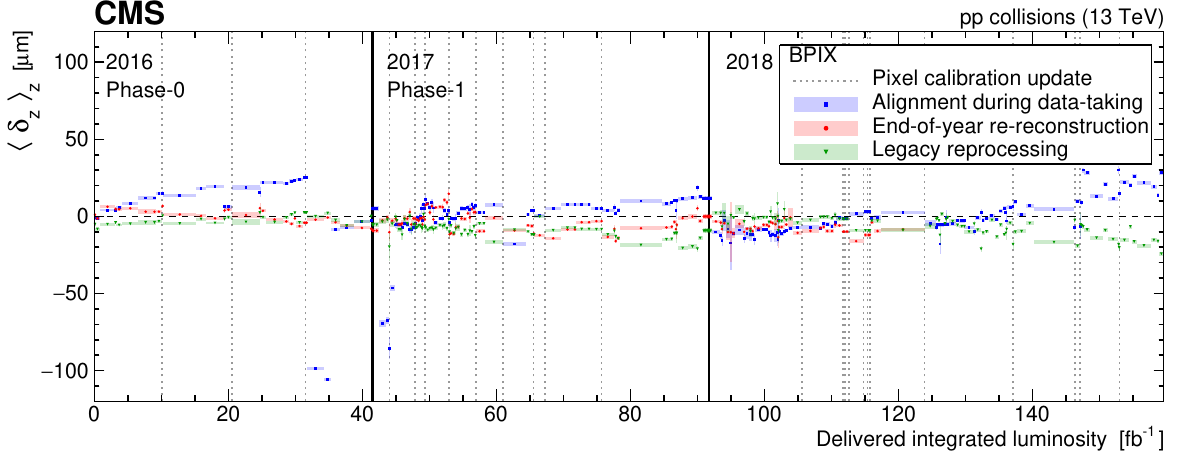}

   \includegraphics[width=\textwidth]{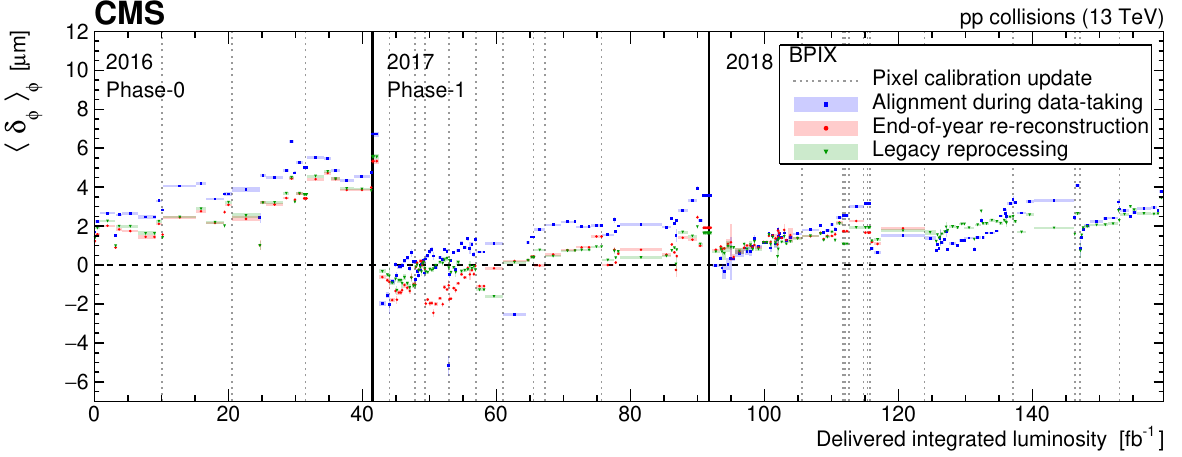}

   \caption{
       The upper (lower) figure shows the mean difference in residuals in the $z$ ($\phi$)~direction for modules overlapping in the $z$ ($\phi$)~direction in the BPIX, $\langle \delta_z \rangle$ ($\langle \delta_\phi \rangle$), as a function of the delivered integrated luminosity. The error bars show the statistical uncertainty in the mean of distribution of the residuals. These residuals are calculated using a sample of data recorded with the inclusive L1 trigger.
   }

   \label{fig:ZZBPIX}\label{fig:ZPhiBPIX} \label{fig:overlapUL}
   
\end{figure*}

The overlap validation was already introduced in Section~\ref{sec:WM}.
Figure~\ref{fig:overlapUL} shows the mean overlap residuals as a function of the delivered integrated luminosity.
The nonzero values, even for the legacy reprocessing, illustrate the limitations of the strategy followed for the alignment fit, where the temporal changes due to the irradiation of the modules are included in the alignment parameters. The biases that were introduced by the radiation damage are artificially absorbed in the alignment constants. Correlations between shifts in the mean overlap in Fig.~\ref{fig:ZZBPIX} and $\Delta\mu$ in Fig.~\ref{fig:DMRtrend} demonstrate that the mean overlap is also an effective measure of the Lorentz angle calibration bias.
Alternatively, if any observed deviations are assumed to be caused purely by systematic distortions, constraints can be placed on the maximal magnitude of the distortions in the tracker. These constraints represent a worst-case misalignment scenario, where no biases are introduced by radiation damage.
In the top row (bottom row) of Fig.~\ref{fig:ZZBPIX}, we observe deviations from zero in the legacy reprocessing ranging from $-21\mum$ to $+2\mum$ (from $-1.9\mum$ to $+5\mum$), which correspond to longitudinal (radial) expansions or contractions ranging from $-105\mum$ to $+10\mum$ at $z = 26\cm$ ($-5.7\mum$ to $+16.5\mum$ at $r = 16\cm$).

\subsubsection{Barycentre of the barrel pixel detector} \label{sec:barycentre}
\begin{figure*}
    \centering

    \includegraphics[width=\textwidth]{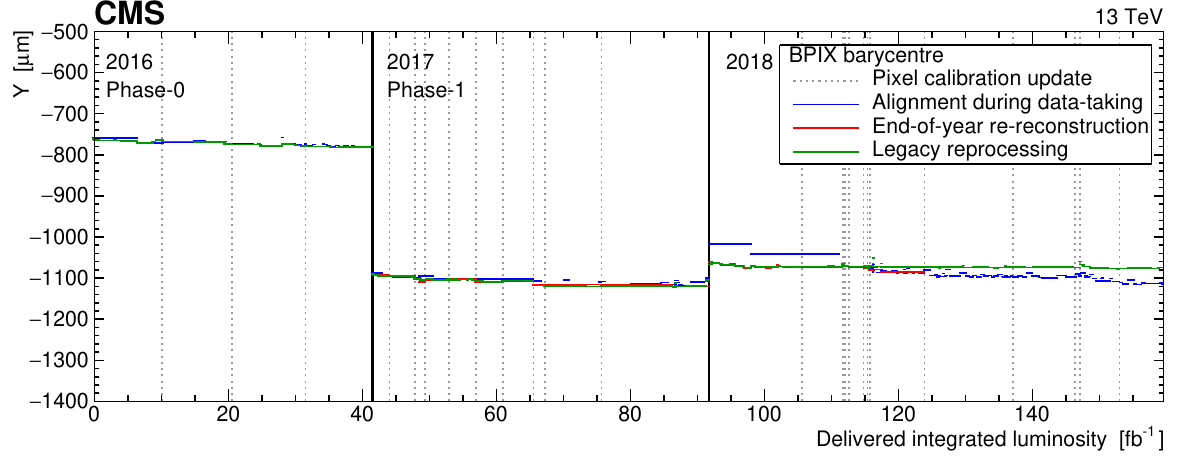}

    \caption{The global $y$~coordinate of the barycentre position of the barrel pixel detector as a function of the delivered integrated luminosity, determined as the centre-of-gravity of the modules in the barrel pixel detector only.
    }
    \label{fig:barycentre}
\end{figure*}

The barycentre position is determined as the centre-of-gravity of the pixel modules, either considering all pixel modules or barrel pixel modules only. The position of the BPIX barycentre is shown in Fig.~\ref{fig:barycentre} as a function of the delivered integrated luminosity.
The barycentre position is extracted directly from the alignment parameters.
The differences between the years illustrate the precision of the mechanical mounting of the detector. During the year, the position is constant within a few \mum, especially for the legacy reprocessing.
Large differences in the position at the beginning of the 2017 and 2018 data-taking periods are caused by the extraction and reinstallation of the pixel detector during the shutdowns. This was done for the Phase-1 upgrade in 2017, and for module replacements in 2018.
The alignment during data taking, which does not include corrections related to the accumulation of the radiation damage, introduces an artificial shift in the barycentre position.
This shift is reduced in the legacy reprocessing, where we have shown that the effects of radiation damage were more successfully included in the alignment procedure.

\subsubsection{Alignment parameter uncertainties} 
The APUs account for the uncertainty in the positions of the modules derived from the alignment fit. 
The difference between the real and measured module positions is equivalent to a residual misalignment that causes a widening of the distributions of the track-hit residuals, which are given by $r = m - f(\mathbf{p},\mathbf{q})$. 
Therefore, to estimate the contribution from the misalignment to the hit residual distribution, the normalized hit residual resolution $\sigma_r$ is calculated. The resolution squared is given by the quadratic sum of the cluster position estimation~(CPE) uncertainty and the uncertainty of the prediction by the track fit, excluding the hit under study:
\begin{equation}
\sigma_r^2 = \sigma_\text{hit}^2 + \sigma_\text{trk}^2.
\end{equation}
The normalized hit residual distribution, $\frac{r}{\sigma_r}$, should have the same width as the design simulation for a perfectly aligned tracker, with a broader and shifted distribution in case of misalignment.

To compensate for the broadening of the hit residual distribution caused by the misalignment, an additional uncertainty $\sigmaalign$ is introduced, which is added in quadrature to the initial resolution:

\begin{equation}
\sigma_{r'}^2 = \sigma_r^2 + \sigmaalign^2.
\end{equation}

Since the additional alignment uncertainty is also included in the tracking covariance matrix and influences the hit association windows used in pattern recognition, $\sigmaalign$ affects the track reconstruction itself.
This is included by calculating $\sigmaalign$ iteratively, where the track candidate is refitted for each iteration, using the previous estimate of $\sigmaalign$. To ensure convergence, 15 iterations are performed, which is sufficient. 

To account for possible incorrectly estimated track fit uncertainties or deficits in the CPE parameterization, the width of the normalized hit residual distribution is compared with values obtained from ideal simulation. The APUs are then determined in such a way that the width of the distribution of $r/\sigma_{r'}$ matches that of the ideal simulation. If the normalized hit residual distribution is not wider than the one for the ideal case, the final $\sigmaalign$ value is set to zero. 

\begin{figure*}[htb!p]
    \centering

    \includegraphics[width=\textwidth]{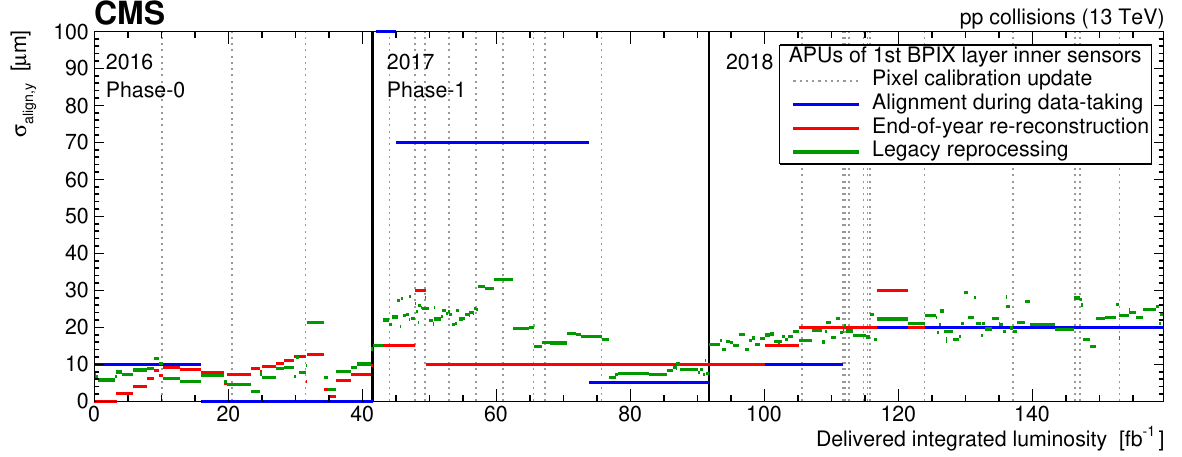}

    \includegraphics[width=\textwidth]{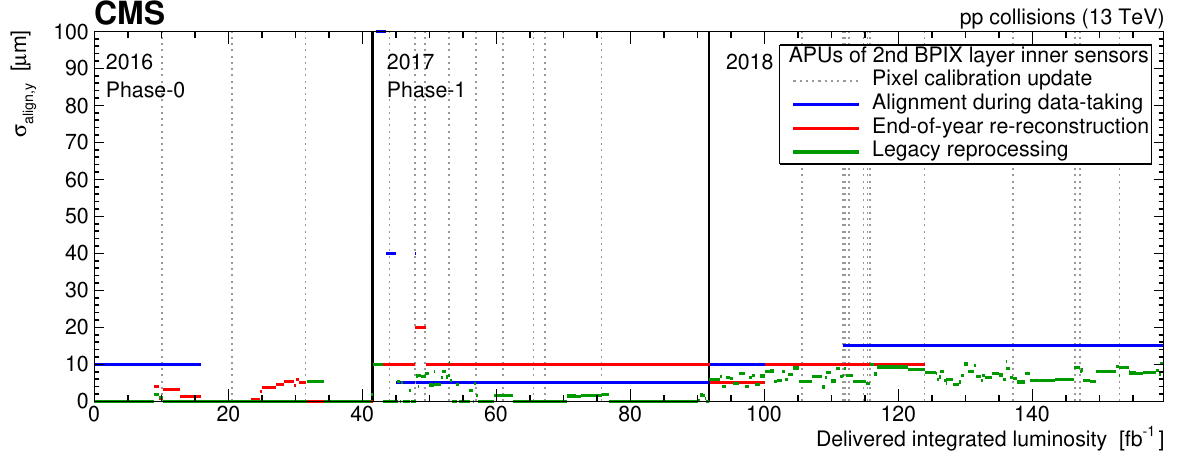}

    \caption{The contribution from the misalignment of the sensors to the total hit resolution for the inner ladders of the first (top) and second (bottom) BPIX layers in the local~$y$ coordinate as a function of the delivered integrated luminosity.
        For the legacy reprocessing, the measurements were performed with a higher granularity than in the other two cases.
    }

    \label{fig:APE_trend1In}\label{fig:APE_trend2In}
\end{figure*}

The contribution from the misalignment of the sensors to the total hit resolution for the inner ladders of the first and second BPIX layers in the local~$y$ coordinate is shown in Fig.~\ref{fig:APE_trend1In} for data taken from 2016 to 2018, as a function of the delivered integrated luminosity.
The alignment uncertainty mainly varies between 0 and 30\mum and is only larger for the alignment during data taking at the beginning of data taking in 2017, because the pixel detector was replaced.
For the legacy reprocessing, the measurements were performed with a higher granularity.
The legacy measurements might have larger values compared with earlier reconstructions because of the higher granularity and because a decrease of the CPE uncertainty can result in an increase of $\sigmaalign$.
The resolution of the second layer is better in comparison with the first layer because of the higher radiation dose in the first layer.

\begin{figure*}[htbp!]
    \centering
    \resizebox{0.98\textwidth}{!}{\includegraphics[height=0.25\textheight]{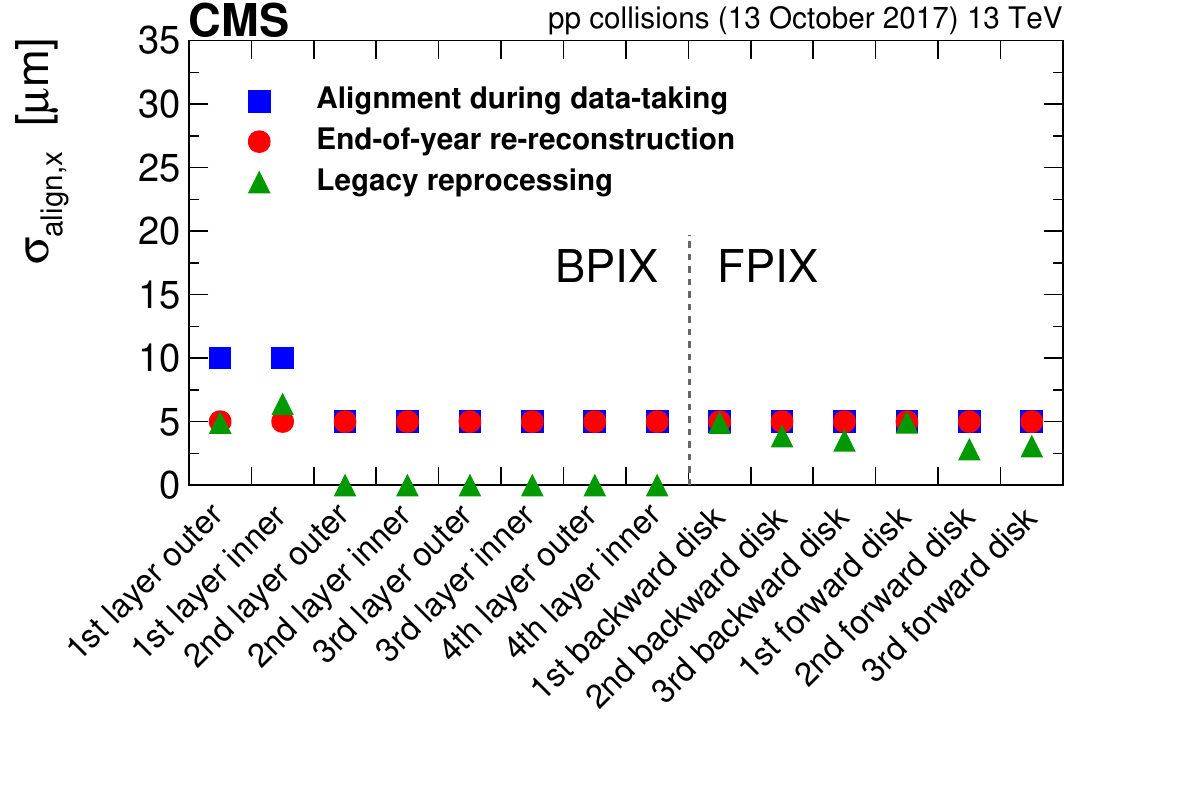}
    \includegraphics[height=0.25\textheight]{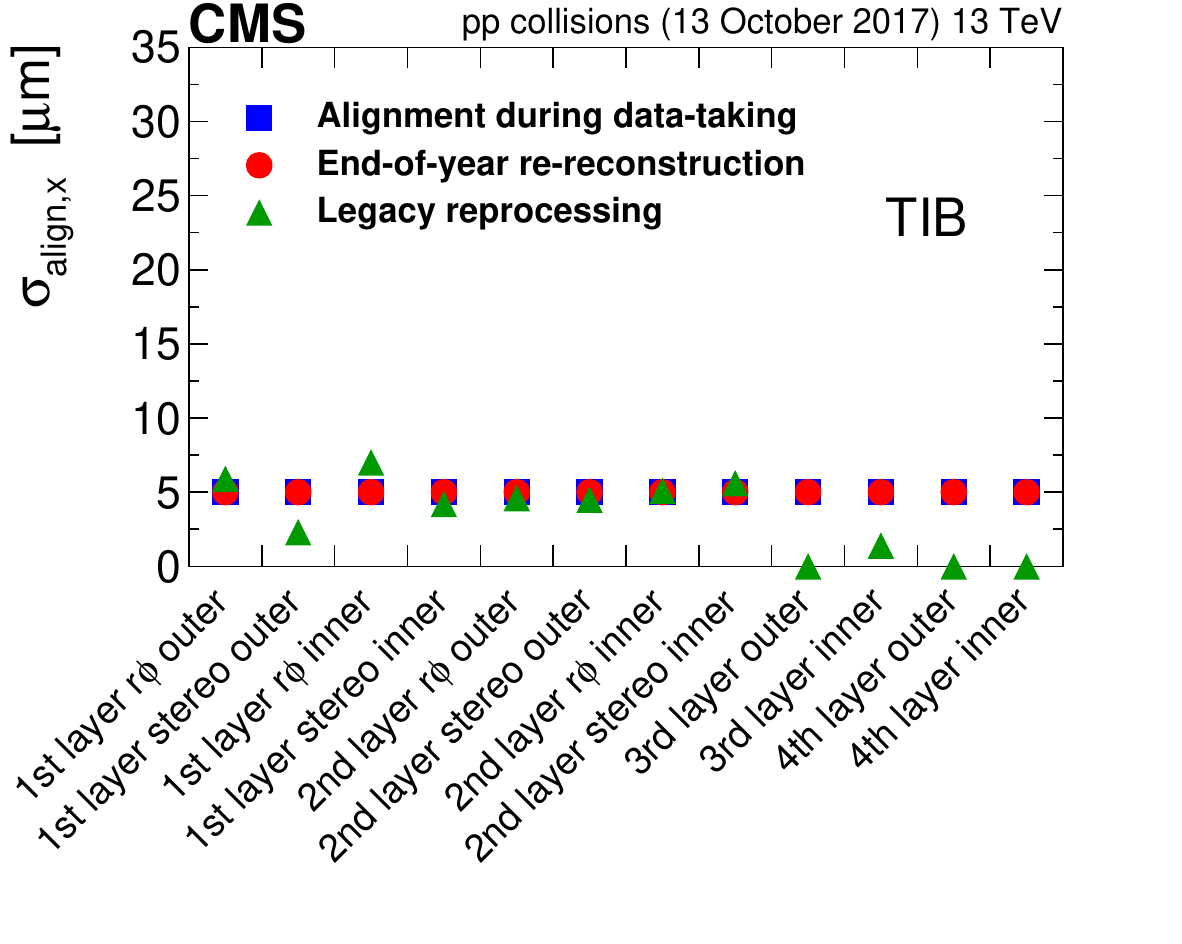}}
    \caption{The contribution from the misalignment of the sensors to the total hit resolution in the local~$x$ coordinate for the tracker pixel detector (left) and the inner barrel region of the strip detector (right). These contributions are shown separately for the different module categories that characterize the hierarchical structure of each subdetector.
    }

    \label{fig:APE_IOV}
\end{figure*}

The contribution from the misalignment of the sensors to the total hit resolution is shown in Fig.~\ref{fig:APE_IOV} for the pixel detector and the inner barrel region of the strip detector.
As an example, a data set recorded in October 2017 was selected as a representative sample showing the performance for the detector geometry in 2017 and 2018. 
The results obtained from the legacy reprocessing are more granular and yield smaller or similar alignment uncertainties than the alignment uncertainties obtained during data taking or from the EOY reconstruction in most module categories.

\begin{figure*} 
    \centering
    \includegraphics[width=0.5\textwidth]{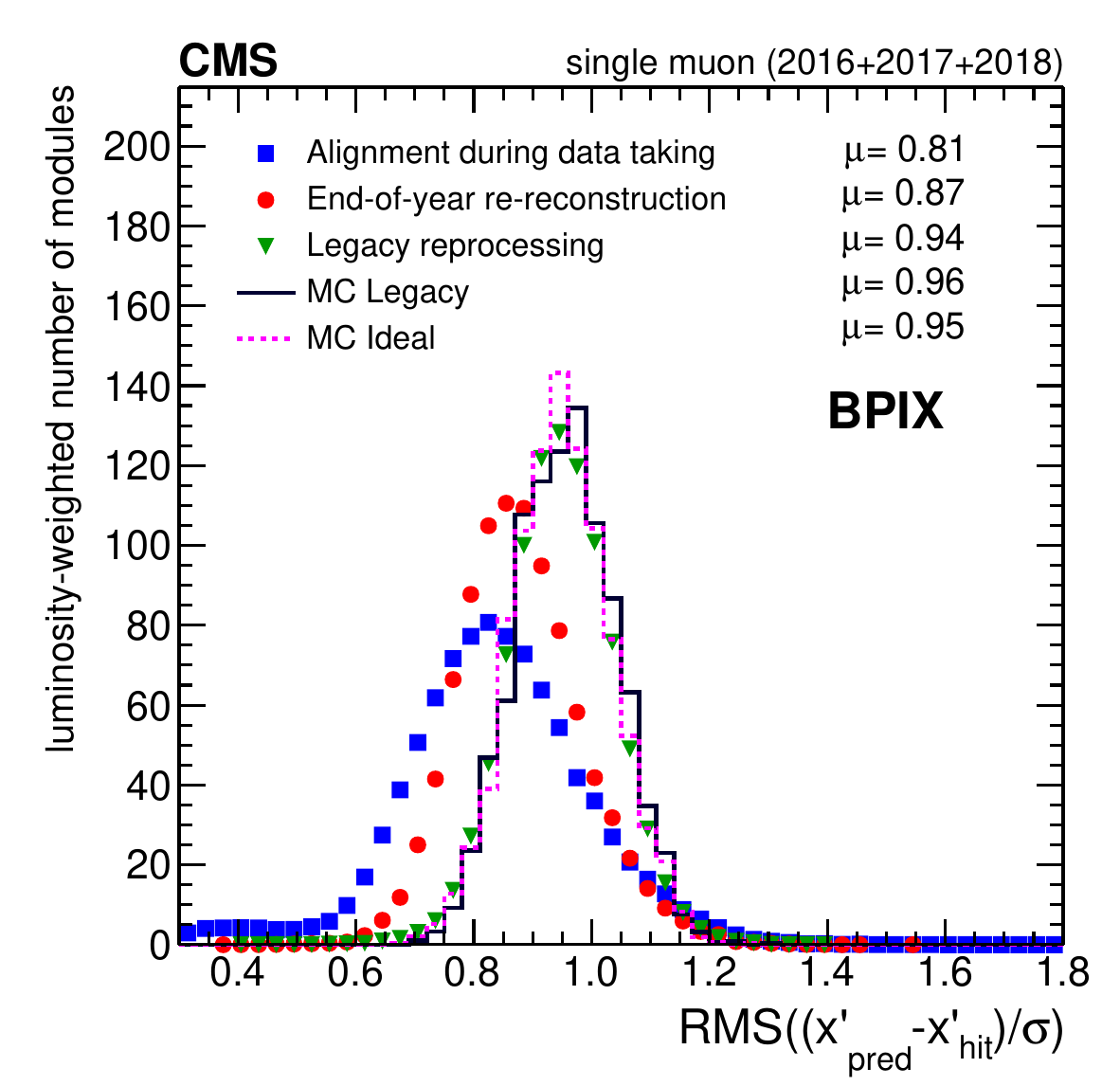}~\includegraphics[width=0.5\textwidth]{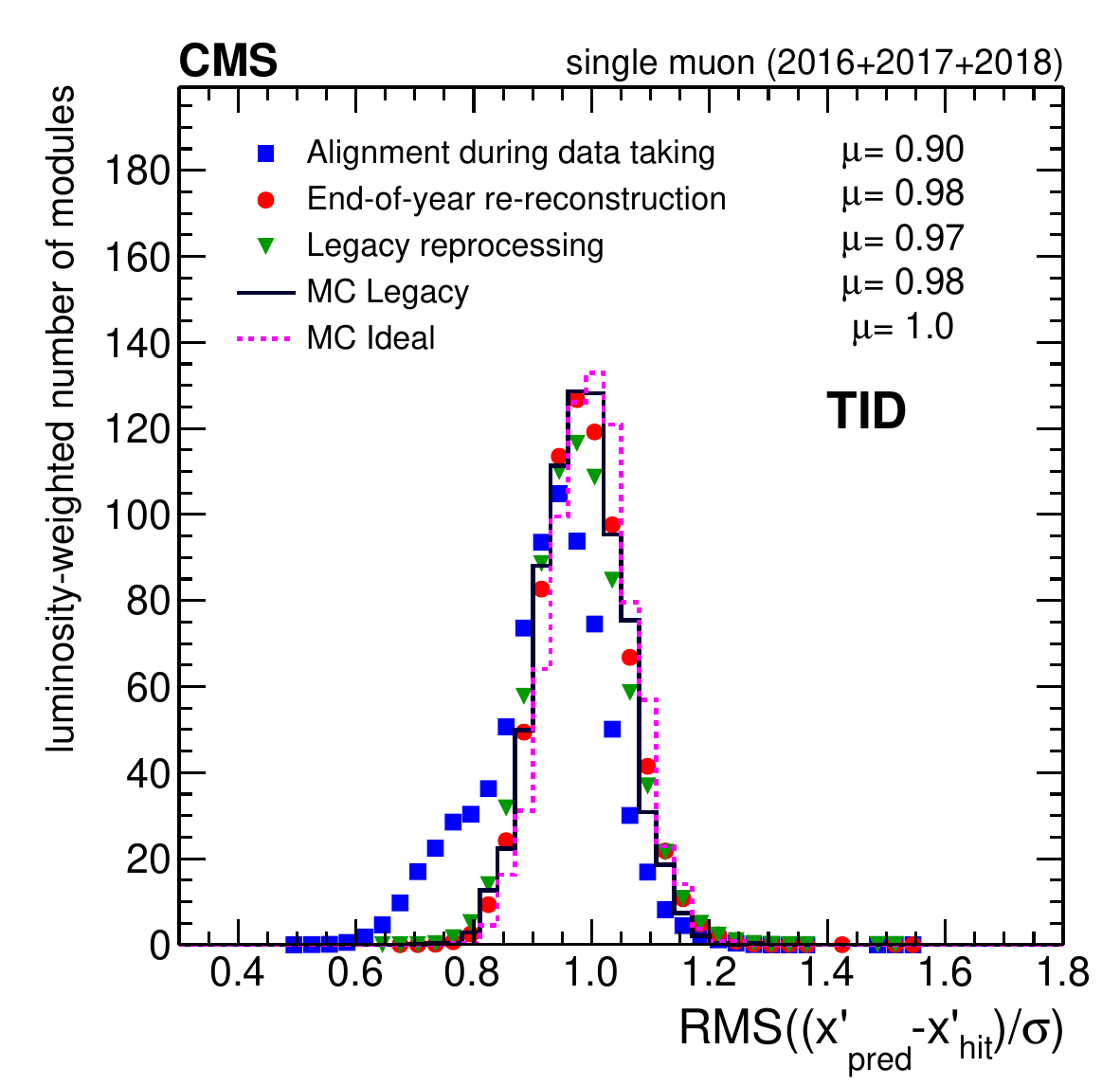}

    \caption{
        The distribution of the RMS of the normalized residuals in the local~$x$ coordinate ($x'$) for modules in the BPIX (left) and in the TID (right).
        The distributions evaluated in data are averaged over all IOVs, weighted by the integrated luminosity delivered in each IOV. The distributions in data are compared with an MC scenario with realistic alignment conditions for the legacy reprocessing and an MC scenario with ideal alignment conditions.
        An improvement is visible in the legacy reprocessing compared with the other two alignments shown; this is quantified by the quoted means $\mu$ of the distributions.
    }
    \label{fig:DRNRs}
\end{figure*}

Finally, the change in the estimation of the alignment accuracy is also visible in Fig.~\ref{fig:DRNRs}, where the distribution of the RMS of the normalized track-hit residuals produced with the single-muon data set is shown for the BPIX and the TID. In this figure, the same high-level structures, and the same condition on the number of hits per module, as in Fig.~\ref{fig:DMRperformance} are used. The distributions in data are averaged over all IOVs, weighted by the corresponding delivered integrated luminosity. These distributions are compared with the realistic and ideal single-muon MC scenarios that will be introduced in Section~\ref{sec:MC}.
The improvement from the legacy reprocessing related to the finer granularity can also be seen, because the means of the distributions shift closer towards unity. The distributions are not expected to be centred exactly around unity, since an incorrect estimation of the track or hit uncertainty causes this centre to deviate from unity. This can be observed in the ideal simulation without misalignment. 

\subsection{Special runs}
The approach used to align the tracker during periods of data taking with lower pileup or a different centre-of-mass energy than usual, as well as during HI collisions, differs from the description in Section \ref{sec:ULstrat}. These data sets are small compared with the standard $\pp$ collision data set and are used independently from it in physics analyses. Short periods of time during which the alignment performance is suboptimal do not have a significant effect on the large standard $\pp$ collision data set. For the much smaller data sets collected during special runs, a short period of suboptimal alignment performance would have a comparatively more important impact. A dedicated alignment strategy is therefore employed to ensure the alignment performance is maintained in these subsets of the data.

\subsubsection{Low-pileup runs and runs at \texorpdfstring{5.02\TeV}{5.02 TeV}}
Several low-pileup runs were taken during Run~2. These runs had an average of three interactions per bunch crossing, whereas under normal operating conditions the average number of interactions per bunch crossing was greater than 20. In 2017 (2018) a data set corresponding to an integrated luminosity of 0.22 (0.03)\fbinv was collected under these low-pileup conditions.
The low-pileup runs in 2017 were included in the global legacy alignment procedure described in Section~\ref{sect:ULstrat}. To improve the performance of the alignment for the low-pileup runs, the result of this global alignment was used as a starting point for an additional alignment fit. This last step used only tracks from the low-pileup runs, and the positions of the pixel detector modules were allowed to vary. The parameters of the strip detector remained fixed to the values determined in the first step. 

In 2017, a data set corresponding to an integrated luminosity of 90\pbinv was collected at a centre-of-mass energy of 5.02\TeV. An alignment fit was performed following the same approach as for the low-pileup runs. However, the runs at 5.02\TeV were not included in the global legacy alignment procedure.

For the low-pileup runs taken during 2018, a different approach was employed. On top of the legacy alignment procedure, a fit using only tracks from the 2018 low-pileup runs was performed. In this step the parameters of the strip detector were fixed to the values obtained in the global procedure and the parameters of the half-barrels and half-cylinders of the pixel detector were allowed to vary. In this case the complexity of the second step is reduced compared with that employed for the low-pileup runs taken in 2017 as a result of the much smaller amount of data collected.

\subsubsection{Heavy ion runs}
The track multiplicity in HI collisions, especially in central ones where the nuclei collide head-on, is much higher than in $\pp$~collisions.
In general, it is not guaranteed that the techniques developed in the context of $\pp$~collision data apply directly to HI runs.

The initial alignment constants used for the HI running period were determined from $\pp$~runs with the same detector settings as used during the HI data-taking period. 
Movements in the high-level structures of the pixel detector were observed by the PCL alignment procedure at the beginning of HI data taking.
The automated alignment procedure triggered by these movements improved the tracking and vertexing performance.

To further improve the performance, a module-level alignment fit was performed in the BPIX and FPIX using the \HIPPY algorithm with 30 iterations. 
As an input for the alignment fit, approximately 100\,000 events from a data set that preferably contained events where the two colliding lead ions only partially overlap were used.
The \HIPPY alignment procedure was validated by investigating the DMRs and impact parameters (presented in Section~\ref{sec:2015}) for the data with and without applying the obtained alignment conditions.
An improvement of the tracking performance in the pixel detector was observed with the new alignment conditions, both in the barrel and in the forward regions.

No additional alignment fits were performed for these data sets as part of the legacy alignment procedure.

\section{Alignment in simulation} \label{sec:MC}
To achieve the best possible modelling of the data, simulated events are processed through the same detector reconstruction chain used for events in the observed data. 
This requires the full set of calibrations, including tracker alignment constants, to be derived for the simulation as well.
The main purpose of such constants is to reproduce, as accurately as possible, the same performance and effects observed in the data. 
This section presents the strategy adopted to derive alignment conditions consistent with the ones for the legacy reprocessing. 
This specific case is most interesting because the alignment constants for the simulation were derived after the final conditions for data reprocessing were available, thus making better tuning possible, as described in Section~\ref{sec:MCstrategy}. The conditions are derived separately for each data-taking year, though no further time dependence is included in the simulation. 
The alignment scenarios for simulation are validated using the same methods used for data. Therefore a direct comparison of the quantities sensitive to alignment effects is possible, as shown in Section~\ref{sec:MCvalidation}.
The good description of the data achieved by the dedicated simulation is also demonstrated in Section~\ref{sec:trackDataMC} in terms of track-related quantities, such as the impact parameters and the $\chi^{2}$ distributions of the track fits. 

\subsection{Derivation of the alignment} \label{sec:MCstrategy}
The general strategy used to derive accurate alignment constants for the simulation relies on reproducing the procedure adopted for the data as closely as possible. For this, a full alignment fit is performed using simulated events. 

The starting geometry for the fit is built starting from the ideal detector geometry, with misalignments applied to reflect the average accuracy of the alignment constants in data after the EOY reconstruction. 
A Gaussian smearing is applied to the ideal module positions in the local $x$, $y$, and $z$~directions. The RMS values for the $x$ and $y$~directions are obtained from the DMR distributions of the EOY reconstruction. This is done separately for each tracker partition.
When appropriate, further systematic shifts of the tracker substructures are also applied, accounting for the presence of systematic misalignments. 
Table~\ref{table:MC} shows the shifts of the module positions applied to construct the scenarios used as a starting point for the MC fits for the 2016 and 2017 legacy alignment conditions. Similarly, a smearing of 10\mum was applied to all tracker substructures in the three local directions for the derivation of the 2018 legacy MC alignment conditions.
\begin{table*}[!ht]
    \centering
    \topcaption{Magnitude of the Gaussian smearing applied to the design geometry to derive the starting geometry for the 2016 and 2017 legacy MC alignments. The adopted values in both years are reported for each substructure and by coordinate.}
    \begin{tabular}{lcrrc}
    Year & Substructure & $\sigma_x$ [\mum] & $\sigma_y$ [\mum] & $\sigma_z$ [\mum] \\
    \hline 
    \multirow{6}{*}{2016} & BPIX & 3.0 & 9.1 & \NA \\
                          & FPIX & 9.0 & 9.0 & \NA \\
                          & TIB & 4.8 & 4.8 & \NA \\
                          & TOB & 11.7 & 11.7 & \NA \\
                          & TID & 3.3 & 3.3 & \NA \\
                          & TEC & 6.9 & 6.9 & \NA \\[\cmsTabSkip]
    \multirow{6}{*}{2017} & BPIX & 6.1 & 17.0 & 5.0 \\
                          & FPIX & 5.3 & 2.7 & 5.0 \\
                          & TIB & 13.7 & 13.7 & 5.0 \\
                          & TOB & 30.9 & 30.9 & 5.0 \\
                          & TID & 6.3 & 6.3 & 5.0 \\
                          & TEC & 13.6 & 13.6 & 5.0 \\[\cmsTabSkip]
    \end{tabular}
    \label{table:MC}
\end{table*}
 
After having determined the starting point for the MC alignment, the fit is performed using the same granularity as for the data. The set of simulated tracks used for the fit is obtained from several samples, reflecting the variety of topologies used in data, as discussed in Section~\ref{sec:datasets}. The composition of the sample used in the data alignment is closely matched using event weights.
 
\subsection{Validation} \label{sec:MCvalidation}
\begin{figure*}[p!]
    \centering

    \includegraphics[width=1.2\textwidth,angle=90]{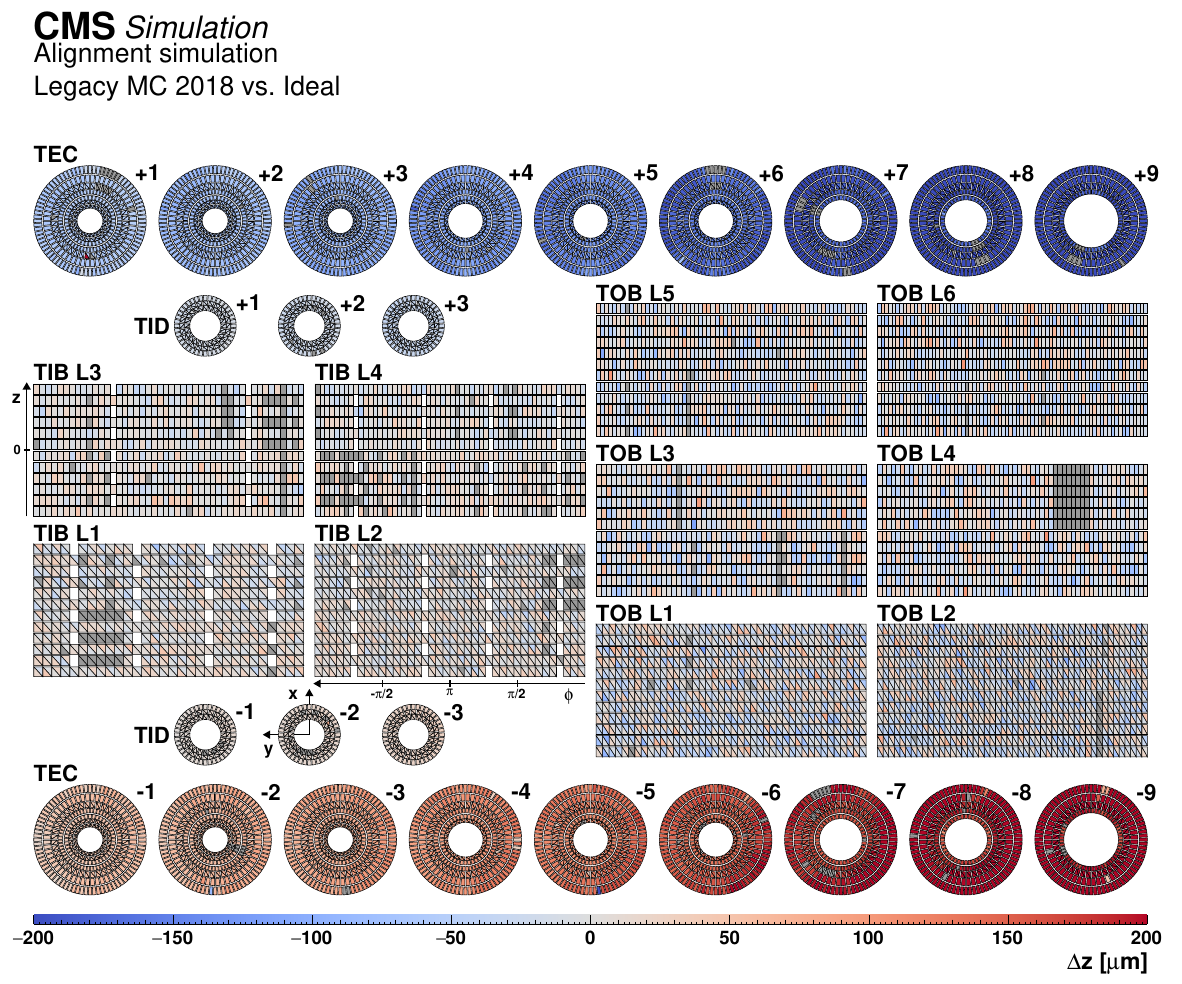}

    \caption{Difference of module positions in the global $z$~coordinate as obtained in the 2018 MC alignment fit with respect to the ideal positions.
        The modules are given a colour corresponding to the value of $\Delta z$ according to the colour bar on the bottom.
        Modules that were inactive during the simulation are indicated in dark grey.
        A pattern typical for the $z$~expansion distortion can be observed in the TEC, with a maximum magnitude of approximately 200\mum. The effect of this systematic misalignment on the alignment performance is expected to be small.
    }  

    \label{fig:TkAlMap_MC2018}
\end{figure*}

The alignment constants derived from the fit are then validated and compared with the data.
First, the tracker geometry obtained from the fit is compared with the ideal one to directly assess the recovery from misalignment and spot any unusual movements or systematic distortions that might be artificially introduced by the fit.  
An example is provided in the right panel of Fig.~\ref{fig:TkAlMap_MC2017}, showing the differences in the module positions in the $\phi$~coordinate for the 2017 MC alignment with respect to the ideal detector geometry. No unexpected large movements or misalignments are observed, providing a first indication that the alignment fit performs well.
Figure~\ref{fig:TkAlMap_MC2018} shows the differences in the module positions in the $z$~coordinate for the 2018 MC alignment. Here, movements characteristic of a $z$~expansion WM are observed in the TEC. These movements are not expected to have a large impact on the alignment performance.  

\begin{figure*}[htbp!]     
    \centering
    \includegraphics[width=0.96\textwidth]{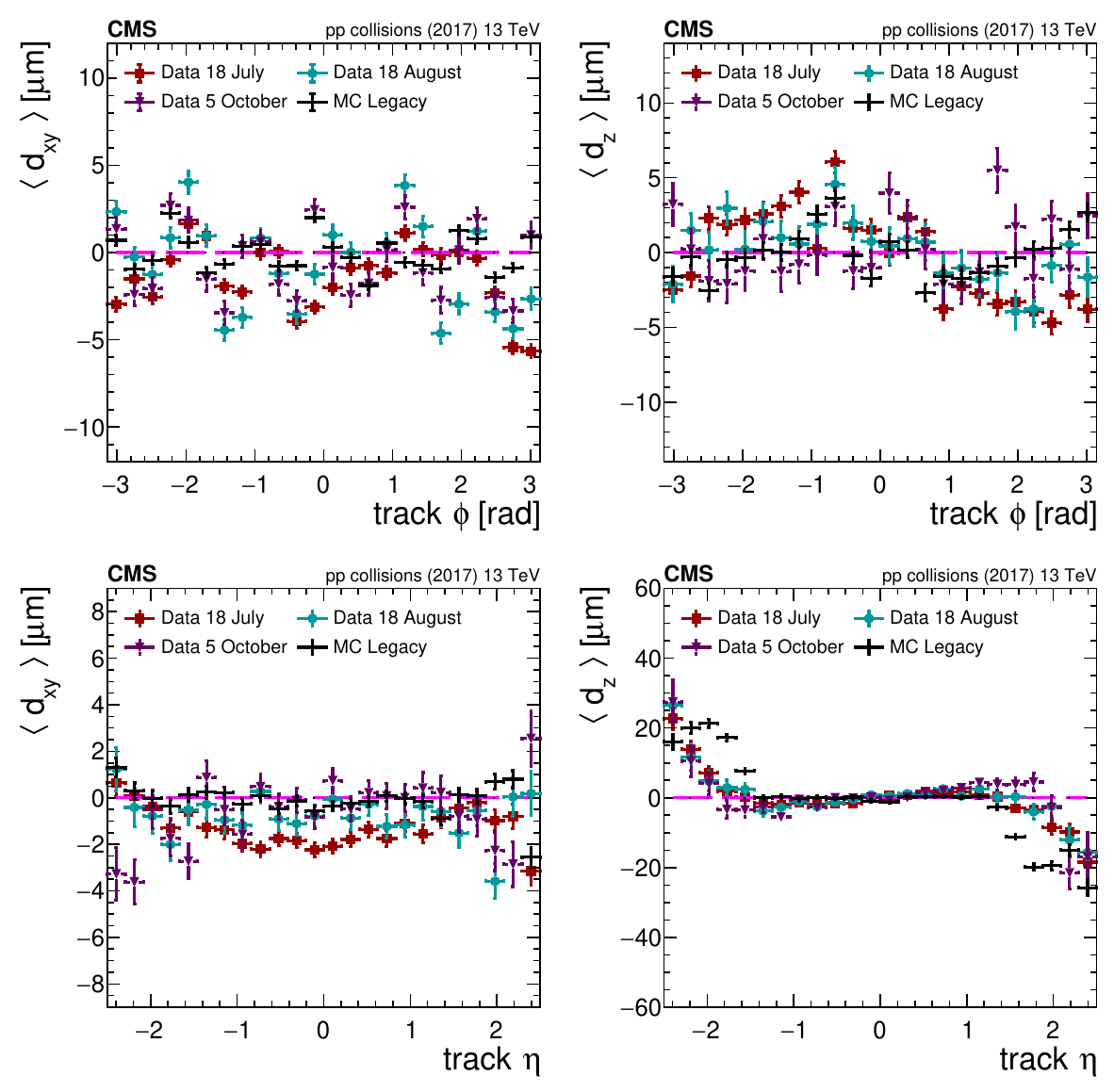}
    \caption{Distribution of the mean impact parameter in the transverse plane (left) and in the longitudinal plane (right) as a function of the track $\phi$ (top) and $\eta$ (bottom) for 2017. The derived MC scenario is compared with three representative IOVs from the year 2017 in data (18~July, 18~August, 5~October) to assess its validity as the final geometry. The error bars show the statistical uncertainty related to the limited number of tracks.}
     \label{fig:MCUL_PV_2017} 
\end{figure*}

\begin{figure*}[htbp!]
    \centering
    \includegraphics[width=0.48\textwidth]{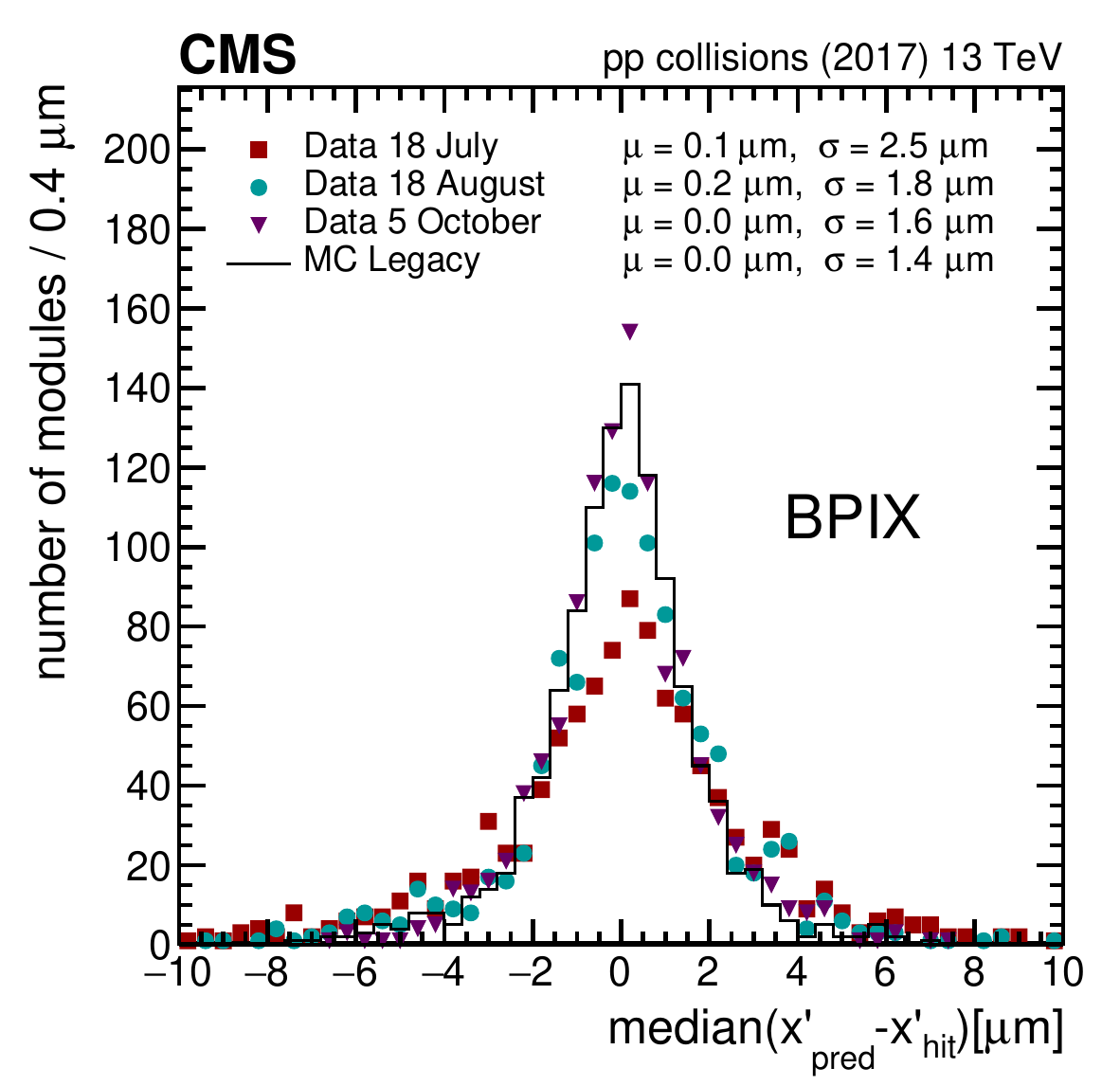}
    \includegraphics[width=0.48\textwidth]{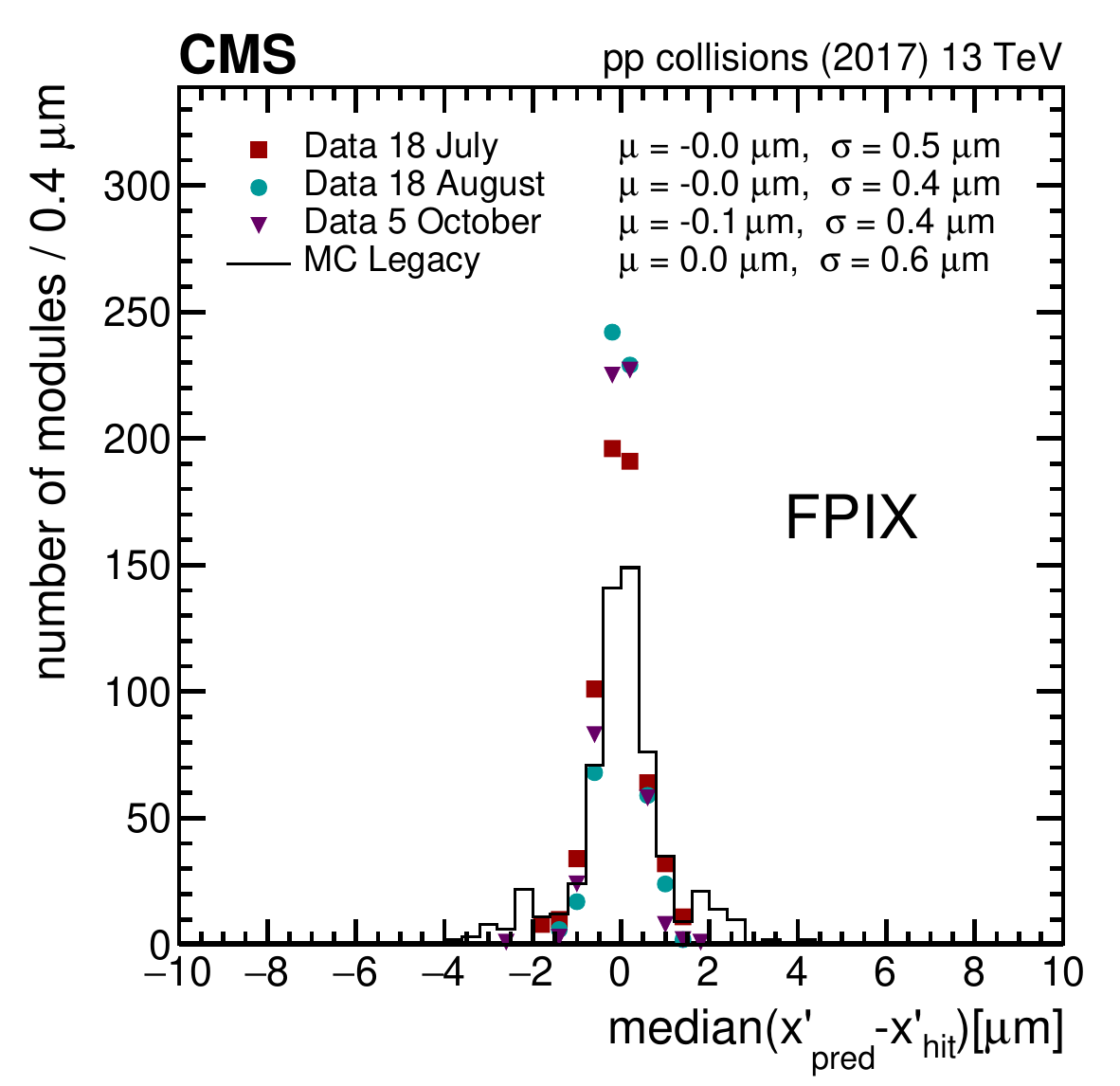}
    \\
    \includegraphics[width=0.48\textwidth]{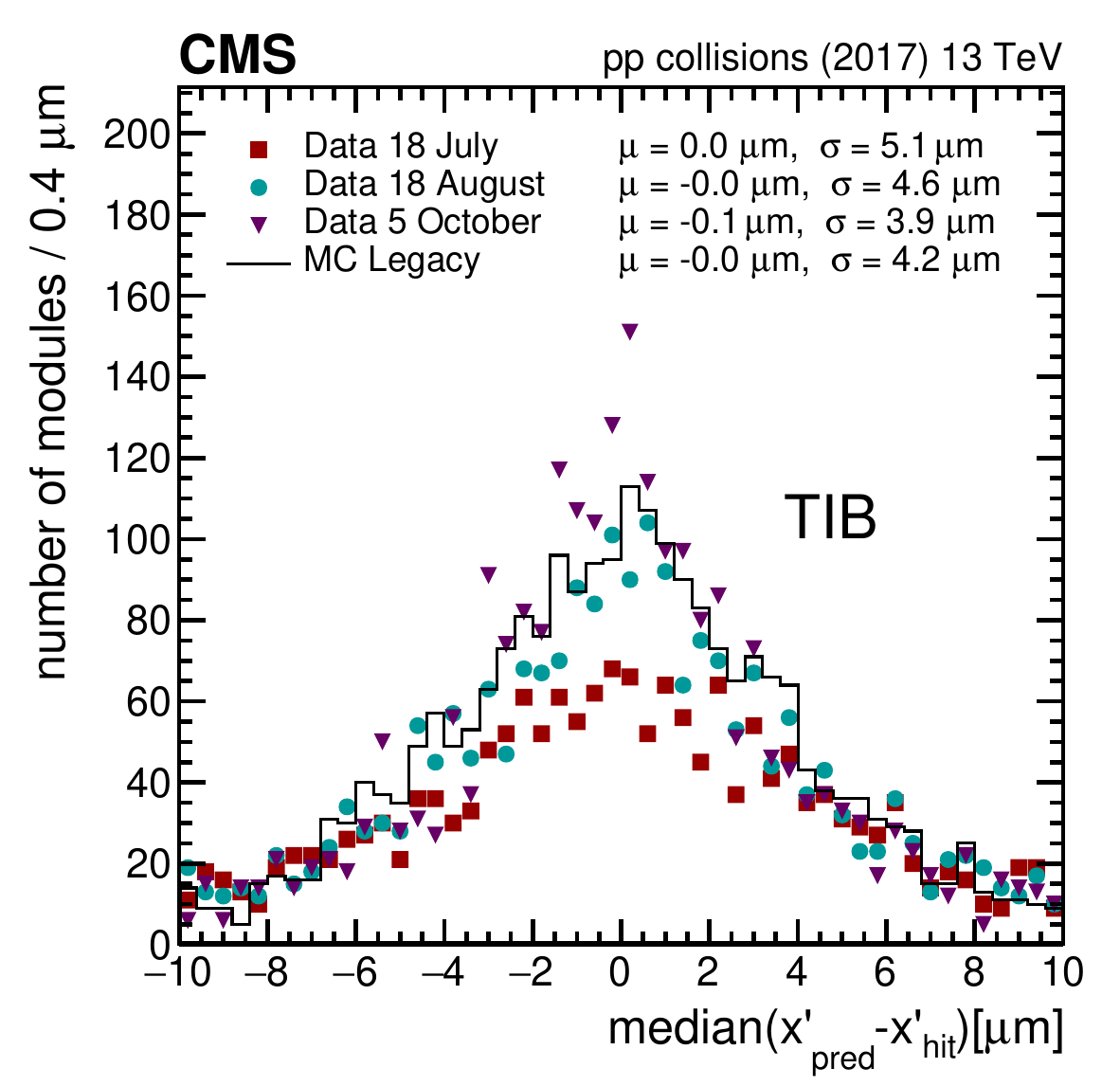}
    \includegraphics[width=0.48\textwidth]{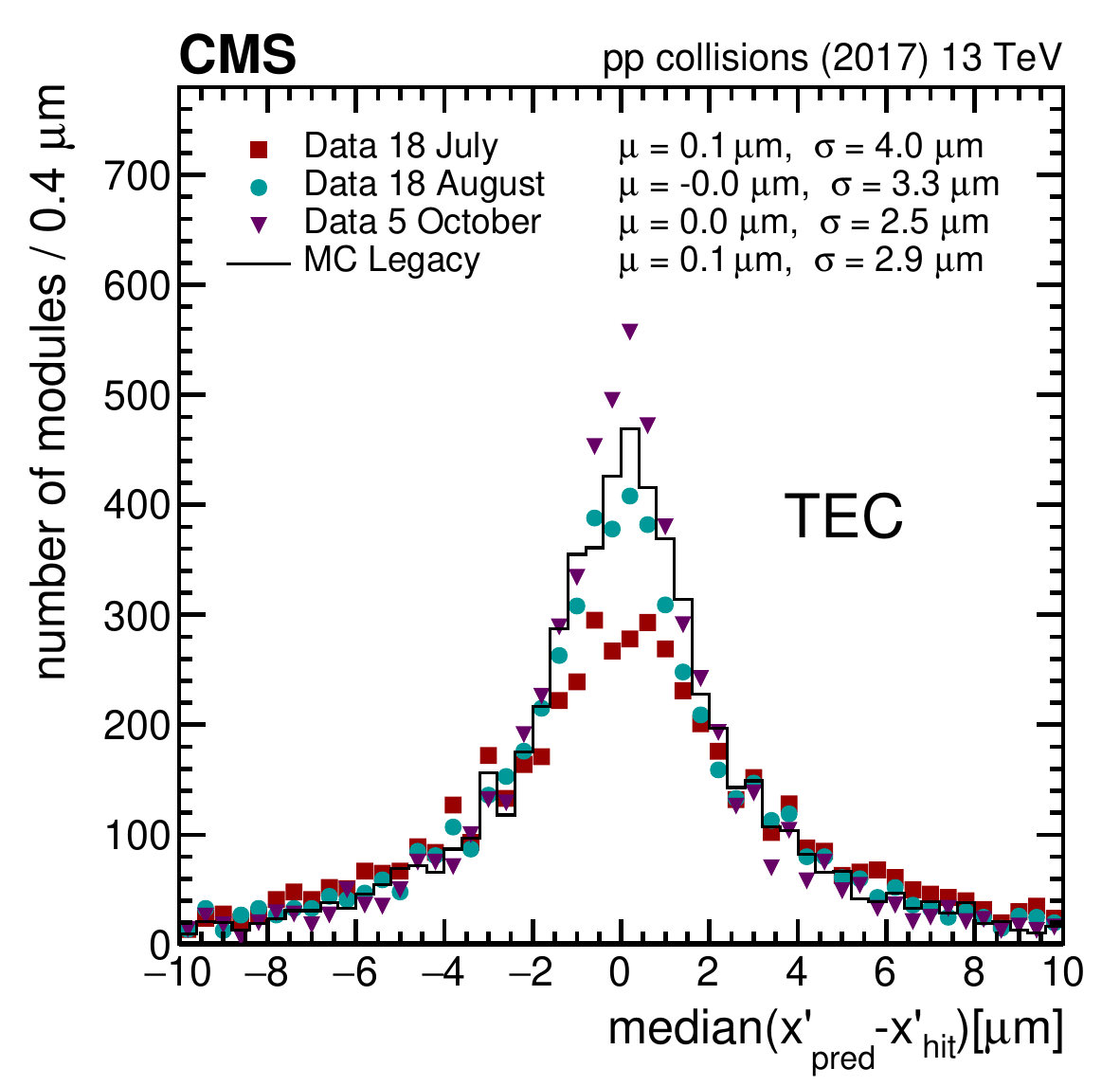}
    \caption{The DMRs in the local~$x$ coordinate for different components of the tracker system. The derived MC scenario is compared with three representative IOVs from the year 2017 in data (18~July, 18~August, 5~October) to assess its validity as the final geometry.}
    \label{fig:MCUL_DMR_2017}
\end{figure*}

\begin{figure*}[htbp!]
    \centering
    \includegraphics[width=0.495\textwidth]{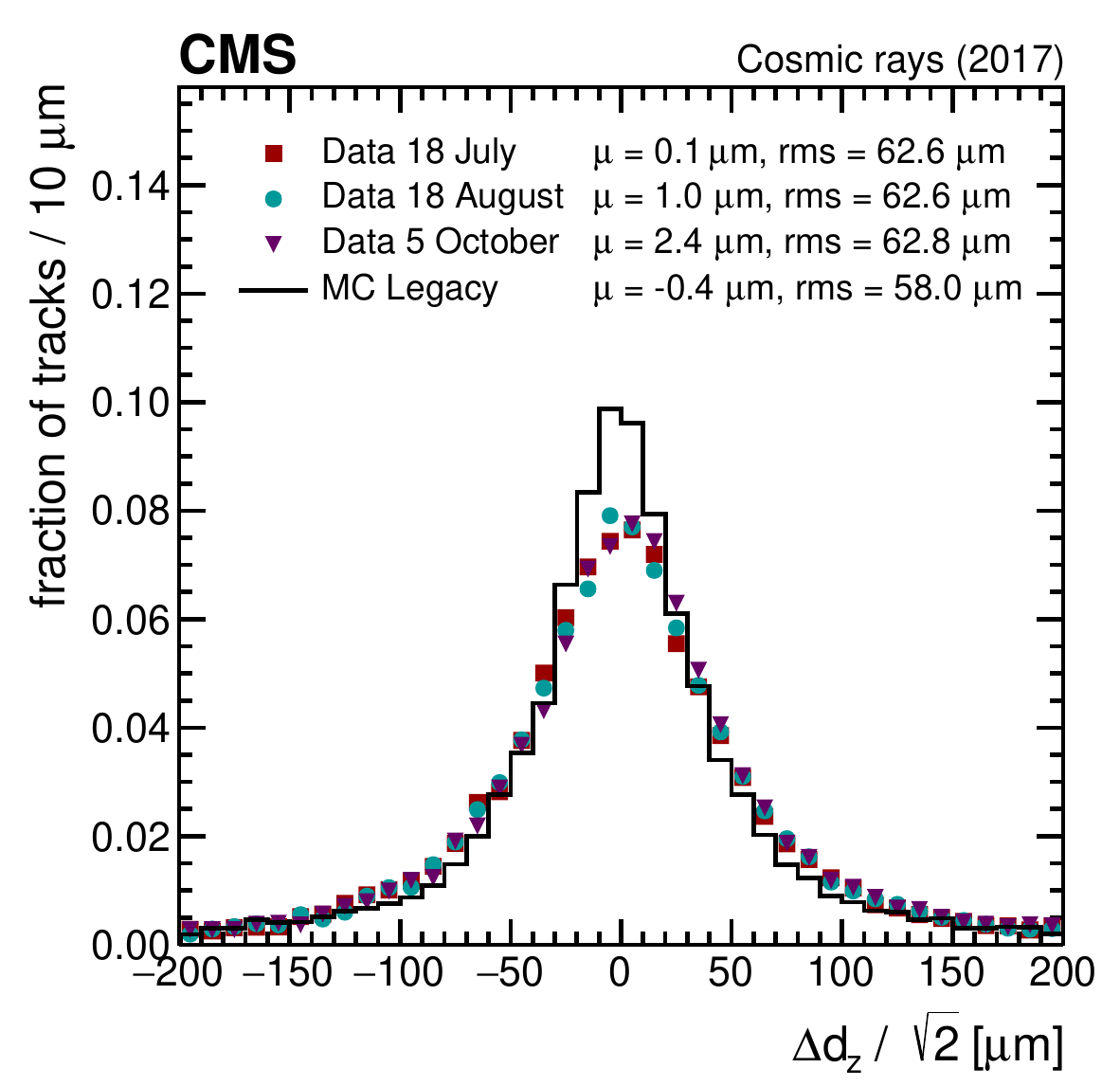}
    \includegraphics[width=0.495\textwidth]{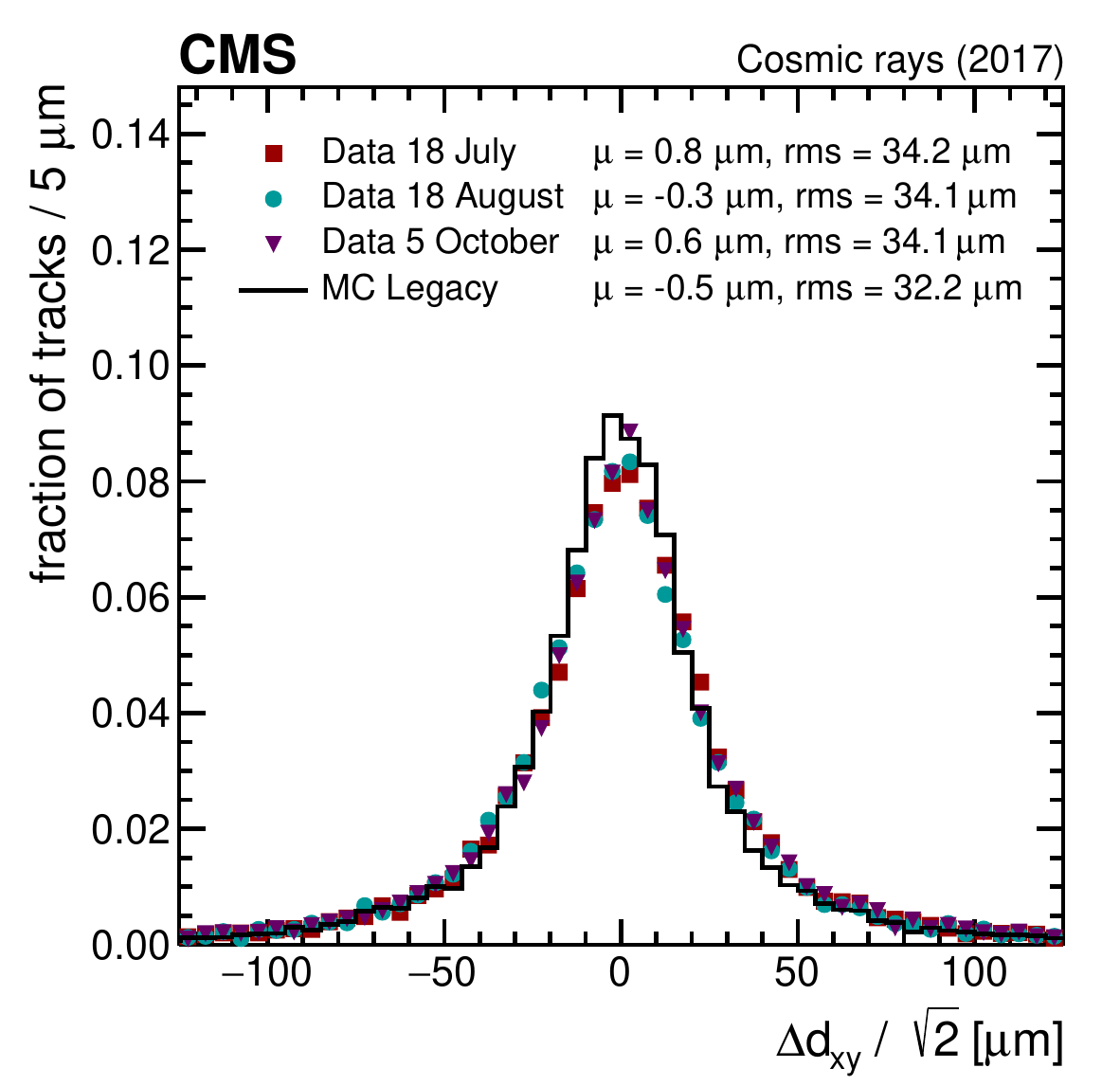}
    \caption{Difference in the impact parameters in the longitudinal (left) and transverse (right) plane as evaluated in the cosmic ray muon track splitting validation. The derived MC scenario is compared with three representative IOVs from the year 2017 in data (18~July, 18~August, 5~October) to assess its validity as the final geometry.}
    \label{fig:MCUL_MTS_2017}
\end{figure*}

Since the performance in data is generally not uniform in time, three representative IOVs are selected and compared with the alignment in simulation. The impact parameter performance for the 2017 MC scenario is shown in Fig.~\ref{fig:MCUL_PV_2017}. An attempt to reproduce the modulations observed at high absolute pseudorapidity for the three selected IOVs was made by applying a coherent shift of 30\mum in the $z$~coordinate to the endcaps of the FPIX. In a similar approach, a 10\mum ladderwise alternation of the bias in the $x$~coordinate was applied to the first layer of the BPIX in the 2018 MC scenario.
The local precision of the MC alignment procedure, which can be estimated from the DMR distributions and compared with the data, is shown in Fig.~\ref{fig:MCUL_DMR_2017} for 2017. The effect of the introduced systematic misalignment can be observed in the distribution corresponding to the FPIX. On average, the performance of the MC alignment matches the performance observed in data. The same observation is supported by the results of the cosmic ray muon track split validation, shown in Fig.~\ref{fig:MCUL_MTS_2017}.

With this set of validations shown for the 2017 MC misalignment scenario, we conclude that the derived scenario is consistent with the conditions observed in the CMS detector. Therefore, it was chosen as the alignment scenario for the 2017 legacy MC simulation. The same procedure was applied for the other two years, and a final candidate consistent with the CMS detector conditions was delivered. 

\begin{figure*}[htbp!]
    \centering
    \includegraphics[width=0.96\textwidth]{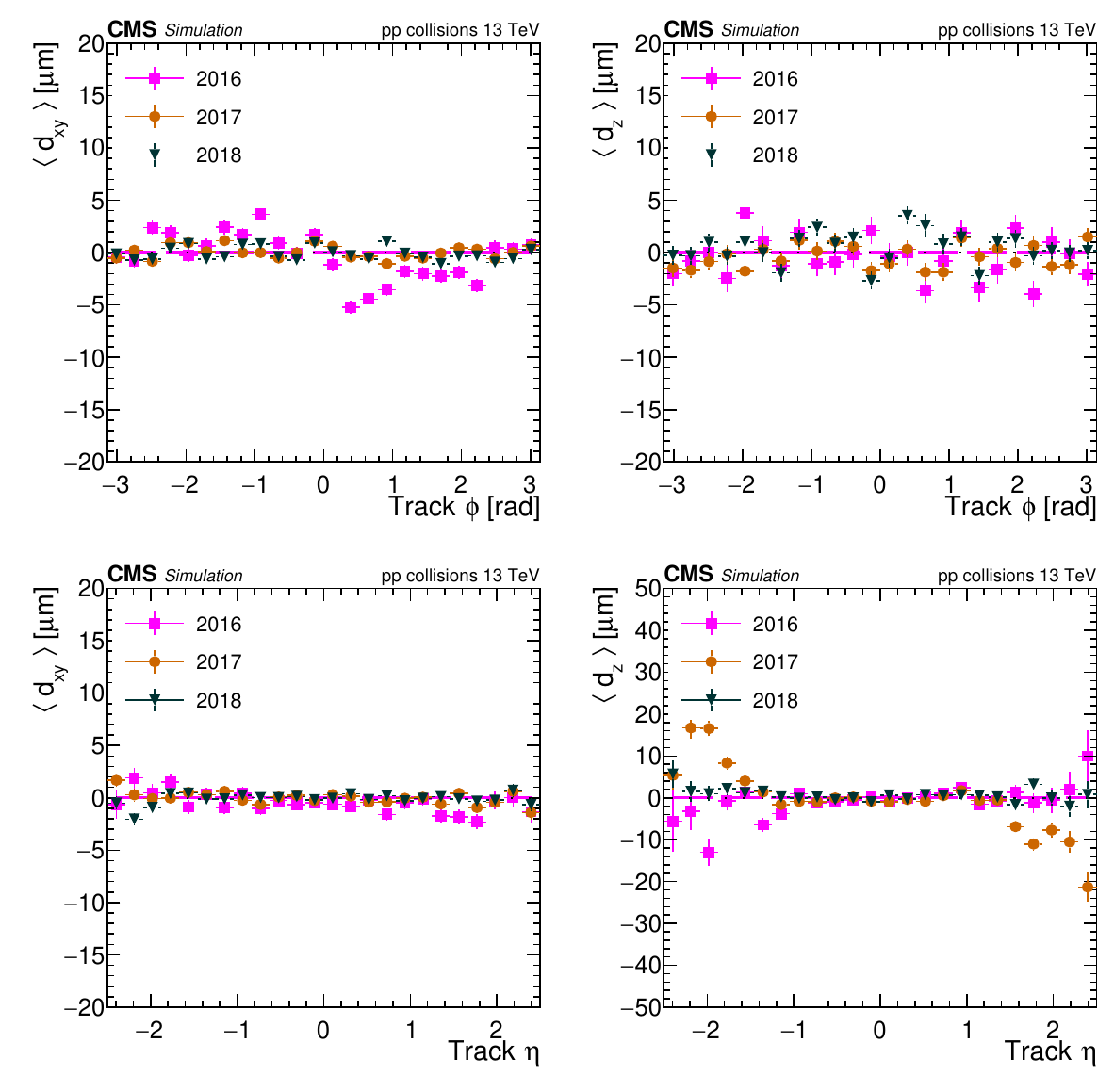}
    \caption{Mean track-vertex impact parameter in the transverse plane (left) and in the longitudinal plane (right), as a function of the track $\phi$ (top) and $\eta$ (bottom). A comparison of the results with the alignment constants derived in the Run~2 legacy MC scenario for 2016, 2017, and 2018 separately is shown. The error bars show the statistical uncertainty related to the limited number of tracks.}
    \label{fig:PV_MC_Legacy_Run2}
\end{figure*}

Finally, the performance of the MC scenario for each year separately is shown in Fig.~\ref{fig:PV_MC_Legacy_Run2}. The performance of the realistic MC scenario scaled to the corresponding delivered integrated luminosity was already shown in Fig.~\ref{fig:PVperformance} to provide a comparison with the average performance in the three considered geometries. The observed residual deviations from the ideal case (mean of 0) should reflect the alignment precision and radiation effects in data, mostly visible for the high-$\abs{\eta}$ region, which cannot be fully fixed by the alignment. Apart from this region, a very similar performance was achieved for the simulation in each year.  

\begin{figure*}[htbp!]    
    \centering
    \includegraphics[width=0.48\textwidth]{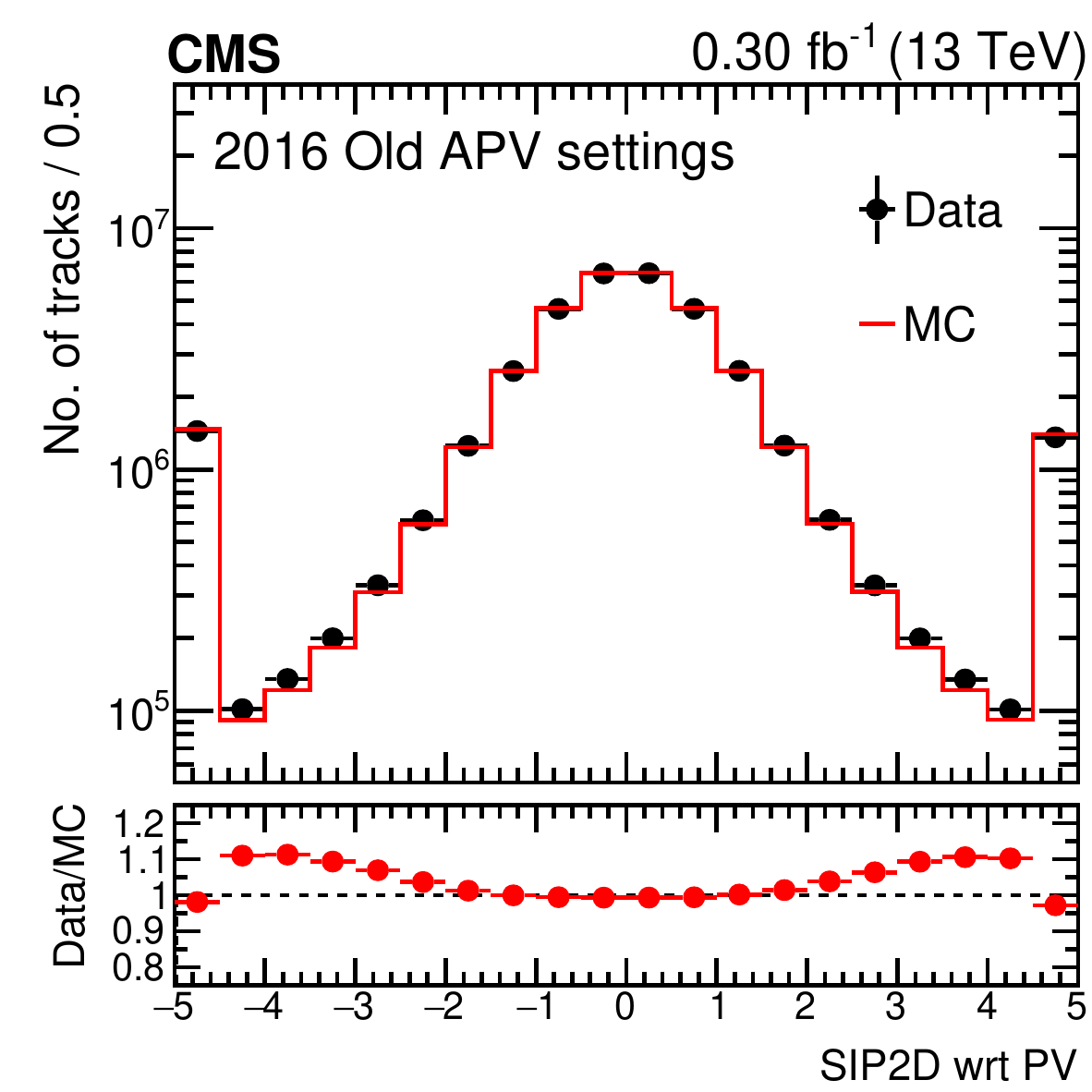}
    \includegraphics[width=0.48\textwidth]{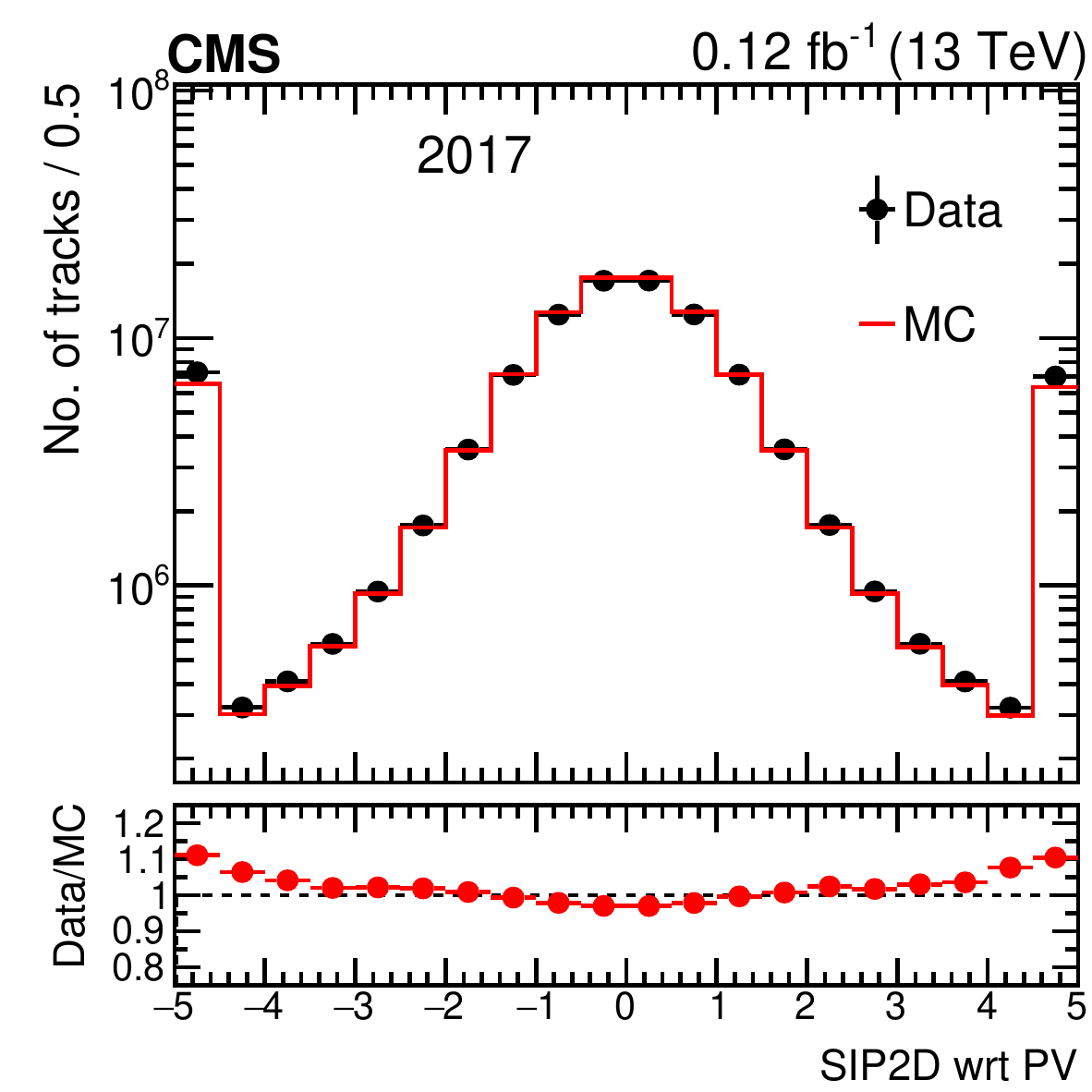}
    \caption{Distribution of the two-dimensional track impact parameter significance~(SIP2D) with respect to the PV.
        Data collected in 2016 (left) and 2017 (right) are compared with the corresponding MC simulation. For 2016, the simulation with old APV~settings is shown. For both data and simulation, the first and last bins include the underflow and overflow, respectively. Vertical error bars represent the statistical uncertainty due to the limited number of tracks; they are smaller than the marker size in both years. Tracks with very low $\chi^{2}$ probability values, related to pattern recognition errors, are not included in these figures.}
     \label{fig:sip3d} 
\end{figure*} 
\begin{figure*}[htbp!]
    \centering
    \includegraphics[width=0.48\textwidth]{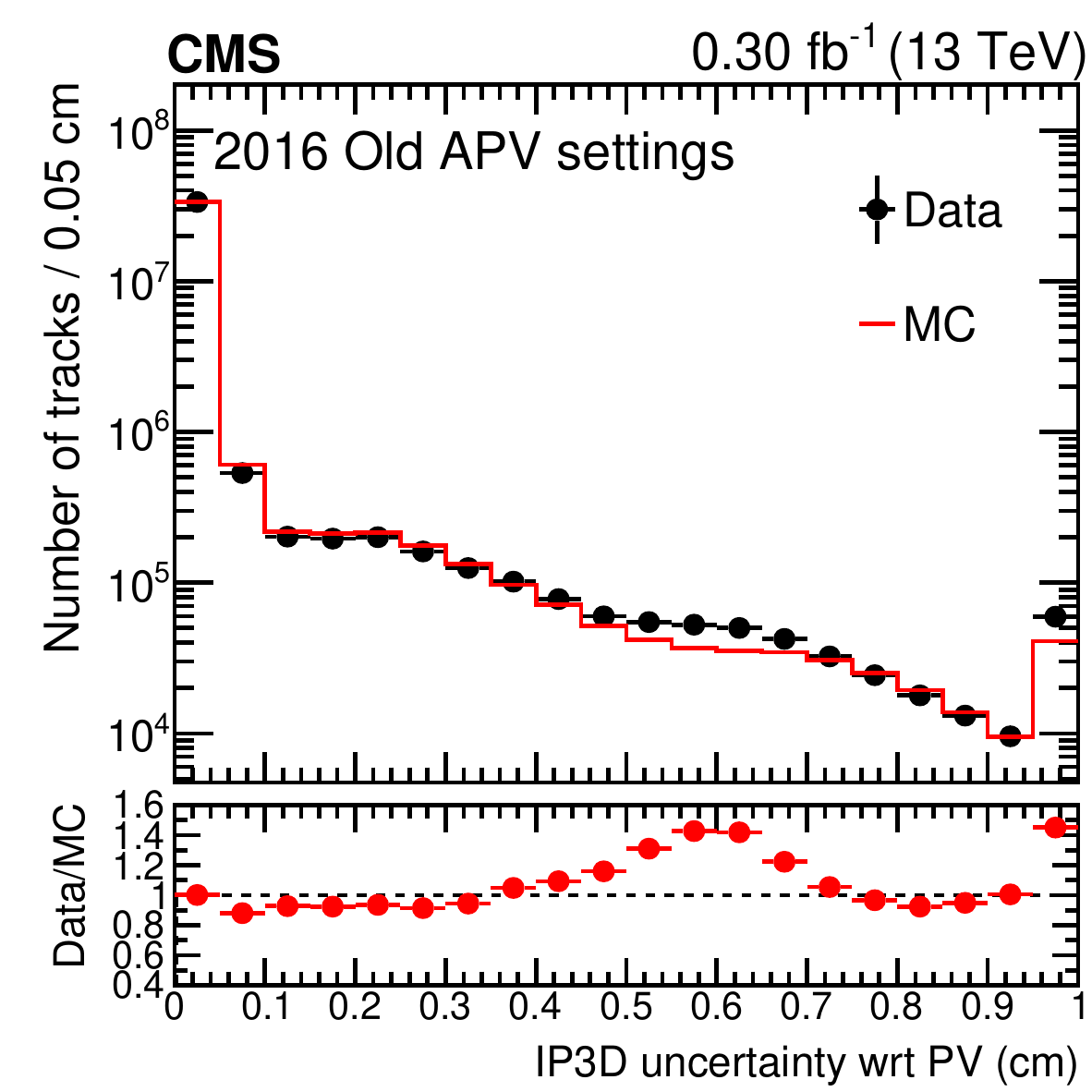}
    \includegraphics[width=0.48\textwidth]{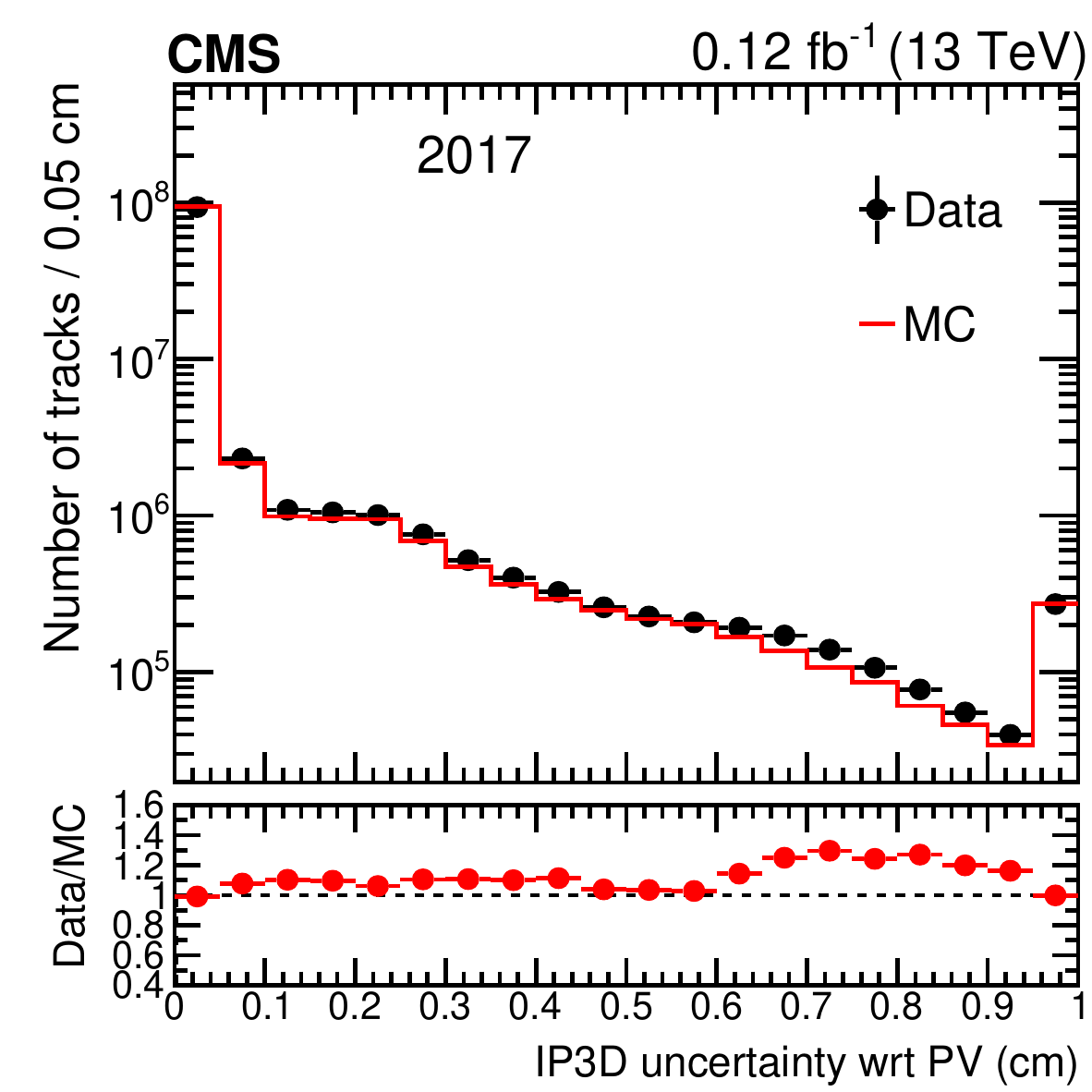}
    \caption{The distribution of the uncertainty in measuring the three-dimensional track impact parameter~(IP3D), with respect to the PV.
        Data collected in 2016 (left) and 2017 (right) are compared with the corresponding MC simulation. For 2016, the simulation with old APV~settings is shown. For both data and simulation, the last bins include the overflow. Vertical error bars represent the statistical uncertainty due to the limited number of tracks; they are smaller than the marker size in both years. Tracks with very low $\chi^{2}$ probability values, related to pattern recognition errors, are not included in these figures.}
    \label{fig:ip3derr}
\end{figure*}
\begin{figure*}[htbp!]
    \centering
    \includegraphics[width=0.48\textwidth]{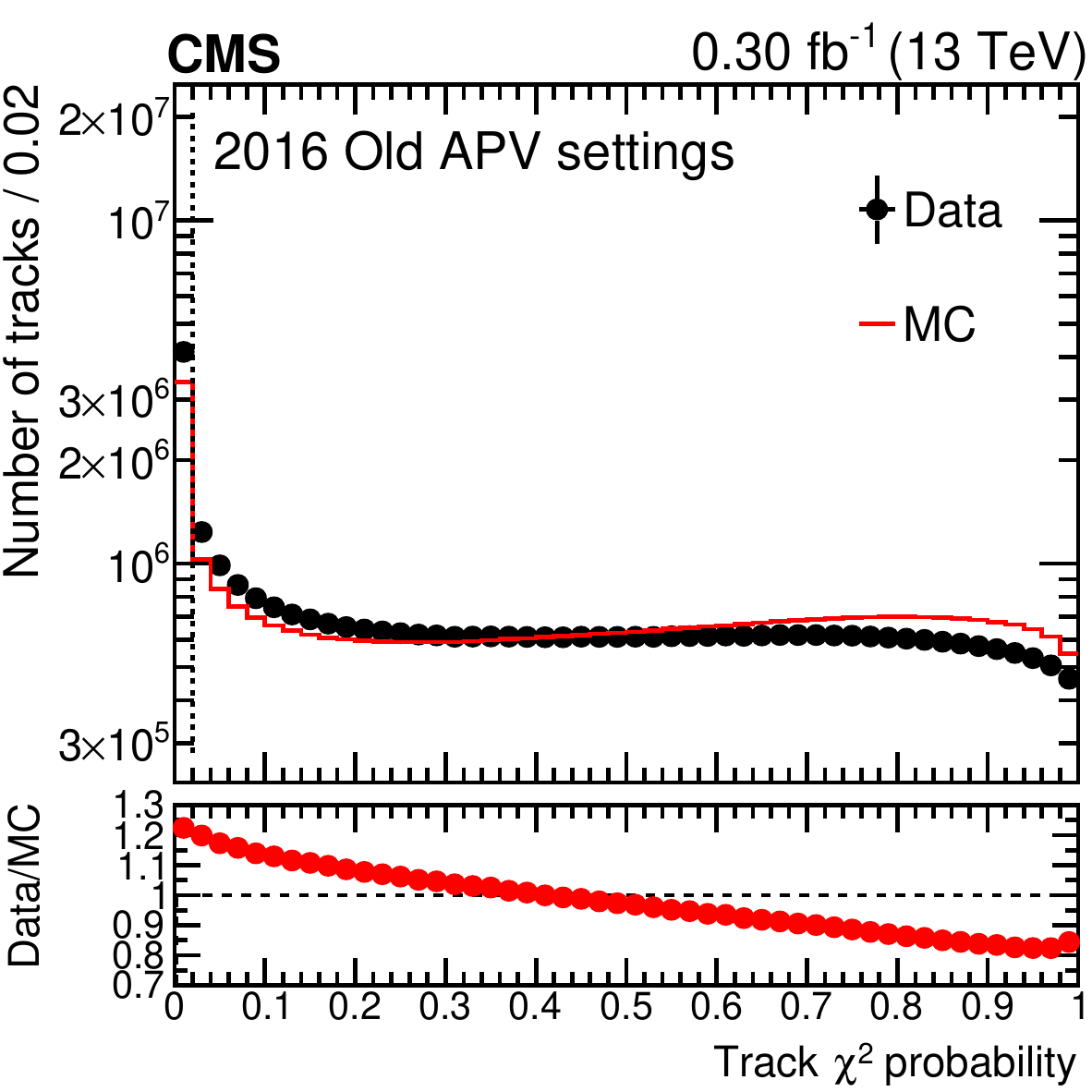}
    \includegraphics[width=0.48\textwidth]{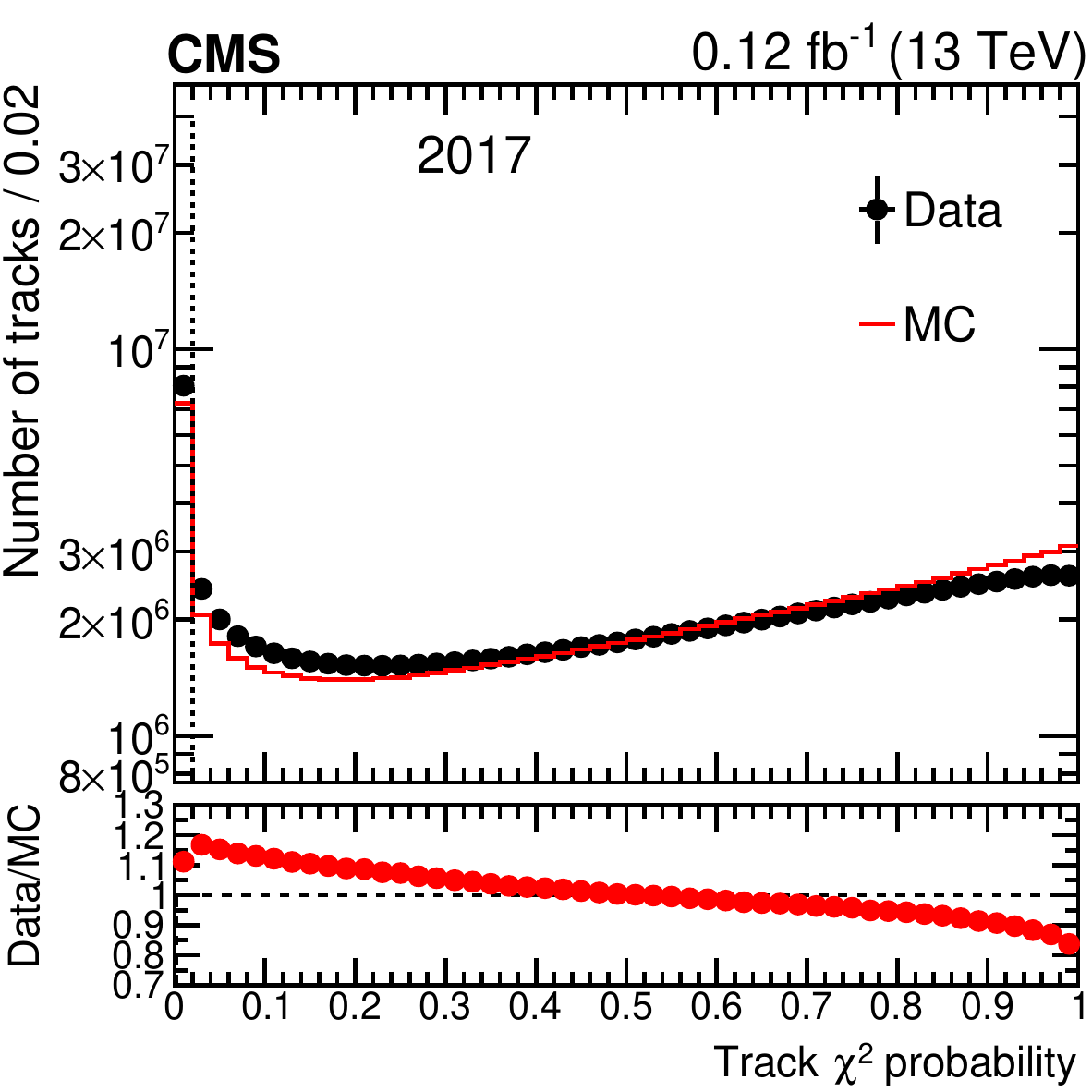}
    \caption{Distribution of the track $\chi^{2}$ probability for the number of degrees of freedom in the track fit.
        Data collected in 2016 (left) and 2017 (right) are compared with the corresponding MC simulation. For 2016, the simulation with old APV~settings is shown. Vertical error bars represent the statistical uncertainty due to the limited number of tracks; they are smaller than the marker size in both years. Some tracks have a very low $\chi^{2}$ probability value, below the dashed line, as a result of pattern recognition errors. Such tracks are excluded from the studies in this section.}
    \label{fig:chi2prob}
\end{figure*}

\subsection{Comparison of data and MC performance} \label{sec:trackDataMC}
A set of track properties is compared between data and MC simulation after the Run~2 legacy reprocessing to check the alignment of the detector. This is done to ensure the data and MC simulation are compatible.
In this study, inclusive L1 trigger $\pp$~collision data reconstructed with the Run~2 legacy reprocessing were compared with a simulated sample of single-neutrino production to produce events containing only pileup and detector noise.
In early 2016, the silicon strip detector observed a decrease in the signal-to-noise ratio, associated with a loss of hits in tracks.
This behaviour was caused by saturation effects~\cite{Butz:2018dum} in the preamplifier circuit used in the strip tracker readout ASIC (APV25) chip~\cite{FRENCH2001359}.
During this period, a data set corresponding to an integrated luminosity of 20\fbinv was impacted by this effect.
It was resolved by optimizing the setting of the parameter governing the drain speed of the preamplifier in the APV25 chip on 13th August 2016 to achieve a recovery of the hit efficiency to the same level as in Run~1.
The period affected by this problem is referred to as having old APV~settings, whereas the 2016 data-taking period after the resolution of the issue is referred to as using new APV~settings.
A legacy MC sample, containing a dedicated description of the APV25 chip dynamic gain~\cite{CMS-DP-2020-045} was made for the first part of the 2016 data-taking period and is referred to as ``MC with APV~simulation''. 

The runs where important updates were made in the legacy reprocessing are used to check the performance.
The study uses 0.30\fbinv of 2016 data with old APV~settings and 0.12\fbinv of 2017 data.
To compare data and simulation, the simulated events are reweighted with respect to the distribution of the number of vertices. In addition, the simulated samples are normalized to the integrated luminosity of the data. 
In general, ``high-purity'' tracks~\cite{Chatrchyan:2014fea} are the default track selection for majority of physics analyses in CMS, unless efficiency is essential and purity is not a major concern. Therefore we consider only good ``high-purity'' tracks to maintain high efficiency (with a typical efficiency of more than 95\%) along with the rejection of a large number of misidentified tracks (the misidentification rate is less than 8\%).
Three track observables that are sensitive to the tracker alignment were used to compare the data and the simulation. For this study we select events in the same phase space as used for the DMRs shown in Fig.~\ref{fig:Run2startup:dmr}, where the alignment should play an important role. Tracks with very low track $\chi^{2}$ probability values ($\chi^{2} < 0.02$) are excluded from the studies in this section. Such low values of the $\chi^{2}$ probability are related to pattern recognition errors.
For all of the years, comparable agreement between the data and MC simulation was observed.
During the 2016 YETS, the Phase-1 upgrade of the pixel detector~\cite{CMS:2012sda,Adam_2021} took place.
To compare data and MC simulation for the Phase-0 detector, a study of 2016 data and MC simulation using old APV~settings, and with a dedicated description corresponding to the APV dynamic gain, was performed. The same was done with 2017 data and MC simulation to make the equivalent comparison for the Phase-1 detector.
 
The two-dimensional track impact parameter significance (SIP2D) with respect to the PV in the transverse plane of the detector is defined as the ratio of the two-dimensional track impact parameter with respect to the PV and its uncertainty. Track impact parameter uncertainties are calculated from the covariance matrix of the fitted track trajectory.
A comparison between data and MC simulation of the distribution of the SIP2D with respect to the PV is shown in Fig.~\ref{fig:sip3d} for the legacy reprocessing of the 2016 and 2017 data.
The symmetric nature of the figures indicates the detector is aligned well in both years.
In 2016, the agreement between data and MC simulation becomes worse for higher values of SIP2D. This discrepancy is not due to misalignment, but is instead caused by the somewhat imperfect simulation of the APV saturation in the case where strip hits contribute more to the tracks in the tails of the distribution.
In 2017, good agreement between data and simulation is seen.
        
The IP3D uncertainty is the uncertainty in measuring the three-dimensional track impact parameter. 
The distribution of the IP3D uncertainty with respect to the PV is shown in Fig.~\ref{fig:ip3derr}, comparing data and simulation.
In both years, despite the steeply falling spectrum for both data and simulation, a notable deviation between data and MC simulation is seen for large values of the IP3D uncertainty. This deviation is possibly due to a residual discrepancy in the simulation of the APV saturation rather than to a misalignment in the detector.
In 2017, the agreement between data and simulation is much improved compared to the 2016 data-taking period with the old APV~settings.

Figure~\ref{fig:chi2prob} shows the track $\chi^{2}$ probability for the number of degrees of freedom of the track fits, in both data and simulation. 
In both years, the peak at very low track $\chi^{2}$ probability values is mainly related to pattern recognition errors. Tracks in this region, indicated by a vertical dashed line in Fig.~\ref{fig:chi2prob}, are excluded from the studies in this section.
Also in both years, the simulation underestimates the data for low probability values, and overestimates the data for high probability values.
By comparing the distribution of the RMS of the normalized residuals (DRNRs) between data and simulation in all the subdetectors for the 2016 and 2017 runs, we find that for most of the subdetectors the DRNR in data is shifted towards larger DRNR values compared with the simulation.
This leads to an underestimation of the simulation, compared with data, for lower values of the $\chi^{2}$ probability.
 
Overall, good agreement between the data and simulation is seen in both the 2016 data-taking period with the old APV~settings and in 2017. 
From the comparison of the data and the simulation, we conclude that the updated alignment conditions for the legacy reprocessing in data lead to consistent good agreement with the MC simulation.
It also confirms a consistent alignment of the tracker during Run~2. 

\section{Summary}\label{sec:Summary}
In this paper, the strategies for and the performance of the alignment of the CMS central tracker during the data-taking period from 2015 to 2018 have been described.

The alignment was determined from a global track fit, where the module parameters were released in addition to the track parameters. Two algorithms based on slightly different approaches were used to perform the minimization of this problem with a large number of parameters, namely \HIPPY and \MILLEPEDE-II.
Improvements of the software introduced for Run~2 were discussed.

Different strategies were applied depending on the number of parameters to determine: at the beginning of the year and with the very limited amount of data available, the modules of the pixel tracker were aligned with respect to the strip tracker; during the year, an automated alignment procedure was performed for each run, correcting for movements of the mechanical structures of the pixel tracker with respect to the strip tracker; finally, in the middle or at the end of the year, once a large amount of data had been recorded (including \eg a sufficient number of muon tracks from \PZ~boson decays), all modules of the strip and pixel trackers were aligned in a single fit accounting for time dependence.
The last strategy leads to the highest precision of all and is preferred for the event reconstruction in the context of physics analyses.

Systematic distortions arising from the ageing of the detector, from internal symmetries of the minimization problem or from external constraints, were monitored using specific distributions. Examples of such distributions are the dimuon invariant mass as a function of the kinematical properties of the outgoing muons and the difference of the mean of the distribution of the median of the residuals for modules pointing in opposite directions.

In general, the increased level of radiation compared with data collected before 2012 required an updated strategy.
As part of the reprocessing of the data recorded during the period from 2016 to 2018 after the end of data taking, the alignment parameters were determined with a higher precision than for the end-of-year reconstruction. This was made possible by artificially moving the ladders and panels in the pixel trackers to absorb the gradually accumulating systematic shift induced by the ageing of the modules and by defining finer intervals of validity.

In addition to the alignment of the modules in real data, scenarios for use in simulation were also derived, aiming to resemble the statistical performance and the systematic effects. A scenario was provided for each year separately, without time dependence in the alignment fit.
Various comparisons of the performance in data and in simulation have been presented, especially in the context of the reprocessing of the data recorded between 2016 and 2018 after the end of data taking.

\begin{acknowledgments}

\hyphenation{Bundes-ministerium Forschungs-gemeinschaft Forschungs-zentren Rachada-pisek} We congratulate our colleagues in the CERN accelerator departments for the excellent performance of the LHC and thank the technical and administrative staffs at CERN and at other CMS institutes for their contributions to the success of the CMS effort. In addition, we gratefully acknowledge the computing centres and personnel of the Worldwide LHC Computing Grid and other centres for delivering so effectively the computing infrastructure essential to our analyses. Finally, we acknowledge the enduring support for the construction and operation of the LHC, the CMS detector, and the supporting computing infrastructure provided by the following funding agencies: the Austrian Federal Ministry of Education, Science and Research and the Austrian Science Fund; the Belgian Fonds de la Recherche Scientifique, and Fonds voor Wetenschappelijk Onderzoek; the Brazilian Funding Agencies (CNPq, CAPES, FAPERJ, FAPERGS, and FAPESP); the Bulgarian Ministry of Education and Science, and the Bulgarian National Science Fund; CERN; the Chinese Academy of Sciences, Ministry of Science and Technology, and National Natural Science Foundation of China; the Ministerio de Ciencia Tecnolog\'ia e Innovaci\'on (MINCIENCIAS), Colombia; the Croatian Ministry of Science, Education and Sport, and the Croatian Science Foundation; the Research and Innovation Foundation, Cyprus; the Secretariat for Higher Education, Science, Technology and Innovation, Ecuador; the Ministry of Education and Research, Estonian Research Council via PRG780, PRG803 and PRG445 and European Regional Development Fund, Estonia; the Academy of Finland, Finnish Ministry of Education and Culture, and Helsinki Institute of Physics; the Institut National de Physique Nucl\'eaire et de Physique des Particules~/~CNRS, and Commissariat \`a l'\'Energie Atomique et aux \'Energies Alternatives~/~CEA, France; the Bundesministerium f\"ur Bildung und Forschung, the Deutsche Forschungsgemeinschaft (DFG), under Germany's Excellence Strategy -- EXC 2121 ``Quantum Universe" -- 390833306, and under project number 400140256 - GRK2497, and Helmholtz-Gemeinschaft Deutscher Forschungszentren, Germany; the General Secretariat for Research and Innovation, Greece; the National Research, Development and Innovation Fund, Hungary; the Department of Atomic Energy and the Department of Science and Technology, India; the Institute for Studies in Theoretical Physics and Mathematics, Iran; the Science Foundation, Ireland; the Istituto Nazionale di Fisica Nucleare, Italy; the Ministry of Science, ICT and Future Planning, and National Research Foundation (NRF), Republic of Korea; the Ministry of Education and Science of the Republic of Latvia; the Lithuanian Academy of Sciences; the Ministry of Education, and University of Malaya (Malaysia); the Ministry of Science of Montenegro; the Mexican Funding Agencies (BUAP, CINVESTAV, CONACYT, LNS, SEP, and UASLP-FAI); the Ministry of Business, Innovation and Employment, New Zealand; the Pakistan Atomic Energy Commission; the Ministry of Science and Higher Education and the National Science Centre, Poland; the Funda\c{c}\~ao para a Ci\^encia e a Tecnologia, grants CERN/FIS-PAR/0025/2019 and CERN/FIS-INS/0032/2019, Portugal; JINR, Dubna; the Ministry of Education and Science of the Russian Federation, the Federal Agency of Atomic Energy of the Russian Federation, Russian Academy of Sciences, the Russian Foundation for Basic Research, and the National Research Center ``Kurchatov Institute"; the Ministry of Education, Science and Technological Development of Serbia; the Secretar\'{\i}a de Estado de Investigaci\'on, Desarrollo e Innovaci\'on, Programa Consolider-Ingenio 2010, Plan Estatal de Investigaci\'on Cient\'{\i}fica y T\'ecnica y de Innovaci\'on 2017--2020, research project IDI-2018-000174 del Principado de Asturias, and Fondo Europeo de Desarrollo Regional, Spain; the Ministry of Science, Technology and Research, Sri Lanka; the Swiss Funding Agencies (ETH Board, ETH Zurich, PSI, SNF, UniZH, Canton Zurich, and SER); the Ministry of Science and Technology, Taipei; the Thailand Center of Excellence in Physics, the Institute for the Promotion of Teaching Science and Technology of Thailand, Special Task Force for Activating Research and the National Science and Technology Development Agency of Thailand; the Scientific and Technical Research Council of Turkey, and Turkish Atomic Energy Authority; the National Academy of Sciences of Ukraine; the Science and Technology Facilities Council, UK; the US Department of Energy, and the US National Science Foundation.

{\tolerance=1600
Individuals have received support from the Marie-Curie programme and the European Research Council and Horizon 2020 Grant, contract Nos.\ 675440, 724704, 752730, 758316, 765710, 824093, 884104, and COST Action CA16108 (European Union) the Leventis Foundation; the Alfred P.\ Sloan Foundation; the Alexander von Humboldt Foundation; the Belgian Federal Science Policy Office; the Fonds pour la Formation \`a la Recherche dans l'Industrie et dans l'Agriculture (FRIA-Belgium); the Agentschap voor Innovatie door Wetenschap en Technologie (IWT-Belgium); the F.R.S.-FNRS and FWO (Belgium) under the ``Excellence of Science -- EOS" -- be.h project n.\ 30820817; the Beijing Municipal Science \& Technology Commission, No. Z191100007219010; the Ministry of Education, Youth and Sports (MEYS) of the Czech Republic; the Lend\"ulet (``Momentum") Programme and the J\'anos Bolyai Research Scholarship of the Hungarian Academy of Sciences, the New National Excellence Program \'UNKP, the NKFIA research grants 123842, 123959, 124845, 124850, 125105, 128713, 128786, and 129058 (Hungary); the Council of Scientific and Industrial Research, India; the Latvian Council of Science; the National Science Center (Poland), contracts Opus 2014/15/B/ST2/03998 and 2015/19/B/ST2/02861; the Funda\c{c}\~ao para a Ci\^encia e a Tecnologia, grant FCT CEECIND/01334/2018; the National Priorities Research Program by Qatar National Research Fund; the Ministry of Science and Higher Education, projects no. 14.W03.31.0026 and no. FSWW-2020-0008, and the Russian Foundation for Basic Research, project No.19-42-703014 (Russia); the Programa de Excelencia Mar\'{i}a de Maeztu, and the Programa Severo Ochoa del Principado de Asturias; the Stavros Niarchos Foundation (Greece); the Rachadapisek Sompot Fund for Postdoctoral Fellowship, Chulalongkorn University, and the Chulalongkorn Academic into Its 2nd Century Project Advancement Project (Thailand); the Kavli Foundation; the Nvidia Corporation; the SuperMicro Corporation; the Welch Foundation, contract C-1845; and the Weston Havens Foundation (USA).
\par}

The \MILLEPEDE-II software is provided by DESY in the framework of the Helmholtz Terascale Alliance.

\end{acknowledgments}

\bibliography{auto_generated}

\cleardoublepage \appendix\section{The CMS Collaboration \label{app:collab}}\begin{sloppypar}\hyphenpenalty=5000\widowpenalty=500\clubpenalty=5000\cmsinstitute{Yerevan~Physics~Institute, Yerevan, Armenia}
A.~Tumasyan
\cmsinstitute{Institut~f\"{u}r~Hochenergiephysik, Vienna, Austria}
W.~Adam\cmsorcid{0000-0001-9099-4341}, J.W.~Andrejkovic, T.~Bergauer\cmsorcid{0000-0002-5786-0293}, D.~Bl\"{o}ch, S.~Chatterjee\cmsorcid{0000-0003-2660-0349}, M.~Dragicevic\cmsorcid{0000-0003-1967-6783}, A.~Escalante~Del~Valle\cmsorcid{0000-0002-9702-6359}, R.~Fr\"{u}hwirth\cmsAuthorMark{1}, V.~Hinger\cmsorcid{0000-0002-2616-4084}, M.~Jeitler\cmsAuthorMark{1}\cmsorcid{0000-0002-5141-9560}, N.~Krammer, L.~Lechner\cmsorcid{0000-0002-3065-1141}, D.~Liko, I.~Mikulec, P.~Paulitsch, F.M.~Pitters, J.~Schieck\cmsAuthorMark{1}\cmsorcid{0000-0002-1058-8093}, R.~Sch\"{o}fbeck\cmsorcid{0000-0002-2332-8784}, D.~Schwarz, H.~Steininger, S.~Templ\cmsorcid{0000-0003-3137-5692}, W.~Waltenberger\cmsorcid{0000-0002-6215-7228}, C.-E.~Wulz\cmsAuthorMark{1}\cmsorcid{0000-0001-9226-5812}
% ...
\vskip\cmsinstskip
% ...
1:~Also at TU~Wien, Wien, Austria\\
% ...
\end{sloppypar}
\end{document}